\pdfoutput=1
\documentclass[12pt,english,british]{phd}
\usepackage[T1]{fontenc}
\usepackage[utf8]{inputenc}
\usepackage[a4paper]{geometry}
\usepackage{verbatim}
\usepackage{float}
\usepackage{amsthm}
\usepackage{amsmath}
\usepackage{amssymb}
\usepackage{wasysym}
\usepackage{graphicx}
\usepackage{setspace}
\usepackage{esint}
\usepackage[authoryear]{natbib}
\makeatletter

\newcommand{\lyxmathsym}[1]{\ifmmode\begingroup\def\b@ld{bold}
  \text{\ifx\math@version\b@ld\bfseries\fi#1}\endgroup\else#1\fi}

\providecommand{\tabularnewline}{\\}

\defcitealias{1980_Wolszczan}{Wo80}
\defcitealias{1988_Lyne}{Ly88}
\defcitealias{1997_Kijak}{Ki97}
\defcitealias{1994_Slane}{Sl94}
\defcitealias{1998_Gotthelf}{Go98}
\defcitealias{1999_Deshpande}{De99}
\defcitealias{2001_Cusumano}{Cu01}
\defcitealias{2001_Kaaret}{Ka01}
\defcitealias{2001_Kaspi}{Ka01}
\defcitealias{2001_Roberts}{Ro01}
\defcitealias{2001_Asgekar}{As01}
\defcitealias{2001_Deshpande}{De01}
\defcitealias{2001_Everett}{Ev01}
\defcitealias{2002_Becker}{Be02}
\defcitealias{2002_Camilo}{Ca02}
\defcitealias{2002_Gotthelf}{Go02}
\defcitealias{2002_Pavlov}{Pa02}
\defcitealias{2002_Petre}{Pe02}
\defcitealias{2003_Gonzales}{Go03}
\defcitealias{2003_Helfand}{He03}
\defcitealias{2003_Hughes}{Hu03}
\defcitealias{2003_Mcgowan}{Mc03}
\defcitealias{2003_Roberts}{Ro03}
\defcitealias{2004_Becker}{Be04}
\defcitealias{2004_Gaensler}{Ga04}
\defcitealias{2004_Hessels}{He04}
\defcitealias{2004_Oosterbroek}{Oo04}
\defcitealias{2004_Roberts}{Ro04}
\defcitealias{2004_Slane}{Sl04}
\defcitealias{2004_Zavlin}{Za04}
\defcitealias{2005_Deluca}{De05}
\defcitealias{2005_Jackson}{Ja05}
\defcitealias{2005_Li}{Li05}
\defcitealias{2005_Mignani}{Mi05}
\defcitealias{2005_Tepedelenl}{Te05}
\defcitealias{2005_Zhang}{Zh05}
\defcitealias{2005_Becker}{Be05}
\defcitealias{2006_Chernyakova}{Ch06}
\defcitealias{2006_DeLaney}{De06}
\defcitealias{2006_Gonzalez}{Go06}
\defcitealias{2006_Kargaltsev}{Ka06}
\defcitealias{2006_Mcgowan}{Mc06}
\defcitealias{2006_Weltevrede}{We06}
\defcitealias{2006_Gil}{Gi06}
\defcitealias{2006_Weltevrede_b}{We06b}
\defcitealias{2007_Gonzalez}{Go07}
\defcitealias{2007_Hui_b}{Hu07b}
\defcitealias{2007_Hui}{Hu07}
\defcitealias{2007_Kargaltsev}{Ka07}
\defcitealias{2007_Lu}{Lu07}
\defcitealias{2007_McGowan}{Mc07}
\defcitealias{2007_Zavlin_b}{Za07b}
\defcitealias{2007_Zavlin}{Za07}
\defcitealias{2007_Rankin}{Ra07}
\defcitealias{2007_Gil_b}{Gi07}
\defcitealias{2007_Herfindal}{He07}
\defcitealias{2008_Campana}{Ca08}
\defcitealias{2008_Gil}{Gi08}
\defcitealias{2008_Misanovic}{Mi08}
\defcitealias{2008_Ng}{Ng08}
\defcitealias{2008_Pavlov}{Pa08}
\defcitealias{2008_VanEtten}{Va08}
\defcitealias{2009_Becker}{Be09}
\defcitealias{2009_Chernyakova}{Ch09}
\defcitealias{2009_Kargaltsev}{Ka09}
\defcitealias{2009_Pavlov}{Pa09}
\defcitealias{2009_Zhu}{Zh09}
\defcitealias{2010_Gotthelf}{Go10}
\defcitealias{2012_Hui}{Hu12}
\defcitealias{2012_Ng}{Ng12}
\defcitealias{2012_Posselt}{Po12}
\makeatother

\usepackage{babel}
\begin{document}
% margins: top:2.5, bottom:2, inner:3, outer:1.5, head sep:0.5, head hight:2, foot skip 0.5
% maximum image size 16cm x 25cm
% images 7.5x6 (2.95276,2.3622 in - double), 16x9 (6.29921,3.54331 in), 15x9.27 (5.90551,3.649606 in - golden), 11x8.25 (4.33071,3.248031 in) 
% 14x8.7 (5.51181,3.4065 in - golden), 9x5.6cm (3.54331,2.1899 in - golden)
% 
\pagestyle{headings}
\fancyhead{}
\fancyfoot{}
\fancyhead[LE,RO]{\thepage}
\fancyhead[RE]{\bf \leftmark}
\fancyhead[LO]{\bf \rightmark}
\renewcommand{\headrulewidth}{0.4pt}
\renewcommand{\footrulewidth}{0pt}
\pagenumbering{alph}
% prevent footnote splitting
\interfootnotelinepenalty=10000

\title{Non-dipolar magnetic field at the polar cap of neutron stars and
the physics of pulsar radiation}

\polishtitle{Niedipolowe pole magnetyczne nad czapą polarną gwiazdy neutronowej
a fizyka promieniowania pulsarów}

\author{Andrzej Szary}

\promoter{prof. dr hab. Giorgi Melikidze}

\city{Zielona Góra}

\date{2013}

\maketitle
\thispagestyle{empty}

\cleardoublepage{}

\thispagestyle{empty} \hbox{} \vfill \vspace{15cm}
\begin{flushright}
	{\it To Natalia, my daughter, and Beata, my wife, for being there... }
\end{flushright}
%empty page
\newpage \thispagestyle{empty} \mbox{}
\newpage

\pagenumbering{roman}
\pagestyle{plain}
\addcontentsline{toc}{chapter}{Contents}

\tableofcontents{}

\clearpage{}

\chapter*{Abstract}

\addcontentsline{toc}{chapter}{Abstract}
%\the\topskip
%\the\textheight
%\the\topmargin
%\the\textwidth

\vspace{0.5cm}

Despite the fact that pulsars have been observed for almost half a
century, until now many questions have remained unanswered. One of
the fundamental problems is describing the physics of pulsar radiation.
By trying to find an answer to this fundamental question we use the
analysis of X-ray observations in order to study the polar cap region
of radio pulsars. The size of the hot spots implies that the magnetic
field configuration just above the stellar surface differs significantly
from a purely dipole one. By using the conservation of the magnetic
flux we can estimate the surface magnetic field as of the order of
$10^{14}\,{\rm G}$. On the other hand, the temperature of the hot
spots is about a few million Kelvins. Based on these two facts the
Partially Screened Gap (PSG) model was proposed to describe the Inner
Acceleration Region (IAR). The PSG model assumes that the temperature
of the actual polar cap is equal to the so-called critical value,
i.e. the temperature at which the outflow of thermal ions from the
surface screens the gap completely. 

We have found that, depending on the conditions above the polar cap,
the generation of high energetic photons in IAR can be caused either
by Curvature Radiation (CR) or by Inverse Compton Scattering (ICS).
Completely different properties of both processes result in two different
scenarios of breaking the acceleration gap: the so-called PSG-off
mode for the gap dominated by CR and the PSG-on mode for the gap dominated
by ICS. The existence of two different mechanisms of gap breakdown
naturally explains the mode-changing phenomenon. Different characteristics
of plasma generated in the acceleration region for both processes
also explain the pulse nulling phenomenon. Furthermore, the mode changes
of the IAR may explain the anti-correlation of radio and X-ray emission
in very recent observations of PSR B0943+10 \citep{2013_Hermsen}.

Simultaneous analysis of X-ray and radio properties have allowed to
develop a model which explains the drifting subpulse phenomenon. According
to this model the drift takes place when the charge density in IAR
differs from the Goldreich-Julian co-rotational density. The proposed
model allows to verify both the radio drift parameters and X-ray efficiency
of the observed pulsars.

\clearpage{}

\vspace*{0.6cm}

\textbf{\huge Streszczenie}{\huge \par}

\vspace*{0.6cm}
Pomimo, że pulsary są badane już od prawie pół wieku, do dzisiaj nie
udało się znaleźć odpowiedzi na wiele pytań. Jednym z fundamentalnych
problemów jest opis fizyki promieniowania pulsarów. Próbując znaleźć
odpowiedź na to fundamentalne pytanie, wykorzystujemy analizę obserwacji
rentgenowskich w celu badania obszaru czapy polarnej pulsarów. Rozmiar
obserwowanych gorących plam wskazuje, że konfiguracja pola magnetycznego
na powierzchni gwiazdy różni się znacznie od pola czysto dipolowego.
Wykorzystując prawo zachowania strumienia magnetycznego możemy oszacować
siłę pola magnetycznego w obszarze czapy polarnej, które dla obserwowanych
pulsarów jest rzędu $10^{14}\,{\rm G}$. Z drugiej strony obserwowana
temperatura gorącej plamy jest rzędu kilku milionów kelwinów. Opierając
się na tych dwóch faktach wykorzystujemy model częściowo-ekranowanej
przerwy akceleracyjnej (z ang. Partially Screened Gap - PSG), aby
opisać wewnętrzną przerwę akceleracyjną (z ang. Inner Acceleration
Region - IAR). Model PSG zakłada, że temperatura czapy polarnej jest
bliska do tak zwanej wartości krytycznej tzn. takiej przy, której
termiczny odpływ jonów z powierzchni w pełni ekranuje przerwę akceleracyjną. 

W zależności od warunków jakie panują w obszarze czapy polarnej, mechanizmem
odpowiedzialnym za generowanie wysokoenergetycznych fotonów w IAR
może być promieniowanie krzywiznowe (z ang. Curvature Radiation -
CR) lub odwrotne rozpraszanie Comptona (z ang. Inverse Compoton Scattering
- ICS). Całkowicie różne właściwości obu tych procesów prowadzą do
sytuacji, w której możemy wyróżnić dwa scenariusze zamknięcia przerwy
akceleracyjnej: tzw. PSG-off dla przerwy zdominowanej przez promieniowanie
CR, oraz tzw. PSG-on dla przerwy zdominowanej przez ICS. Istnienie
dwóch różnych mechanizmów zamknięcia przerwy w naturalny sposób tłumaczy
zjawisko zmiany trybu promieniowania pulsarów (z ang. mode-changing).
Różna charakterystyka plazmy generowanej w obszarze akceleracyjnym
dla obu tych trybów tłumaczy zjawisko sporadycznego braku pojedynczych
pulsów (z ang. pulse nulling) w obserwacjach radiowych. Co więcej
zmiana trybu w jakim pracuje przerwa akceleracyjna może zostać powiązana
z antykorelacją promieniowania radiowego i rentgenowskiego wykazaną
w ostatnich obserwacjach PSR~B0943+10 \citep{2013_Hermsen}.

Jednoczesna analiza właściwości promieniowania rentgenowskiego i radiowego
pozwoliła na opracowanie modelu dryfujących składowych pulsu pojedynczego
(z ang. subpulses). Model ten zakłada, że dryf jest wynikiem różnicy
gęstości ładunku w IAR w stosunku do gęstości korotacji. Proponowany
model pozwala zarówno na weryfikację wyznaczonych parametrów dryfu
oraz na weryfikację np. efektywności promieniowania rentgenowskiego.

\chapter*{Introduction}

\addcontentsline{toc}{chapter}{Introduction}

The history of neutron stars began in the early 1930s when Subrahmanyan
Chandrasekhar calculated the critical mass for a white dwarf. As soon
as the mass of a white dwarf exceeds the critical value (e.g. due
to accretion of matter from a companion star) it collapses and a neutron
star is formed. Chandrasekhar estimated that the critical mass was
approximately $1.4$ solar masses (${\rm M}_{\odot}$). Even before
James Chadwick's discovery of neutrons \citeyearpar{1932_Chadwick},
Lev Landau anticipated the existence of neutron stars by writing about
stars in which ``atomic nuclei come in close contact, forming one
gigantic nucleus''. In 1934 \citeauthor{1934_Baade} proposed that
the ``supernova process represents the transition of an ordinary
star into a neutron star''. Five years later \citet{1939_Oppenheimer},
using the work of \citet{1939_Tolman}, computed an upper bound on
the mass of a star composed of neutron-degenerate matter. They assumed
that the neutrons in a neutron star form a cold degenerate Fermi gas
which leads to an upper bound of approximately $0.7\,{\rm M}_{\odot}$.
Modern estimates of the critical mass for neutron stars range from
approximately $1.5\,{\rm M}_{\odot}$ to $3\,{\rm M}_{\odot}$ \citep{1996_Bombaci}.
This uncertainty reflects the fact that the equation of state for
extremely dense matter is not well known. Let us note that the radius
of a neutron star should be $R\approx10{\rm \, km}$. On the other
hand nobody expected to detect any emission from neutron stars due
to their small size and the lack of theoretical predictions about
any radiation processes, except for thermal radiation. Thus, it took
almost forty years to detect emission from a neutron star.

The breakthrough came on 28 November 1967 with the radio observations
that were performed by Jocelyn Bell-Burnell and Anthony Hewish. They
observed radio pulses separated by $1.33$ seconds. The world ``pulsar''
was adopted to reflect the specific property of these celestial objects.
The suggestion that pulsars were rotating neutron stars was put forth
independently by \citet{1968_Gold} and \citet{1968_Pacini}, and
was soon proved beyond a reasonable doubt by the discovery of a pulsar
with a very short ($33$-millisecond) pulse period in the Crab nebula.
It was suggested that this pulsar powers the activity of the nebula
\citep{1968_Pacini}. Nearly 2000 pulsars have been found so far.
Observations of pulsars provide valuable information about neutron
star physics, general relativity, the interstellar medium, celestial
mechanics, planetary physics, the Galactic gravitational potential,
the magnetic field and even cosmology. Studying neutron stars is therefore
a very broad issue and it is beyond the scope of this thesis to describe
the current status of the theory of neutron stars or pulsar population
studies in detail. We rather refer the reader to the literature \citep{1991_Michel,1992_Meszaros,1996_Glendenning,1999_Weber}
and provide only a basic theoretical background that is relevant to
the subject of this thesis.

Following the ideas of \citet{1968_Pacini} and \citet{1968_Gold}
radio pulsars can be interpreted as rapidly spinning, strongly magnetised
neutron stars radiating at the expense of their rotational energy.
Neutron stars consist of compressed matter with density in its core
exceeding nuclear density $\rho_{{\rm nuc}}=2.8\times10^{14}{\rm \, g\, cm^{-3}}$.
Direct and accurate mass measurements come from timing observations
of binary pulsars and are consistent with a typically assumed neutron
star mass $M\approx1.4\,{\rm M}_{\odot}$. Most models predict a radius
of $R\sim10\,{\rm km}$, which is consistent with the theoretical
upper and lower limits. However, the measurements of neutron star
radii are much less reliable than the mass measurements. Therefore,
the moment of inertia for these canonical values ($M=1.4\,{\rm M}_{\odot}$,
$R=10{\rm \, km}$) $I\approx\left(2/5\right)MR^{2}\approx10^{45}{\rm \, g\, cm^{2}}$
may be uncertain by $\sim70\%$. The increase rate of a pulsar period,
$\dot{P}={\rm d}P/{\rm d}t$, is related to the rate of rotational
kinetic energy loss (spin-down luminosity) $\dot{E}=L_{{\rm SD}}=4\pi^{2}I\dot{P}P^{-3}$.
In most cases only a tiny fraction of $\dot{E}$ can be converted
into radio emission. The efficiency, $\chi_{{\rm radio}}=L_{{\rm radio}}/\dot{E}$,
in the radio bands is typically in the range of $\sim10^{-7}-10^{-5}$.
It is assumed that the bulk of the rotational energy is converted
into magnetic dipole radiation. The expected evolution of the angular
velocity ($\Omega=2\pi/P$) of a rotating magnetic dipole can be described
as $\dot{\Omega}\sim\Omega^{n}$, and the breaking index is $n=3$
for the pure dipole radiation. Indeed, the observed values of the
breaking index (e.g. \citealp{2009_Becker}) confirm the above statement,
e.g.: for the Crab $n=2.515\pm0.005$, for PSR B1509-58 $n=2.8\text{\ensuremath{\pm}}0.2$,
for PSR B0540-69 $n=2.28\text{\ensuremath{\pm}}0.02$, for PSR J1911-6127
$n=2.91\pm0.05$, for PSR J1846-0258 $n=2.65\text{\ensuremath{\pm}}0.01$,
and for the Vela pulsar $n=1.4\text{\ensuremath{\pm}}0.2$. On the
other hand the observations of pulsar wind nebulae suggest that a
significant fraction of the pulsar rotational energy is carried away
by a pulsar wind. Furthermore, recent observations of high energy
radiation from pulsars show that significantly more energy is radiated
in the form of X-rays and $\gamma$-rays than in the form of radio
emission (e.g. \citealp{2010_Abdo}). Thus, pure magnetic breaking
does not provide full information about the physical processes that
take place in the pulsar magnetosphere. 

Despite the fact that pulsars have been observed for almost half a
century, many questions still remain unanswered. One of the fundamental
problems concerns the physics of pulsar radiation. Radio observations
alone cannot point to the model (e.g. vacuum gap, slot gap, outer
gap, free outflow, etc.) that correctly describes the source of pulsar
activity. Observations carried out by relatively new high-energy instruments,
e.g. \textit{Chandr}a and \textit{XMM-Newton}, significantly extended
the spectra over which we can study pulsars and their environments.
There is no consensus about the origin of pulsar X-ray emission \citep{1991_Michel}.
We can distinguish two main types of models: the polar gap and the
outer gap. The polar gap models suggest that the emission region is
located in the vicinity of the neutron star polar caps, while the
outer gap models assume that particle acceleration and X-ray emission
take place close to the pulsar light cylinder%
\footnote{The light cylinder with radius $R_{{\rm LC}}=cP/2\pi$ is defined
as a place where the azimuthal velocity of the co-rotating magnetic
field lines is equal to the speed of light ($c$)%
}. In both types of models high energy radiation is generated by relativistic
particles accelerated in charge-depleted regions, while the high energy
photons are emitted by means of Curvature Radiation (CR), Synchrotron
Radiation (SR) and Inverse Compton Scattering (ICS). Both models are
able to interpret existing observational data.

In this thesis we will use the Partially Screened Gap (PSG) model
\citep{2007_Gil}. The PSG model assumes the existence of the Inner
Acceleration Region (IAR) above the polar cap (a region penetrated
by the open field lines) where the electric field has a component
along the magnetic field. In this region particles (electrons and
positrons) are accelerated in both directions: outward and toward
the stellar surface. Consequently, outflowing particles are responsible
for generation of magnetospheric emission (radio and high-frequency)
while the backflowing particles heat the surface and provide the required
energy for thermal emission. The PSG model is an extension of the
Standard Model developed by \citet{1975_Ruderman} and takes into
account the thermionic ion flow from the stellar surface heated up
to a high temperature (a few million Kelvins) by the backstreaming
particles. In such a scenario an analysis of X-ray radiation is an
excellent method of obtaining insight into the most intriguing region
of the neutron star.

\chapter{X-ray emission from Radio Pulsars\label{chap:x-ray_emission}}

\pagenumbering{arabic}
\thispagestyle{headings}
\pagestyle{headings}

\section{Brief historical overview}

X-ray photons can only be detected by telescopes operating at high
altitudes or above the Earth's atmosphere, thus detectors should be
mounted on high-flying balloons, rockets or satellites. The first
(i.e. carried out from space) X-ray observations were performed by
a team led by Herbert Friedman in 1948. The team estimated the luminosity
of X-ray radiation from the solar corona. They found that X-ray luminosity
is weaker by a factor of $10^{6}$ than luminosity in the optical
wave range. Up until the early 1960s it was widely believed that all
other stars should be so faint in the X-rays that their observations
would be hopeless. The situation changed in 1962 when a team led by
Bruno Rossi and Riccardo Giacconi, when trying to find fluorescent
X-ray photons from the moon, accidentally detected X-rays from Sco
X-1. Subsequent flights launched to confirm these first results detected
Tau X-1, a source in the constellation Taurus which coincided with
the Crab supernova remnant \citep{1964_Bower}. The search for similar
sources became a source of strong motivation for the further development
of X-ray astronomy.

Before the first direct detection of a neutron star by \citet{1968_Hewish},
it was predicted that neutron stars could be powerful sources of thermal
X-ray emission due to a high surface temperature ($T_{{\rm s}}$).
The expected value of the surface temperature was estimated as $T_{{\rm s}}\sim1\,{\rm MK}$
\citep{1964_Chiu,1964_Tsuruta}. The first X-ray observations of isolated
neutron stars %
\footnote{The term ''isolated'' is omitted hereafter in the text however all
X-ray observations presented in this thesis concern isolated neutron
stars%
} were initiated by the \textit{Einstein Observatory,} which was launched
by NASA in 1978. Using a high-resolution imaging camera sensitive
in the $0.2-3.5\,{\rm keV}$ energy range provided unprecedented levels
of sensitivity (hundreds of times better than had previously been
achieved). The \textit{Einstein} detected X-ray emission from a number
of neutron stars (mainly as compact sources in supernova remnants)
such as the middle-aged radio pulsars B0656+14, B1055-52 and the old
pulsar B0950+08. The \textit{Einstein} observatory re-entered the
Earth's atmosphere and burned up on 25 March 1982. The next ''decade
of space science'' was opened in the 1990s with the launch of the
\textit{ROSAT} mission that was sensitive in the $0.1-2.4\,{\rm keV}$
energy range. One of the major results achieved with the \textit{ROSAT}
was the identification of the $\gamma$-ray source Geminga as a pulsar,
hence a neutron star \citep{1992_Halpern}.

The current era of X-ray observations of neutron stars was begun with
the launch of two satellites: the \textit{XMM-Newton} owned by the
European Space Agency and the \textit{Chandra} owned by the National
Aeronautics and Space Administration. These two grazing-incidence
X-ray telescopes were placed in orbit in 1999. They were equipped
with cameras and high-resolution spectrometers sensitive to low-energy
X-rays: from $0.08$ to $10\,{\rm keV}$ for the \textit{Chandra}
and from $0.1$ to $15\,{\rm keV}$ for the \textit{XMM-Newton}. While
the two observatories have similar designs, they are not identical.
The \textit{XMM-Newton} observatory has three X-ray telescopes that
provide six times the collecting area and a broader spectral range
in images than the \textit{Chandra}, while the \textit{Chandra} has
a much finer spatial resolution and a broader spectral range in its
high-resolution spectroscopy than does the \textit{XMM-Newton}. Both
observatories are in a highly-elliptical orbit that permits continuous
observations of up to 40 hours. The \textit{Chandra} and \textit{XMM-Newton}
have greatly increased the quality and availability of observations
of X-ray thermal radiation from neutron star surfaces. The total number
of isolated neutron stars of different types detected in X-rays is
hard to find since not all data have been published. Some authors
estimate that about one hundred rotation-powered pulsars were detected
in the X-rays \citep{2007_Zavlin,2009_Becker}.

\section{X-ray emission from isolated neutron stars}

X-ray emission is a common feature of all kinds of neutron stars.
Furthermore, X-ray observations have led to the discovery of other
types of neutron stars that for various reasons were missed in the
standard searches for radio pulsars. These new classes, such as X-ray
Dim Isolated Neutron Stars, Central Compact Objects in supernovae
remnants, Anomalous X-ray Pulsars, and Soft Gamma-ray Repeaters, are
only a small fraction of the whole number of observed pulsars but
provide valuable information on the diversity of the neutron star
population.

X-ray radiation from an isolated neutron star can in general consist
of two distinguishable components: thermal and nonthermal emissions.
The thermal emission can originate either from the entire surface
of a cooling neutron star or from spots around the magnetic poles
on the stellar surface (polar caps and adjacent areas). The temperature
of a neutron star at the moment of its formation is extremely high
- its value is even as high as $10^{10}-10^{11}\,{\rm K}$. Such a
high initial temperature leads to very fast cooling, and after several
minutes the temperature of the star interior falls to $10^{9}-10^{10}{\rm \, K}$.
After $10-100\,{\rm yr}$ the neutron star will cool down to a few
times $10^{6}\,{\rm K}$. At this point, depending on the still poorly
known properties of super-dense matter, the temperature evolution
can follow two different scenarios. The standard cooling scenario
predicts that the temperature decreases gradually, down to $\sim\left(0.3-1\right)\times10^{6}\,{\rm K}$
by the end of the neutrino cooling era and then falls exponentially
to temperatures lower than $\sim10^{5}\,{\rm K}$ in $\sim10^{7}\,{\rm yr}$.
In the accelerated cooling scenario, which implies higher central
densities (up to $10^{15}\,{\rm g\, cm^{-3}}$) and/or exotic interior
composition (e.g. quark plasma), at the age of $\sim10-100\,{\rm yr}$
the temperature decreases rapidly down to $\sim\left(0.3-0.5\right)\times10^{6}\,{\rm K}$
and is followed by a more gradual decrease down to the same $\sim10^{5}\,{\rm K}$
in $\sim10^{7}\,{\rm yr}$ \citep{2009_Becker}. The thermal evolution
of neutron stars is very sensitive to the composition (and structure)
of their interiors, therefore, measuring surface temperatures is an
important tool in studying super-dense matter. In addition to a thermal
component emitted from the entire surface, other thermal components
can also be seen. One of these additional components could be related
to the reheating of the polar cap region by relativistic backflowing
particles (electron and/or positrons) created and accelerated in the
so-called polar gaps (see Chapter \ref{chap:psg}). The temperature
of these hot spots does not obey the same age dependence as the thermal
evolution of neutron stars. Thus, depending on the pulsar age the
thermal radiation may be dominated by either the entire surface (for
younger neutron stars) or the hot spot components (for older neutron
stars). The nonthermal component is usually attributed to the emission
produced by Synchrotron Radiation (SR) and/or Inverse Compton Scattering
(ICS) of charged relativistic particles accelerated in the pulsar
magnetosphere. As the energy of these particles follows a power-law
distribution, nonthermal emission is also characterised by power-law
spectra.

The X-ray spectrum of a neutron star (thermal and nonthermal) depends
on many factors, e.g. the age of the star ($\tau$), inclination angle,
strength and geometry of the magnetic field, etc. In most of the very
young pulsars ($\tau\sim1\,{\rm kyr}$) the nonthermal component dominates,
thus making it impossible to accurately measure the thermal flux --
only the upper limits for the surface temperature can be derived.
As a pulsar becomes older, its activity (nonthermal luminosity) decreases
roughly proportionally to its spin-down luminosity $L_{{\rm SD}}$.
A spin-down luminosity generally decreases with the increasing star
age, as $L_{{\rm SD}}\propto\tau^{-m}$, where $m\simeq2-4$ depends
on the pulsar dipole breaking index \citep{2007_Zavlin_b}. With the
increase of the pulsar age the luminosity of the surface thermal radiation
decreases more slowly than the luminosity of the nonthermal one. Thus,
the thermal radiation from an entire stellar surface can dominate
a soft X-ray spectrum of middle-aged ($\tau\sim100\,{\rm kyr}$) and
some younger ($\tau\sim10\,{\rm kyr}$) pulsars. For the old neutron
stars ($\tau>1\,{\rm Myr}$), a surface temperature $T_{{\rm s}}<0.1\,{\rm MK}$
makes it impossible to detect the thermal radiation from the entire
surface by available observatories. However, most of the pulsar models
predict the heating up of polar caps to very high temperatures ($T_{{\rm s}}\apprge1\,{\rm MK}$)
by relativistic particles which are created in the pulsar acceleration
zones. Conventionally, it is assumed that the polar cap radius is
$R_{{\rm dp}}=\sqrt{2\pi R^{3}/cP}$. 

\begin{figure}[th]
\begin{centering}
\includegraphics{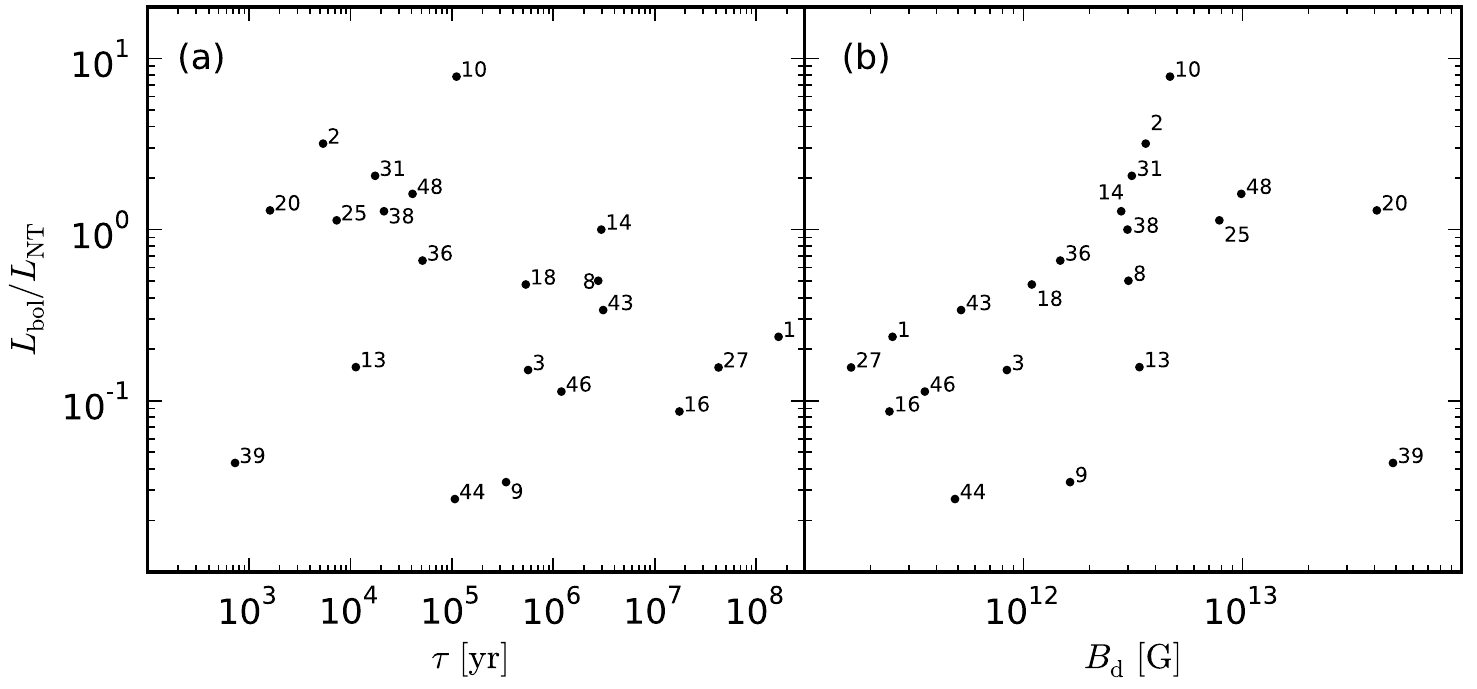}
\par\end{centering}

\centering{}\caption[Nonthermal and thermal components in the X-ray spectra of neutron
stars]{Ratio of X-ray luminosities (thermal and nonthermal components) as
a function of $\tau$ (panel a) and $B_{{\rm d}}$ (panel b). The
plots contain only those pulsars for which the BB+PL (Black-Body plus
Power-Law) spectral fit exists. The number labels at the points correspond
to the pulsar numbers in Table \ref{tab:pulsars}. \label{fig:x-ray_age_field}}
\end{figure}

Since the spin-down luminosity $L_{{\rm SD}}$ is the source for both
nonthermal (magnetospheric) and thermal (polar cap) components, it
is hard to predict which one would prevail in the X-ray flux of old
neutron stars. Figure \ref{fig:x-ray_age_field}a shows the ratio
of a thermal luminosity to a nonthermal one as a function of the pulsar
age. Since calculating this ratio is possible only for pulsars with
blackbody plus power-law fit, only these pulsars are included in the
Figure. There is also a significant number of pulsars ($16$) with
the spectra dominated by nonthermal components. Let us note that it
is impossible to determine the thermal components for these pulsars.
Most of them are young neutron stars $\sim10^{3}-10^{4}\,{\rm yr}$,
but there are also much older ones ($\sim10^{6}\,{\rm yr}$). In addition,
there is a group of $4$ pulsars with the spectra dominated by thermal
components (without a visible nonthermal component). Their age also
varies in quite a wide range $10^{4}-10^{6}\,{\rm yr}$. 

\begin{comment}
http://localhost:9090/pulsars/graphs/

\textasciitilde{}/Html/pulsar/media/images/xray\_age\_field.svg
\end{comment}

As it follows from the left panel of Figure \ref{fig:x-ray_age_field},
there is no obvious relation between pulsar age and the ratio of luminosities.
The spectra of pulsars with a similar age may be dominated either
by nonthermal (e.g. PSR B1951+32, PSR B1046-58) or thermal (e.g. PSR
B0656+14, PSR J0538+2817) components. It is difficult to provide a
more detailed analysis because, on the one hand, the observational
errors are large and, on the other hand, a separation of thermal and
nonthermal components is often not possible. The ratio of luminosities
also does not show any correlation with the strength of the dipolar
magnetic field (see the right panel of Figure \ref{fig:x-ray_age_field}).
Let us note that the value of the dipolar magnetic field is conventionally
calculated by adopting that the spin-down luminosity is equal to the
power of magneto-dipole radiation (neglecting the influence of a pulsar
wind). Then, assuming a dipolar structure of the neutron star magnetic
field down to the stellar surface, we estimate its strength (measured
in Gauss) at the pole as 
\begin{equation}
B_{{\rm d}}=2.02\times10^{12}\left(P\dot{P}_{-15}\right)^{0.5}.
\end{equation}
Here $P$ is a period in seconds and $\dot{P}_{-15}=\dot{P}\times10^{15}$.
The actual strength of the surface magnetic field can greatly exceed
the above value (see Chapter \ref{chap:model}). 

Table \ref{tab:pulsars} presents the basic parameters of the 48 pulsars
that we use in this thesis, while the results of the X-ray observations
of these pulsars are listed in Tables \ref{tab:x-ray_nonthermal}
and \ref{tab:x-ray_thermal}.

\begin{comment}
http://localhost:9090/pulsars/table\_psrs/

\textasciitilde{}/Html/pulsar/download/data/table\_psrs.tex
\end{comment}

\begin{table}[H]
\caption[Parameters of rotation-powered normal pulsars with detected X-ray
radiation]{ Parameters of rotation powered normal pulsars with detected X-ray
radiation. The individual columns are as follows: (1) Pulsar name,
(2) Barycentric period $P$ of the pulsar, (3) Time derivative of
barycentric period $\dot{P}$, (4) Canonical value of the dipolar
magnetic field $B_{{\rm d}}$ at the poles, (5) Spin-down energy loss
rate $L_{{\rm SD}}$ (spin-down luminosity) , (6) Dispersion measure
$DM$, (7) Best estimate of pulsar distance $D$ (used in all calculations),
(8) Best estimate of pulsar age or spin-down age $\tau=P/\left(2\dot{P}\right)$,
(9) Pulsar number (used in the Figures). Parameters of the radio pulsar
have been taken from the ATNF catalogue. \label{tab:pulsars} }

\centering{}%
\begin{tabular}{|l|c|c|c|c|c|c|c|c|}
\hline 
 &  &  &  &  &  &  &  & \tabularnewline
Name  & $P$  & $\dot{P}$  & $B_{{\rm d}}$  & $\log L_{{\rm SD}}$  & $DM$  & $D$  & $\tau$  & No. \tabularnewline
 & {\scriptsize $\left({\rm s}\right)$}  & {\scriptsize $\left(10^{-15}\right)$}  & {\scriptsize $\left(10^{12}\,{\rm G}\right)$}  & {\scriptsize $\left({\rm erg\, s^{-1}}\right)$}  & {\scriptsize $\left({\rm cm^{-3}\, pc}\right)$}  & {\scriptsize $\left({\rm kpc}\right)$}  &  & \tabularnewline
\hline 
 &  &  &  &  &  &  &  & \tabularnewline
J0108--1431  & $0.808$  & $0.077$  & $0.504$  & $30.76$  & $2.38$  & $0.18$  & $166$ Myr  & 1 \tabularnewline
J0205+6449  & $0.066$  & $193.9$  & $7.210$  & $37.43$  & $141$  & $3.20$  & $5.37$ kyr  & 2 \tabularnewline
B0355+54  & $0.156$  & $4.397$  & $1.675$  & $34.65$  & $57.1$  & $1.04$  & $564$ kyr  & 3 \tabularnewline
B0531+21  & $0.033$  & $422.8$  & $7.555$  & $38.66$  & $56.8$  & $2.00$  & $1.24$ kyr  & 4 \tabularnewline
J0537--6910  & $0.016$  & $51.78$  & $1.846$  & $38.69$  & --  & $47.0$  & $4.93$ kyr  & 5 \tabularnewline
 &  &  &  &  &  &  &  & \tabularnewline
J0538+2817  & $0.143$  & $3.669$  & $1.464$  & $34.69$  & $39.6$  & $1.20$  & $30.0$ kyr  & 6 \tabularnewline
B0540--69  & $0.050$  & $478.9$  & $9.934$  & $38.18$  & $146$  & $55.0$  & $1.67$ kyr  & 7 \tabularnewline
B0628--28  & $1.244$  & $7.123$  & $6.014$  & $32.18$  & $34.5$  & $1.45$  & $2.77$ Myr  & 8 \tabularnewline
J0633+1746  & $0.237$  & $10.97$  & $3.258$  & $34.51$  & --  & $0.16$  & $342$ kyr  & 9 \tabularnewline
B0656+14  & $0.385$  & $55.00$  & $9.294$  & $34.58$  & $14.0$  & $0.29$  & $111$ kyr  & 10 \tabularnewline
 &  &  &  &  &  &  &  & \tabularnewline
J0821--4300  & $0.113$  & $1.200$  & $0.743$  & $34.52$  & --  & $2.20$  & $3.7$ kyr  & 11 \tabularnewline
B0823+26  & $0.531$  & $1.709$  & $1.924$  & $32.65$  & $19.5$  & $0.34$  & $4.92$ Myr  & 12 \tabularnewline
B0833--45  & $0.089$  & $125.0$  & $6.750$  & $36.84$  & $68.0$  & $0.21$  & $11.3$ kyr  & 13 \tabularnewline
B0834+06  & $1.274$  & $6.799$  & $5.945$  & $32.11$  & $12.9$  & $0.64$  & $2.97$ Myr  & 14 \tabularnewline
B0943+10  & $1.098$  & $3.493$  & $3.956$  & $32.00$  & $15.4$  & $0.63$  & $4.98$ Myr  & 15 \tabularnewline
 &  &  &  &  &  &  &  & \tabularnewline
B0950+08  & $0.253$  & $0.230$  & $0.487$  & $32.75$  & $2.96$  & $0.26$  & $17.5$ Myr  & 16 \tabularnewline
B1046--58  & $0.124$  & $96.32$  & $6.972$  & $36.30$  & $129$  & $2.70$  & $20.3$ kyr  & 17 \tabularnewline
B1055--52  & $0.197$  & $5.833$  & $2.166$  & $34.48$  & $30.1$  & $0.75$  & $535$ kyr  & 18 \tabularnewline
J1105--6107  & $0.063$  & $15.83$  & $2.020$  & $36.40$  & $271$  & $7.00$  & $63.3$ kyr  & 19 \tabularnewline
J1119--6127  & $0.408$  & $4022$  & $81.80$  & $36.36$  & $707$  & $8.40$  & $1.61$ kyr  & 20 \tabularnewline
 &  &  &  &  &  &  &  & \tabularnewline
\hline 
\multicolumn{9}{|r|}{\emph{Continued on next page}}\tabularnewline
\hline 
\end{tabular}
\end{table}

\begin{table}[H]
\centering{}Table \ref{tab:pulsars} - continued from previous page
\begin{tabular}{|l|c|c|c|c|c|c|c|c|}
\hline 
 &  &  &  &  &  &  &  & \tabularnewline
Name  & $P$  & $\dot{P}$  & $B_{{\rm d}}$  & $\log L_{{\rm SD}}$  & $DM$  & $D$  & $\tau$  & No. \tabularnewline
 & {\scriptsize $\left({\rm s}\right)$}  & {\scriptsize $\left(10^{-15}\right)$}  & {\scriptsize $\left(10^{12}\,{\rm G}\right)$}  & {\scriptsize $\left({\rm erg\, s^{-1}}\right)$}  & {\scriptsize $\left({\rm cm^{-3}\, pc}\right)$}  & {\scriptsize $\left({\rm kpc}\right)$}  &  & \tabularnewline
\hline 
\hline 
 &  &  &  &  &  &  &  & \tabularnewline
J1124--5916  & $0.135$  & $747.1$  & $20.31$  & $37.08$  & $330$  & $6.00$  & $2.87$ kyr  & 21 \tabularnewline
B1133+16  & $1.188$  & $3.734$  & $4.254$  & $31.94$  & $4.86$  & $0.36$  & $5.04$ Myr  & 22 \tabularnewline
J1210--5226  & $0.424$  & $0.066$  & $0.338$  & $31.53$  & --  & $2.45$  & $102$ Myr  & 23 \tabularnewline
B1259--63  & $0.048$  & $2.276$  & $0.666$  & $35.91$  & $147$  & $2.00$  & $332$ kyr  & 24 \tabularnewline
J1357--6429  & $0.166$  & $360.2$  & $15.62$  & $36.49$  & $128$  & $2.50$  & $7.31$ kyr  & 25 \tabularnewline
 &  &  &  &  &  &  &  & \tabularnewline
J1420--6048  & $0.068$  & $83.17$  & $4.810$  & $37.00$  & $360$  & $8.00$  & $13.0$ kyr  & 26 \tabularnewline
B1451--68  & $0.263$  & $0.098$  & $0.325$  & $32.32$  & $8.60$  & $0.48$  & $42.5$ Myr  & 27 \tabularnewline
J1509--5850  & $0.089$  & $9.170$  & $1.824$  & $35.71$  & $138$  & $2.56$  & $154$ kyr  & 28 \tabularnewline
B1509--58  & $0.151$  & $1537$  & $30.73$  & $37.26$  & $252$  & $4.18$  & $1.55$ kyr  & 29 \tabularnewline
J1617--5055  & $0.069$  & $135.1$  & $6.183$  & $37.20$  & $467$  & $6.50$  & $8.13$ kyr  & 30 \tabularnewline
 &  &  &  &  &  &  &  & \tabularnewline
B1706--44  & $0.102$  & $92.98$  & $6.235$  & $36.53$  & $75.7$  & $2.50$  & $17.5$ kyr  & 31 \tabularnewline
B1719--37  & $0.236$  & $10.85$  & $3.234$  & $34.52$  & $99.5$  & $1.84$  & $345$ kyr  & 32 \tabularnewline
J1747--2958  & $0.099$  & $61.32$  & $4.972$  & $36.40$  & $102$  & $5.00$  & $25.5$ kyr  & 33 \tabularnewline
B1757--24  & $0.125$  & $127.9$  & $8.075$  & $36.41$  & $289$  & $5.00$  & $15.5$ kyr  & 34 \tabularnewline
B1800--21  & $0.134$  & $134.1$  & $8.551$  & $36.34$  & $234$  & $4.00$  & $15.8$ kyr  & 35 \tabularnewline
 &  &  &  &  &  &  &  & \tabularnewline
J1809--1917  & $0.083$  & $25.54$  & $2.936$  & $36.26$  & $197$  & $3.50$  & $51.3$ kyr  & 36 \tabularnewline
J1811--1925  & $0.065$  & $44.00$  & $3.407$  & $36.81$  & --  & $5.00$  & $23.3$ kyr  & 37 \tabularnewline
B1823--13  & $0.101$  & $75.06$  & $5.575$  & $36.45$  & $231$  & $4.00$  & $21.4$ kyr  & 38 \tabularnewline
J1846--0258  & $0.326$  & $7083$  & $97.02$  & $36.91$  & --  & $6.00$  & $0.73$ kyr  & 39 \tabularnewline
B1853+01  & $0.267$  & $208.4$  & $15.08$  & $35.63$  & $96.7$  & $2.60$  & $20.3$ kyr  & 40 \tabularnewline
 &  &  &  &  &  &  &  & \tabularnewline
B1916+14  & $1.181$  & $212.4$  & $31.99$  & $33.71$  & $27.2$  & $2.10$  & $88.1$ kyr  & 41 \tabularnewline
J1930+1852  & $0.137$  & $750.6$  & $20.47$  & $37.08$  & $308$  & $5.00$  & $2.89$ kyr  & 42 \tabularnewline
B1929+10  & $0.227$  & $1.157$  & $1.034$  & $33.59$  & $3.18$  & $0.36$  & $3.10$ Myr  & 43 \tabularnewline
B1951+32  & $0.040$  & $5.845$  & $0.971$  & $36.57$  & $45.0$  & $2.00$  & $107$ kyr  & 44 \tabularnewline
J2021+3651  & $0.104$  & $95.60$  & $6.361$  & $36.53$  & $371$  & $10.0$  & $17.2$ kyr  & 45 \tabularnewline
 &  &  &  &  &  &  &  & \tabularnewline
J2043+2740  & $0.096$  & $1.270$  & $0.706$  & $34.75$  & $21.0$  & $1.80$  & $1.20$ Myr  & 46 \tabularnewline
B2224+65  & $0.683$  & $9.659$  & $5.187$  & $33.08$  & $36.1$  & $2.00$  & $1.12$ Myr  & 47 \tabularnewline
B2334+61  & $0.495$  & $191.7$  & $19.69$  & $34.79$  & $58.4$  & $3.10$  & $40.9$ kyr  & 48 \tabularnewline
 &  &  &  &  &  &  &  & \tabularnewline
\hline 
\end{tabular}
\end{table}

\section{Nonthermal X-ray radiation}

The nonthermal emission, which is generally observed from radio to
$\gamma$-ray frequencies, should be generated by charged particles
accelerated at the expense of rotational energy in the magnetosphere
of the neutron star. Nonthermal X-ray radiation is characterised by
highly anisotropic emission patterns, which give rise to large pulsed
fractions. The pulse profiles often show narrow (often double) peaks,
however, in many cases nearly sinusoidal profiles are observed. As
the X-ray efficiency is strongly correlated with $L_{{\rm SD}}$,
the most X-ray luminous sources (among rotationally powered pulsars)
are the Crab pulsar and two young pulsars in the Large Magellanic
Cloud, which are the only pulsars with $L_{{\rm SD}}>10^{38}\,{\rm erg\, s^{-1}}$
\citep{2011_Mereghetti}.

\begin{figure}[h]
\begin{centering}
\includegraphics{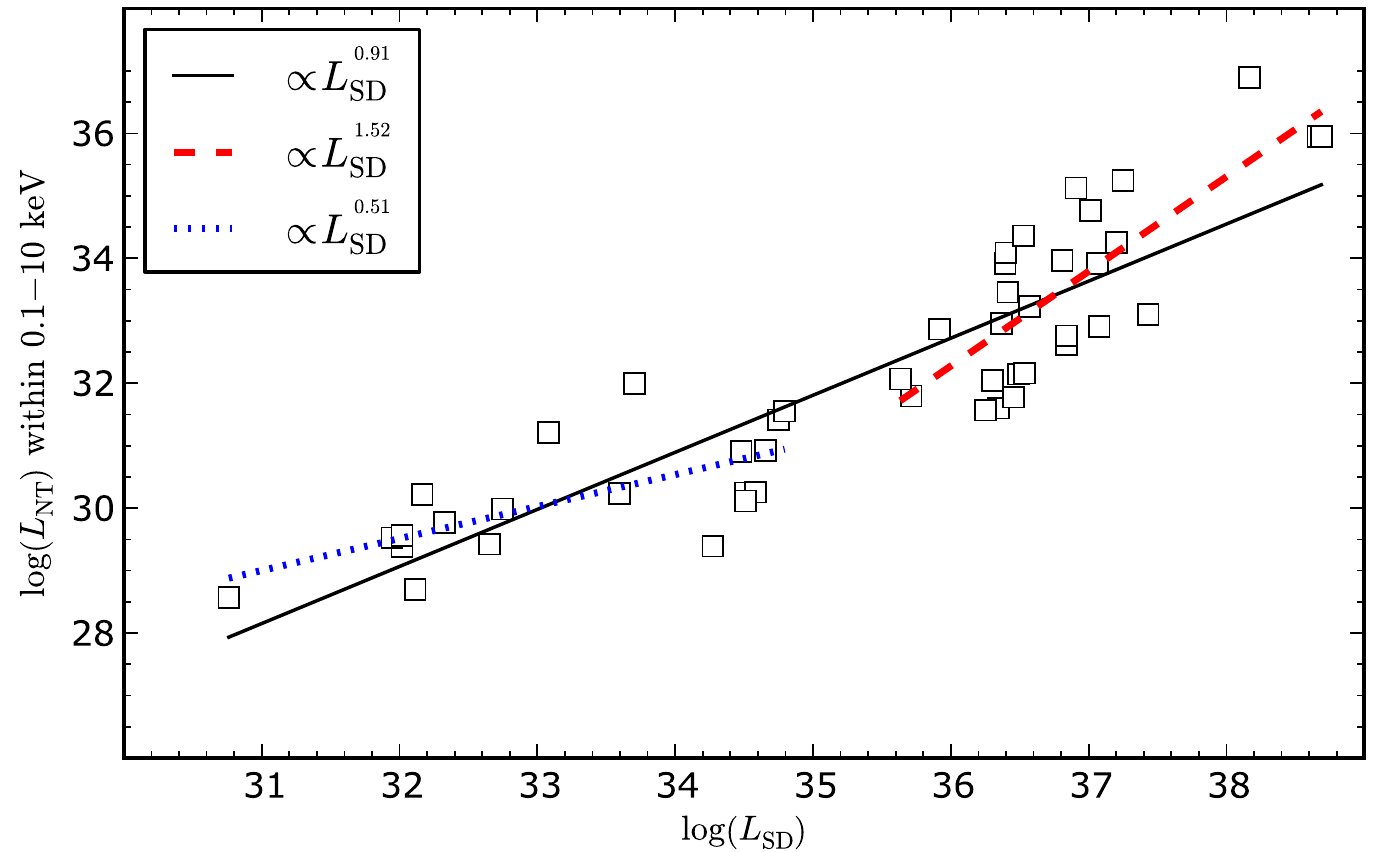}
\par\end{centering}

\centering{}\caption[Nonthermal luminosity within the $0.1-10\,{\rm keV}$ band vs spin-down
luminosity]{Nonthermal luminosity within the $0.1-10\,{\rm keV}$ band ($L_{{\rm NT}}$)
vs spin-down luminosity ($L_{{\rm SD}}$). The black solid line corresponds
to the linear fitting for all pulsars, while the blue dotted and red
dashed lines correspond to the linear fit for less luminous ($L_{{\rm SD}}<10^{35}\,{\rm erg\, s^{-1}}$)
and more luminous ($L_{{\rm SD}}>10^{35}\,{\rm erg\, s^{-1}}$) pulsars,
respectively. \label{fig:x-ray_sd} }
\end{figure}

\citet{1997_Becker} suggested that in the $0.1\lyxmathsym{–}2.4\,{\rm keV}$
band \textit{ROSAT} sources that are identified as rotation-powered
pulsars exhibit an X-ray efficiency which can be approximated as a
linear function $L_{{\rm X}}=\xi L_{{\rm SD}}$, where the total X-ray
efficiency \linebreak{}
$\xi=\xi_{_{{\rm BB}}}+\xi_{_{{\rm NT}}}\approx10^{-3}$, here $\xi_{_{{\rm BB}}}$
and $\xi_{_{{\rm NT}}}$ are efficiencies of the thermal (without
the cooling component) and nonthermal X-ray emission, respectively.
The higher sensitivity of both the \emph{Chandra} and \emph{XMM-Newton
}allows detection of less efficient ($\xi<10^{-3}$) X-ray pulsars
(see Figure \ref{fig:x-ray_sd}). \citet{2009_Becker} suggested that
for these faint pulsars the orientation of the magnetic/rotation axes
to the observer's line of sight might not be optimal. We believe that
the efficiency of spin-down energy conversion processes is mostly
affected by the strength and structure of the surface magnetic field.
The variation of $\xi$ is rather due to the nature of physical processes
than the geometrical effects. Let us note that the nonthermal X-ray
luminosities presented in Figure \ref{fig:x-ray_sd} are calculated
assuming an isotropic radiation pattern. In general, the X-ray emission
pattern differs quite essentially from the isotropic one. Thus, one
should introduce a beaming factor as the ratio of the opening angle
of the radiation cone to the full solid angle $4\pi$. Since a beaming
factor is generally unknown, the actual X-ray efficiency may differ
by up to an order of magnitude (or even more) than we have presented.

\begin{comment}
http://localhost:9090/pulsars/graphs/ (generate data, \textasciitilde{}/Html/pulsar/media/images/xray\_sd.dat)

cp \textasciitilde{}/Html/pulsar/media/images/xray\_sd.dat \textasciitilde{}/Programs/studies/phd/xray\_sd/. 

\textasciitilde{}/Programs/studies/phd/xray\_sd/x\_ray.py

cp \textasciitilde{}/Programs/studies/phd/xray\_sd/x\_ray.svg \textasciitilde{}/Documents/studies/phd/images/x-ray/.
\end{comment}

Various fitting parameters and efficiencies of nonthermal X-ray radiation
suggest that the efficiency of processes responsible for the generation
of nonthermal X-ray radiation should highly depend on the pulsar parameters
(see Figure \ref{fig:x-ray_sd}). The fitting parameters for the data
of all pulsars show a linear trend with $\xi\approx10^{-3}$, however,
if we divide them into two groups of less and more luminous pulsars,
we can see that the fitting parameters for these two groups differ
from one another. The efficiency of less luminous X-ray pulsars depends
on $L_{{\rm SD}}$ to a lesser extent than is the case for more luminous
pulsars.

As we mentioned in the Introduction, there are two main types of models:
the polar cap models and the outer gap models. The outer gap model
was proposed to explain the bright $\gamma$-ray emission from the
Crab and Vela pulsars \citep{1986_Cheng_a,1986_Cheng_b}. Placing
a $\gamma$-ray emission zone at the light cylinder, where the magnetic
field strength is considerably reduced to $B_{{\rm LC}}=B_{{\rm d}}\left(R/R_{{\rm LC}}\right)^{3}$,
provides higher $\gamma$-ray emissivities that are in somewhat better
agreement with the observations. The observational data can be interpreted
with any of the two models, although under completely different assumptions
about pulsar parameters.

\subsection{Observations}

Generally, the X-ray spectrum of relatively young ($\tau<10\,{\rm kyr}$)
and middle-aged \linebreak{}
($\tau<10\,{\rm kyr}$) pulsars is dominated by the nonthermal component.
However, it is not possible to find an exact correlation between $\tau$
and the type of spectra, i.e. which component, thermal or nonthermal,
dominates the spectrum (see the left panel of Figure \ref{fig:x-ray_age_field}).
As we mentioned above, it is quite often impossible to resolve the
components. The Crab pulsar ($\tau=958\,{\rm yrs}$) is the most characteristic
example of a young pulsar. The upper limit for X-ray luminosity of
the Crab pulsar (one of the strongest known X-ray radio pulsars) is
about $L_{{\rm NT}}^{{\rm ^{max}}}=8.9\times10^{35}\,{\rm erg\, s^{-1}}$.
This value is calculated assuming an isotropic radiation pattern,
however, even if we assume an angular anisotropy of the radiation
(beaming factor $\approx1/4\pi$), the lower limit of its luminosity
$L_{{\rm NT}}^{^{{\rm min}}}=7.1\times10^{34}\,{\rm erg\, s^{-1}}$
continues to be very high. The luminosities calculated above correspond
to the following X-ray efficiencies: ${\displaystyle \xi_{_{{\rm NT}}}^{^{{\rm max}}}=10^{-2.71}}$
(isotropic radiation pattern) and $\xi_{_{{\rm NT}}}^{{\rm ^{min}}}=10^{-3.81}$
(anisotropic radiation pattern). Although ${\displaystyle \xi_{_{{\rm NT}}}}$
is quite small, the nonthermal component still obscures all the thermal
ones. To obtain a similar efficiency of the thermal radiation from
the entire stellar surface, its temperature should be $T_{{\rm s}}=5.9\times10^{6}\,{\rm K}$
(assuming $R=10\,{\rm km}$), which vastly exceeds the upper limit
($T_{{\rm s}}<2.3\times10^{6}\,{\rm K}$). Furthermore, the temperature
of the polar caps should be about $2.5\times10^{7}\,{\rm K}$ to obtain
a comparable luminosity.

The Vela-like pulsars compose another characteristic group of pulsars.
This group consists of pulsars with high spin-down luminosities but
considerably low X-ray efficiencies $\xi_{_{{\rm NT}}}\apprle10^{-4}$.
A characteristic age of the Vela is about $1.1\times10^{4}\,{\rm yrs}$
($10$ times older than the Crab), but it can still be classified
as a very young pulsar. The nonthermal luminosity of a Vela pulsar
is $L_{{\rm NT}}^{^{{\rm max}}}=4.2\times10^{32}$ and efficiency
$\xi_{_{{\rm NT}}}^{^{{\rm max}}}=10^{-4.22}$. Some of the Vela-like
pulsars (like the Vela itself) also exhibit a thermal component, which
in some cases can be comparable to the nonthermal component. The thermal
efficiency of the Vela $\xi_{_{{\rm BB}}}=10^{-4.72}$ is quite similar
to $\xi_{_{{\rm NT}}}^{^{{\rm max}}}=10^{-4.22}$, but if we assume
an anisotropic radiation pattern of the nonthermal component than
$\xi_{_{{\rm NT}}}^{^{{\rm min}}}=10^{-5.32}$, thus even less than
$\xi_{_{{\rm BB}}}$.

The third group includes pulsars with low spin-down luminosity $L_{{\rm SD}}\apprle10^{35}$.
In most cases, the X-ray spectra of such pulsars (e.g. PSR 9050+08,
PSR B1929+10) have both thermal and nonthermal components, with similar
efficiencies. Thus, the spectrum fitting procedure is more complicated.
The nonthermal X-ray efficiencies of these pulsars, $\xi_{_{{\rm NT}}}\sim10^{-3}$,
are considerably higher than those of the Vela-like pulsars. Note
that even when the observed spectra are dominated by nonthermal radiation,
we cannot rule out a situation that the thermal component is stronger
than the nonthermal one, but due to unfavourable geometry we cannot
observe it.

Even with the improved quality of X-ray observations performed by
both the \emph{Chandra} and \emph{XMM-Newton}, the available data
do not allow us to fully discriminate between the~different emission
scenarios. However, these data can be used to verify whether the~proposed
model of X-ray emission meets all the requirements. Table \ref{tab:x-ray_nonthermal}
presents the~observed spectral properties of pulsars showing nonthermal
components.

\begin{comment}
http://127.0.0.1:9090/pulsars/table\_pl/ (then rename citet to citetalias
after lyx import of \textasciitilde{}/Html/pulsar/download/data/table\_pl.tex)
\end{comment}

\begin{landscape} 
\begin{table}[H]
\thispagestyle{fancy} \caption[{Observed X-ray spectral properties of rotation-powered pulsars {[}nonthermal{]}}]{Observed spectral properties of rotation-powered pulsars with X-ray
spectrum showing the nonthermal (power-law) component. The individual
columns are as follows: (1) Pulsar name, (2) Additional information,
(3) Spectral components required to fit the observed spectra, PL:
power law, BB: blackbody, (4) Pulse phase average photon index, (5)
Maximum nonthermal luminosity $L_{{\rm NT}}$, (6) Maximum nonthermal
X-ray efficiency $\xi_{_{{\rm NT}}}^{^{{\rm max}}}$, (7) Minimum
nonthermal X-ray efficiency $\xi_{_{{\rm NT}}}^{{\rm ^{min}}}$, (8)
Total thermal luminosity $L_{{\rm BB}}$, (9) Thermal efficiency $\xi_{_{{\rm BB}}}$,
(10) References, (11) Number of the pulsar. Both nonthermal luminosities
and efficiencies were calculated in the $0.1-10\,{\rm keV}$ band.
The maximum value was calculated with the assumption that the X-ray
radiation is isotropic while the minimum value was calculated assuming
strong angular anisotropy of the radiation ($\xi_{_{{\rm NT}}}^{{\rm ^{min}}}\approx1/\left(4\pi\right)\cdot\xi_{_{{\rm NT}}}^{{\rm ^{max}}}$).
Pulsars are sorted by nonthermal X-ray luminosity (5). \label{tab:x-ray_nonthermal} }

\centering{}%
\begin{tabular}{|l|c|c|c|c|c|c|c|c|c|c|}
\hline 
%\multicolumn{11}{c}{} \\
 &  &  &  &  &  &  &  &  &  & \tabularnewline
Name  & Comment  & Spectrum  & Photon-Index  & $\log L_{{\rm NT}}$  & $\log\xi_{_{{\rm NT}}}^{{\rm ^{max}}}$  & $\log\xi_{_{{\rm NT}}}^{^{{\rm min}}}$  & $\log L_{{\rm BB}}$  & $\log\xi_{_{{\rm BB}}}$  & Ref.  & No. \tabularnewline
 &  &  &  & {\scriptsize $\left({\rm erg\, s^{-1}}\right)$}  &  &  & {\scriptsize $\left({\rm erg\, s^{-1}}\right)$}  &  &  & \tabularnewline
\hline 
\hline 
 &  &  &  &  &  &  &  &  &  & \tabularnewline
B0540--69  & N158A, LMC  & PL  & $1.92_{-0.11}^{+0.11}$  & $36.90$  & $-1.27$  & $-2.37$  & --  & --  & \citetalias{2001_Kaaret}, \citetalias{2008_Campana}  & 7 \tabularnewline
B0531+21  & Crab  & PL  & $1.63_{-0.07}^{+0.07}$  & $35.95$  & $-2.71$  & $-3.81$  & --  & --  & \citetalias{2009_Becker}  & 4 \tabularnewline
J0537--6910  & N157B, LMC  & PL  & $1.80_{-0.10}^{+0.10}$  & $35.95$  & $-2.74$  & $-3.84$  & --  & --  & \citetalias{2005_Mignani}  & 5 \tabularnewline
B1509--58  & Crab-like pulsar  & PL  & $1.19_{-0.04}^{+0.04}$  & $35.24$  & $-2.00$  & $-3.10$  & --  & --  & \citetalias{2001_Cusumano}, \citetalias{2006_DeLaney}, \citetalias{2009_Becker}  & 29 \tabularnewline
J1846--0258  & Kes 75  & BB + PL  & $1.90_{-0.10}^{+0.10}$  & $35.13$  & $-1.78$  & $-2.88$  & $34.06$  & $-2.85$  & \citetalias{2008_Ng}, \citetalias{2003_Helfand}  & 39 \tabularnewline
 &  &  &  &  &  &  &  &  &  & \tabularnewline
J1420--6048  &  & PL  & $1.60_{-0.40}^{+0.40}$  & $34.77$  & $-2.25$  & $-3.35$  & --  & --  & \citetalias{2001_Roberts}  & 26 \tabularnewline
J2021+3651  &  & PL, BB  & $1.70_{-0.20}^{+0.30}$  & $34.36$  & $-2.17$  & $-3.27$  & $33.78$  & $-2.75$  & \citetalias{2008_VanEtten},\citetalias{2004_Hessels}  & 45 \tabularnewline
J1617--5055  & Crab-like pulsar  & PL  & $1.14_{-0.06}^{+0.06}$  & $34.25$  & $-2.95$  & $-4.05$  & --  & --  & \citetalias{2009_Kargaltsev}, \citetalias{2002_Becker}  & 30 \tabularnewline
J1747--2958  & Mouse  & PL, BB  & $1.80_{-0.08}^{+0.08}$  & $34.09$  & $-2.31$  & $-3.41$  & --  & --  & \citetalias{2004_Gaensler}  & 33 \tabularnewline
J1811--1925  & G11.2-0.3  & PL  & $0.97_{-0.32}^{+0.39}$  & $33.97$  & $-2.84$  & $-3.94$  & --  & --  & \citetalias{2003_Roberts}, \citetalias{2004_Roberts}  & 37 \tabularnewline
 &  &  &  &  &  &  &  &  &  & \tabularnewline
J1930+1852  & Crab-like pulsar  & PL  & $1.20_{-0.20}^{+0.20}$  & $33.92$  & $-3.15$  & $-4.25$  & --  & --  & \citetalias{2007_Lu}, \citetalias{2002_Camilo}  & 42 \tabularnewline
\hline 
\multicolumn{11}{|r|}{\emph{Continued on next page}}\tabularnewline
\hline 
\end{tabular}
\end{table}

\end{landscape}

\begin{landscape} 
\begin{table}[H]
\thispagestyle{fancy} 

\centering{}Table \ref{tab:x-ray_nonthermal} - continued from previous
page %
\begin{tabular}{|l|c|c|c|c|c|c|c|c|c|c|}
\hline 
 &  &  &  &  &  &  &  &  &  & \tabularnewline
Name  & Comment  & Spectrum  & Photon-Index  & $\log L_{{\rm NT}}$  & $\log\xi_{_{{\rm NT}}}^{^{max}}$  & $\log\xi_{_{{\rm NT}}}^{^{min}}$  & $\log L_{{\rm BB}}$  & $\log\xi_{_{{\rm BB}}}$  & Ref.  & No. \tabularnewline
 &  &  &  & {\scriptsize $\left({\rm erg\, s^{-1}}\right)$}  &  &  & {\scriptsize $\left({\rm erg\, s^{-1}}\right)$}  &  &  & \tabularnewline
\hline 
\hline 
 &  &  &  &  &  &  &  &  &  & \tabularnewline
J1105--6107  &  & PL  & $1.80_{-0.40}^{+0.40}$  & $33.91$  & $-2.48$  & $-3.58$  & --  & --  & \citetalias{1998_Gotthelf}  & 19 \tabularnewline
B1757--24  & Duck  & PL  & $1.60_{-0.50}^{+0.60}$  & $33.46$  & $-2.95$  & $-4.05$  & --  & --  & \citetalias{2001_Kaspi}  & 34 \tabularnewline
B1951+32  & CTB 80  & BB + PL  & $1.63_{-0.05}^{+0.03}$  & $33.22$  & $-3.35$  & $-4.45$  & $31.95$  & $-4.62$  & \citetalias{2005_Li}  & 44 \tabularnewline
J0205+6449  & 3C58  & BB + PL  & $1.78_{-0.04}^{+0.02}$  & $33.10$  & $-4.33$  & $-5.43$  & $33.60$  & $-3.83$  & \citetalias{2004_Slane}  & 2 \tabularnewline
J1119--6127  & G292.2-0.5  & BB + PL  & $1.50_{-0.20}^{+0.30}$  & $32.95$  & $-3.42$  & $-4.51$  & $33.37$  & $-3.00$  & \citetalias{2007_Gonzalez}, \citetalias{2012_Ng}  & 20 \tabularnewline
 &  &  &  &  &  &  &  &  &  & \tabularnewline
J1124--5916  & Vela-like pulsar  & PL  & $1.60_{-0.10}^{+0.10}$  & $32.91$  & $-4.17$  & $-5.27$  & --  & --  & \citetalias{2003_Hughes},\citetalias{2003_Gonzales}  & 21 \tabularnewline
B1259--63  & Be-star bin  & PL  & $1.69_{-0.04}^{+0.04}$  & $32.87$  & $-3.05$  & $-4.15$  & --  & --  & \citetalias{2009_Chernyakova}, \citetalias{2006_Chernyakova}  & 24 \tabularnewline
B0833--45  & Vela  & BB + PL  & $2.70_{-0.40}^{+0.40}$  & $32.62$  & $-4.22$  & $-5.32$  & $32.12$  & $-4.72$  & \citetalias{2007_Zavlin_b}  & 13 \tabularnewline
B1706--44  & G343.1-02.3  & BB + PL  & $2.00_{-0.50}^{+0.50}$  & $32.16$  & $-4.37$  & $-5.47$  & $32.78$  & $-3.76$  & \citetalias{2002_Gotthelf}  & 31 \tabularnewline
J1357--6429  &  & BB + PL  & $1.30_{-0.20}^{+0.20}$  & $32.15$  & $-4.35$  & $-5.44$  & $32.50$  & $-3.99$  & \citetalias{2007_Zavlin}  & 25 \tabularnewline
 &  &  &  &  &  &  &  &  &  & \tabularnewline
B1853+01  & W44  & PL  & $1.28_{-0.48}^{+0.48}$  & $32.07$  & $-3.57$  & $-4.66$  & --  & --  & \citetalias{2002_Petre}  & 40 \tabularnewline
B1046--58  & Vela-like pulsar  & PL  & $1.70_{-0.20}^{+0.40}$  & $32.04$  & $-4.26$  & $-5.36$  & --  & --  & \citetalias{2006_Gonzalez}  & 17 \tabularnewline
B1916+14  &  & BB, PL  & $3.50_{-0.70}^{+1.60}$  & $32.00$  & $-1.71$  & $-2.81$  & $31.07$  & $-2.63$  & \citetalias{2009_Zhu}  & 41 \tabularnewline
J1509--5850  & MSH 15-52  & PL  & $1.00_{-0.30}^{+0.20}$  & $31.80$  & $-3.92$  & $-5.02$  & --  & --  & \citetalias{2007_Hui}  & 28 \tabularnewline
B1823--13  & Vela-like  & BB + PL  & $1.70_{-0.70}^{+0.70}$  & $31.78$  & $-4.67$  & $-5.77$  & $32.19$  & $-4.27$  & \citetalias{2008_Pavlov}  & 38 \tabularnewline
 &  &  &  &  &  &  &  &  &  & \tabularnewline
B1800--21  & Vela-like pulsar  & PL + BB  & $1.40_{-0.60}^{+0.60}$  & $31.60$  & $-4.74$  & $-5.84$  & --  & --  & \citetalias{2007_Kargaltsev}  & 35 \tabularnewline
\hline 
\multicolumn{11}{|r|}{\emph{Continued on next page}}\tabularnewline
\hline 
\end{tabular}
\end{table}

\end{landscape}

\begin{landscape} 
\begin{table}[H]
\thispagestyle{fancy} 

\centering{}Table \ref{tab:x-ray_nonthermal} - continued from previous
page %
\begin{tabular}{|l|c|c|c|c|c|c|c|c|c|c|}
\hline 
 &  &  &  &  &  &  &  &  &  & \tabularnewline
Name  & Comment  & Spectrum  & Photon-Index  & $\log L_{{\rm NT}}$  & $\log\xi_{_{{\rm NT}}}^{^{max}}$  & $\log\xi_{_{{\rm NT}}}^{^{min}}$  & $\log L_{{\rm BB}}$  & $\log\xi_{_{{\rm BB}}}$  & Ref.  & No. \tabularnewline
 &  &  &  & {\scriptsize $\left({\rm erg\, s^{-1}}\right)$}  &  &  & {\scriptsize $\left({\rm erg\, s^{-1}}\right)$}  &  &  & \tabularnewline
\hline 
\hline 
 &  &  &  &  &  &  &  &  &  & \tabularnewline
J1809--1917  &  & BB + PL  & $1.23_{-0.62}^{+0.62}$  & $31.57$  & $-4.68$  & $-5.78$  & $31.69$  & $-4.56$  & \citetalias{2007_Kargaltsev}  & 36 \tabularnewline
B2334+61  &  & BB + PL  & $2.20_{-1.40}^{+3.00}$  & $31.55$  & $-3.24$  & $-4.34$  & $32.06$  & $-2.73$  & \citetalias{2006_Mcgowan}  & 48 \tabularnewline
J2043+2740  &  & BB + PL  & $2.80_{-0.80}^{+1.00}$  & $31.41$  & $-3.34$  & $-4.44$  & $30.77$  & $-3.98$  & \citetalias{2004_Becker}  & 46 \tabularnewline
B2224+65  & Guitar  & PL, BB  & $2.20_{-0.30}^{+0.20}$  & $31.21$  & $-1.87$  & $-2.97$  & $30.51$  & $-2.57$  & \citetalias{2012_Hui}, \citetalias{2007_Hui_b}  & 47 \tabularnewline
B0355+54  &  & BB + PL  & $1.00_{-0.20}^{+0.20}$  & $30.92$  & $-3.73$  & $-4.83$  & $30.40$  & $-4.25$  & \citetalias{2007_McGowan},\citetalias{1994_Slane}  & 3 \tabularnewline
 &  &  &  &  &  &  &  &  &  & \tabularnewline
B1055--52  &  & BB+BB+PL  & $1.70_{-0.10}^{+0.10}$  & $30.91$  & $-3.57$  & $-4.67$  & $32.63$  & $-1.85$  & \citetalias{2005_Deluca}  & 18 \tabularnewline
B0656+14  &  & BB+BB+PL  & $2.10_{-0.30}^{+0.30}$  & $30.26$  & $-4.33$  & $-5.42$  & $32.77$  & $-1.81$  & \citetalias{2005_Deluca}  & 10 \tabularnewline
J0633+1746  & Geminga  & BB+BB+PL  & $1.68_{-0.06}^{+0.06}$  & $30.24$  & $-4.27$  & $-5.37$  & $31.67$  & $-2.84$  & \citetalias{2005_Jackson}  & 9 \tabularnewline
B1929+10  &  & BB + PL  & $1.73_{-0.66}^{+0.46}$  & $30.23$  & $-3.36$  & $-4.46$  & $30.06$  & $-3.53$  & \citetalias{2008_Misanovic}  & 43 \tabularnewline
B0628--28  &  & BB + PL  & $2.98_{-0.65}^{+0.91}$  & $30.22$  & $-1.94$  & $-3.04$  & $30.22$  & $-1.94$  & \citetalias{2005_Tepedelenl} , \citetalias{2005_Becker} & 8 \tabularnewline
 &  &  &  &  &  &  &  &  &  & \tabularnewline
B0950+08  &  & BB + PL  & $1.31_{-0.14}^{+0.14}$  & $29.99$  & $-2.76$  & $-3.86$  & $28.92$  & $-3.82$  & \citetalias{2004_Zavlin}  & 16 \tabularnewline
B1451--68  &  & BB + PL  & $1.40_{-0.50}^{+0.50}$  & $29.77$  & $-2.56$  & $-3.66$  & $29.27$  & $-3.06$  & \citetalias{2012_Posselt}  & 27 \tabularnewline
B1133+16  &  & BB, PL  & $2.51_{-0.33}^{+0.36}$  & $29.52$  & $-2.42$  & $-3.52$  & $28.56$  & $-3.38$  & \citetalias{2006_Kargaltsev}  & 22 \tabularnewline
B0823+26  &  & PL  & $1.58_{-0.33}^{+0.43}$  & $29.42$  & $-3.23$  & $-4.33$  & --  & --  & \citetalias{2004_Becker}  & 12 \tabularnewline
B0943+10  & Chameleon & BB, PL  & $2.60_{-0.50}^{+0.70}$  & $29.38$  & $-2.64$  & $-3.74$  & $28.40$  & $-3.62$  & \citetalias{2005_Zhang},\citetalias{2006_Kargaltsev}  & 15 \tabularnewline
 &  &  &  &  &  &  &  &  &  & \tabularnewline
B0834+06  &  & BB + PL  & -- & $28.70$  & $-3.41$  & $-4.51$  & $28.70$  & $-3.41$  & \citetalias{2008_Gil}  & 14 \tabularnewline
J0108--1431  &  & BB + PL &  $3.10_{-0.20}^{+0.50}$ & $28.57$ & $-2.19$ & $-3.29$ & $27.94$ & $-2.82$ & \citetalias{2012_Posselt}, \citetalias{2009_Pavlov} & 1\tabularnewline
\hline 
\end{tabular}
\end{table}

\end{landscape}

\section{Thermal X-ray radiation\label{sec:x-ray.thermal}}

\subsection{Modelling of thermal radiation from a neutron star}

Thermal X-ray emission seems to be quite a common feature of radio
pulsars. The blackbody fit to the observed thermal spectrum of a neutron
star allows us to obtain the redshifted effective temperature $T^{\infty}$
and redshifted total bolometric flux $F^{\infty}$ (measured by a
distant observer). To estimate the actual (unredshifted) parameters,
one should take into account the gravitational redshift, $g_{{\rm r}}=\sqrt{1-2GM/Rc^{2}}$,
determined by the neutron star mass $M$ and radius $R$, here $G$
is the gravitational constant. Then the actual effective temperature
and actual total bolometric flux can be written as: 
\begin{equation}
\begin{split}T & =g_{{\rm r}}^{-1}T^{\infty},\\
F & =g_{{\rm r}}^{-2}F^{\infty}.
\end{split}
\label{eq:x-ray.infty_eff}
\end{equation}

Knowing the distance to the neutron star, $D$, we can use the effective
temperature and total bolometric flux to calculate the size of the
radiating region. If we assume that the radiation is isotropic (same
in all directions, e.g. radiation from the entire stellar surface)
then the radius of the radiating sphere (star) can be calculated as
\citep{2007_Zavlin} 
\begin{equation}
R_{\perp}^{\infty}=D\sqrt{\frac{F^{\infty}}{\sigma T^{\infty4}}}=g_{{\rm r}}^{-1}R_{\perp},\label{eq:x-ray.r_infty}
\end{equation}
where $\sigma\approx5.6704\times10^{-5}{\rm \, erg\, cm^{-2}\, s^{-1}\, K^{-4}}$
is the Stefan-Boltzmann constant.

Knowing that $L_{{\rm BB}}=4\pi D^{2}F$ and using Equations \ref{eq:x-ray.infty_eff}
and \ref{eq:x-ray.r_infty}, we can write that

\begin{equation}
L_{{\rm BB}}=g_{{\rm r}}^{-2}L_{{\rm BB}}^{\infty}.
\end{equation}

The modelling of thermal radiation is more complicated if we assume
that it comes from the hot spot on the stellar surface. One should
take into account such factors as: time-averaged cosine of the angle
between the magnetic axis and the line of sight $\left<\cos i\right>$,
gravitational bending of light, as well as whether the radiation comes
from two opposite poles of the star or from one hot spot only. In
general, the observed luminosity of the hot spot can be written as:
\begin{equation}
L_{{\rm hs}}^{\infty}=A_{{\rm hs}}^{\infty}\sigma T^{\infty4},\label{x-ray.lbol_spot}
\end{equation}
 where $A_{{\rm hs}}^{\infty}=\pi R_{{\rm hs}}^{\infty2}$ is the
observed area of the radiating region.

The observed area of the radiating spot is also influenced by the
geometrical factor $f$. This geometrical factor depends on following
angles: $\zeta$ between the line of sight and the spin axis, and
$\alpha$ between the spin and magnetic axes, as well as on $g_{{\rm r}}$
and whether the radiation comes from the star's two opposite poles
or from a single hot spot only: 
\begin{equation}
\begin{split}A_{{\rm hs}}^{\infty} & =g_{{\rm r}}^{-2}fA_{{\rm hs}},\\
R_{{\rm hs}} & =g_{{\rm r}}f^{-1/2}R_{{\rm hs}}^{\infty}.
\end{split}
\label{x-ray.infty_eff2}
\end{equation}

Finally, the hot spot luminosity can be calculated as

\begin{equation}
L_{{\rm hs}}=g_{{\rm r}}^{-2}f^{-1}L_{{\rm hs}}^{\infty}.\label{x-ray.lbol_spot-1}
\end{equation}

The luminosity of a radiating sphere with radius $R_{\perp}$ can
be calculated as\linebreak{}
 $L_{{\rm sp}}=4A_{\perp}\sigma T^{4}=4\pi R_{\perp}^{2}\sigma T^{4}$.
On the other hand, if we assume that the radiation originates only
from one hot spot we can calculate the luminosity as $L_{{\rm hs}}=A_{{\rm hs}}\sigma T^{4}$.
If the hot spot size is small compared to the star radius ($R_{{\rm hs}}\ll R$)
then the area of the spot can be calculated as $A_{{\rm hs}}\approx\pi R_{\perp}^{2}$.
Thus, we have to remember that the luminosity calculated assuming
a spherical source will be four times higher than the actual luminosity
of a radiating hot spot $L_{{\rm hs}}=1/4\cdot L_{{\rm sp}}$ (see
the next section for details).

\subsection{Thermal radiation of hot spots\label{sub:x-ray.hot_spots}}

\begin{figure}[H]
\begin{centering}
\includegraphics[width=8.5cm]{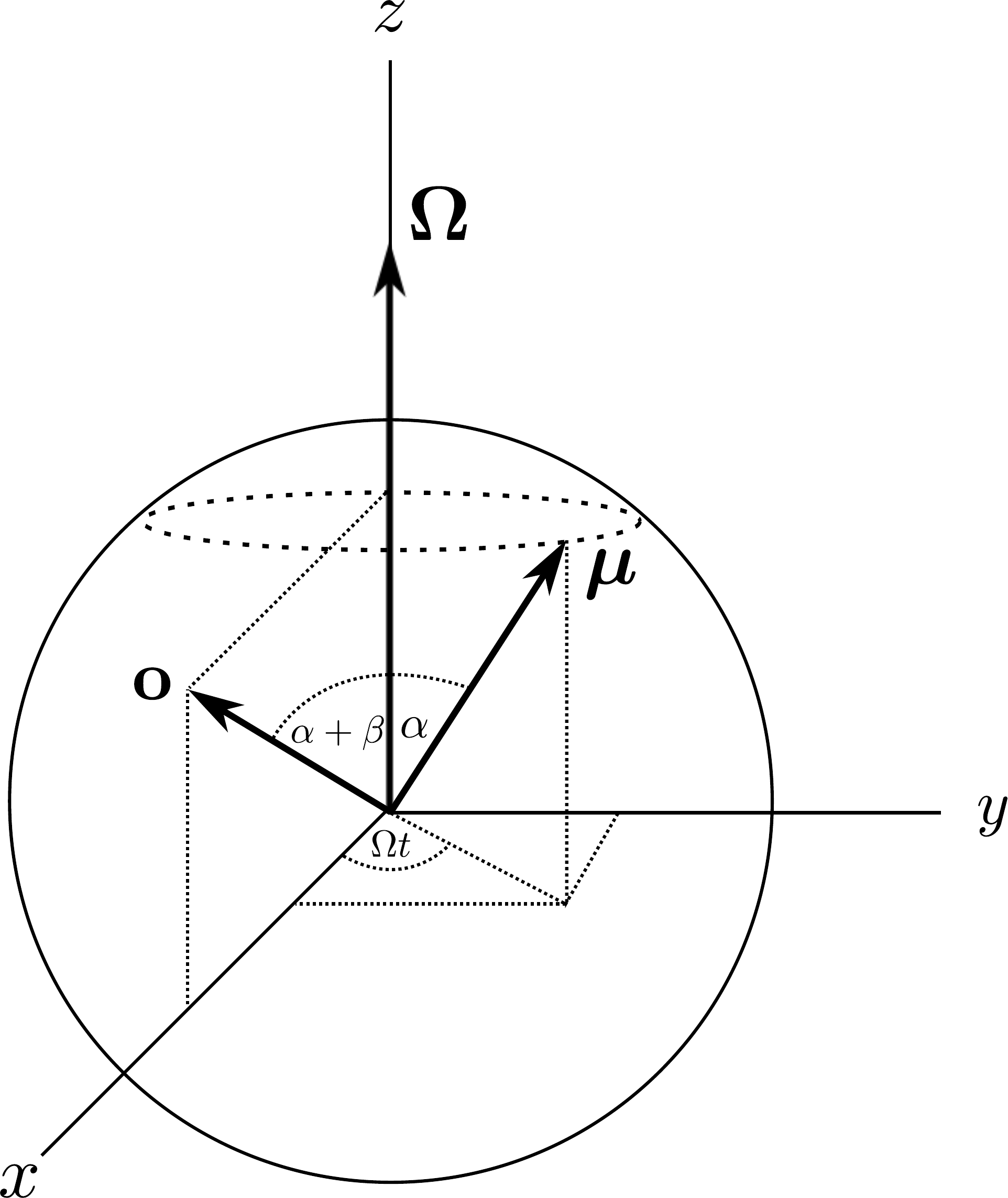}
\par\end{centering}

\centering{}\caption[Coordinate system co-rotating with a star]{Coordinate system co-rotating with a star. The system was chosen
so that the z-axis is along ${\bf \Omega}$ (the angular velocity)
and ${\bf o}$ lies in the x-z plane (fiducial plane, i.e. at longitude
zero). Here, $\boldsymbol{\hat{\mu}}$ is a unit vector in the direction
of the magnetic axis and $\alpha$ is the angle between ${\bf \Omega}$
and $\boldsymbol{\hat{\mu}}$, $\beta$ is the impact parameter. \label{fig:x-ray_coordinates}}
\end{figure}

Let us consider a neutron star with two antipodal hot spots associated
with polar caps of a stellar magnetic field. For simplicity's sake
we assume that the spot size is small compared to the star radius
$R$. If the magnetic axis $\boldsymbol{\hat{\mu}}$ is inclined to
the spin axis by an angle $\alpha\leq90^{\circ}$, the spots periodically
change their position and inclination with respect to a distant observer.
To compute the radiation fluxes from the primary (closer to the observer)
as well as the antipodal spot, we need to know their inclinations:
$\cos i_{1}={\bf n\cdot o}$ and $\cos i_{2}={\bf \bar{n}\cdot o}=-\cos i_{1}$,
where ${\bf n}$ and ${\bf \bar{n}}=-{\bf n}$ are normal vectors
to spots surfaces, and ${\bf o}$ is the unit vector pointing toward
the observer. In the calculations we use a coordinate system co-rotating
with a star. The z-axis is along ${\bf \Omega}$ (the angular velocity)
and ${\bf o}$ lies in the x-z plane (see Figure \ref{fig:x-ray_coordinates}). 

In the chosen coordinate system we can write that the spherical coordinates
of vectors have the following components: 

\begin{equation}
\begin{array}{ccc}
{\bf \Omega} & = & \left(\Omega,\,0,\,0\right);\\
{\bf o} & = & \left(1,\,\alpha+\beta,\,0\right);\\
\boldsymbol{\hat{\mu}} & = & \left(1,\,\alpha,\,\Omega t\right).
\end{array}
\end{equation}

Here the impact parameter $\beta$ represents the closest approach
of the line of sight to the magnetic axis. Note that $\boldsymbol{\hat{\mu}}={\bf n}$
and ${\bf \bar{{\bf n}}}=-\boldsymbol{\hat{\mu}}$; thus, we can write
the following components of Cartesian coordinates:

\begin{equation}
\begin{array}{ccc}
{\bf {\bf o}} & = & \left(\sin\left(\alpha+\beta\right),\,0,\,\cos\left(\alpha+\beta\right)\right);\\
{\bf n} & = & \left(\sin\alpha\cos\Omega t,\,\sin\Omega t\sin\alpha,\,\cos\alpha\right).
\end{array}
\end{equation}

\begin{comment}
$\left(\sin\left(\pi-\alpha\right)\cos\left(\pi+\Omega t\right),\,\sin\left(\pi+\Omega t\right)\cdot\sin\left(\pi-\alpha\right),\,\cos\left(\pi-\alpha\right)\right)=\left(-\sin\alpha\cos\Omega t,\,-\sin\Omega t\cdot\sin\alpha,\,-\cos\alpha\right)$
\end{comment}

Finally, the inclination angle for both primary and antipodal hot
spots can be calculated as

\begin{equation}
\begin{array}{ccc}
\cos i_{1} & = & \sin\alpha\cdot\cos\Omega t\cdot\sin\left(\alpha+\beta\right)+\cos\alpha\cdot\cos\left(\alpha+\beta\right);\\
\cos i_{2} & = & -\cos i_{1}=-\sin\alpha\cdot\cos\Omega t\cdot\sin\left(\alpha+\beta\right)-\cos\alpha\cdot\cos\left(\alpha+\beta\right).
\end{array}
\end{equation}

We can estimate the contributions of the primary and antipodal spots
to the observed X-ray flux by calculating the time-averaged cosine
of the angle between the magnetic axis and the line of sight. Note
that we should take into account only positive values of $\cos i$
since for larger angles ($i>90^{\circ}$) the spot is not visible
(at least in this approximation, see Section \ref{sub:x-ray:gravitational}
for more details). Thus, the contribution of the primary spot can
be calculated as follows:

\begin{equation}
\begin{array}{c}
\left\langle \cos i_{1}\right\rangle =\begin{cases}
\begin{split}\int_{0}^{P}\cos\left(i_{1}\right){\rm d}t\end{split}
 & {\rm if}\ \frac{1}{\tan\alpha\tan\left(\alpha+\beta\right)}<-1\ {\rm or}\ \frac{1}{\tan\alpha\tan\left(\alpha+\beta\right)}>1,\\
\begin{split}\int_{0}^{t_{-}}\cos\left(i_{1}\right){\rm d}t\end{split}
+\begin{split}\int_{t_{+}}^{2\pi}\cos\left(i_{1}\right){\rm d}t\end{split}
 & {\rm if}\ -1<\frac{1}{\tan\alpha\tan\left(\alpha+\beta\right)}<1,
\end{cases}\end{array}
\end{equation}
where integration limits are

\begin{equation}
t_{\pm}=\frac{P}{2}\pm\frac{P}{2\pi}\arccos\left[\frac{1}{\tan\alpha\tan\left(\alpha+\beta\right)}\right].
\end{equation}

On the other hand, the contribution of the antipodal spot can be calculated
as

\begin{equation}
\begin{array}{c}
\left\langle \cos i_{2}\right\rangle =\begin{cases}
\begin{split}0\end{split}
 & {\rm if}\ \frac{1}{\tan\alpha\tan\left(\alpha+\beta\right)}<-1\ {\rm or}\ \frac{1}{\tan\alpha\tan\left(\alpha+\beta\right)}>1,\\
\begin{split}\int_{t_{-}}^{t_{+}}\cos\left(i_{2}\right){\rm d}t\end{split}
 & {\rm if}\ -1<\frac{1}{\tan\alpha\tan\left(\alpha+\beta\right)}<1.
\end{cases}\end{array}
\end{equation}

\begin{figure}[H]

\begin{centering}
\includegraphics[height=8.7cm]{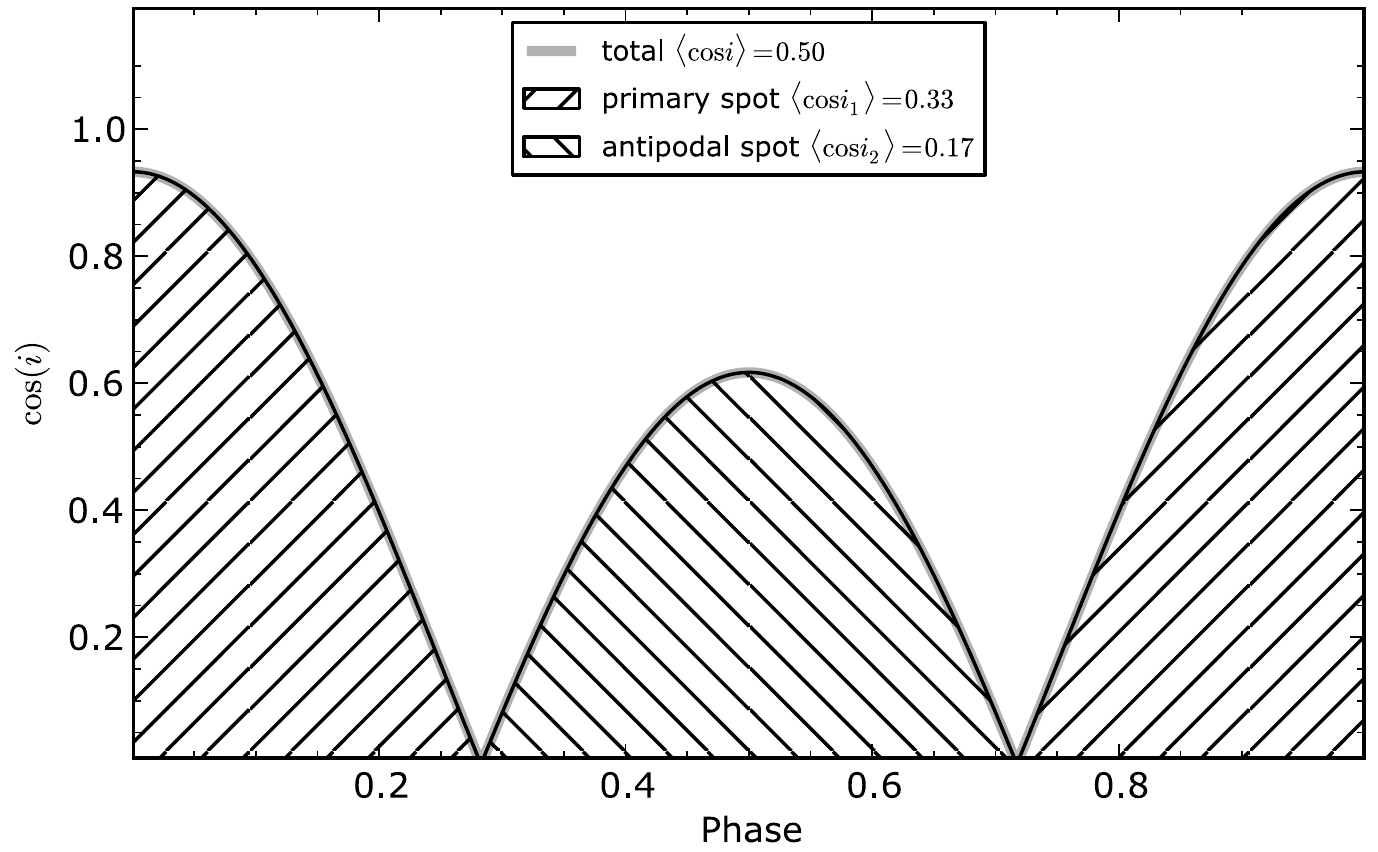}
\par\end{centering}

\caption[{Cosine of the hot spots' inclination angle {[}PSR B0950+08{]}}]{Cosine of the hot spots' inclination angle as a function of the pulsar
phase for PSR~B0950+08. The following parameters were used: $\alpha=105.46^{\circ}$,
$\beta=21.1^{\circ}$. For this geometry the thermal radiation of
both primary and antipodal spots has a significant influence on the
observed thermal flux. \label{fig:x-ray_B0950+08}}
\end{figure}

\begin{figure}[H]
\begin{centering}
\includegraphics[height=8.7cm]{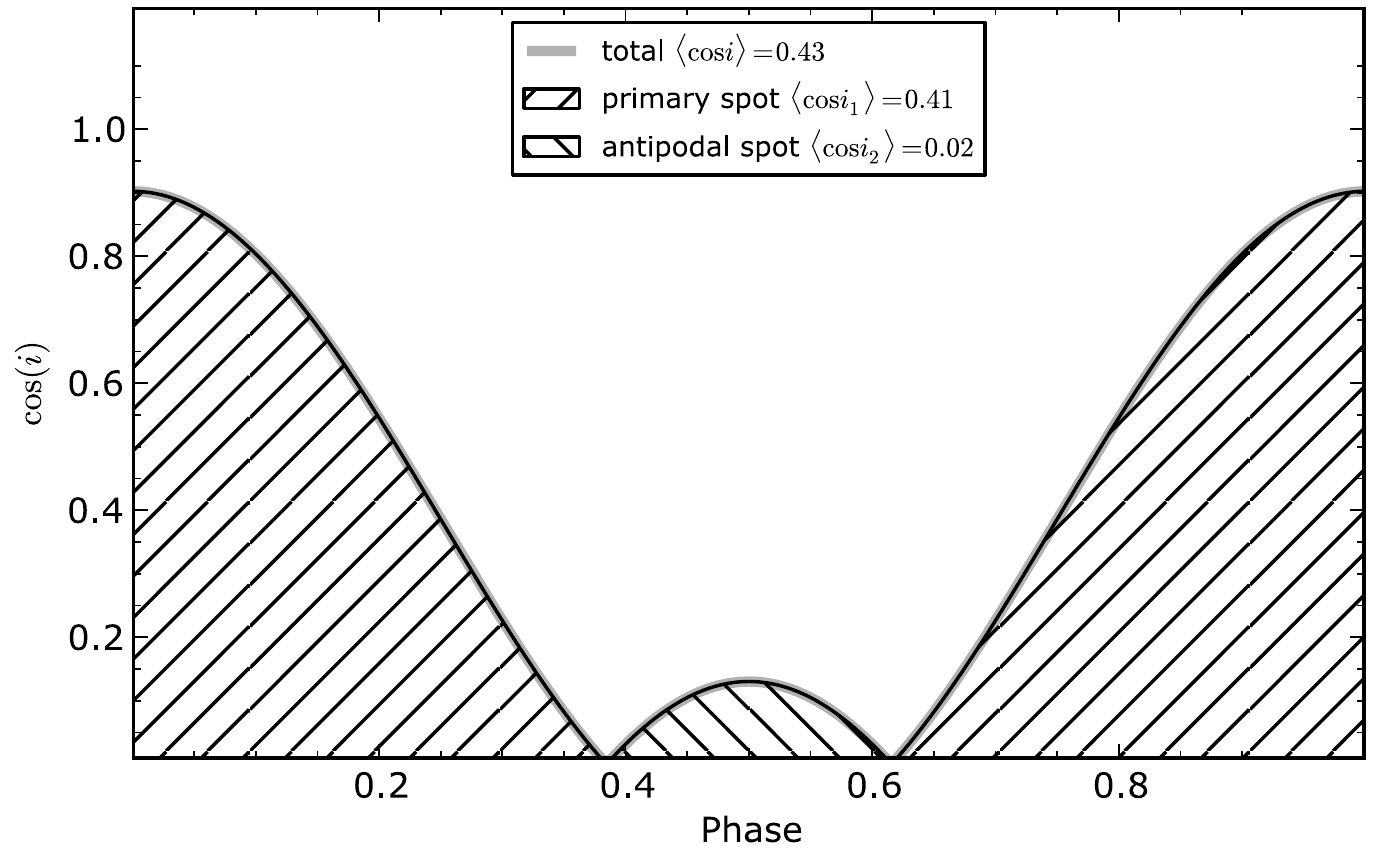}
\par\end{centering}

\caption[{Cosine of the hot spots' inclination angle {[}PSR B1929+10{]}}]{Cosine of the hot spots' inclination angle as a function of the pulsar
phase for PSR~B1929+10. The following parameters were used: $\alpha=35.97$,
$\beta=25.55$. For this geometry there is only a small contribution
from the antipodal spot.\label{fig:x-ray_B1929+10}}
\end{figure}

\begin{figure}[H]
\begin{centering}
\includegraphics[height=8.7cm]{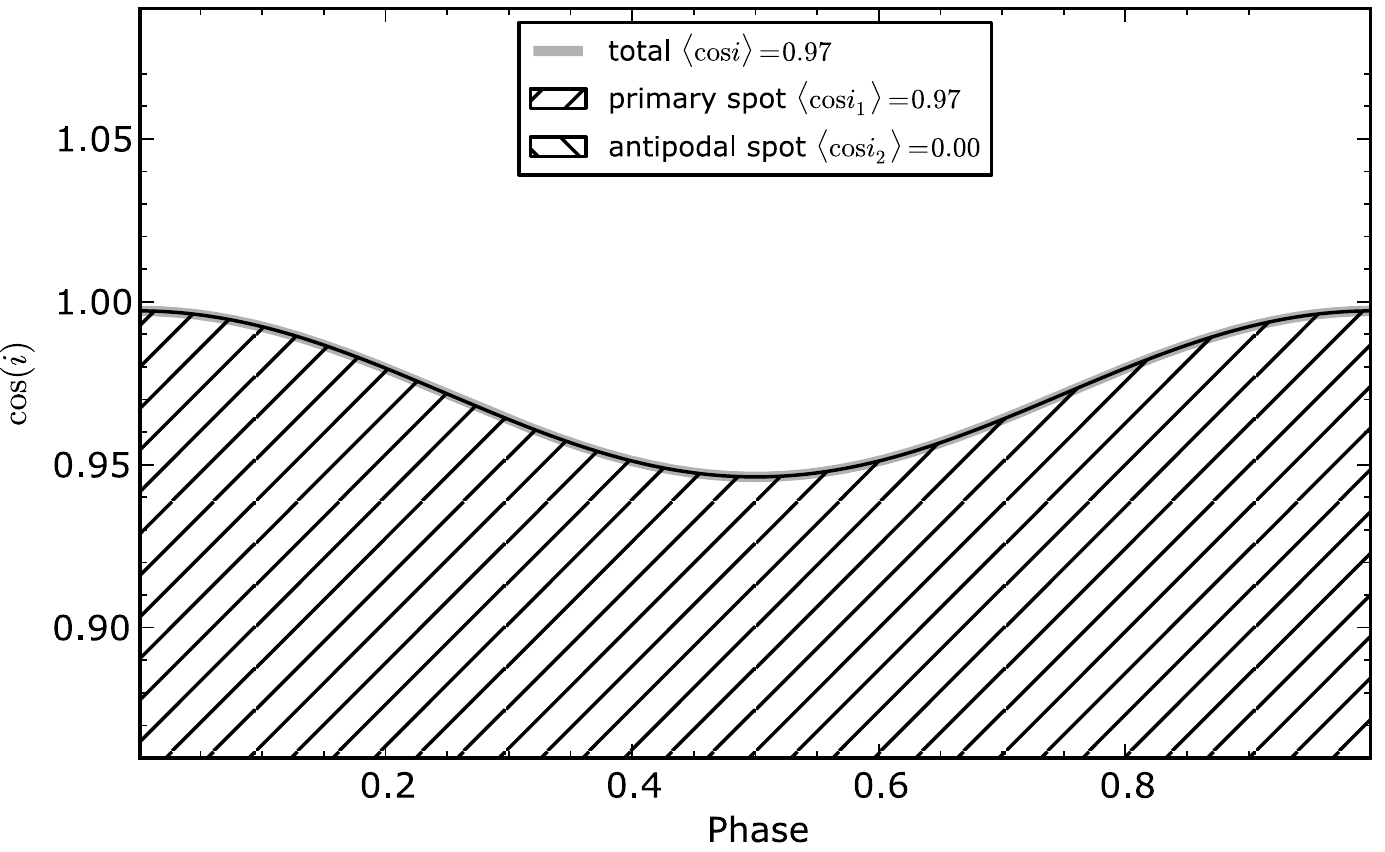}
\par\end{centering}

\caption[{Cosine of the hot spots' inclination angle {[}PSR B0943+10{]}}]{Cosine of the hot spots' inclination angle as a function of the pulsar
phase for PSR~B0943+10. The following parameters were used: $\alpha=11.58^{\circ}$,
$\beta=-4.29^{\circ}$. For this geometry only the primary hot spot
is visible.\label{fig:x-ray_B0943+10}}
\end{figure}

Depending on the orientation of ${\bf \Omega}$, ${\bf o}$ and $\boldsymbol{\hat{\mu}}$,
the thermal radiation may originate from: (1) both the primary and
antipodal hot spots (see Figure \ref{fig:x-ray_B0950+08}); (2) mainly
the primary spot but with a small contribution from the antipodal
spot (see Figure \ref{fig:x-ray_B1929+10}); (3) the primary spot
only (see Figure \ref{fig:x-ray_B0943+10}).

\begin{comment}
\textasciitilde{}/Documents/studies/phd/data/hot\_spot

\textasciitilde{}/Programs/studies/phd/hot\_spots/hot\_spots.pu (dim
14x8.7 cm)
\end{comment}

\subsection{Gravitational bending of light near stellar surface \label{sub:x-ray:gravitational}}

The radius of a neutron star is only a few times larger than the Schwarzschild
radius. The approach presented in the previous section does not include
the gravitational bending effect, which is very strong in neutron
stars. A strong gravitational field just above the stellar surface
causes the bending of light. A photon emitted near a neutron star
surface at an angle $\delta$ with respect to the radial direction
escapes to infinity at a different angle $\delta^{\prime}>\delta$.
As a consequence, even when the spot inclination angle to the line
of sight is $i\gtrsim90^{\circ}$ we can still observe thermal radiation
from this spot. For a Schwarzschild metric we can calculate an observed
flux fraction from the primary $f_{1}=F_{1}/F_{0}$ and antipodal
$f_{2}=F_{2}/F_{0}$ spots. Here $F_{0}$ is the maximum possible
flux that is observed when the primary spot is viewed face-on. The
primary and antipodal fluxes are given by \citep{2002_Beloborodov}

\begin{equation}
\begin{array}{ccc}
f_{1} & = & \cos\left(i\right)\left(1-\frac{r_{{\rm g}}}{R}\right)+\frac{r_{{\rm g}}}{R},\\
f_{2} & = & -\cos\left(i\right)\left(1-\frac{r_{{\rm g}}}{R}\right)+\frac{r_{{\rm g}}}{R},
\end{array}
\end{equation}
here $r_{{\rm g}}=2GM/c^{2}$ is the Schwarzschild radius. The primary
spot is visible when\linebreak{}
 $\cos i_{1}>-r_{{\rm g}}/\left(R-r_{{\rm g}}\right)$ and the antipodal
spot when $\cos i_{2}>-r_{{\rm g}}/\left(R-r_{{\rm g}}\right)$. Consequently,
both spots are seen when $-r_{{\rm g}}/\left(R-r_{{\rm g}}\right)<\cos i<r_{{\rm g}}/\left(R-r_{{\rm g}}\right)$,
and then the observed flux fraction is

\begin{equation}
f_{{\rm min}}=f_{1}+f_{2}=\frac{2r_{{\rm g}}}{R}.
\end{equation}

\begin{figure}[H]
\begin{centering}
\includegraphics[height=8.7cm]{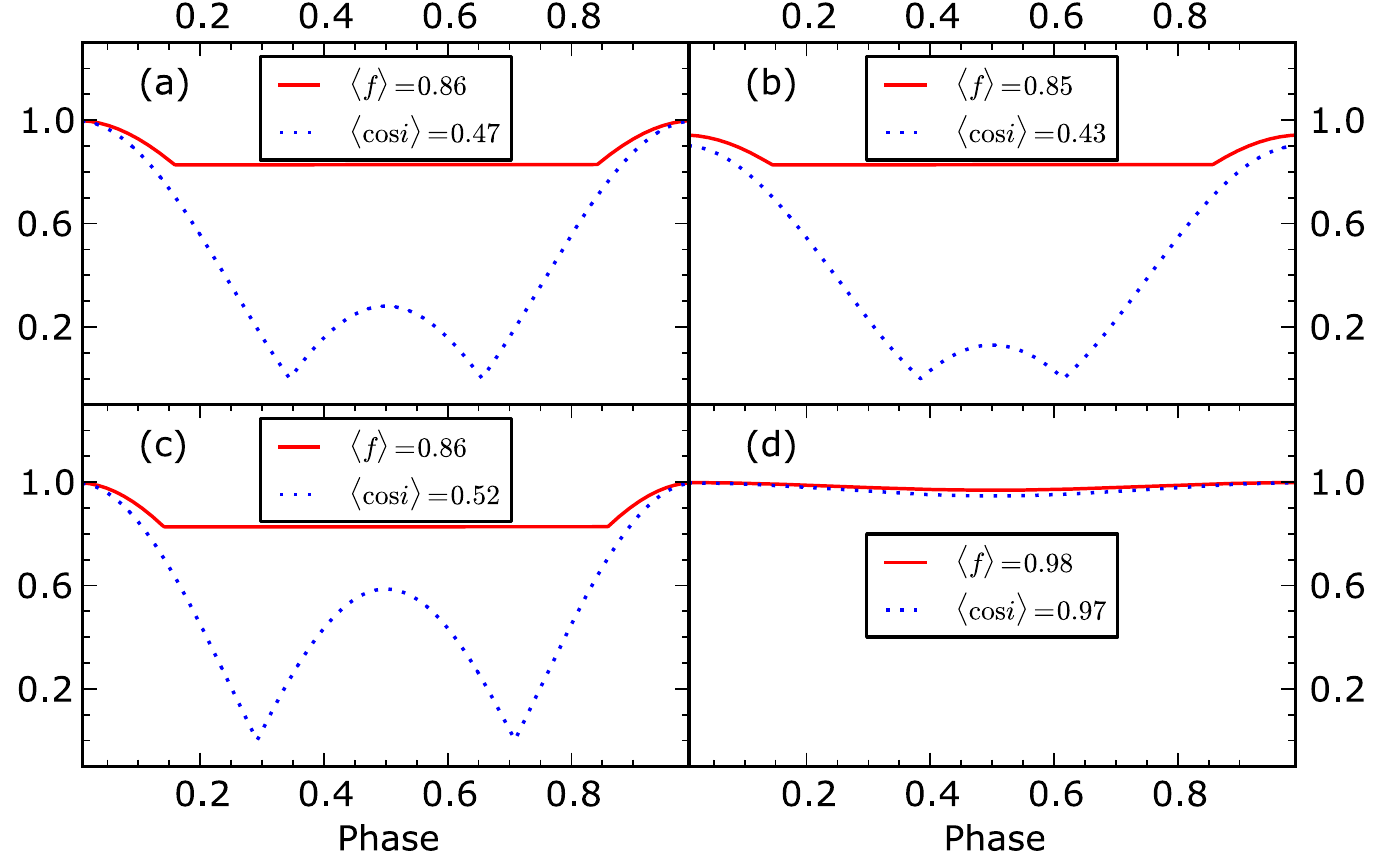}
\par\end{centering}

\caption[Comparison of the observed flux fractions for geometric effect only
and for geometric effect with the inclusion of a gravitational bending
of light]{Comparison of the observed flux fraction for geometric effect only
(blue dotted line) and for geometric effect with the inclusion of
a gravitational bending of light (red solid line). Individual panels
correspond to the following pulsars: (a) PSR B1133+16, (b) PSR B1929+10,
(c) PSR B0834+06 (d) PSR B0943+10. Parameters used in the calculations
are presented in Table \ref{tab:x-ray_inclination}. \label{fig:x-ray.PSRS_quad}}
\end{figure}

Hence, the blackbody pulse of primary and antipodal spots must display
a plateau whenever both spots are in sight. Depending on the geometry
of a pulsar we can distinguish four classes \citep{2002_Beloborodov}.
Class I: when the antipodal spot is never seen and the primary spot
is visible all the time (see the bottom right panel of Figure \ref{fig:x-ray.PSRS_quad}).
For such pulsars the blackbody pulse has a perfect sinusoidal shape.
Class II: when the primary spot is seen all the time and the antipodal
spot is also in the visible zone for some time (see panels a, b and
c of Figure \ref{fig:x-ray.PSRS_quad}). For these pulsars the sinusoidal
pulse shape is interrupted by the plateau. Class III: the primary
spot is not visible for a fraction of the period and during this time
only the antipodal spot is seen. The primary sinusoidal profile of
such pulsars is interrupted by the plateau, and the plateau is interrupted
by a weaker sinusoidal subpulse from the antipodal spot. Class IV:
both spots are seen at any time. The observed blackbody flux of such
pulsars is constant.

\begin{figure}[H]
\begin{centering}
\includegraphics[height=8.7cm]{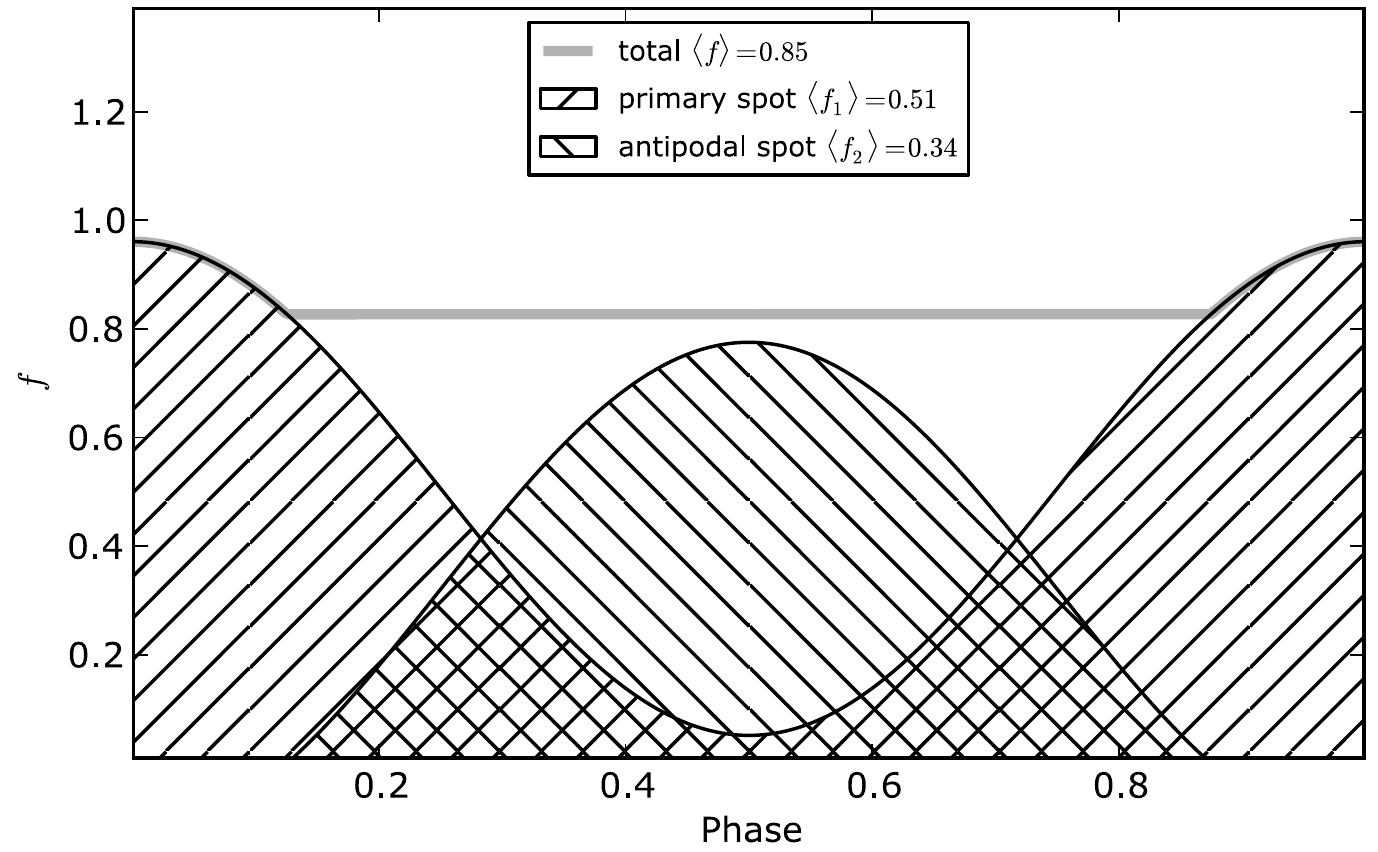}
\par\end{centering}

\caption[{Observed flux fraction as a function of the rotation phase {[}PSR
B0950+08{]}}]{Observed flux fraction $f$ as a function of the rotation phase for
PSR B0950+08. The following parameters were used: $\alpha=105.46^{\circ}$,
$\beta=21.1^{\circ}$, $M=1.4\,{\rm M}_{\odot}$, $R=10\,{\rm km}$.
The gravitational bending of light increases the flux ratio of the
antipodal to primary spots almost two times ($0.85/0.5=1.7$) and
also increases the antipodal to the primary flux ratio ($\sim1.3$).
\label{fig:x-ray_B0950+08_f}}
\end{figure}

\begin{figure}[H]
\begin{centering}
\includegraphics[height=8.7cm]{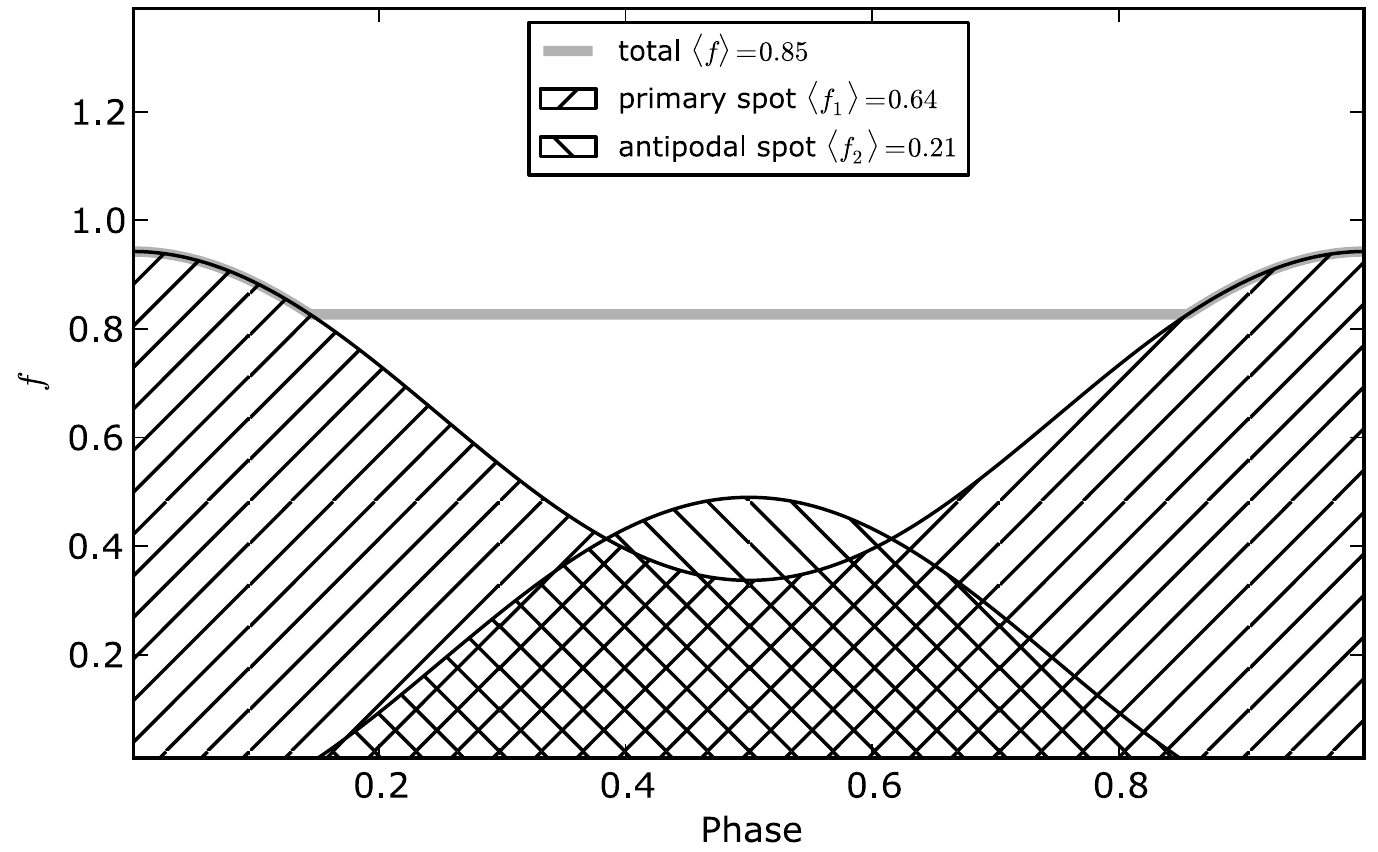}
\par\end{centering}

\caption[{Observed flux fraction as a function of the rotation phase {[}PSR
B1929+10{]}}]{Observed flux fraction $f$ as a function of the rotation phase for
PSR B1929+10. The following parameters were used: $\alpha=35.97$,
$\beta=25.55^{\circ}$, $M=1.4\,{\rm M}_{\odot}$, $R=10\,{\rm km}$.
The gravitational bending of light increases the observed flux fraction
two times ($\left\langle f\right\rangle /\left\langle \cos i\right\rangle =1.98$)
and also increases the flux ratio of the antipodal to primary spots
almost seven times ($\sim6.7$).\label{fig:x-ray_B1929+10_f}}
\end{figure}

The gravitational bending of light can significantly increase the
visibility of a pulsar (i.e. the observed flux, compare Figures \ref{fig:x-ray_B0950+08}
and \ref{fig:x-ray_B0950+08_f}). For some specific geometry the gravitational
effects can also drastically change primary to the antipodal flux
ratio (compare Figures \ref{fig:x-ray_B1929+10_f} and \ref{fig:x-ray_B1929+10}).
Our calculations show that for canonical values $M=1.4\,{\rm M}_{\odot}$
and $R=10\,{\rm km}$ the gravitational effect is quite strong and
the observed flux fraction is in the range of $0.85-1$, while the
geometric approach results in the $0.43-1$ range (see Table \ref{tab:x-ray_inclination}).

\begin{comment}
http://localhost:9090/pulsars/table\_inclination/ 

\textasciitilde{}/Html/pulsar/download/data/table\_inclination.tex
\end{comment}

\begin{table}[H]
\caption[Viewing geometry of pulsars]{Viewing geometry of pulsars. The individual columns are as follows:
(1) Pulsar name, (2) Inclination angle with respect to the rotation
axis $\alpha$, (3) Opening angle $\rho$, (4) Impact parameter $\beta$,
(5) Total flux correction factor (including gravitational bending
of light) $\left\langle f\right\rangle $ , (6) Flux correction factor
of the primary spot $\left\langle f_{1}\right\rangle $, (7) Flux
correction factor of the antipodal spot $\left\langle f_{2}\right\rangle $,
(8, 9, 10) Time-averaged cosine of the angle between the magnetic
axis and the line of sight: $\left<\cos i\right>$ (the total value),
$\left<\cos i_{1}\right>$(the primary spot), $\left<\cos i_{2}\right>$
(the antipodal spot), (10) Number of the pulsar. The gravitational
bending effect was calculated using $M=1.4\,{\rm M}_{\odot}$ and
$R=10\,{\rm km}$. \label{tab:x-ray_inclination} }

\centering{}%
\begin{tabular}{|c|c|c|c|c|c|c|c|c|c|c|}
\hline 
 &  &  &  &  &  &  &  &  &  & \tabularnewline
Name  & $\alpha$  & $\beta$  & $\rho$ & $\left\langle f\right\rangle $  & $\left\langle f_{1}\right\rangle $  & $\left\langle f_{2}\right\rangle $  & $\left<\cos i\right>$  & $\left<\cos i_{1}\right>$  & $\left<\cos i_{2}\right>$  & No. \tabularnewline
 & {\scriptsize $\left({\rm deg}\right)$}  & {\scriptsize $\left({\rm deg}\right)$}  & {\scriptsize $\left({\rm deg}\right)$}  &  &  &  &  &  &  & \tabularnewline
\hline 
\hline 
 &  &  &  &  &  &  &  &  &  & \tabularnewline
B0628--28  & $70.0$ & $-12.0$  & $19.6$ & $0.86$  & $0.52$  & $0.34$  & $0.52$  & $0.35$  & $0.17$  & 8 \tabularnewline
B0834+06  & $60.7$  & $4.5$  & $7.1$ & $0.86$  & $0.53$  & $0.32$  & $0.52$  & $0.36$  & $0.16$  & 14 \tabularnewline
B0943+10  & $11.6$  & $-4.3$  & $4.5$ & $0.98$  & $0.98$  & $0.00$  & $0.97$  & $0.97$  & $0.00$  & 15 \tabularnewline
B0950+08  & $105.4$  & $22.1$  & $25.6$ & $0.85$  & $0.51$  & $0.34$  & $0.50$  & $0.33$  & $0.17$  & 16 \tabularnewline
B1133+16  & $52.5$  & $4.5$  & $8.1$ & $0.86$  & $0.61$  & $0.25$  & $0.48$  & $0.40$  & $0.07$  & 22 \tabularnewline
 &  &  &  &  &  &  &  &  &  & \tabularnewline
B1451--68  & $37.0$  & $-6.0$  & -- & $0.88$  & $0.81$  & $0.06$  & $0.68$  & $0.68$  & $0.00$  & 27 \tabularnewline
B1929+10  & $36.0$  & $25.6$  & $26.8$ & $0.85$  & $0.64$  & $0.21$  & $0.43$  & $0.41$  & $0.02$  & 43 \tabularnewline
 &  &  &  &  &  &  &  &  &  & \tabularnewline
\hline 
\end{tabular}
\end{table}

\subsection{Observations \label{sub:x-ray.thermal_observations}}

As we have shown in the previous sections, the blackbody fit to the
X-ray observations allows us to directly obtain the surface temperature
$T_{{\rm s}}$. Using the distance to pulsar $D$ and the luminosity
of thermal emission $L_{{\rm BB}}$ we can estimate the area of spot
$A_{{\rm bb}}$. In most cases, $A_{{\rm bb}}$ differs from the conventional
polar cap area $A_{{\rm dp}}\approx6.2\times10^{4}P^{-1}\,{\rm m^{2}}$.
We use parameter $b=A_{{\rm dp}}/A_{{\rm bb}}$ to describe the difference
between $A_{{\rm dp}}$ and $A_{{\rm bb}}$.

\subsubsection*{Entire surface radiation and warm spot component (b<1) \label{sec:x-ray.blt}}

In most cases the observed spot area $A_{{\rm bb}}$ is larger than
the conventional polar cap area (see Table \ref{tab:x-ray_thermal}).
We can distinguish two types of pulsars in this group, with $b\ll1$
and $b\lesssim1$.

The first type is associated with observations of a thermal emission
from the entire stellar surface and can be used to test cooling models.
Although the entire surface radiation is strongest for young pulsars
($\tau\lesssim10$ kyr ), observation of this radiation is very difficult
due to the strong nonthermal component. A common practice is to separately
fit the nonthermal (PL) and thermal (BB) components. However, the
temperature obtained in such a BB fit (without the PL component) is
most likely overestimated (e.g. see PSR~J2021+3651 in Table \ref{tab:x-ray_thermal}).
The nonthermal luminosity of an aging neutron star decreases proportionally
to its spin-down luminosity $L_{{\rm SD}}$, which is thought to drop
with the star age as $L_{{\rm SD}}\propto\tau^{-m}$, where $m\simeq2-4$
depends on the pulsar dipole breaking index \citep{2007_Zavlin_b}.
As a pulsar becomes older, its surface temperature decreases. Depending
on the model, a predicted temperature decrease in the early stages
is gradual (the standard model) or rapid (the accelerated cooling
scenario). For a number of middle-aged ($\tau\sim100\,{\rm kyr}$)
and some younger ($\tau\sim10\,{\rm kyr}$) pulsars the thermal radiation
from the entire stellar surface dominates the radiation at soft X-ray
energies (e.g. PSR J0633+1746, PSR B1055-52, PSR J0821-4300, PSR B0656+14,
PSR J0205+6449, PSR J2021+3651). However, the sample of pulsars is
not sufficient to unambiguously identify the cooling scenario.

\begin{figure}[H]
\begin{centering}
\includegraphics[height=8cm]{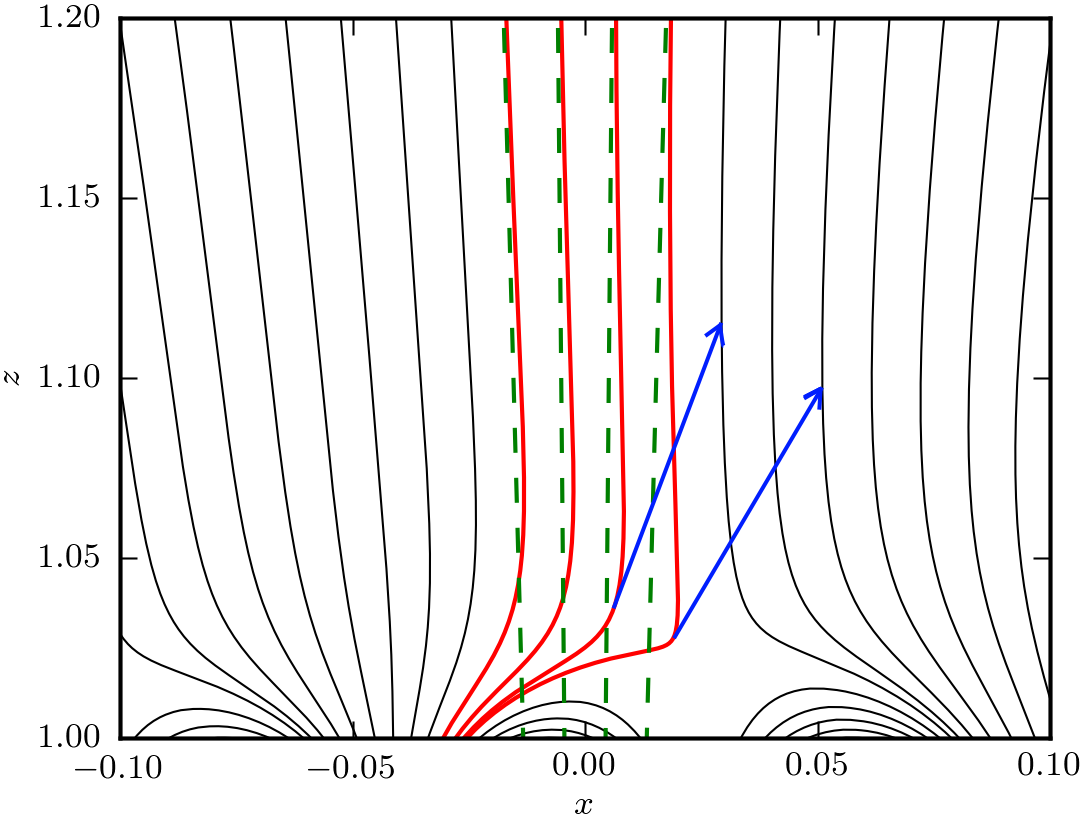}
\par\end{centering}

\centering{}\caption[Cartoon of the magnetic field lines in the polar cap region]{Cartoon of the magnetic field lines in the polar cap region. Red
lines are open field lines and green dashed lines correspond to the
dipole field. The blue arrows show the direction of the curvature
photon emission.\label{fig:x-ray.non_dipolar_heat} }
\end{figure}

The second type is associated with observations of the warm spot area
that is larger than the conventional polar cap area but still significantly
less than the area of the star ($b\lesssim1$). The age of pulsars
in this group varies from very young ($\tau\sim1\,{\rm kyr}$) to
middle-aged ($\tau\sim100\,{\rm kyr}$) neutron stars. There is one
exception, namely PSR J1210-5226, which is very old ($\tau=105\,{\rm Myr}$)
and can still be classified as a pulsar with the large warm spot component.
Note, however, that the age of this pulsar is estimated using a characteristic
value and if the pulsar period at birth is comparable with the current
period then the age is highly overestimated (see, e.g. PSR J0821-4300).
Furthermore, the fit to the X-ray spectrum was performed using only
one thermal component and assuming no nonthermal radiation (PL). We
believe that in many cases the size of the warm spot component and
its temperature are overestimated by neglecting other sources of X-ray
radiation, i.e. the nonthermal component and the hot spot radiation.
The small number of observed X-ray photons in some cases prevents
a full spectrum fit with all thermal and nonthermal components. Therefore,
we need observations with better statistics so that the spectrum fit
can be extended using more spectral components. The non-dipolar structure
of the surface magnetic field may cause significant deviations from
the spherical symmetry of the transport processes in the crust. The
magnetic field slightly enhances heat transport along the magnetic
lines, but strongly suppresses it in the perpendicular direction \citep{1983_Greenstein}.
Hence, the non-isothermality of the crust strongly depends on the
geometry of the magnetic field \citep{2004_Geppert}. The drastic
difference of the crustal transport process causes significant differences
in the surface temperature distribution \citep{2006_Page}. Thus,
the non-dipolar structure of the surface magnetic field can explain
the existence of large warm spot components for young and middle-aged
pulsars. We also suggested a mechanism of heating the surface adjacent
to the polar cap \citep{2011_Szary}. The model of such heating is
also based on the assumption that the pulsar magnetic field near the
stellar surface differs significantly from the pure dipole one. The
calculations show that it is natural to obtain such a geometry of
the magnetic field lines that allows pair creation in the closed field
line region (see Figure \ref{fig:x-ray.non_dipolar_heat}). 

The pairs move along the closed magnetic field lines and heat the
surface beyond the polar cap on the opposite side of the star. In
such a scenario the heating energy is generated in IAR, and hence
the luminosity of such a warm spot is limited by the power of the
outflowing particles (for more details see Section \ref{sec:radiation.thermal_reheating}).
In most cases the large size of the emitting area and its high temperature
make it unlikely that the warm spot is related to the particles accelerated
in IAR and is rather connected with the non-isothermality of the crust
(e.g. PSR J1210-5226, PSR J1119-6127).

\subsubsection*{The hot spot component (b > 1)\label{sec:x-ray.bgt1}}

In many cases the observed hot spot area $A_{{\rm bb}}$ is less than
the conventional polar cap area ($b>1$). The temperature of the emitting
area of these pulsars is usually higher than the temperature of the
emitting area of pulsars with a warm spot component ($b<1$). The
hot spot component is a natural consequence of the non-dipolar structure
of the surface magnetic field (see Figure \ref{fig:x-ray.non_dipolar_heat}).
In order to define an actual polar cap we need to follow the open
field lines from the light cylinder up to the stellar surface by taking
into account the non-dipolar structure of the surface magnetic field
(see Figure \ref{fig:x-ray.non_dipolar_heat}), which can be estimated
by the magnetic flux conservation law as $b=A_{{\rm dp}}/A_{{\rm bb}}$
= $B_{{\rm s}}/B_{{\rm d}}$. Thus, if $b\gg1$ then $B_{{\rm s}}\gg B_{{\rm d}}$. 

\begin{figure}[H]
\begin{centering}
\includegraphics{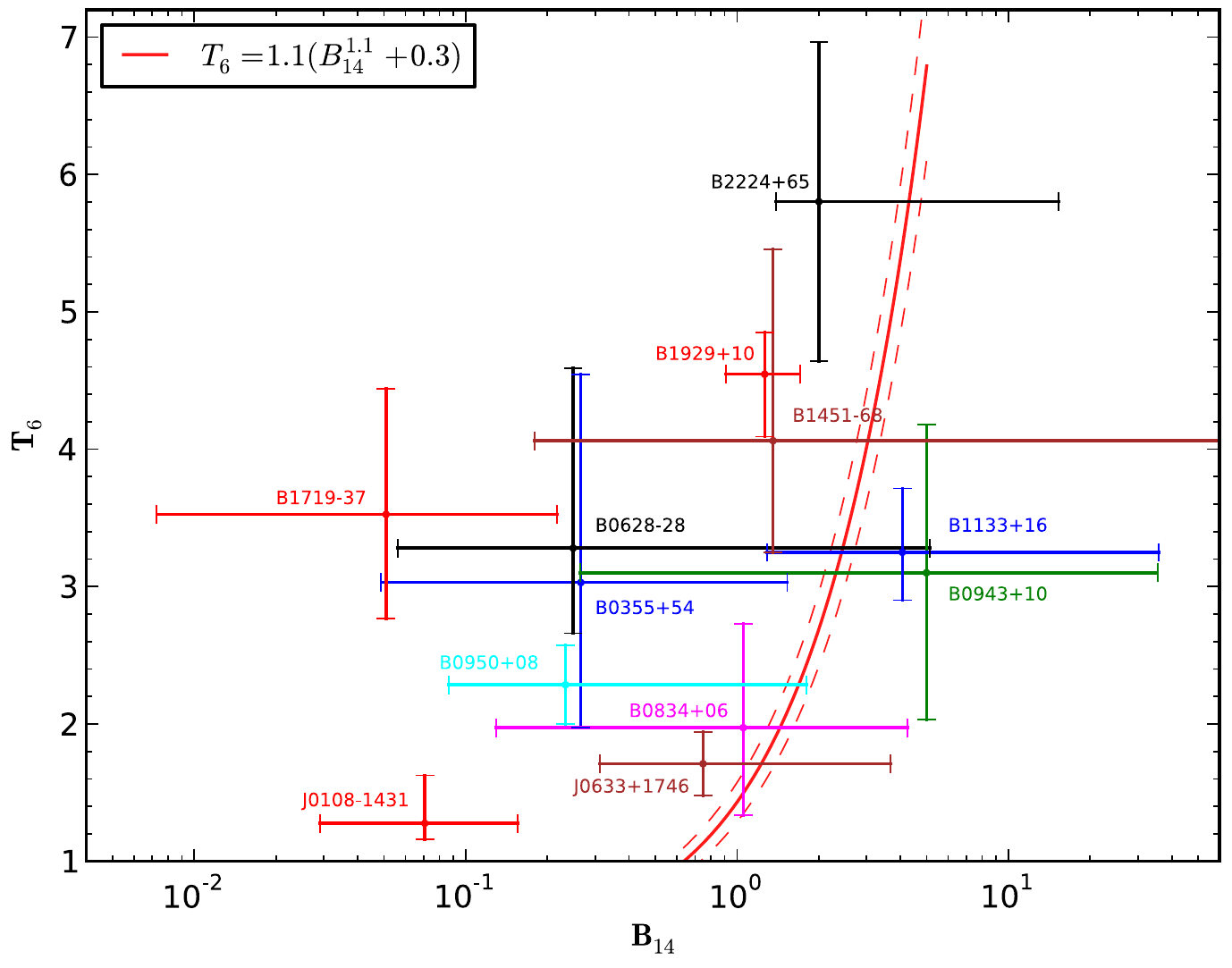}
\par\end{centering}

\centering{}\caption[Diagram of the surface temperature vs. the surface magnetic field]{Diagram of the surface temperature ($T_{6}=T_{{\rm s}}/\left(10^{6}\,{\rm K}\right)$)
vs. the surface magnetic field ($B_{14}=B_{{\rm s}}/\left(10^{14}\,{\rm G}\right)$).
The red line represents the dependence of $T_{{\rm crit}}$ on $B_{14}$
according to \citet{2008_Medin} and the dashed lines correspond to
uncertainties in the calculations. The diagram includes all pulsars
with $b>1$ with the exception of PSR J2043+2740, for which the blackbody
fit was performed using a fixed radius (estimation of the surface
magnetic field is not possible). Error bars correspond to $1\sigma$.\label{fig:x-ray.medin_lai}}
\end{figure}

In neutron stars with positively charged polar caps (${\bf \Omega}\cdot{\bf B}<0$),
the outflow of iron ions depends on the surface temperature and the
surface binding energy (the so-called cohesive energy) \citep{1980_Cheng,1986_Jones,1991_Abrahams,2003_Gil}.
The cohesive energy of condensed matter increases with magnetic field
strength \citep{2007_Medin}. If for a given strength of the surface
magnetic field the temperature is below the so-called critical temperature
$T_{{\rm crit}}$ the ions can tightly bind to the condensed surface
and a polar gap can form (see Chapter \ref{chap:psg} for details).
\citet{2008_Medin} calculated the dependence of the critical temperature
(for a vacuum gap formation) on the strength of the surface magnetic
field. In Figure \ref{fig:x-ray.medin_lai} we present the positions
of pulsars with derived surface temperature $T_{{\rm s}}$ and hot
spot area $A_{{\rm bb}}$ on the $B_{{\rm s}}-T_{{\rm s}}$ diagram,
where $B_{{\rm s}}$ is estimated as $B_{{\rm s}}=bB_{{\rm d}}$.
The red line represents the dependence of the critical temperature
$T_{{\rm crit}}$ on $B_{{\rm s}}$. We can see that in most cases
the pulsars' positions follow the $B_{{\rm s}}-T_{{\rm crit}}$ theoretical
curve. Note that the Figure includes only pulsars with a visible hot
spot component (old pulsars). For younger pulsars (with warm spot
components) it is not possible to estimate the surface magnetic field.
There are a few cases which do not coincide with the theoretical curve.
We believe that they correspond to the observations of warm spot component
but with the area of radiation smaller than the conventional polar
cap area (e.g. due to reheating of the surface beyond the polar cap,
see Section \ref{sec:x-ray.blt}).

According to our model the actual surface temperature is almost equal
to the critical value $T_{{\rm s}}\approx T_{{\rm crit}}$, which
leads to the formation of the Partially Screened Gap (PSG) above the
polar caps of a neutron star \citep{2003_Gil}. The hot spot parameters
derived from X-ray observations of isolated neutron stars are presented
in Table \ref{tab:x-ray_thermal}.

\begin{comment}
http://localhost:9090/pulsars/graphs/ {[}\textasciitilde{}/Html/pulsars/data/models.py
(t6\_b14\_log\_zoom){]}

cp \textasciitilde{}/Html/pulsar/media/images/t6\_b14\_log\_zoom.svg
\textasciitilde{}/Documents/studies/phd/images/x-ray/observations\_zoom.svg
\end{comment}

\begin{comment}
http://localhost:9090/pulsars/table\_bb\_age/ (PL instead of age in
table\_bb.tex)

\textasciitilde{}/Html/pulsar/download/data/table\_bb\_age.tex (import
in lyx and replace citet with citetalias)
\end{comment}

\begin{landscape} 
\begin{table}[H]
\thispagestyle{fancy} \caption[{Observed X-ray spectral properties of rotation-powered pulsars {[}thermal{]}}]{Spectral properties of rotation-powered pulsars with detected blackbody
X-ray components. The individual columns are as follows: (1) Pulsar
name, (2) Spectral components required to fit the observed spectra,
PL: power law, BB: blackbody, (3) Radius of the spot obtained from
the blackbody fit $R_{{\rm bb}}$, (4) Surface temperature $T_{{\rm s}}$,
(5) Surface magnetic field strength $B_{{\rm s}}$, (6) $b=A_{{\rm dp}}/A_{{\rm bb}}=B_{{\rm s}}/B_{{\rm d}}$,
$A_{{\rm dp}}$ - conventional polar cap area, $A_{{\rm bb}}$ - actual
polar cap area, (7) Bolometric luminosity of blackbody component $L_{{\rm BB}}$,
(8) Bolometric efficiency $\xi_{_{{\rm BB}}}$, (9) Maximum nonthermal
luminosity $L_{{\rm NT}}^{^{{\rm max}}}$, (10) Maximum nonthermal
X-ray efficiency $\xi_{_{{\rm NT}}}^{^{{\rm max}}}$, (11) Best estimate
of pulsar age or spin down age, (12) References, (13) Number of the
pulsar. Nonthermal luminosity and efficiency were calculated in the
$0.1-10\,{\rm keV}$ band. The maximum value was calculated with the
assumption that the X-ray nonthermal radiation is isotropic. Pulsars
are sorted by $b$ parameter (6). \label{tab:x-ray_thermal} }

\centering{}%
\begin{tabular}{|l|c|c|c|c|c|c|c|c|c|c|c|c|}
\hline 
 &  &  &  &  &  &  &  &  &  &  &  & \tabularnewline
Name  & Spectrum & $R_{{\rm bb}}$  & $T_{{\rm s}}$  & $B_{{\rm s}}$  & $b$  & $\log L_{{\rm BB}}$  & $\log\xi_{_{{\rm BB}}}$  & $\log L_{{\rm X}}$  & $\log\xi_{_{{\rm NT}}}^{^{{\rm max}}}$  & $\tau$  & Ref.  & No. \tabularnewline
 &  &  & {\scriptsize $\left(10^{6}{\rm K}\right)$}  & {\scriptsize $\left(10^{14}{\rm G}\right)$}  &  & {\scriptsize $\left({\rm erg\, s^{-1}}\right)$} &  & {\scriptsize $\left({\rm erg\, s^{-1}}\right)$}  &  &  &  & \tabularnewline
\hline 
\hline 
 &  &  &  &  &  &  &  &  &  &  &  & \tabularnewline
B1451--68  & {\scriptsize BB + PL}  & $14_{-12.3}^{+24.2}$ m  & $4.1_{-0.81}^{+1.39}$  & $1.36_{-1.18}^{+114}$  & $418$  & $29.27$  & $-3.06$  & $29.77$  & $-2.56$  & $42.5$ Myr  & \citetalias{2012_Posselt}  & 27 \tabularnewline
B0943+10  & {\scriptsize BB, PL}  & $12_{-7.7}^{+41.2}$ m  & $3.1_{-1.07}^{+1.08}$  & $4.99_{-4.72}^{+30.45}$  & $126$  & $28.40$  & $-3.62$  & $29.38$  & $-2.64$  & $4.98$ Myr  & \citetalias{2005_Zhang},\citetalias{2006_Kargaltsev}  & 15 \tabularnewline
B1929+10  & {\scriptsize BB + PL}  & $28_{-3.8}^{+4.9}$ m  & $4.5_{-0.45}^{+0.30}$  & $1.26_{-0.35}^{+0.44}$  & $122$  & $30.06$  & $-3.53$  & $30.23$  & $-3.36$  & $3.10$ Myr  & \citetalias{2008_Misanovic}  & 43 \tabularnewline
B1133+16  & {\scriptsize BB, PL}  & $14_{-9.0}^{+10.5}$ m  & $3.2_{-0.35}^{+0.46}$  & $4.06_{-2.77}^{+31.79}$  & $95.5$  & $28.56$  & $-3.38$  & $29.52$  & $-2.42$  & $5.04$ Myr  & \citetalias{2006_Kargaltsev}  & 22 \tabularnewline
B0950+08  & {\scriptsize BB + PL}  & $42_{-26.6}^{+26.6}$ m  & $2.3_{-0.29}^{+0.29}$  & $0.23_{-0.15}^{+1.57}$  & $47.9$  & $28.92$  & $-3.82$  & $29.99$  & $-2.76$  & $17.5$ Myr  & \citetalias{2004_Zavlin}  & 16 \tabularnewline
 &  &  &  &  &  &  &  &  &  &  &  & \tabularnewline
B2224+65  & {\scriptsize PL, BB}  & $28_{-18.0}^{+5.6}$ m  & $5.8_{-1.16}^{+1.16}$  & $2.00_{-0.61}^{+13.31}$  & $38.6$  & $30.51$  & $-2.57$  & $31.21$  & $-1.87$  & $1.12$ Myr  & \citetalias{2012_Hui}, \citetalias{2007_Hui_b}  & 47 \tabularnewline
J0633+1746  & {\scriptsize BB+BB+PL}  & $62_{-34.0}^{+34.0}$ m  & $1.7_{-0.23}^{+0.23}$  & $0.75_{-0.44}^{+2.92}$  & $23.0$  & $29.07$  & $-5.44$  & $30.24$  & $-4.27$  & $342$ kyr  & \citetalias{2005_Jackson}  & 9 \tabularnewline
------ &  & $11.17_{-1}^{+1}$ km  & $0.5_{-0.1}^{+0.1}$  &  &  & $31.67$  & $-2.84$  &  &  &  &  & \tabularnewline
B0834+06  & {\scriptsize BB + PL}  & $30_{-15.3}^{+56.4}$ m  & $2.0_{-0.64}^{+0.75}$  & $1.05_{-0.92}^{+3.19}$  & $17.7$  & $28.70$  & $-3.41$  & $28.70$  & $-3.41$  & $2.97$ Myr  & \citetalias{2008_Gil}  & 14 \tabularnewline
B0355+54  & {\scriptsize BB + PL}  & $92_{-53.6}^{+122.5}$ m  & $3.0_{-1.06}^{+1.51}$  & $0.27_{-0.22}^{+1.27}$  & $15.9$  & $30.40$  & $-4.25$  & $30.92$  & $-3.73$  & $564$ kyr  & \citetalias{2007_McGowan},\citetalias{1994_Slane}  & 3 \tabularnewline
J0108--1431  & {\scriptsize BB + PL}  & $43_{-14.0}^{+24.0}$ m  & $1.3_{-0.12}^{+0.35}$  & $0.07_{-0.04}^{+0.08}$  & $14.0$  & $27.94$  & $-2.82$  & $28.57$  & $-2.19$  & $166$ Myr  & \citetalias{2012_Posselt}, \citetalias{2009_Pavlov}  & 1 \tabularnewline
 &  &  &  &  &  &  &  &  &  &  &  & \tabularnewline
\hline 
\multicolumn{13}{|r|}{\emph{Continued on next page}}\tabularnewline
\hline 
\end{tabular}
\end{table}

\end{landscape}

\begin{landscape} 
\begin{table}[H]
\thispagestyle{fancy} 

\centering{}Table \ref{tab:x-ray_thermal} - continued from previous
page %
\begin{tabular}{|l|c|c|c|c|c|c|c|c|c|c|c|c|}
\hline 
 &  &  &  &  &  &  &  &  &  &  &  & \tabularnewline
Name  & Spectrum & $R_{{\rm bb}}$  & $T_{{\rm s}}$  & $B_{{\rm s}}$  & $b$  & $\log L_{{\rm BB}}$  & $\log\xi_{_{{\rm BB}}}$  & $\log L_{{\rm X}}$  & $\log\xi_{_{{\rm NT}}}^{^{{\rm max}}}$  & $\tau$  & Ref.  & No. \tabularnewline
 &  &  & {\scriptsize $\left(10^{6}{\rm K}\right)$}  & {\scriptsize $\left(10^{14}{\rm G}\right)$}  &  & {\scriptsize $\left({\rm erg\, s^{-1}}\right)$} &  & {\scriptsize $\left({\rm erg\, s^{-1}}\right)$}  &  &  &  & \tabularnewline
\hline 
\hline 
 &  &  &  &  &  &  &  &  &  &  &  & \tabularnewline
B0628--28  & {\scriptsize BB + PL}  & $64_{-49.7}^{+70.3}$ m  & $3.3_{-0.62}^{+1.31}$  & $0.25_{-0.19}^{+4.88}$  & $4.14$  & $30.22$  & $-1.94$  & $30.22$  & $-1.94$  & $2.77$ Myr  & \citetalias{2005_Tepedelenl}, \citetalias{2005_Becker}  & 8 \tabularnewline
J2043+2740  & {\scriptsize BB + PL}  & $358_{-153.2}^{+153.2}$ m  & $1.9_{-0.45}^{+0.45}$  & $0.01_{-0.01}^{+0.02}$  & $1.70$  & $30.77$  & $-3.98$  & $31.41$  & $-3.34$  & $1.20$ Myr  & \citetalias{2004_Becker}  & 46 \tabularnewline
B1719--37  & {\scriptsize BB}  & $237_{-122.5}^{+390.6}$ m  & $3.5_{-0.76}^{+0.91}$  & $0.05_{-0.04}^{+0.17}$  & $1.57$  & $31.19$  & $-3.32$  & --  & --  & $345$ kyr  & \citetalias{2004_Oosterbroek}  & 32 \tabularnewline
J1846--0258  & {\scriptsize BB + PL}  & $306_{-153.2}^{+153.2}$ m  & $13.6_{-3.03}^{+3.03}$  & --  & $0.686$  & $34.06$  & $-2.85$  & $35.13$  & $-1.78$  & $0.73$ kyr  & \citetalias{2008_Ng}, \citetalias{2003_Helfand}  & 39 \tabularnewline
B1055--52  & {\scriptsize BB+BB+PL}  & $460_{-60.0}^{+60.0}$ m  & $1.8_{-0.06}^{+0.06}$  & --  & $0.503$  & $30.89$  & $-3.59$  & $30.91$  & $-3.57$  & $535$ kyr  & \citetalias{2005_Deluca}  & 18 \tabularnewline
 &  & $12.30_{-1}^{+2}$ km  & $0.8_{-0.03}^{+0.03}$  &  &  & $32.62$  & $-1.86$  &  &  &  &  & \tabularnewline
 &  &  &  &  &  &  &  &  &  &  &  & \tabularnewline
J0538+2817  & {\scriptsize BB}  & $666_{-38.3}^{+38.3}$ m  & $2.8_{-0.04}^{+0.05}$  & --  & $0.330$  & $31.97$  & $-2.73$  & --  & --  & $30.0$ kyr  & \citetalias{2003_Mcgowan}  & 6 \tabularnewline
J1809--1917  & {\scriptsize BB + PL}  & $951_{-693.3}^{+920.2}$ m  & $2.0_{-0.35}^{+0.35}$  & --  & $0.280$  & $31.69$  & $-4.56$  & $31.57$  & $-4.68$  & $51.3$ kyr  & \citetalias{2007_Kargaltsev}  & 36 \tabularnewline
J0821--4300  & {\scriptsize BB + BB}  & $1.22_{-0.13}^{+0.13}$ km  & $6.3_{-0.19}^{+0.19}$  & --  & $0.125$  & $33.61$  & $-0.91$  & --  & --  & $3.70$ kyr  & \citetalias{2010_Gotthelf}  & 11 \tabularnewline
------ &  & $6.02_{-0.4}^{+0.4}$ km  & $3.2_{-0.10}^{+0.10}$  &  &  & $33.86$  & $-0.66$  &  &  &  &  & \tabularnewline
B1951+32  & {\scriptsize BB + PL}  & $2.20_{-0.80}^{+1.40}$ km  & $1.5_{-0.23}^{+0.23}$  & --  & $0.110$  & $31.95$  & $-4.62$  & $33.22$  & $-3.35$  & $107$ kyr  & \citetalias{2005_Li}  & 44 \tabularnewline
B0833--45  & {\scriptsize BB + PL}  & $1.61_{-0.15}^{+0.15}$ km  & $1.9_{-0.05}^{+0.05}$  & --  & $0.091$  & $32.12$  & $-4.72$  & $32.62$  & $-4.22$  & $11.3$ kyr  & \citetalias{2007_Zavlin_b}  & 13 \tabularnewline
 &  &  &  &  &  &  &  &  &  &  &  & \tabularnewline
J1357--6429  & {\scriptsize BB + PL}  & $1.91_{-0.38}^{+0.38}$ km  & $2.2_{-0.26}^{+0.26}$  & --  & $0.034$  & $32.50$  & $-3.99$  & $32.15$  & $-4.35$  & $7.31$ kyr  & \citetalias{2007_Zavlin}  & 25 \tabularnewline
J1210--5226  & {\scriptsize BB}  & $1.23$ km  & $3.8$  & --  & $0.033$  & $33.04$  & $1.51$  & --  & --  & $102$ Myr  & \citetalias{2002_Pavlov}  & 23 \tabularnewline
B1823--13  & {\scriptsize BB + PL}  & $2.52_{-0.00}^{+0.00}$ km  & $1.6_{-0.07}^{+0.10}$  & --  & $0.032$  & $32.19$  & $-4.27$  & $31.78$  & $-4.67$  & $21.4$ kyr  & \citetalias{2008_Pavlov}  & 38 \tabularnewline
B1916+14  & {\scriptsize BB, PL}  & $800_{-100.0}^{+100.0}$ m  & $1.5_{-0.12}^{+0.12}$  & --  & $0.028$  & $31.07$  & $-2.63$  & $32.00$  & $-1.71$  & $88.1$ kyr  & \citetalias{2009_Zhu}  & 41 \tabularnewline
B1706--44  & {\scriptsize BB + PL}  & $2.76_{-0.69}^{+0.69}$ km  & $2.2_{-0.20}^{+0.22}$  & --  & $0.027$  & $32.78$  & $-3.76$  & $32.16$  & $-4.37$  & $17.5$ kyr  & \citetalias{2002_Gotthelf}  & 31 \tabularnewline
 &  &  &  &  &  &  &  &  &  &  &  & \tabularnewline
\hline 
\end{tabular}
\end{table}

\end{landscape}

\begin{landscape} 
\begin{table}[H]
\thispagestyle{fancy} 

\centering{}Table \ref{tab:x-ray_thermal} - continued from previous
page %
\begin{tabular}{|l|c|c|c|c|c|c|c|c|c|c|c|c|}
\hline 
 &  &  &  &  &  &  &  &  &  &  &  & \tabularnewline
Name  & Spectrum & $R_{{\rm bb}}$  & $T_{{\rm s}}$  & $B_{{\rm s}}$  & $b$  & $\log L_{{\rm BB}}$  & $\log\xi_{_{{\rm BB}}}$  & $\log L_{{\rm X}}$  & $\log\xi_{_{{\rm NT}}}^{^{{\rm max}}}$  & $\tau$  & Ref.  & No. \tabularnewline
 &  &  & {\scriptsize $\left(10^{6}{\rm K}\right)$}  & {\scriptsize $\left(10^{14}{\rm G}\right)$}  &  & {\scriptsize $\left({\rm erg\, s^{-1}}\right)$} &  & {\scriptsize $\left({\rm erg\, s^{-1}}\right)$}  &  &  &  & \tabularnewline
\hline 
\hline 
 &  &  &  &  &  &  &  &  &  &  &  & \tabularnewline
B2334+61  & {\scriptsize BB + PL}  & $1.27_{-0.30}^{+0.45}$ km  & $2.1_{-0.76}^{+0.46}$  & --  & $0.026$  & $32.06$  & $-2.73$  & $31.55$  & $-3.24$  & $40.9$ kyr  & \citetalias{2006_Mcgowan}  & 48 \tabularnewline
B0656+14  & {\scriptsize BB+BB+PL}  & $1.80_{-0.15}^{+0.15}$ km  & $1.2_{-0.03}^{+0.03}$  & --  & $0.017$  & $31.45$  & $-3.13$  & $30.26$  & $-4.33$  & $111$ kyr  & \citetalias{2005_Deluca}  & 10 \tabularnewline
------ &  & $20.90_{-4}^{+3}$ km  & $0.7_{-0.01}^{+0.01}$  &  &  & $32.74$  & $-1.84$  &  &  &  &  & \tabularnewline
J1119--6127  & {\scriptsize BB + PL}  & $2.60_{-0.23}^{+1.38}$ km  & $3.1_{-0.26}^{+0.39}$  & --  & $0.008$  & $33.37$  & $-3.00$  & $32.95$  & $-3.42$  & $1.61$ kyr  & \citetalias{2007_Gonzalez}, \citetalias{2012_Ng} & 20 \tabularnewline
J0205+6449  & {\scriptsize BB + PL}  & $8.1$ km  & $1.7$  & --  & $0.005$  & $33.60$  & $-3.83$  & $33.10$  & $-4.33$  & $5.37$ kyr  & \citetalias{2004_Slane}  & 2 \tabularnewline
J2021+3651  & {\scriptsize PL, BB}  & $7.00_{-1.70}^{+4.00}$ km  & $2.4_{-0.30}^{+0.30}$  & --  & $0.004$  & $33.78$  & $-2.75$  & $34.36$  & $-2.17$  & $17.2$ kyr  & \citetalias{2008_VanEtten},\citetalias{2004_Hessels}  & 45 \tabularnewline
 &  &  &  &  &  &  &  &  &  &  &  & \tabularnewline
\hline 
\end{tabular}
\end{table}

\end{landscape}

\chapter{Model of a non-dipolar surface magnetic field \label{chap:model}}

\thispagestyle{headings}

\section{The magnetic field of neutron stars}

Generally, the properties of pulsar radio emission support the assumption
that the magnetic field of pulsars is purely dipolar at least in the
radio emission region \citep{1969_Radhakrishnan}. However, radio
emission is generated at altitudes $R_{{\rm em}}$ of more than several
stellar radii (e.g. \citet{1997_Kijak}, \citet{1998_Kijak}, \citet{2009_Krzeszowski}
and references therein). Thus, radio observations do not provide information
about the structure of the magnetic field at the surface of the neutron
star. On the other hand, strong non-dipolar surface magnetic fields
have long been thought to be a necessary condition for pulsar activities,
e.g. the vacuum gap model proposed by \citet{1975_Ruderman} implicitly
assumes that the radius of curvature of field lines above the polar
cap should be about $10^{6}\,{\rm cm}$ in order to sustain pair production.
This curvature is approximately $100$ times higher than that expected
from a global dipolar magnetic field. Furthermore, to explain radiation
from the Crab Nebula, the Crab pulsar should provide quite a dense
stellar wind, as such a high particle multiplicity is not possible
in a purely dipolar magnetic field.

There are several theoretical studies concerning the formation and
evolution of the non-dipolar magnetic fields of neutron stars (e.g.
\citealt{1983_Blandford,1991_Krolik,1991_Ruderman,1993_Arons,1993_Chen,1994_Geppert,1999_Mitra,2006_Page}).
According to \citet{1964_Woltjer}, the magnetic field in neutron
stars results from the fossil field of the progenitor stars which
is amplified during the collapse and remains anchored in the superfluid
core of the neutron star. Several authors also noted that during the
collapse (or shortly after) there is possible magnetic field generation
in the external crust, for instance, by a mechanism like thermomagnetic
instabilities \citep{1983_Blandford}. \citet{1986_Urpin} also showed
that it is possible to form small-scale magnetic field anomalies in
the neutron star crust with a typical size of the order of $100$
meters.

The soft X-ray observations of pulsars presented in Chapter \ref{chap:x-ray_emission}
show non-uniform surface temperatures which can be attributed to small-scale
magnetic anomalies in the crust. Further observational arguments in
favour of the non-dipolar nature of the surface magnetic field can
be found in many articles (e.g. \citealt{1992_Bulik,1995_Thompson,1995_Bulik,1996_Page,1996_Thompson,1997_Becker,1998_Cheng,1999_Rudak,1999_Cheng,1999_Murakami,2001_Tauris,2012_Maciesiak}).

\section{Modelling of the surface magnetic field}

\begin{figure}[H]
\begin{centering}
\includegraphics[width=7cm]{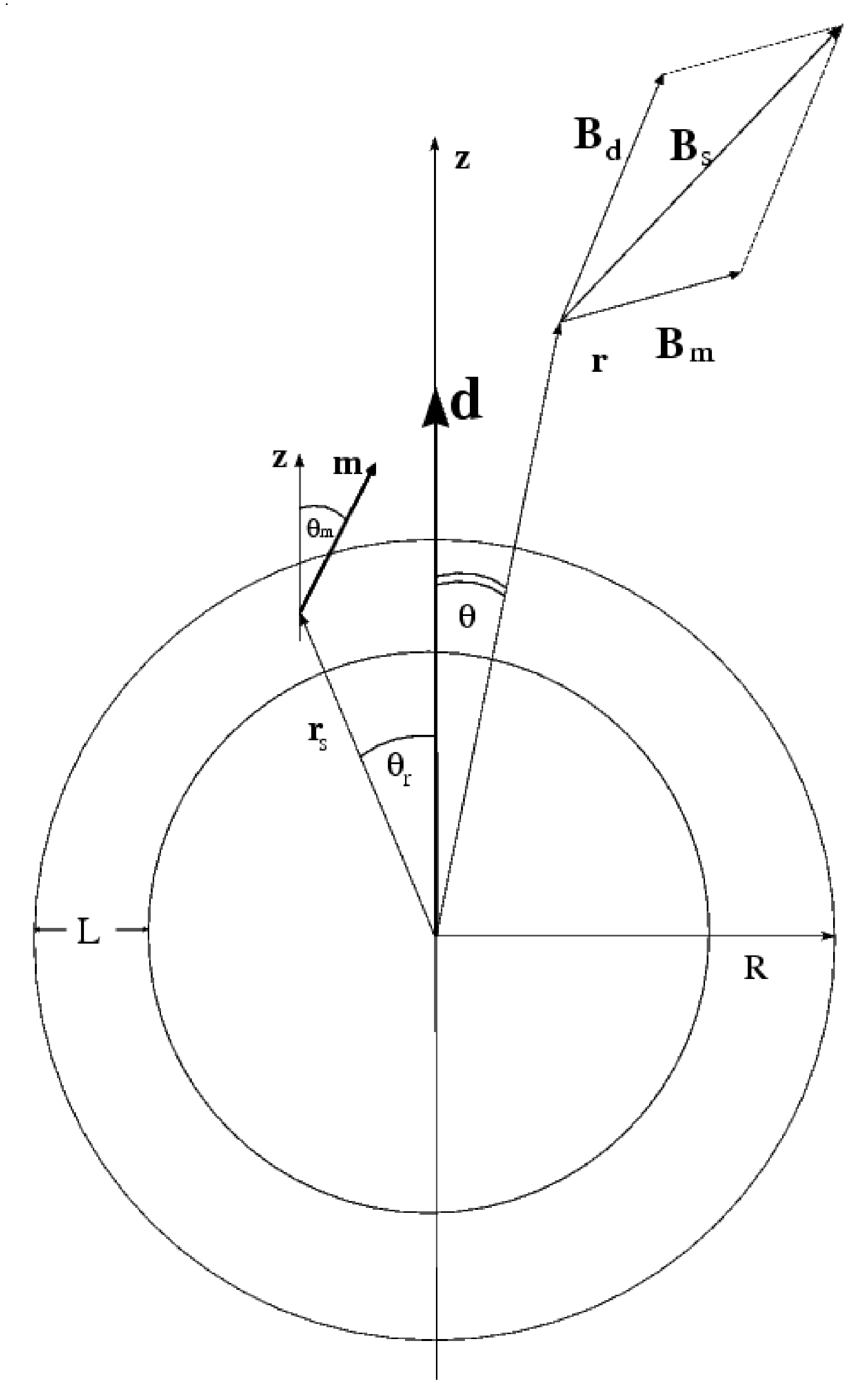}
\par\end{centering}

\centering{}\caption[Model of a non-dipolar surface magnetic field]{Superposition of the star-centred global magnetic dipole $\mathbf{d}$
and crust-anchored local dipole anomaly $\mathbf{m}$ located at $\mathbf{r_{s}}=(r_{s}\sim R,\,\theta=\theta_{r})$
and inclined to the $z$-axis by an angle $\theta_{m}$. The actual
surface magnetic field at radius vector $\mathbf{r}=(r,\,\theta)$
is $\mathbf{B_{s}}=\mathbf{B_{d}}+\mathbf{B_{{\rm m}}}$, where $B_{d}=2d/r^{3}$,
$B_{{\rm m}}=2m/|\mathbf{r}\mathbf{r_{s}}|^{3}$, $r$ is the radius
and $\theta$ - is the polar angle. $R$ is the radius of the neutron
star and L is the external crust thickness. \citet{2002_Gil}\label{fig:model.magnetic_field} }
\end{figure}

In order to model a surface magnetic field we used the scenario proposed
by \citet{2002_Gil}. In this scenario the magnetic field at the neutron
star's surface is non-dipolar in nature, which is due to superposition
of the fossil field in the core and crustal field structures. To calculate
the actual surface magnetic field described by superposition of the
star-centred global dipole $\mathbf{d}$ and the crust-anchored dipole
moment $\mathbf{m}$, let us consider the general situation presented
in Figure \ref{fig:model.magnetic_field}

The actual surface magnetic field is a sum of the global magnetic
dipole and crust-anchored local anomalies 
\begin{equation}
\mathbf{B_{s}}=\mathbf{B_{d}}+\mathbf{B_{{\rm m}}}+...\label{eq:model.field}
\end{equation}
Using the star-centred spherical coordinates with the $z$-axis directed
along the global magnetic dipole moment we obtain: 
\begin{equation}
\mathbf{B_{d}}=\left(\frac{2d\cos\theta}{r^{3}},\,\frac{d\sin\theta}{r^{3}},\,0\right),\label{eq:model.b_d}
\end{equation}

\begin{equation}
\mathbf{B_{{\rm m}}}=\frac{3(\mathbf{r}-\mathbf{r_{s}})(\mathbf{m}\cdot(\mathbf{r}-\mathbf{r_{s}}))-\mathbf{m}|\mathbf{r}-\mathbf{r_{s}}|^{2}}{|\mathbf{r}-\mathbf{r_{s}}|^{5}}.
\end{equation}
Here $\mathbf{r_{s}}=(r_{s},\,\theta_{r},\,\phi_{r})$, $\mathbf{m}=(m,\,\theta_{m},\,\phi_{m})$
and the spherical components of $\mathbf{B_{{\rm m}}}$ are explicitly
given in Equation \ref{eq:model.b_m}.

The global magnetic moment can be written as 
\begin{equation}
d=\frac{1}{2}B_{{\rm p}}R^{3},
\end{equation}
where $B_{{\rm p}}=6.4\times10^{19}\left(P\dot{P}\right){}^{1/2}\,{\rm G}$
is the dipole component at the pole derived from pulsar spin-down
energy loss.

The crust-anchored local dipole moment is 
\begin{equation}
m=\frac{1}{2}B_{{\rm m}}\Delta R^{3},
\end{equation}
where $\Delta R\sim0.05R<L$ and $L\sim10^{5}\,{\rm cm}$ is the characteristic
crust dimension (for\linebreak{}
 $R=10^{6}\,{\rm cm}$). For these values a local anomaly can significantly
influence the surface magnetic field ($B_{{\rm m}}>B_{{\rm d}}$)
if $m/d>10^{-4}$.

The system of differential equations for a field line of the vector
field $\mathbf{B=}\left(B_{r},\, B_{\theta},\, B_{\phi}\right)$ in
spherical coordinates can be written as 
\begin{equation}
\begin{cases}
\frac{{\rm d}\theta}{{\rm d}r} & =\frac{B_{\theta}}{rB_{r}}\\
\frac{{\rm d}\phi}{{\rm d}r} & =\frac{B_{\phi}}{r\sin(\theta)B_{r}}.
\end{cases}\label{eq:model.diff_eqs}
\end{equation}
The solution of these equations, with the initial conditions $\theta_{0}=\theta(r=R)$
and $\mbox{\ensuremath{\phi_{0}}=\ensuremath{\phi}(r=R)}$ determining
a given field line at the stellar surface, describes the parametric
equation of the magnetic field lines. The spherical components of
$\mathbf{B_{{\rm m}}}$ can be written in the following form

\begin{equation}
\begin{split}B_{r}^{m} & =-\frac{1}{D^{2.5}}\left(3Tr_{r}^{s}-3Tr+Dm_{r}\right),\\
B_{\theta}^{m} & =-\frac{1}{D^{2.5}}\left(3Tr_{\theta}^{s}+Dm_{\theta}\right),\\
B_{\phi}^{m} & =-\frac{1}{D^{2.5}}\left(3Tr_{\phi}^{s}+Dm_{\phi}\right).
\end{split}
\label{eq:model.b_m}
\end{equation}
Here 
\begin{equation}
D=r_{s}^{2}+r^{2}-2r_{s}r\left(\sin\theta_{r}\sin\theta\cos\left(\phi-\phi_{r}\right)+\cos\theta_{r}\cos\theta\right),
\end{equation}
and 
\begin{equation}
T=m_{r}r-\left(m_{r}r_{r}^{s}+m_{\theta}r_{\theta}^{s}\right).
\end{equation}

According to the geometry presented in Figure \ref{fig:model.magnetic_field},
the components of the radius vector of the origin of the crust-anchored
local dipole anomaly can be written as

\begin{equation}
\begin{split}r_{r}^{s} & =r_{s}\left(\sin\theta_{r}\sin\theta\cos\left(\phi-\phi_{r}\right)+\cos\theta_{r}\cos\theta\right),\\
r_{\theta}^{s} & =r_{s}\left(\sin\theta_{r}\cos\theta\cos\left(\phi-\phi_{r}\right)+\cos\theta_{r}\sin\theta\right),\\
r_{\phi}^{s} & -r_{s}\sin\theta_{r}\sin\left(\phi-\phi_{r}\right).
\end{split}
\end{equation}
The components of the local dipole anomaly are 
\begin{equation}
\begin{split}m_{r} & =m\left(\sin\theta_{m}\sin\theta\cos\left(\phi-\phi_{m}\right)+\cos\theta_{m}\cos\theta\right),\\
m_{\theta} & =m\left(\sin\theta_{m}\cos\theta\cos\left(\phi-\phi_{m}\right)+\cos\theta_{m}\sin\theta\right),\\
m_{\phi} & =-m\sin\theta_{m}\sin\left(\phi-\phi_{m}\right).
\end{split}
\end{equation}

Finally, we obtain the system of differential equations from Equation
\ref{eq:model.diff_eqs} by substitutions $B_{r}=B_{r}^{d}+B_{r}^{m}$,
$B_{\theta}=B_{\theta}^{d}+B_{\theta}^{m}$ and $B_{\phi}=B_{\phi}^{d}+B_{\phi}^{m}$
(Equations \ref{eq:model.b_d} and \ref{eq:model.b_m})

\begin{equation}
\frac{{\rm d}\theta}{{\rm d}r}=\frac{B_{\theta}^{d}+B_{\theta}^{m}}{r\left(B_{r}^{d}+B_{r}^{m}\right)}\equiv\Theta_{1},\label{model.line_diff}
\end{equation}

\begin{equation}
\frac{{\rm d}\phi}{{\rm d}r}=\frac{B_{\phi}^{m}}{r\left(B_{r}^{d}+B-r^{m}\right)\sin\theta}\equiv\Phi_{1}.\label{model.line_diff2}
\end{equation}

\section{Curvature of magnetic field lines \label{sec:model.curvature}}

As Curvature Radiation (CR) may play a decisive role in radiation
processes, it is important to calculate the curvature (or curvature
radius) for each field line. The curvature $\rho_{c}=1/\Re$ of field
lines (where $\Re$ is the radius of curvature) is calculated as \citep{2002_Gil}
\begin{equation}
\rho_{c}=\left(\frac{{\rm d}s}{{\rm d}r}\right)^{-3}\left|\left(\frac{{\rm d}^{2}\mathbf{r}}{{\rm d}r^{2}}\frac{{\rm d}s}{{\rm d}r}-\frac{{\rm d}\mathbf{r}}{{\rm d}r}\frac{{\rm d}^{2}s}{{\rm d}r^{2}}\right)\right|,
\end{equation}
where 
\begin{equation}
\frac{{\rm d}s}{{\rm d}r}=\sqrt{\left[1+r^{2}\Theta_{1}^{2}+r^{2}\Phi_{1}^{2}\sin^{2}(\theta)\right]}.
\end{equation}

Thus, the curvature can be written in the form 
\begin{equation}
\rho_{c}=\left(S_{1}\right)^{-3}\left(J_{1}^{2}+J_{2}^{2}+J_{3}^{2}\right)^{1/2},
\end{equation}
where 
\begin{equation}
\begin{split}J_{1}= & X_{2}S_{1}-X_{1}S_{2},\\
J_{2}= & Y_{2}S_{1}-Y_{1}S_{2},\\
J_{3}= & Z_{2}S_{1}-Z_{1}S_{2},\\
X_{1}= & \sin\theta\cos\phi+r\Theta_{1}\cos\theta\cos\phi-r\Phi_{1}\sin\theta\sin\phi,\\
Y_{1}= & \sin\theta\sin\phi+r\Theta_{1}\cos\theta\sin\phi-r\Phi_{1}\sin\theta\cos\phi,\\
Z_{1}= & \cos\theta-r\Theta_{1}\sin\theta,\\
X_{2}= & \left(2\Theta_{1}+r\Theta_{2}\right)\cos\theta\cos\phi-\left(2\Phi_{1}+r\Phi_{2}\right)\sin\theta\sin\phi-\\
 & r\left(\Theta_{1}^{2}+\Phi_{1}^{2}\right)\sin\theta\cos\phi+2r\Theta_{1}\Phi_{1}\cos\theta\sin\phi,\\
Y_{2}= & \left(2\Theta_{1}+r\Theta_{2}\right)\cos\theta\sin\phi+\left(2\Phi_{1}+r\Phi_{2}\right)\sin\theta\cos\phi-\\
 & r\left(\Theta_{1}^{2}+\Phi_{1}^{2}\right)\sin\theta\sin\phi+2r\Theta_{1}\Phi_{1}\cos\theta\cos\phi,\\
Z_{2}= & -\Theta_{1}\sin\theta-\Theta_{1}\sin\theta-r\Theta_{2}\sin\theta-r\Theta_{1}^{2}\cos\theta,\\
S_{1}= & \sqrt{1+r^{2}\Theta_{1}^{2}+r^{2}\Phi_{1}^{2}\sin^{2}\theta},\\
S_{2}= & S_{1}^{-1}\left(t\Theta_{1}^{2}+r^{2}\Theta_{1}\Theta_{2}+r\Phi_{1}^{2}\sin^{2}\theta+r^{2}\Phi_{1}\Phi_{2}\sin^{2}\theta+r^{2}\Theta_{1}\Phi_{1}^{2}\sin\theta\cos\theta\right),\\
\Theta_{2}\equiv & \frac{{\rm d}\Theta_{1}}{{\rm d}r},\\
\Phi_{2}\equiv & \frac{{\rm d}\Phi_{1}}{{\rm d}r}.
\end{split}
\end{equation}

\subsection{Numerical calculation of the curvature}

\begin{figure}[H]
\begin{centering}
\includegraphics[height=9cm]{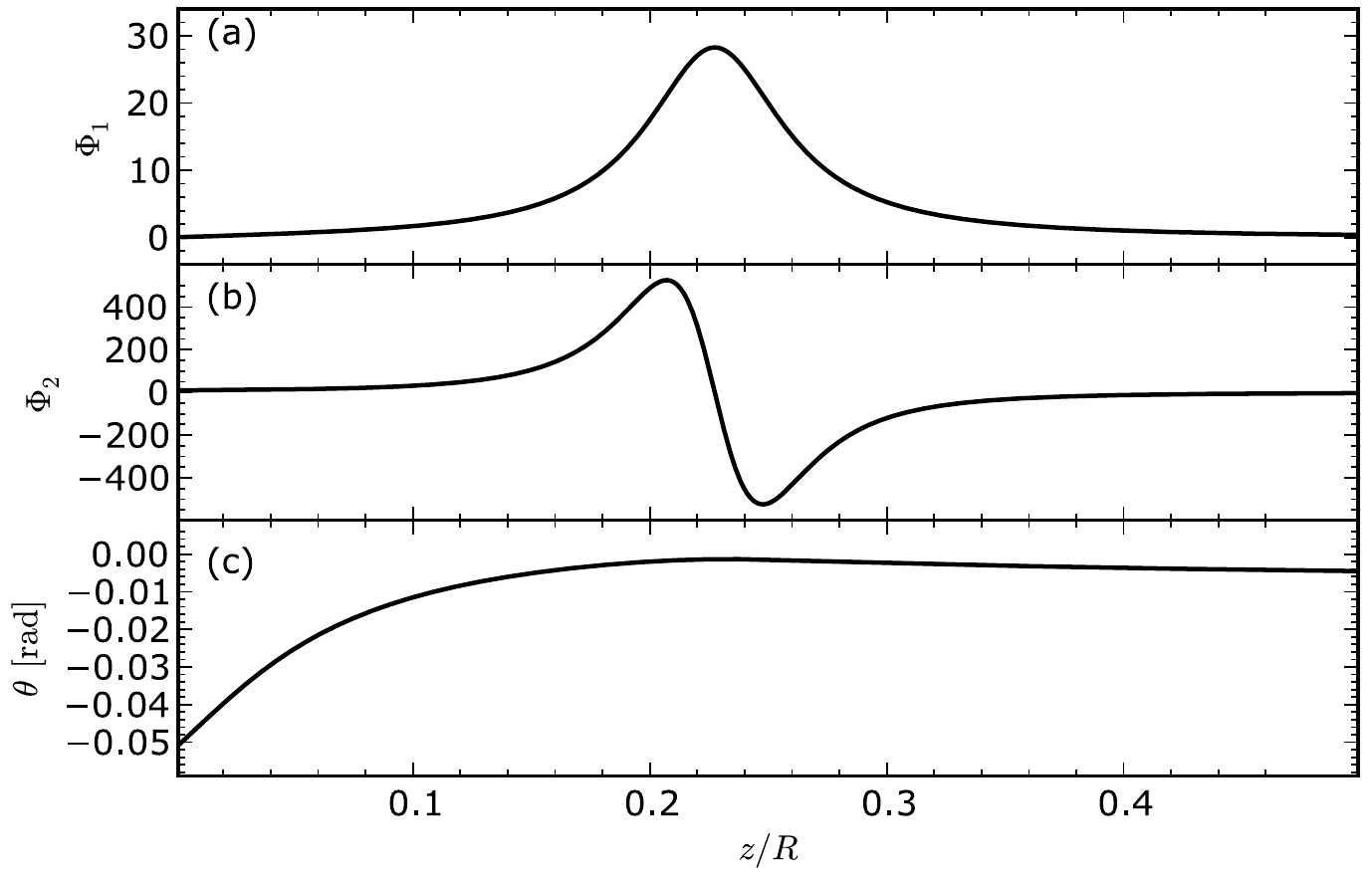}
\par\end{centering}

\centering{}\caption[First and second derivative of the $\phi$-coordinate of the magnetic
field line]{Plot of the first and second derivative of the $\phi$-coordinate
of the magnetic field line with respect to the $r$-coordinate vs.
the distance from the stellar surface. Values were calculated using
the approach described in Section \ref{sec:model.curvature}. Panels
(a) and (b) show the first and second derivative of the $\phi$-coordinate
while panel (c) shows the $\theta$-coordinate of the magnetic field
line. \label{fig:model.an_curva}}
\end{figure}

Let us note that when evaluating Equation \ref{eq:model.diff_eqs}
it was assumed that $\sin\theta\neq0$. Thus $\Phi_{1}$ and $\Phi_{2}$
are undefined for $\theta=0$. Figure \ref{fig:model.an_curva} presents
the first and second derivative ($\Phi_{1}$, $\Phi_{2}$) of the
$\phi$-coordinate of a magnetic field line with respect to the $r$-coordinate
for a magnetic field structure with $B_{\phi}\neq0$.

The singularity in Equation \ref{eq:model.diff_eqs} may result in
an overestimation of curvature of field lines that cross the $\theta=0$
plane for a complex structure of the surface magnetic field. To solve
this problem, and in addition to the analytical approach, the numerical
calculation of the curvature of magnetic field lines was implemented. 

\begin{figure}[H]
\begin{centering}
\includegraphics[height=9cm]{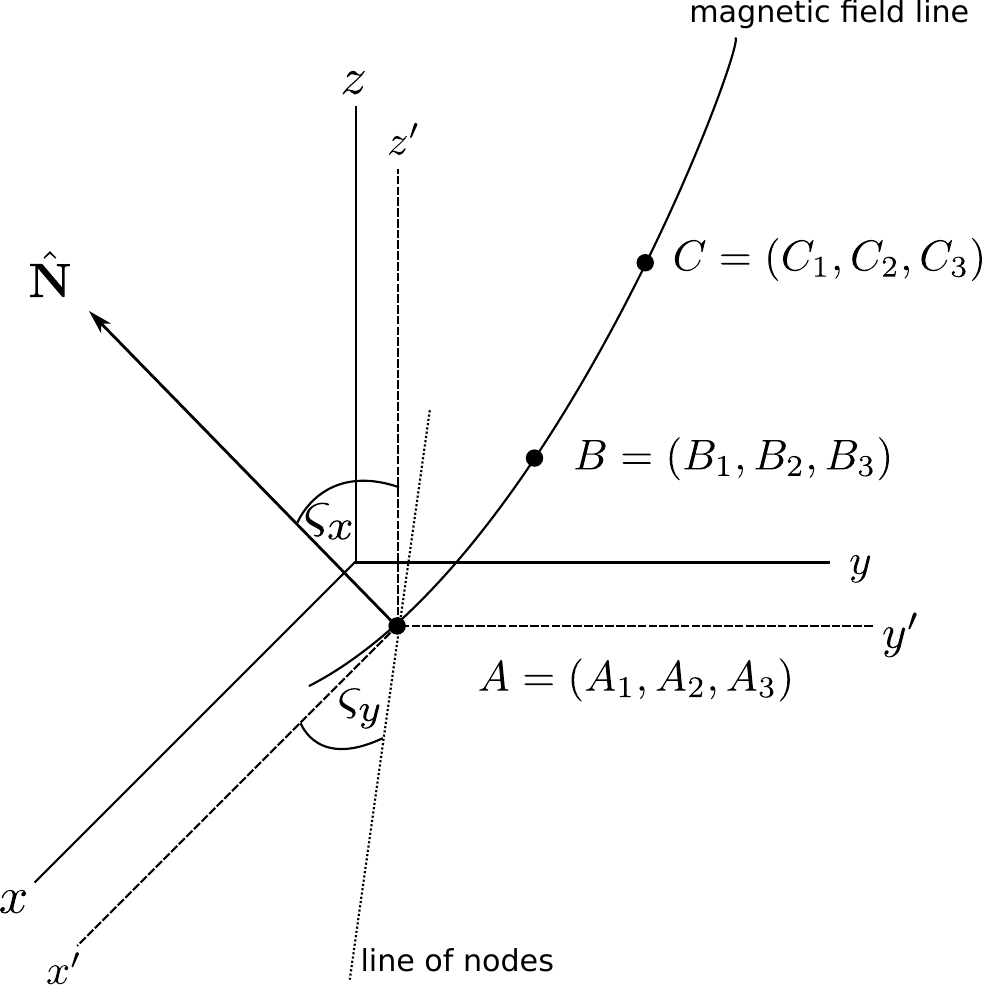}
\par\end{centering}

\centering{}\caption[Curvature of magnetic field lines (numerical approach)]{For any given three points ($A$, $B$, $C$) we can always find
a common plane. We use the following transformations to achieve this:
(I) shift the origin of the system to point $A$ (prime), (II) rotate
the shifted system by an angle $\varsigma_{y}$ around the $y^{\prime}$--axis
and by an angle $\varsigma_{x}$ around the $x^{\prime\prime}$-axis
(double prime). After these transformations the $z^{\prime\prime}$-axis
will be aligned with normal vector $\hat{{\bf N}}$ and all points
will lie in the $x^{\prime\prime}y^{\prime\prime}$-plane of such
a system of coordinates. \label{fig:model.curvature_model}}
\end{figure}

Let us consider three consecutive points of the given magnetic field
line $A$, $B$, $C$ (see Figure \ref{fig:model.curvature_model}). 

\begin{comment}
\textasciitilde{}/Programs/studies/phd/curvature/curvature.py (show\_angles)
with 99 data set
\end{comment}

To calculate curvature (or radius of curvature) in point $B$ we can
use the following procedure:
\begin{itemize}
\item simplify the 3-D problem to 2-D by finding a common plane for all
three points

\begin{itemize}
\item move the origin of the coordinate system to point $A$
\item rotate the coordinate system to align the z-axis with the normal vector
to the common plane of all three points ($A$, $B$, $C$)
\end{itemize}
\item calculate the radius of the circle passing through all three points
(in a 2-D coordinate system)
\end{itemize}

\subsubsection*{3-D to 2-D transition}

To simplify the calculations we shift the origin of the coordinate
system so that point $A$ will be the origin of the new system:

\begin{eqnarray}
A^{\prime} & = & \left(0,\,0,\,0\right);\nonumber \\
B^{\prime} & = & \left(B_{1}-A_{1},\, B_{2}-A_{2},\, B_{3}-A_{3}\right);\\
C^{\prime} & = & \left(C_{1}-A_{1},\, C_{2}-A_{2},\, C_{3}-A_{3}\right).\nonumber 
\end{eqnarray}

The unit normal vector to the plane enclosing all three points ($A^{\prime}$,
$B^{\prime}$, $C^{\prime}$) can be calculated as

\begin{equation}
\hat{{\bf N}}=\frac{{\bf b}\times{\bf c}}{\left|{\bf b}\times{\bf c}\right|}=\left(N_{1},\, N_{2},\, N_{3}\right),
\end{equation}
where ${\bf b}=\left(B_{1}^{\prime},\, B_{2}^{\prime},\, B_{3}^{\prime}\right)$
and ${\bf c}=\left(C_{1}^{\prime},\, C_{2}^{\prime},\, C_{3}^{\prime}\right)$.

The next step is to rotate the shifted coordinate system to align
the $z^{\prime}$-axis with normal vector $\hat{{\bf N}}$. In the
new system all three points will lie in the $x^{\prime\prime}y^{\prime\prime}$-plane.
In our calculations we rotate the shifted system by an angle $\varsigma_{y}$
around the $y^{\prime}$-axis, $R_{y}\left(\varsigma_{y}\right)$,
and a rotation by an angle $\varsigma_{x}$ around the $x^{\prime\prime}$-axis,
$R_{x}\left(\varsigma_{x}\right)$. The final rotation matrix can
be written as

\begin{equation}
R_{yx}=R_{y}\left(\varsigma_{y}\right)R_{x}\left(\varsigma_{x}\right)=\left[\begin{array}{ccc}
\cos\varsigma_{y} & \sin\varsigma_{x}\sin\varsigma_{y} & \sin\varsigma_{y}\cos\varsigma_{x}\\
0 & \cos\varsigma_{x} & -\sin\varsigma_{x}\\
-\sin\varsigma_{y} & \cos\varsigma_{y}\sin\varsigma_{x} & \cos\varsigma_{y}\cos\varsigma_{x}
\end{array}\right]
\end{equation}

The Euler angles for these rotations can be calculated as

\begin{center}
$\varsigma_{x}={\rm atan2}\left(N_{2},N_{3}\right).$%
\footnote{where ${\rm atan2}\left(y,\, x\right)$ equals: (1) $\arctan\left(y/x\right)$
if $x>0$; (2) $\arctan\left(y/x\right)+\pi$ if $y\ge0$ and $x<0$;
(3) $\arctan\left(y/x\right)-\pi$ if $y<0$ and $x<0$; (4) $\pi/2$
if $y>0$ and $x=0$; (5) $-\pi/2$ if $y<0$ and $x=0$; (6) is undefined
if $y=0$ and $x=0$. This function is available in many programming
languages.\label{fn:model.atan2}%
}
\par\end{center}

\begin{equation}
\begin{array}{c}
\varsigma_{y}=\begin{cases}
\arctan\left(-\frac{N_{1}}{N_{3}}\cos\varsigma_{x}\right) & {\rm if\ }N_{3}\neq0\\
\arctan\left(-\frac{N_{1}}{N_{2}}\sin\varsigma_{x}\right) & {\rm if\ }N_{2}\neq0\\
\frac{\pi}{2} & {\rm if\ }N_{2}=0\ {\rm and}\ N_{3}=0
\end{cases}\end{array}
\end{equation}

\begin{comment}
where $\arctan2\left(x,\, y\right)$ is the arc tangent of the two
variables $x$ and $y$. It is similar to calculating the arc tangent
of $x/y$, except that the signs of both arguments are used to determine
the quadrant of the result, which lies in the range $\left[-\pi,\,\pi\right]$.
This function is available in many programming languages and often
is called atan2.
\end{comment}

Finally, we can write the components of all three points in our new
(shifted and double-rotated) system of coordinates as follows

\begin{eqnarray}
A^{\prime\prime} & = & R_{yx}A^{\prime}=\left(0,\,0,\,0\right);\nonumber \\
B^{\prime\prime} & = & R_{yx}B^{\prime}=\left(B_{1}^{\prime\prime},\, B_{2}^{\prime\prime},\,0\right);\\
C^{\prime\prime} & = & R_{yx}C^{\prime}=\left(C_{1}^{\prime\prime},\, C_{2}^{\prime\prime},\,0\right).\nonumber 
\end{eqnarray}

\subsubsection*{Circle passing through 3 points \citep{2012_Bourne}}

Finding the radius of the circle passing through three consecutive
points of a given magnetic field line ($A=\left(0,\,0\right)$, $B=\left(B_{1},\, B_{2}\right)$,
$C=\left(C_{1},\, C_{2}\right)$) is an exact method for finding the
radius of curvature $\Re$ and hence the curvature $\rho=1/\Re$ of
this line. Note that for simplicity's sake we hereafter describe points
without double prime notation but they refer to coordinates in the
shifted and double-rotated system of coordinates (e.g. $B=\left(B_{1},\, B_{2}\right)=\left(B_{1}^{\prime\prime},\, B_{2}^{\prime\prime}\right)$).

\begin{figure}[H]
\begin{centering}
\includegraphics{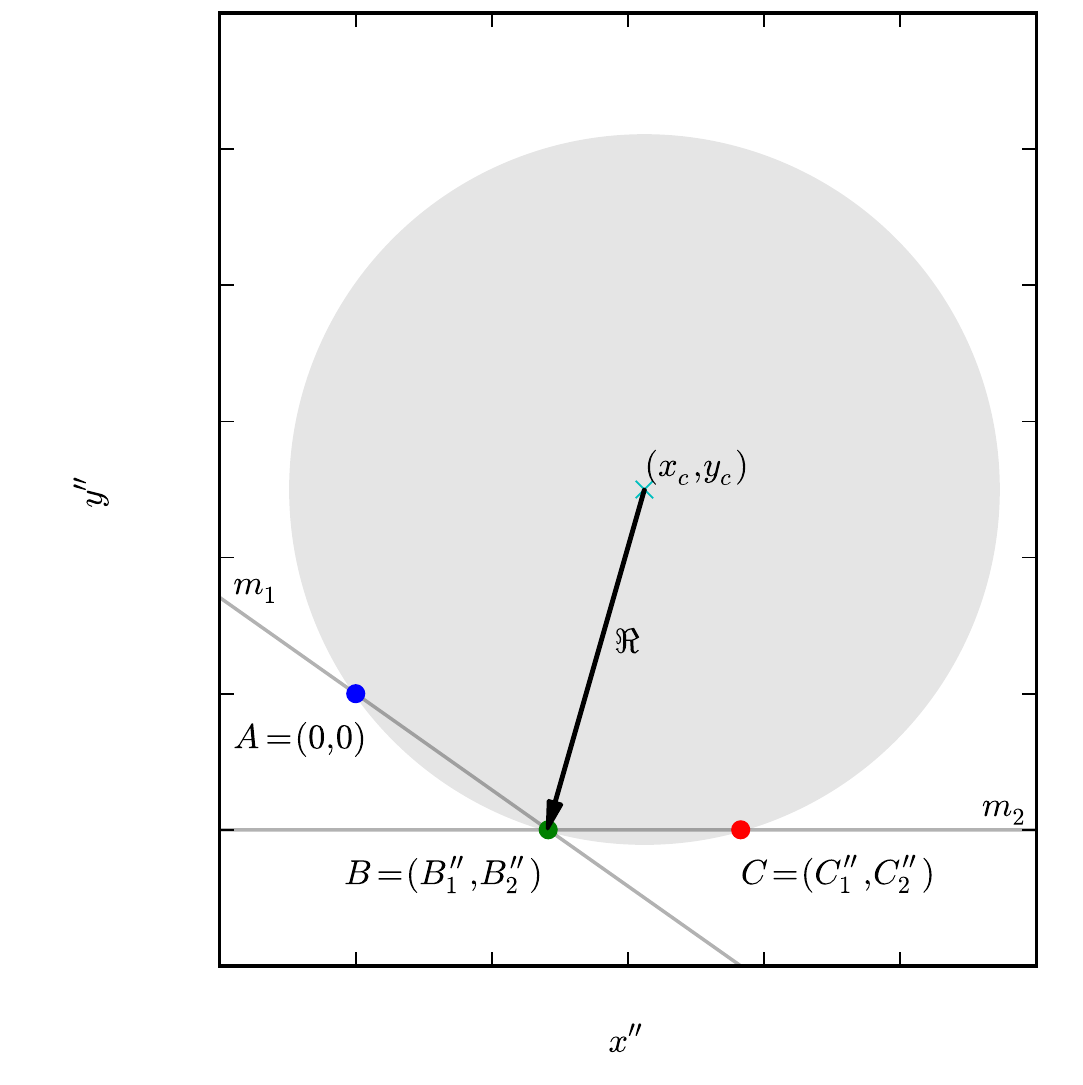}
\par\end{centering}

\centering{}\caption[The radius of curvature of the magnetic field line]{ The radius of curvature $\Re$ of the magnetic field line at a given
point $B$ can be calculated as the radius of the circle passing through
this and the neighbouring two points ($A$, $C$). The system of coordinates
was moved and double-rotated so that $A$ is in its origin and all
points lie in the $x^{\prime\prime}y^{\prime\prime}$-plane. The slopes
of the lines joining $A$ to $B$ and $B$ to $C$ are described by
Equation \ref{eq:model.slopes}. \label{fig:model.curva_circle}}
\end{figure}

Slope $m_{1}$ of the line joining $A$ to $B$ and slope $m_{2}$
of the line joining $B$ to $C$ (see Figure \ref{fig:model.curva_circle})
are given by 

\begin{equation}
\begin{split}m_{1}= & \frac{\Delta y}{\Delta x}=\frac{B_{2}}{B_{1}},\\
m_{2}= & \frac{\Delta y}{\Delta x}=\frac{C_{2}-B_{2}}{C_{1}-B_{1}}.
\end{split}
\label{eq:model.slopes}
\end{equation}

In general, the centre of the circle passing through our points is
given by

\begin{equation}
\begin{split}x_{c}= & \frac{m_{1}m_{2}\left(A_{2}-C_{2}\right)+m_{2}\left(A_{1}-B_{1}\right)-m_{1}\left(B_{1}-C_{1}\right)}{2\left(m_{2}-m_{1}\right)},\\
y_{c}= & \frac{1}{m_{1}}\left(x_{c}-\frac{A_{1}+B_{1}}{2}\right)+\frac{A_{2}+B_{2}}{2}.
\end{split}
\end{equation}

\begin{comment}
\textasciitilde{}/Programs/magnetic/magnetic/src/model/curvature.py
(plot\_phd)

cp \textasciitilde{}/Programs/magnetic/magnetic/src/model/curvature\_circle.pdf
\textasciitilde{}/Documents/studies/phd/images/model/.
\end{comment}

Since point $A$ is in the centre of the coordinate system we can
simplify these formulas as follows

\begin{equation}
\begin{split}x_{c}= & \frac{2\, B_{1}^{2}B_{2}-2\, B_{1}B_{2}C_{1}+B_{2}C_{1}^{2}-B_{2}C_{2}^{2}-\left(B_{1}^{2}-B_{2}^{2}\right)C_{2}}{2\,\left(B_{1}C_{2}-B_{2}C_{1}\right)},\\
y_{c}= & \frac{B_{2}^{2}-\left(B_{2}-2\, x_{c}\right)B_{1}}{2\, B_{2}}.
\end{split}
\end{equation}

Finally, we can calculate the radius of curvature simply by finding
the distance between the centre of the circle and any of the points
on the circle (we have chosen point $A$)

\begin{equation}
\Re=\frac{1}{\rho}=\sqrt{x_{c}^{2}+y_{c}^{2}}.
\end{equation}

\begin{figure}[H]
\begin{centering}
\includegraphics[height=9.5cm]{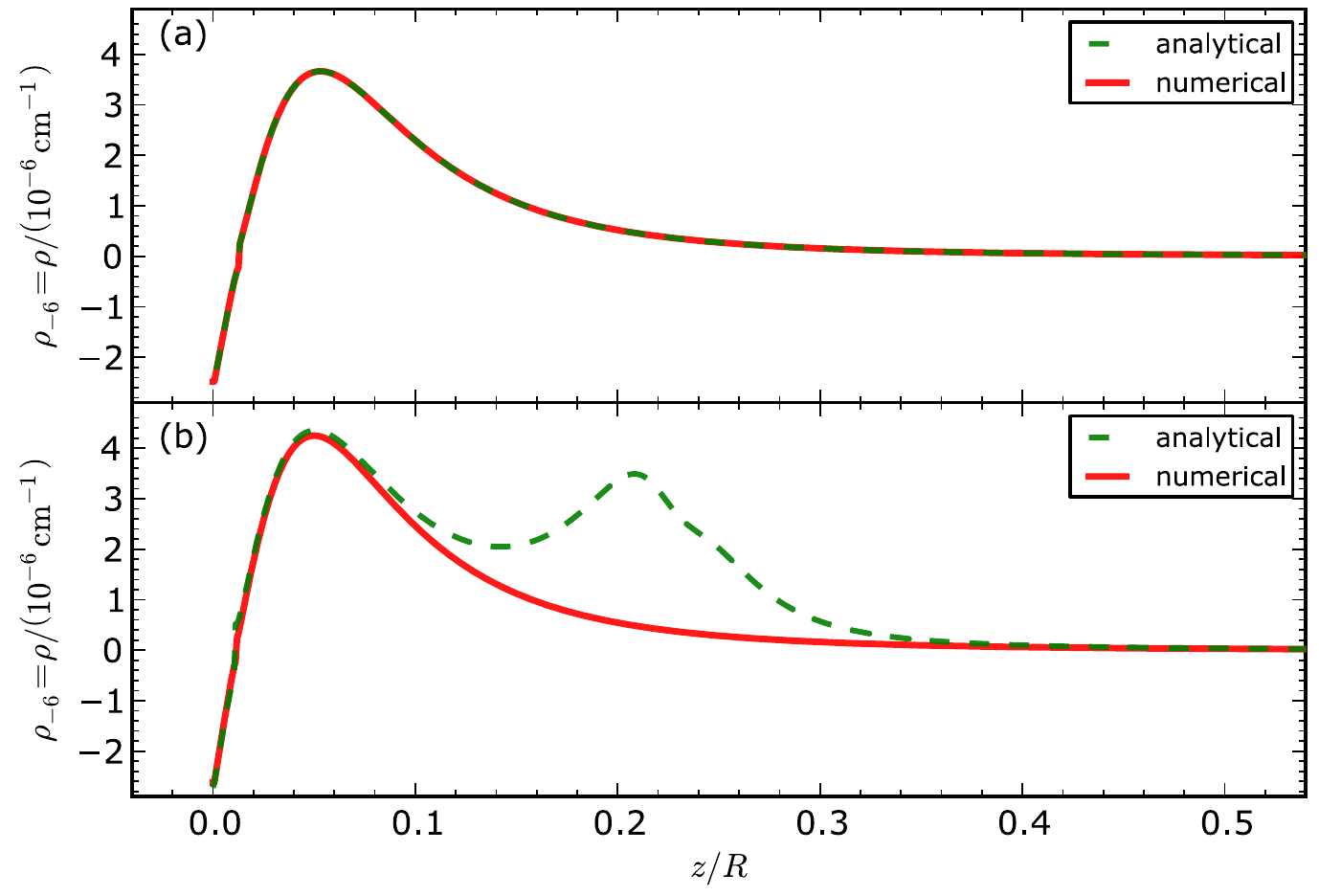}
\par\end{centering}

\centering{}\caption[Curvature of the magnetic field lines vs. the height above the stellar
surface]{Curvature of the magnetic field lines vs. the height above the stellar
surface calculated using the analytical approach described in Section
{[}\ref{sec:model.curvature}{]} (green lines) and the numerical approach
presented above (red lines). Panel (a) corresponds to the magnetic
field line which has no $\phi$ component, while panel (b) corresponds
to a more general scenario i.e. the nonzero $\phi$ component of the
magnetic field line. As can be seen, the analytical approach is not
valid for every case. This is caused by the undefined value of the
$\phi$ derivative for $\theta=0$ (see Equation {[}\ref{eq:model.diff_eqs}{]}).
Here $\rho_{-6}=1/\Re_{6}=\rho/\left(10^{-6}\,{\rm cm}^{-1}\right)$
and $\Re_{6}=\Re/\left(10^{6}\,{\rm cm}\right)$. \label{fig:model.an_num_curva} }
\end{figure}

In this thesis we consider complex structures of the surface magnetic
field, thus the numerical method presented above was used in all the
calculations of curvature. The analytical approach may result in an
overestimation of curvature for points with $\theta\approx0$ (see
Figure \ref{fig:model.an_num_curva}).

\begin{comment}
\textasciitilde{}/Programs/studies/phd/curvature/curvature.py (show\_curva)
with 99 data set 
\end{comment}

\section{Simulation results}

In this section we model the surface non-dipolar magnetic field structure
for some pulsars. Note that we can estimate the size of the polar
cap and the strength of the surface magnetic field only for pulsars
with an observed hot spot (see Section \ref{sec:x-ray.blt}). Here
we present only pulsars listed in Table \ref{tab:psg.psg_top}.

\begin{figure}[H]
\begin{centering}
\includegraphics[height=8.5cm]{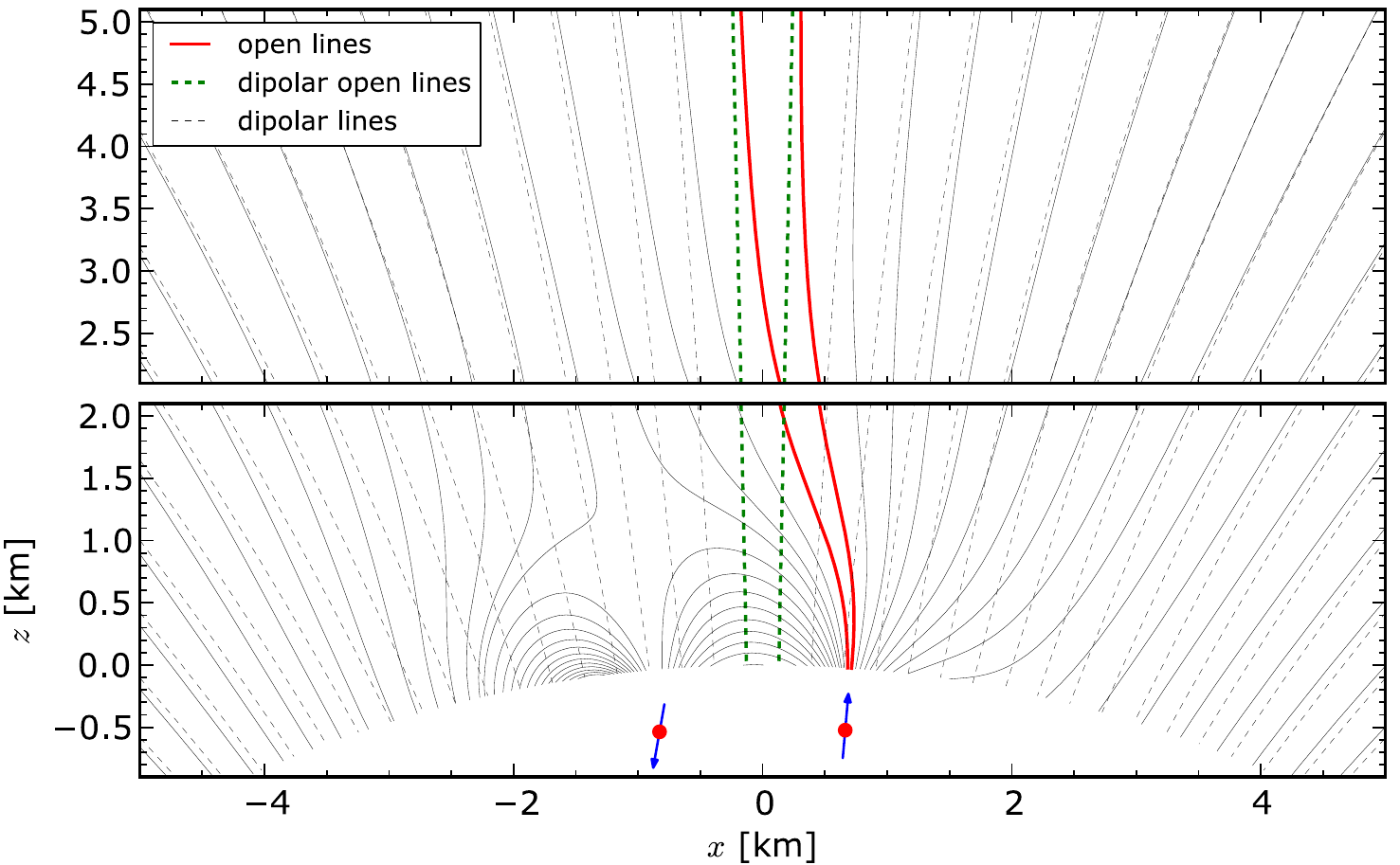}
\par\end{centering}

\caption[{Possible non-dipolar structure of the magnetic field lines {[}PSR
B0628-28{]}}]{Possible non-dipolar structure of the magnetic field lines of PSR
B0628-28.\protect \linebreak{}
The structure was obtained using two crust anchored anomalies located
at: \protect \linebreak{}
${\bf r_{1}}=\left(0.95R,\,4^{\circ},\,0^{\circ}\right)$, ${\bf r_{2}}=\left(0.95R,\,5^{\circ},\,180^{\circ}\right)$,
with the dipole moments\protect \linebreak{}
 ${\bf m_{1}}=\left(4.5\times10^{-3}d,\,5^{\circ},\,0^{\circ}\right)$,
${\bf m_{2}}=\left(4.5\times10^{-3}d,\,170^{\circ},\,180^{\circ}\right)$
respectively (blue arrows). The influence of the anomalies is negligible
at distances $D\gtrsim2R$, where $B_{{\rm m}}/B_{{\rm d}}\approx m/d=4.5\times10^{-3}$
(top panel). For more details on the polar cap region see Figure \ref{fig:model.b0628_zoom}.\label{fig:model.b0628}}
\end{figure}

\begin{figure}[H]
\begin{centering}
\includegraphics{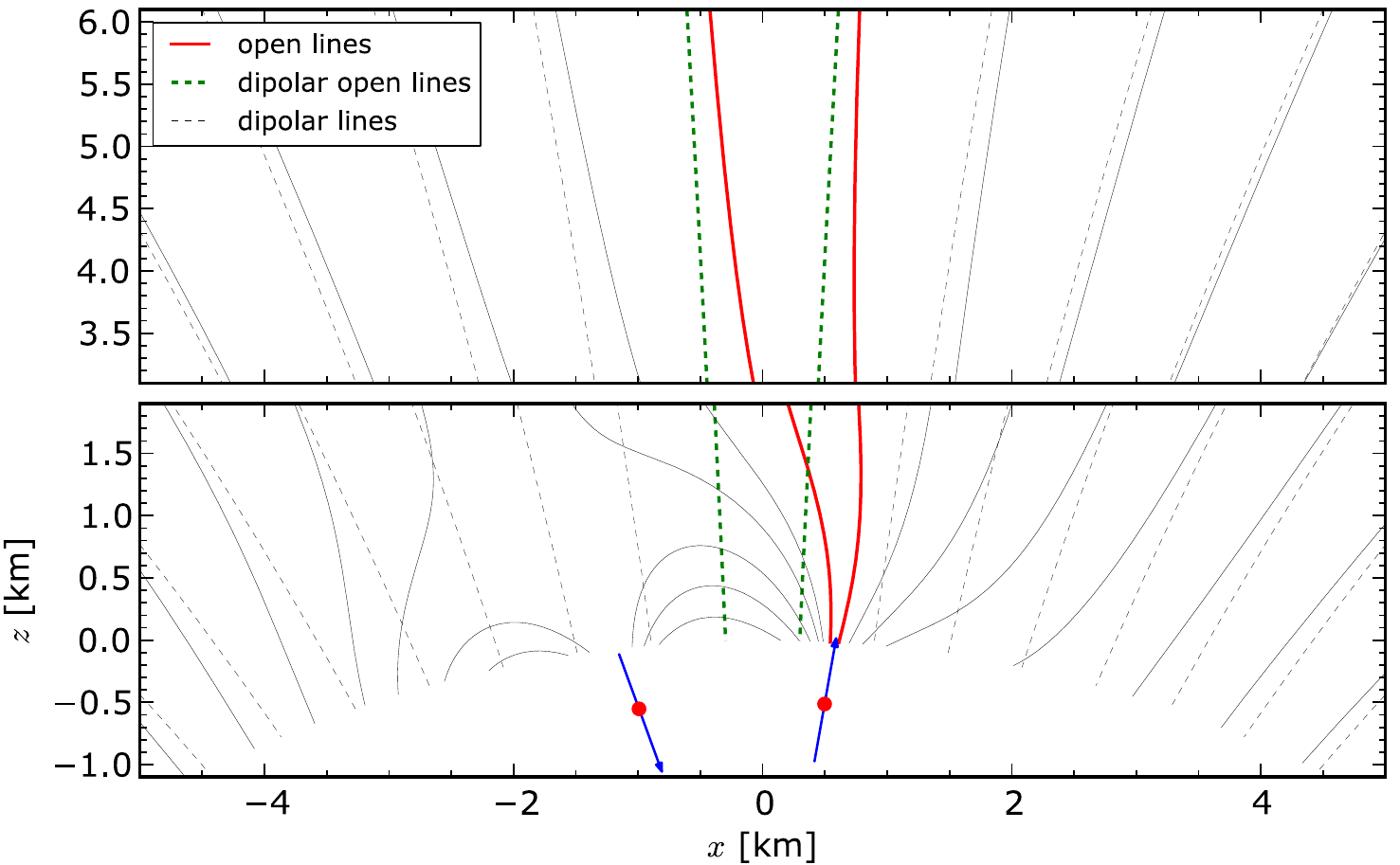}
\par\end{centering}

\caption[{Possible non-dipolar structure of the magnetic field lines {[}PSR
J0633+1746{]}}]{Possible non-dipolar structure of the magnetic field lines of PSR
J0633+1746.\protect \linebreak{}
 The structure was obtained using two crust anchored anomalies located
at:\protect \linebreak{}
 ${\bf r_{1}}=\left(0.95R,\,3^{\circ},\,0^{\circ}\right)$, ${\bf r_{2}}=\left(0.95R,\,6^{\circ},\,180^{\circ}\right)$,
with the dipole moments\protect \linebreak{}
${\bf m_{1}}=\left(5.5\times10^{-3}d,\,10^{\circ},\,0^{\circ}\right)$,
${\bf m_{2}}=\left(5.5\times10^{-3}d,\,160^{\circ},\,0^{\circ}\right)$
respectively (blue arrows). The influence of the anomalies is negligible
at distances $D\gtrsim3.1R$, where $B_{{\rm m}}/B_{{\rm d}}\approx m/d=5.5\times10^{-3}$
(top panel). For more details on the polar cap region see Figure \ref{fig:model.j0633_zoom}.\label{fig:model.j0633}}
\end{figure}

\begin{figure}[H]
\begin{centering}
\includegraphics{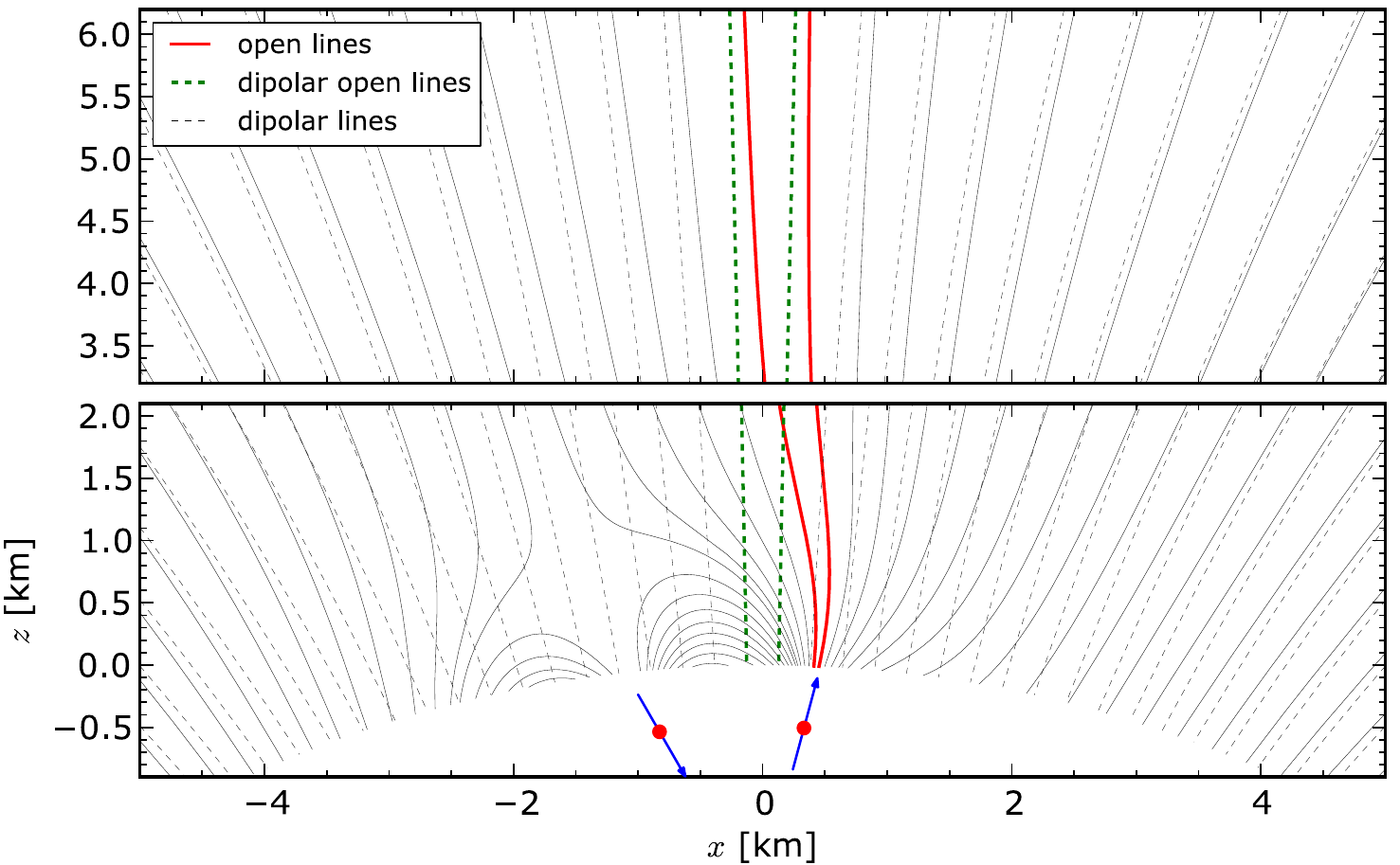}
\par\end{centering}

\caption[{Possible non-dipolar structure of the magnetic field lines {[}PSR
B0834+06{]}}]{Possible non-dipolar structure of the magnetic field lines of PSR
B0834+06.\protect \linebreak{}
 The structure was obtained using two crust anchored anomalies located
at:\protect \linebreak{}
${\bf r_{1}}=\left(0.95R,\,2^{\circ},\,0^{\circ}\right)$, ${\bf r_{2}}=\left(0.95R,\,5^{\circ},\,180^{\circ}\right)$,
with the dipole moments\protect \linebreak{}
${\bf m_{1}}=\left(3\times10^{-3}d,\,15^{\circ},\,0^{\circ}\right)$,
${\bf m_{2}}=\left(3\times10^{-3}d,\,150^{\circ},\,0^{\circ}\right)$
respectively (blue arrows). The influence of the anomalies is negligible
at distances $D\gtrsim3.2R$, where $B_{{\rm m}}/B_{{\rm d}}\approx m/d=3\times10^{-3}$
(top panel). For more details on the polar cap region see Figure \ref{fig:model.b0834_zoom}.\label{fig:model.b0834}}
\end{figure}

We use spherical coordinates $\left(r,\,\theta,\,\phi\right)$ to
describe the location and orientation of crust-anchored local anomalies.
The parameters of anomalies are as follows: ${\bf r_{a}}=\left(r_{a},\,\theta_{a},\,\phi_{a}\right)$
is a radius vector which points to the location of the anomaly and
${\bf m_{a}}=\left(m_{a},\,\theta_{a}\,,\phi_{a}\right)$ is its dipole
moment. The value of $m_{a}$ is measured in units of the global dipole
moment $d$, i.e. the moment which corresponds to the pulsar's global
magnetic field. In the Figures showing a possible non-dipolar structure
(e.g. Figure \ref{fig:model.b0628}, \ref{fig:model.j0633}, \ref{fig:model.b0834})
the dashed lines correspond to the dipolar configuration of the magnetic
field lines, while the solid lines correspond to the actual magnetic
field lines (taking into account the crust-anchored anomalies). Green
and red lines represent the open magnetic field lines for dipolar
and non-dipolar structures, respectively.

\subsection{PSR B0628-28\label{sec:model.0628}}

Pulsar B0628-28, a bright radio pulsar, was discovered by \citet{1969_Large}
during a pulsar search at 408 MHz. The pulsar period $P\approx1.24\,{\rm s}$
and its first derivative $\dot{P}_{-15}\approx7.1$ result in a dipolar
component of magnetic field $B_{{\rm d}}=6\times10^{12}\,{\rm G}$
and a characteristic age $\tau_{c}\approx2.8\,{\rm Myr}$, which makes
it a typical, old pulsar. The large distance to this pulsar $D=1.44\,{\rm kpc}$
(evaluated using the Galactic free electron density model of \citealp{2002_Cordes})
makes it impossible to use the parallax method to determine the distance
with better accuracy. 

PSR B0628-28 is one of the longest period pulsars among those detected
in X-rays. The pulsar was first detected in the X-ray band by \textit{ROSAT}
and then later observed with both the \textit{Chandra} and \textit{XMM-Newton}.
Observations with the \textit{Chandra} revealed no pulsations, while
the\textit{ XMM-Newton} observations revealed pulsations with a period
consistent with the period of radio emission \citep{2005_Tepedelenl}.
The inconsistency of the observations is a reflection of the fact
that the pulsar is detectable just at the threshold of sensitivity
of both the observatories. The two-component spectral fit (BB+PL)
shows that both the nonthermal and thermal components have a comparable
luminosity (at least if we assume that the nonthermal radiation is
isotropic, see Table \ref{tab:x-ray_nonthermal}). PSR~$\mbox{B0628-28}$
is characterised by one of the largest X-ray efficiencies among the
observed pulsars $\xi_{{\rm BB}}\approx\xi_{{\rm NT}}^{^{{\rm max}}}\approx10^{-2}$.

\begin{comment}
\textasciitilde{}/Programs/studies/phd/lines/lines.py (plot\_b0628),
400, 401, data sets
\end{comment}

\begin{comment}
\textasciitilde{}/Programs/studies/phd/lines/lines.py (plot\_b0628\_zoom),
400 data set
\end{comment}

\begin{figure}[H]
\begin{centering}
\includegraphics{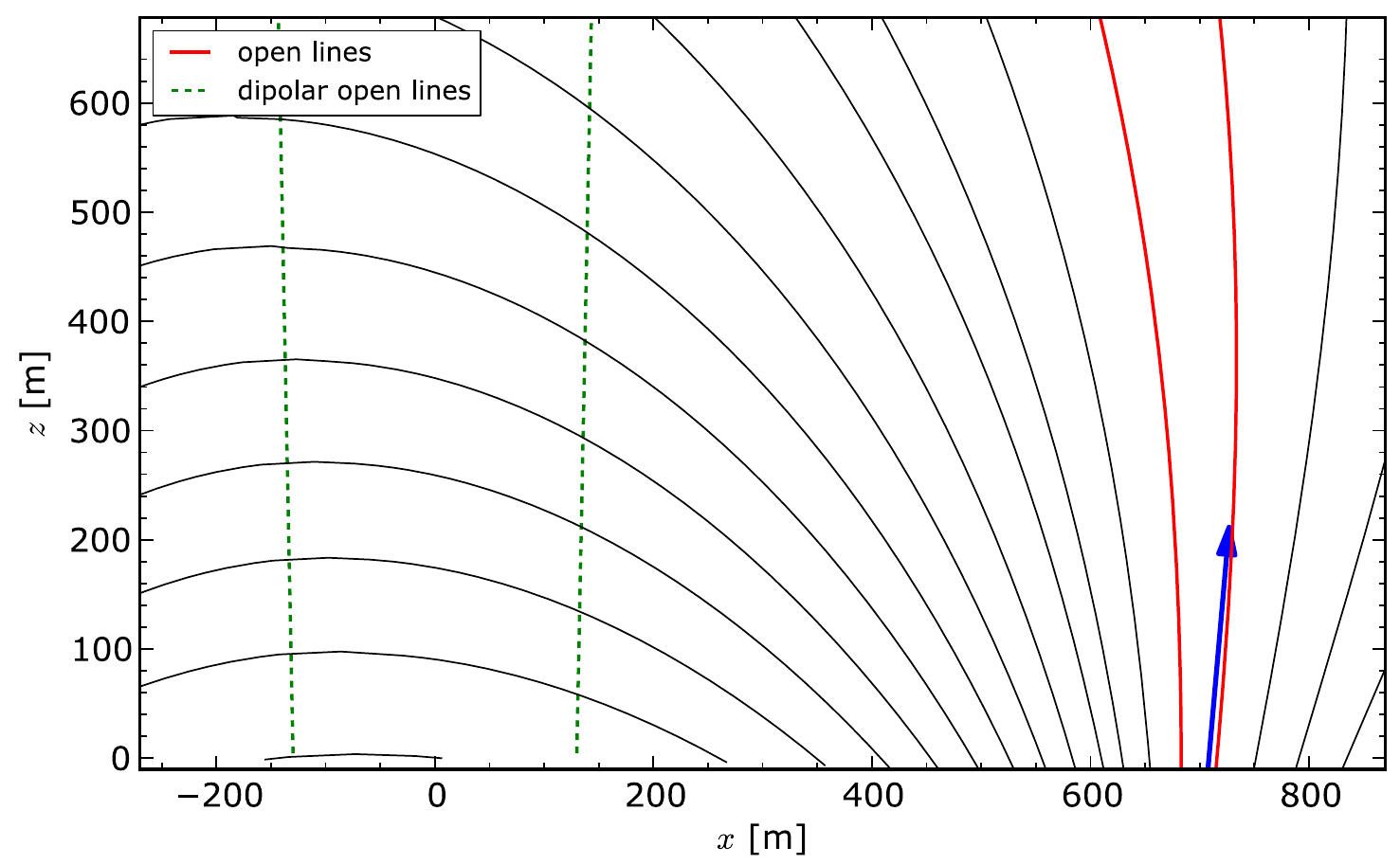}
\par\end{centering}

\caption[{Zoom of the polar cap region {[}PSR B0628-28{]}}]{Zoom of the polar cap region of PSR B0628-28. See Figure \ref{fig:model.b0628}
for a description. \label{fig:model.b0628_zoom}}
\end{figure}

\vspace*{0.5cm}

\begin{comment}
\textasciitilde{}/Programs/studies/phd/lines/lines.py (curvature\_b0628),
403 data set
\end{comment}

\begin{figure}[H]
\begin{centering}
\includegraphics{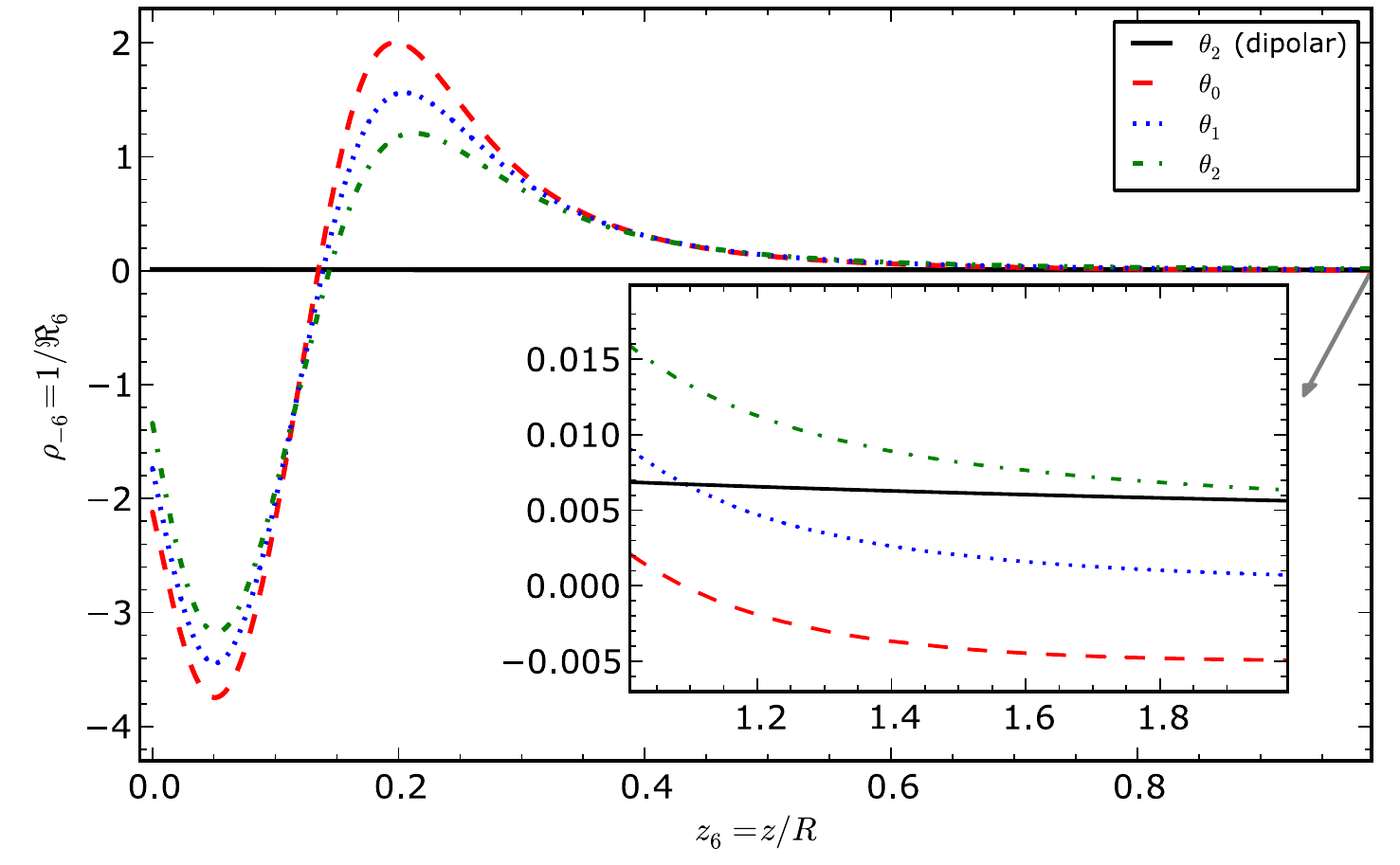}
\par\end{centering}

\caption[{Curvature of the open magnetic field lines {[}PSR B0628-28{]}}]{Dependence of a curvature of the open magnetic field lines on the
distance from the stellar surface for PSR B0628-28. The distance is
in units of the stellar radius $z_{6}=z/R$ and the curvature of the
magnetic field lines is $\rho_{-6}=1/\Re_{6}=\rho/\left(10^{-6}\,{\rm cm}^{-1}\right)$.
\label{fig:model.b0628_curva}}
\end{figure}

\clearpage{}

\subsection{PSR J0633+1746\label{sec:model.0633}}

Geminga was discovered in 1972 as a $\gamma$-ray source by \citet{1975_Fichtel}.
The visual magnitude of the pulsar was estimated by \citet{1987_Bignami}
to be of the order of $\sim25.5^{^{{\rm mag}}}$. The pulse modulation
was discovered in X-rays \citep{1992_Halpern}, in $\gamma$-rays,
and at optical wavelengths \citep{1998_Shearer}. Geminga has been
determined to be a relatively old ($\tau=342\,{\rm kyr}$) radio-quiet
pulsar with a period $P=237\,{\rm ms}$. The distance to the pulsar
$D=0.16\,{\rm kpc}$, evaluated using the parallax method, makes it
the closest pulsar with available X-ray data. 

The pulsar exhibits one of the weakest radio luminosities known and
a cutoff at frequencies higher than about $100\,{\rm MHz}$. The model
presented by \citet{1998_Gil} explains this weak radio emission with
absorption by the magnetised relativistic plasma inside the light
cylinder. As the exact model of radio emission is still unknown (see
Section \ref{sec:radiation.radio_emission}), it is difficult to verify
if this weak radio emission is a result of absorption or the absence
of coherent radio emission.

The three-component fit to the X-ray spectrum (PL+BB+BB, see Table
\ref{tab:x-ray_thermal}) reveals the hot spot component with a size
that is considerably smaller than the conventional polar cap size
($b\approx23$). The entire surface temperature $T_{{\rm s}}=0.5\,{\rm MK}$
is consistent with the theoretical value predicted by the cooling
model. 

\begin{comment}
\textasciitilde{}/Programs/studies/phd/lines/lines.py (plot\_j0633),
370, 371data sets
\end{comment}

\vspace*{0.5cm}

\begin{comment}
\textasciitilde{}/Programs/studies/phd/lines/lines.py (plot\_j0633\_zoom),
373 data set
\end{comment}

\begin{figure}[H]
\begin{centering}
\includegraphics{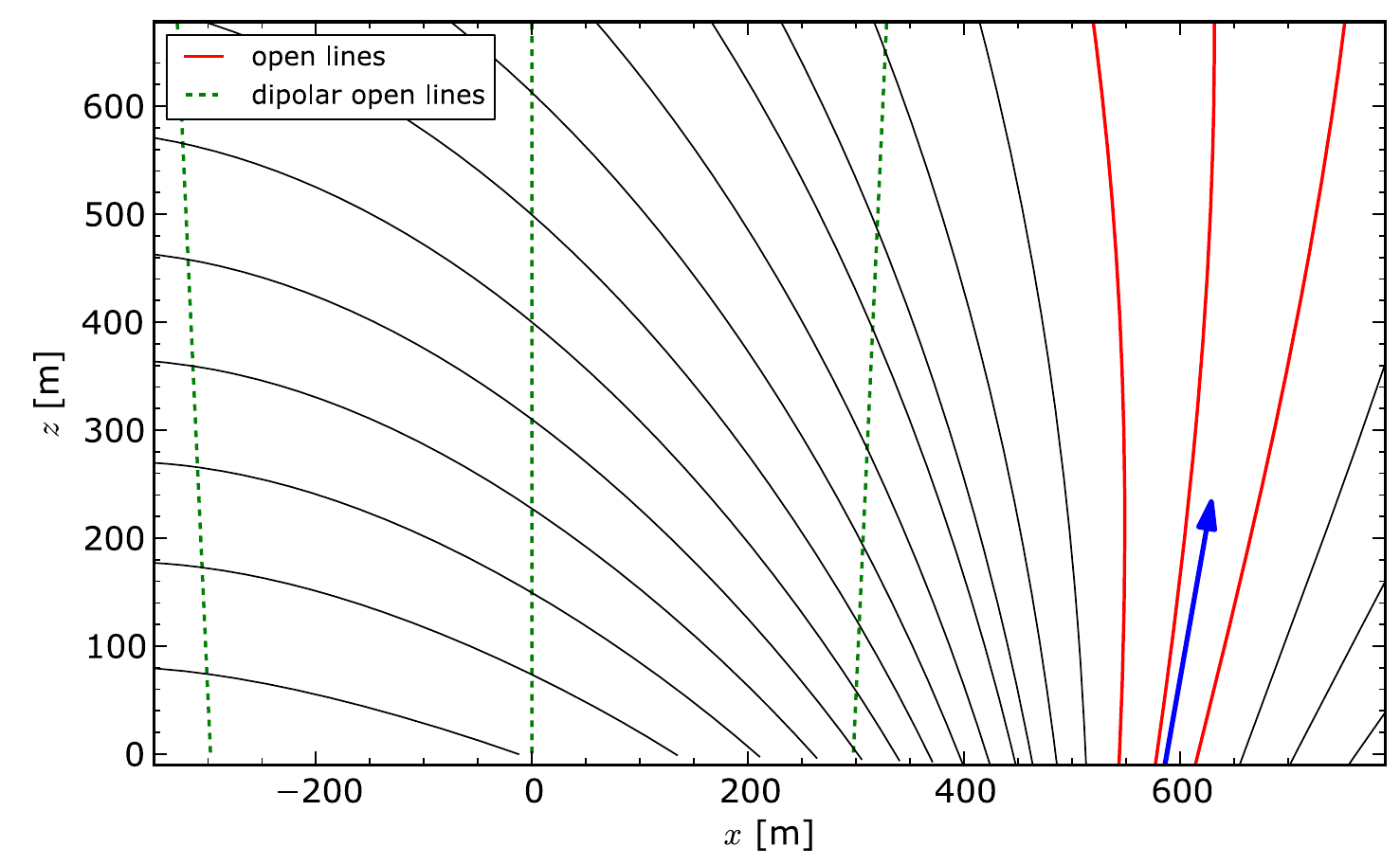}
\par\end{centering}

\caption[{Zoom of the polar cap region {[}PSR J0633+1746{]}}]{Zoom of the polar cap region of PSR J0633+1746. See Figure \ref{fig:model.j0633}
for a description. \label{fig:model.j0633_zoom}}
\end{figure}

\vspace*{0.5cm}

\begin{comment}
\textasciitilde{}/Programs/studies/phd/lines/lines.py (curvature\_0633),
data set
\end{comment}

\begin{figure}[H]
\begin{centering}
\includegraphics{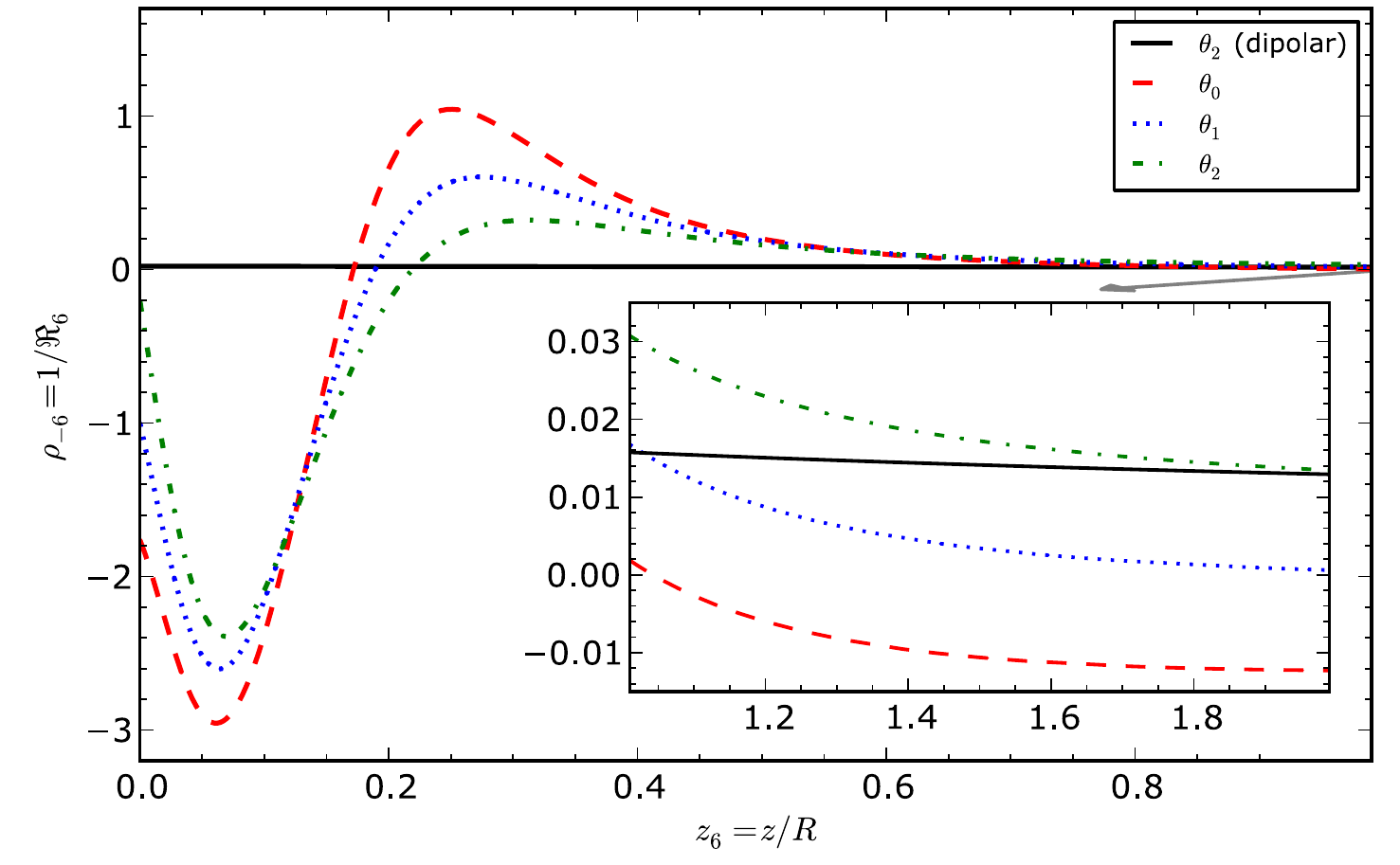}
\par\end{centering}

\caption[{Curvature of the open magnetic field lines {[}PSR J0633+1746{]}}]{Dependence of a curvature of the open magnetic field lines on the
distance from the stellar surface for PSR J0633+1746. The distance
is in units of the stellar radius $z_{6}=z/R$ and the curvature of
the magnetic field lines is $\rho_{-6}=1/\Re_{6}=\rho/\left(10^{-6}\,{\rm cm}^{-1}\right)$.
\label{fig:model.j0633_curva}}
\end{figure}

\clearpage{}

\subsection{PSR B0834+06}

The bright radio emission of PSR B0834+06 shows frequent nulls (nearly
$9\%$ of the pulses is absent, see \citealp{2008_Rankin}) . With
a relatively long rotational period $P=1.27\,{\rm s}$ and $\dot{P}_{-15}\approx7.1$
\citep{2000_Taylor}, its inferred physical properties, e.g. $B_{{\rm d}}=3\times10^{12}\,{\rm G}$,
are close to the average. The characteristic age $\tau_{c}=2.97\,{\rm Myr}$
implies that the pulsar should be categorised as an old pulsar. The
distance to the pulsar, estimated as $D=0.64\,{\rm kc}$, was derived
from its dispersion measure using the Galactic free-electron density
model of \citealp{2002_Cordes}. \citet{2006_Weltevrede} suggest
a drift of subpulses, but the estimated value of a subpulse separation
is larger than the pulse width. Despite the fact that the geometry
based on the carousel model could be fitted to the observations, there
is no clear evidence for a drift of emission between the components
of the pulsar \citep{2007_Rankin}.

The pulsar was detected in X-ray by \citet{2008_Gil} with a total
of $70$ counts from over $50\,{\rm ks}$ exposure time. Because of
the low statistical quality of the X-ray data, it was not possible
to constrain the absorbing column density $N_{H}$. The two-component
spectral fit (BB + PL), as presented in this thesis, was performed
using the assumption that both the thermal and nonthermal fluxes are
of the same order.

\begin{comment}
\textasciitilde{}/Programs/studies/phd/lines/lines.py (plot\_b0834),
380, 381 data sets
\end{comment}

\vspace*{0.5cm}

\begin{comment}
\textasciitilde{}/Programs/studies/phd/lines/lines.py (plot\_b0834\_zoom),
380 set
\end{comment}

\begin{figure}[H]
\begin{centering}
\includegraphics{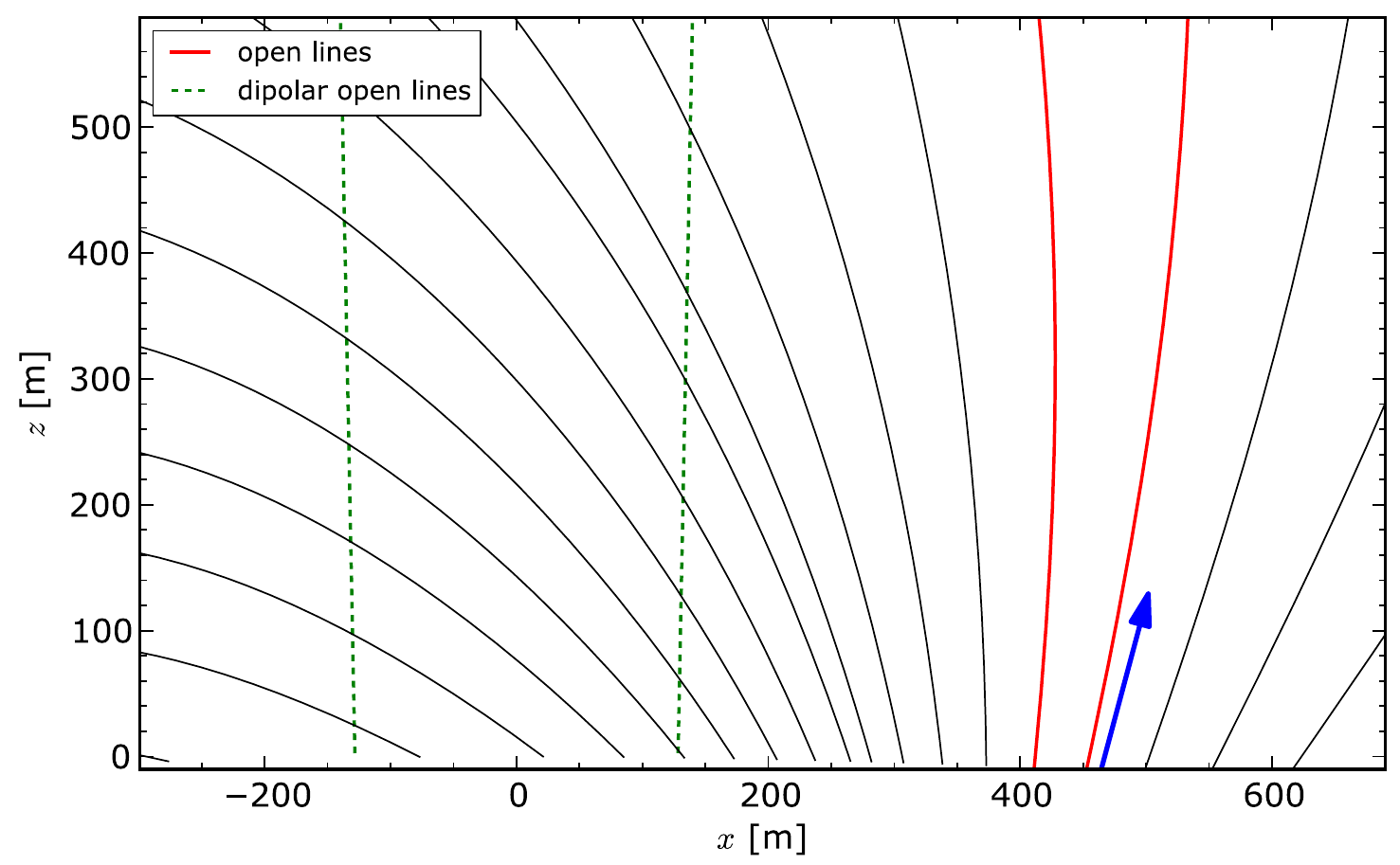}
\par\end{centering}

\caption[{Zoom of the polar cap region {[}PSR B0834+06{]}}]{Zoom of the polar cap region of PSR B0834+06. See Figure \ref{fig:model.b0834}
for a description. \label{fig:model.b0834_zoom}}
\end{figure}

\vspace*{0.5cm}

\begin{comment}
\textasciitilde{}/Programs/studies/phd/lines/lines.py (curvature\_b0834),
data set
\end{comment}

\begin{figure}[H]
\begin{centering}
\includegraphics{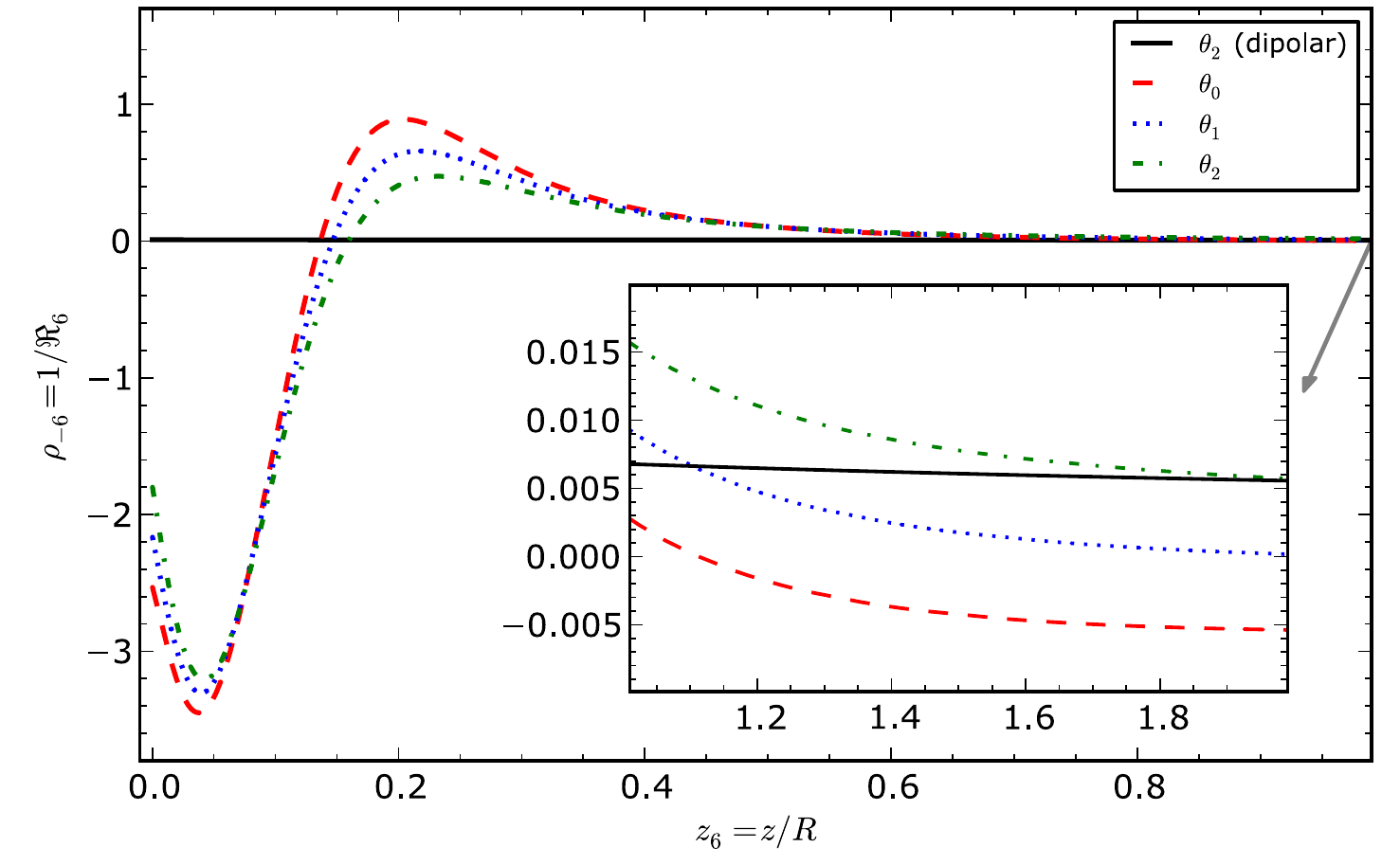}
\par\end{centering}

\caption[{Curvature of the open magnetic field lines {[}PSR B0834+06{]}}]{Dependence of a curvature of the open magnetic field lines on the
distance from the stellar surface for PSR B0834+06. The distance is
in units of stellar radius ($z_{6}=z/R$) and the curvature of the
magnetic field lines is $\rho_{-6}=1/\Re_{6}=\rho/\left(10^{-6}\,{\rm cm}^{-1}\right)$.
\label{fig:model.b0834_curva}}
\end{figure}

\clearpage{}

\subsection{PSR B0943+10\label{sec:model.b0943}}

Pulsar B0943+10 is a relatively old pulsar with a characteristic age
of $\tau_{c}=4.98\,{\rm Myr}$. The pulsar period $P=1.1\,{\rm s}$
and its first derivative $\dot{P}_{-15}\approx3.5$ result in the
dipolar component of a magnetic field $B_{{\rm d}}=4.0\times10^{12}\,{\rm G}$.
Using the Galactic free-electron density model of \citealp{2002_Cordes},
we can estimate the distance to the pulsar $D=0.63\,{\rm kpc}$.

PSR B0943+10 is a well-known example of a pulsar exhibiting both the
mode changing and subpulse drifting phenomenon. Strong, regular subpulse
drifting is observed only in radio-bright mode, and only hints of
the modulation feature have been found in the radio-quiescent mode.
Very recent results presented by \citet{2013_Hermsen} show synchronous
switching in the radio and X-ray emission properties. When the pulsar
is in a radio-bright mode, the X-rays exhibit only an unpulsed component.
On the other hand, when the pulsar is in a radio-quiet mode, the flux
of X-rays is doubled and a pulsed component is also visible.

\begin{comment}
\textasciitilde{}/Programs/studies/phd/lines/lines.py (plot\_0943),
910 data sets
\end{comment}

\begin{figure}[H]
\begin{centering}
\includegraphics{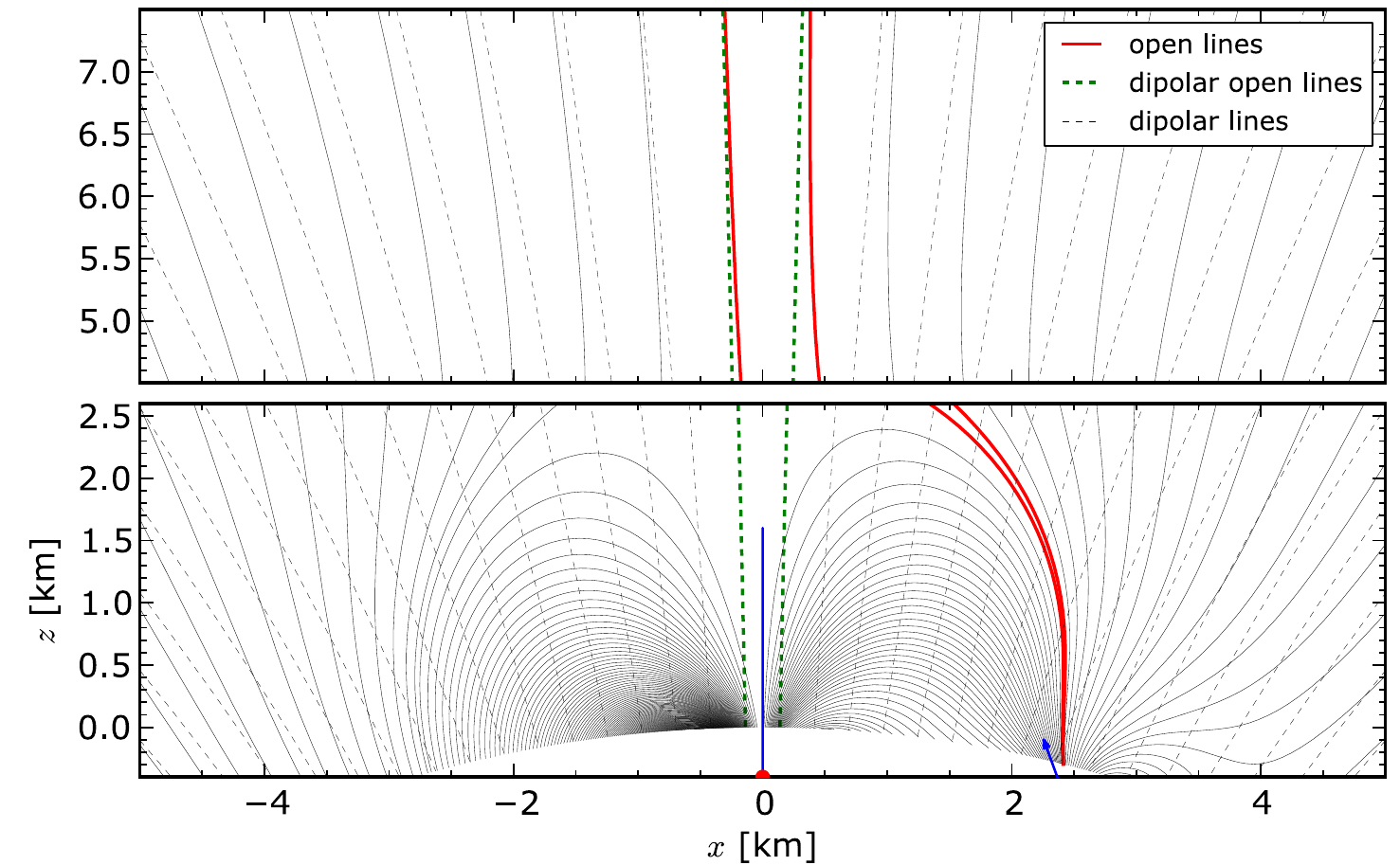}
\par\end{centering}

\caption[{Possible non-dipolar structure of the magnetic field lines {[}PSR
B0943+10{]}}]{Possible non-dipolar structure of the magnetic field lines of PSR
B0943+10.\protect \linebreak{}
The structure was obtained using two crust anchored anomalies located
at:\protect \linebreak{}
${\bf r_{1}}=\left(0.96R,\,0^{\circ},\,0^{\circ}\right)$, ${\bf r_{2}}=\left(0.96R,\,15^{\circ},\,0^{\circ}\right)$,
with the dipole moments \protect \linebreak{}
${\bf m_{1}}=\left(2.0\times10^{-2}d,\,180^{\circ},\,0^{\circ}\right)$,
${\bf m_{2}}=\left(6\times10^{-3}d,\,20^{\circ},\,180^{\circ}\right)$
respectively (blue arrows). The influence of the anomalies is negligible
at distances $D\gtrsim4.5R$, where $B_{{\rm m}}/B_{{\rm d}}\approx m/d=2\times10^{-2}$
(top panel). For more details on the polar cap region see Figure \ref{fig:model.b0943_zoom}.\label{fig:model.b0943}}
\end{figure}

\vspace*{0.5cm}

\begin{comment}
\textasciitilde{}/Programs/studies/phd/lines/lines.py (plot\_0943),
910 data sets
\end{comment}

\begin{figure}[H]
\begin{centering}
\includegraphics{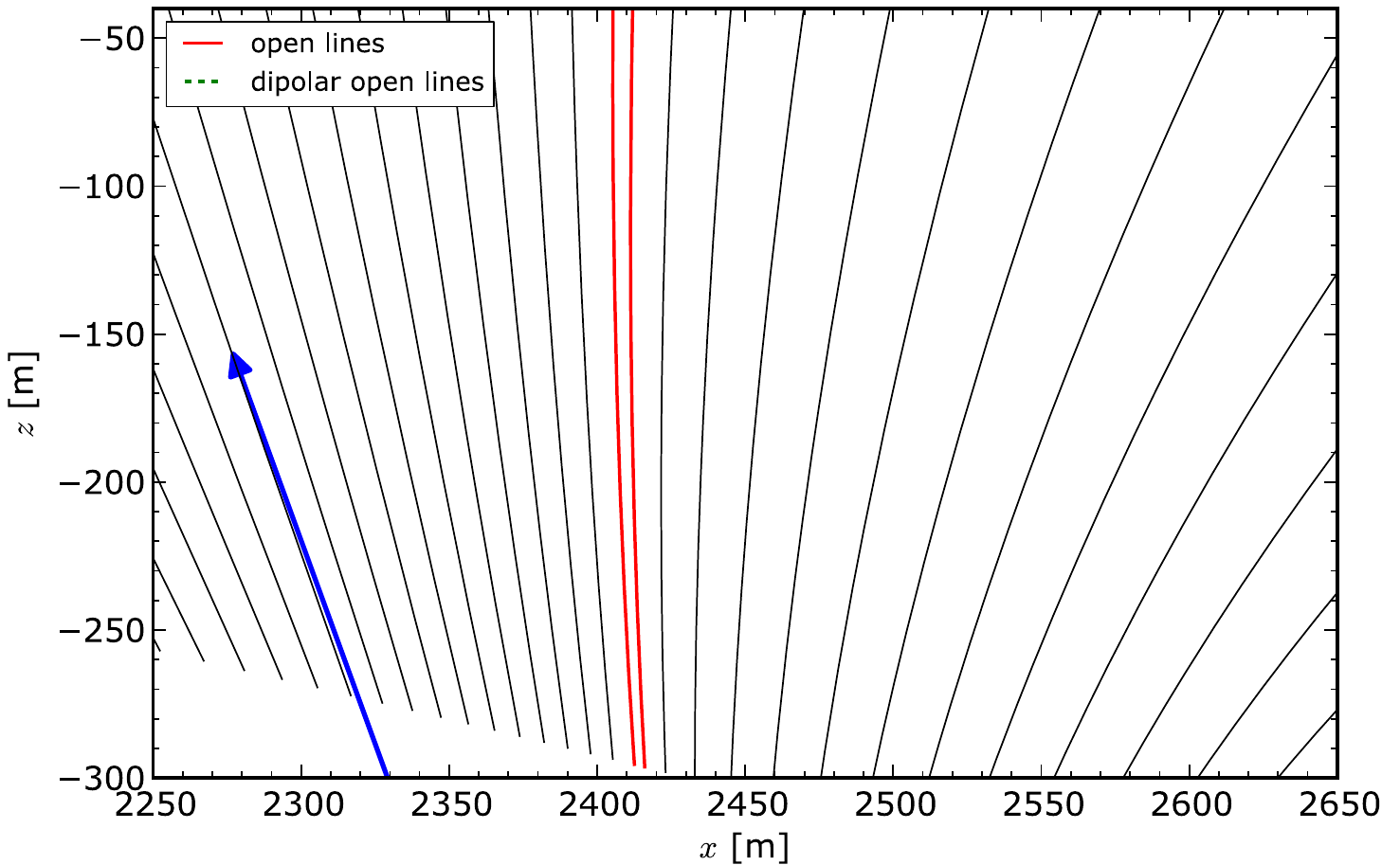}
\par\end{centering}

\caption[{Zoom of the polar cap region {[}PSR B0943+10{]}}]{Zoom of the polar cap region of PSR B0943+10. See Figure \ref{fig:model.b0943}
for a description. \label{fig:model.b0943_zoom}}
\end{figure}

\vspace*{0.5cm}

\begin{comment}
\textasciitilde{}/Programs/studies/phd/lines/lines.py (curvature\_0943),
912 data set
\end{comment}

\begin{figure}[H]
\begin{centering}
\includegraphics{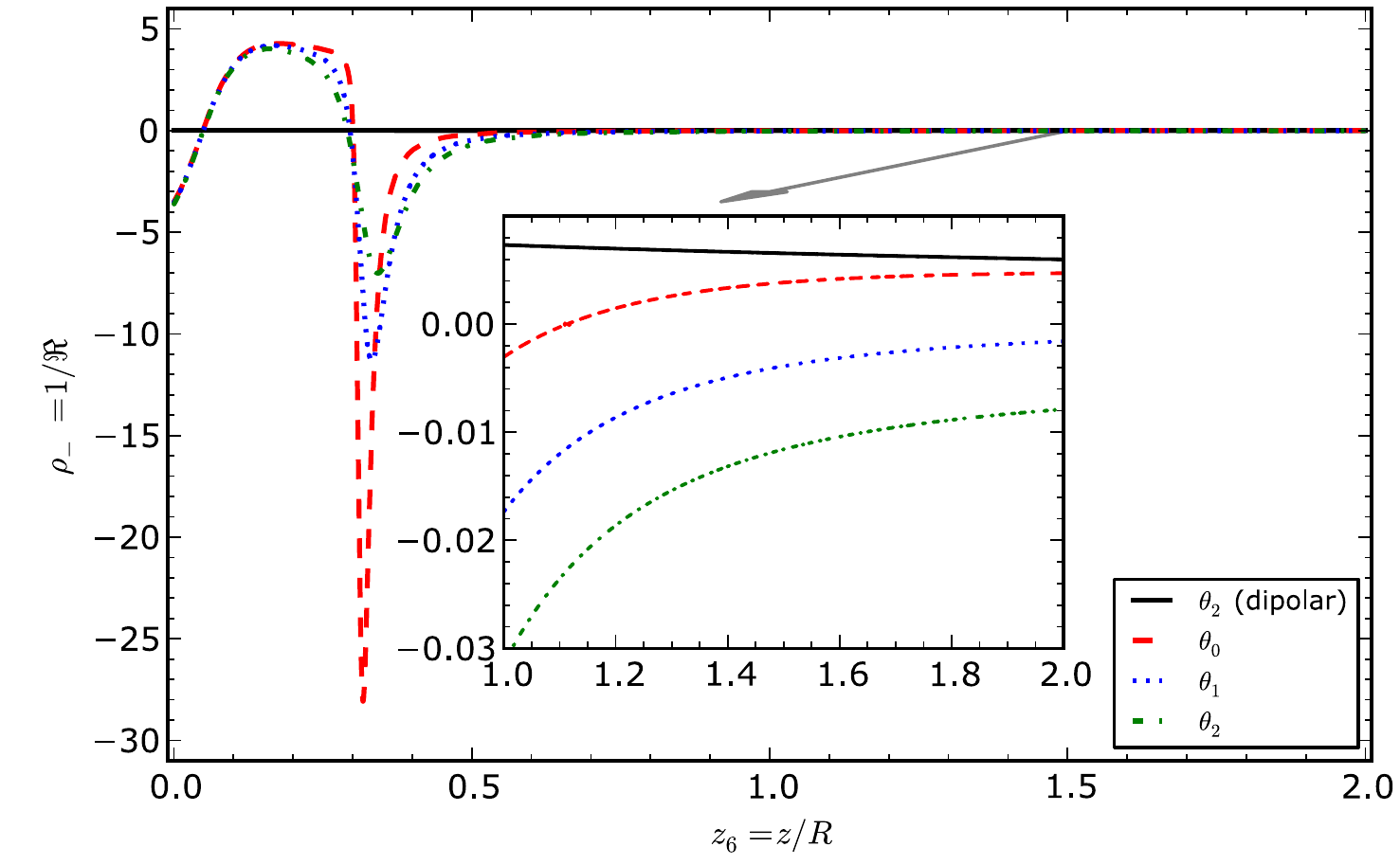}
\par\end{centering}

\caption[{Curvature of the open magnetic field lines {[}PSR B0943+10{]}}]{Dependence of a curvature of the open magnetic field lines on the
distance from the stellar surface for PSR B0943+10. The distance is
in units of stellar radius ($z_{6}=z/R$) and the curvature of the
magnetic field lines is $\rho_{-6}=1/\Re_{6}=\rho/\left(10^{-6}\,{\rm cm}^{-1}\right)$.
\label{fig:model.b0943_curva}}
\end{figure}

\clearpage{}

\subsection{PSR B0950+08\label{sec:model.0950}}

Pulsar B0950+08 is one of the strongest pulsed radio sources in the
metre wavelength range. The pulsar radiation also exhibits an interpulse
located at $152^{\circ}$ from the main pulse \citep{1992_Smirnova}.
Based on the period $P=1.1\,{\rm s}$ and its first derivative $\dot{P}_{-15}\approx3.5$,
we can estimate the pulsar's characteristic age $\tau_{c}=17.5\,{\rm Myr}$.
PSR B0950+08 has a relatively weak dipolar component of magnetic field
$B_{{\rm d}}=0.5\times10^{12}$. For this pulsar the distance $D=0.26\,{\rm kpc}$
was estimated using the parallax method.

PSR B0950+08 was detected in the ultraviolet-optical range ($2400-4600\,\AA$)
by \citet{1996_Pavlov} with the \textit{Hubble Space Telescope}.
Further observations suggest that the optical radiation of the pulsar
is most likely of a nonthermal origin \citep{2002_Mignani,2004_Zarikov}.

X-rays from PSR B0950+08 were first detected with the \textit{ROSAT}
by \citet{1994_Manning} ($\sim55$ source counts). Further X-ray
observations revealed pulsations of the X-ray flux at the radio period
of the pulsar \citep{2004_Zavlin}. The X-ray spectrum manifests two
components (thermal and nonthermal). Which of the two components dominates
the spectrum depends on the radiation pattern of the nonthermal component
(isotropic or anisotropic). Due to the poor quality of the X-ray data,
the connection of the optical and X-ray spectra remained unclear.

\begin{comment}
\textasciitilde{}/Programs/studies/phd/lines/lines.py (plot\_b0950),
355, 356 data sets
\end{comment}

\begin{figure}[H]
\begin{centering}
\includegraphics{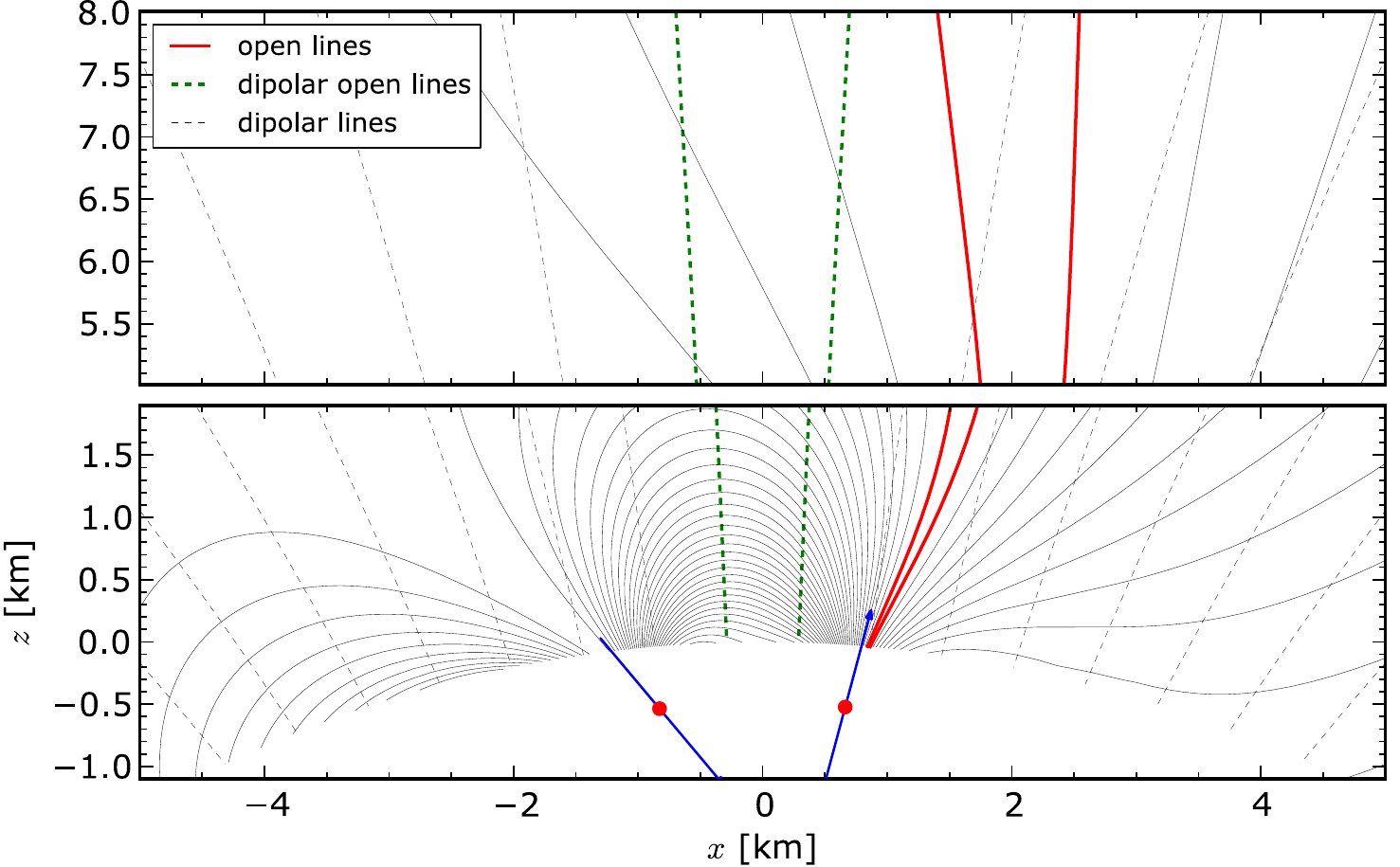}
\par\end{centering}

\caption[{Possible non-dipolar structure of the magnetic field lines {[}PSR
B0950+08{]}}]{Possible non-dipolar structure of the magnetic field lines of PSR
B0950+08. The structure was obtained using two crust anchored anomalies
located at: ${\bf r_{1}}=\left(0.95R,\,4^{\circ},\,0^{\circ}\right)$,
${\bf r_{2}}=\left(0.95R,\,5^{\circ},\,180^{\circ}\right)$, with
the dipole moments ${\bf m_{1}}=\left(5.9\times10^{-2}d,\,15^{\circ},\,0^{\circ}\right)$,
${\bf m_{2}}=\left(5.9\times10^{-2}d,\,140^{\circ},\,0^{\circ}\right)$
respectively (blue arrows). The influence of the anomalies is negligible
at distances $D\gtrsim5.0R$, where $B_{{\rm m}}/B_{{\rm d}}\approx m/d=5.9\times10^{-2}$
(top panel). For more details on the polar cap region see Figure \ref{fig:model.b0950_zoom}.\label{fig:model.b0950}}
\end{figure}

\vspace*{0.5cm}

\begin{comment}
\textasciitilde{}/Programs/studies/phd/lines/lines.py (plot\_b0950\_zoom),
355 data set
\end{comment}

\begin{figure}[H]
\begin{centering}
\includegraphics{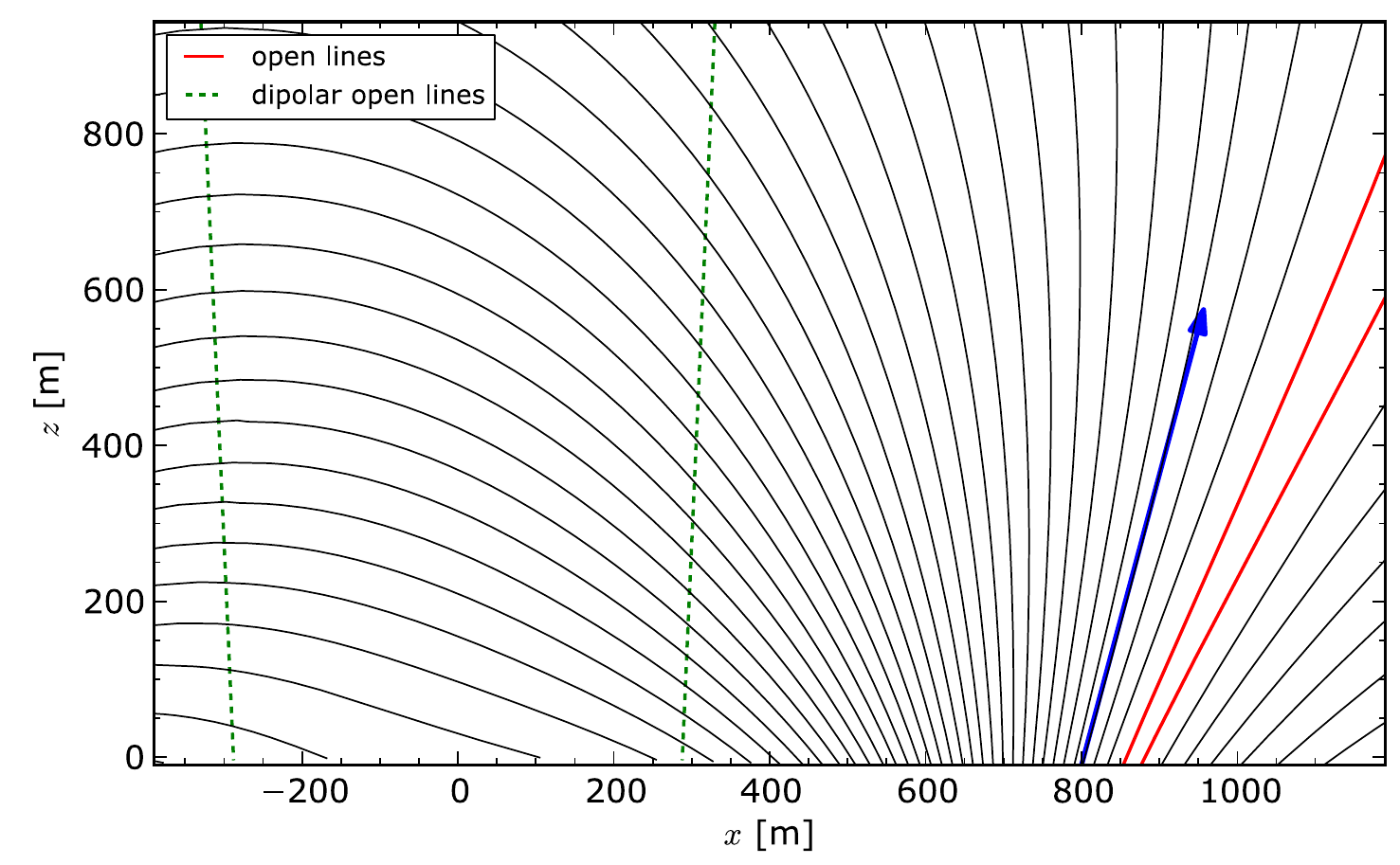}
\par\end{centering}

\caption[{Zoom of the polar cap region {[}PSR B0950+08{]}}]{Zoom of the polar cap region of PSR B0950+08. See Figure \ref{fig:model.b0950}
for a description. \label{fig:model.b0950_zoom}}
\end{figure}

\vspace*{0.5cm}

\begin{comment}
\textasciitilde{}/Programs/studies/phd/lines/lines.py (curvature\_b0959),
357 data set
\end{comment}

\begin{figure}[H]
\begin{centering}
\includegraphics{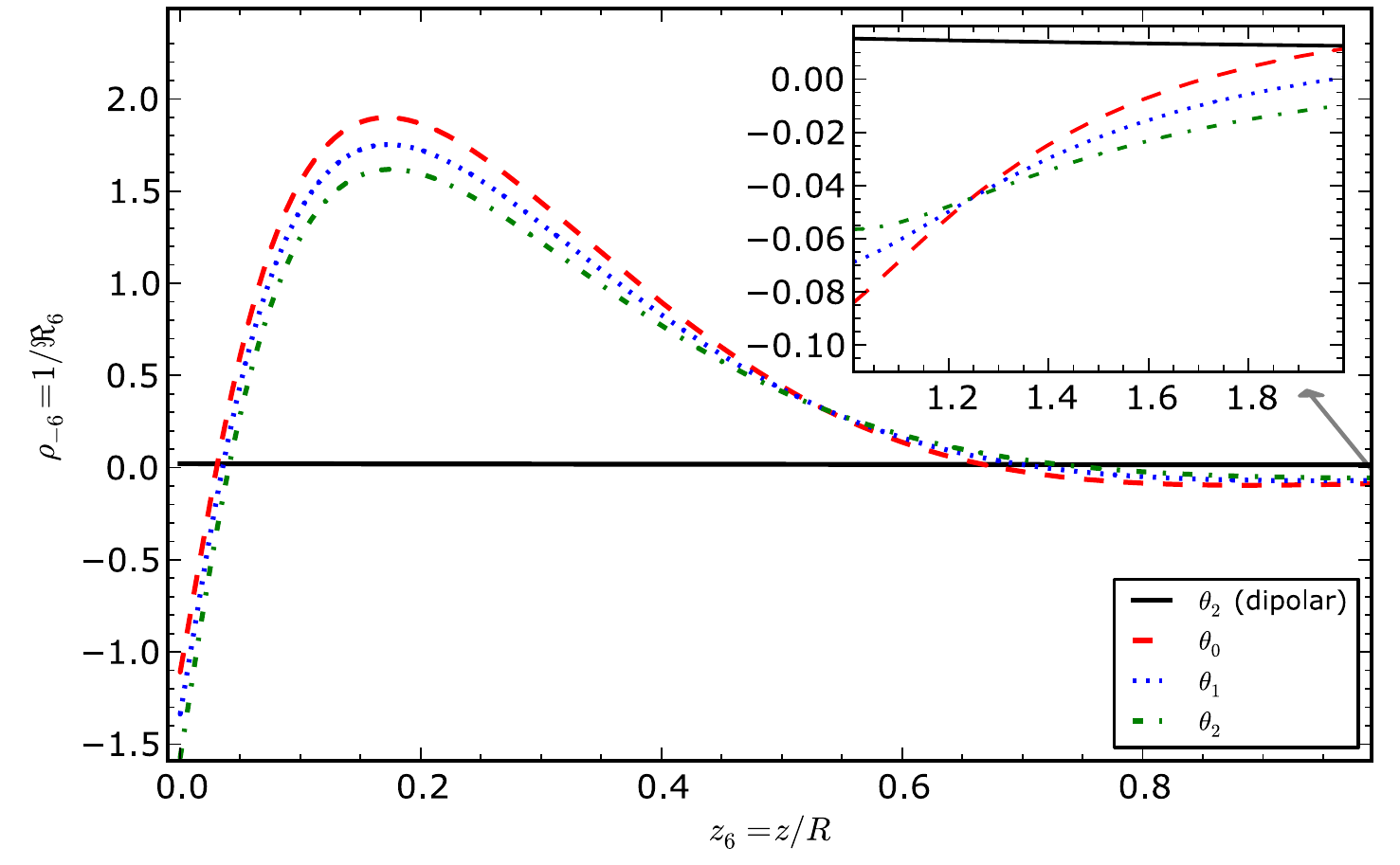}
\par\end{centering}

\caption[{Curvature of the open magnetic field lines {[}PSR B0950+08{]}}]{Dependence of a curvature of the open magnetic field lines on the
distance from the stellar surface for PSR B0950+08. The distance is
in units of stellar radius ($z_{6}=z/R$) and the curvature of the
magnetic field lines is $\rho_{-6}=1/\Re_{6}=\rho/\left(10^{-6}\,{\rm cm}^{-1}\right)$.
\label{fig:model.b0950_curva}}
\end{figure}

\clearpage{}

\subsection{PSR B1133+16\label{sec:model.1133}}

Pulsar B1133+16 is one of the brightest pulsating radio sources in
the Northern hemisphere \citep{2000_Maron}. The relatively long pulse
period $P=1.19\,{\rm s}$ and its first derivative $\dot{P}_{-15}\approx3.5$
result in the following inferred physical properties: $B_{{\rm d}}=4.3\times10^{12}\,{\rm G}$,
$\tau_{c}=5.04\,{\rm Myr}$. The pulsar profile exhibits a classic
double peak along with the usual S-shaped polarisation-angle traverse.
The pulsar also shows the phenomenon of drifting subpulses but only
for some finite time-spans, outside of which the behaviour of individual
pulses is chaotic \citep{2012_Honnappa}. 

PSR B1133+16 is located at a high galactic latitude, thus implying
a low interstellar extinction \citep{1998_Schlegel}. \citet{2008_Zharikov}
suggested a possible optical counterpart with brightness $B=28^{^{{\rm mag}}}$.

X-ray observations performed by \citet{2005_Kargaltsev} with the
\textit{Chandra} result in a small number of counts ($33$ counts
from over $17\,{\rm ks}$), thus the X-ray spectrum can be described
by various models. The photon statistics are so low that they allowed
only separate fits for the thermal (BB) and nonthermal (PL) components.

\begin{comment}
\textasciitilde{}/Programs/studies/phd/lines/lines.py (plot\_1133),
340, 341 data sets
\end{comment}

\begin{figure}[H]
\begin{centering}
\includegraphics{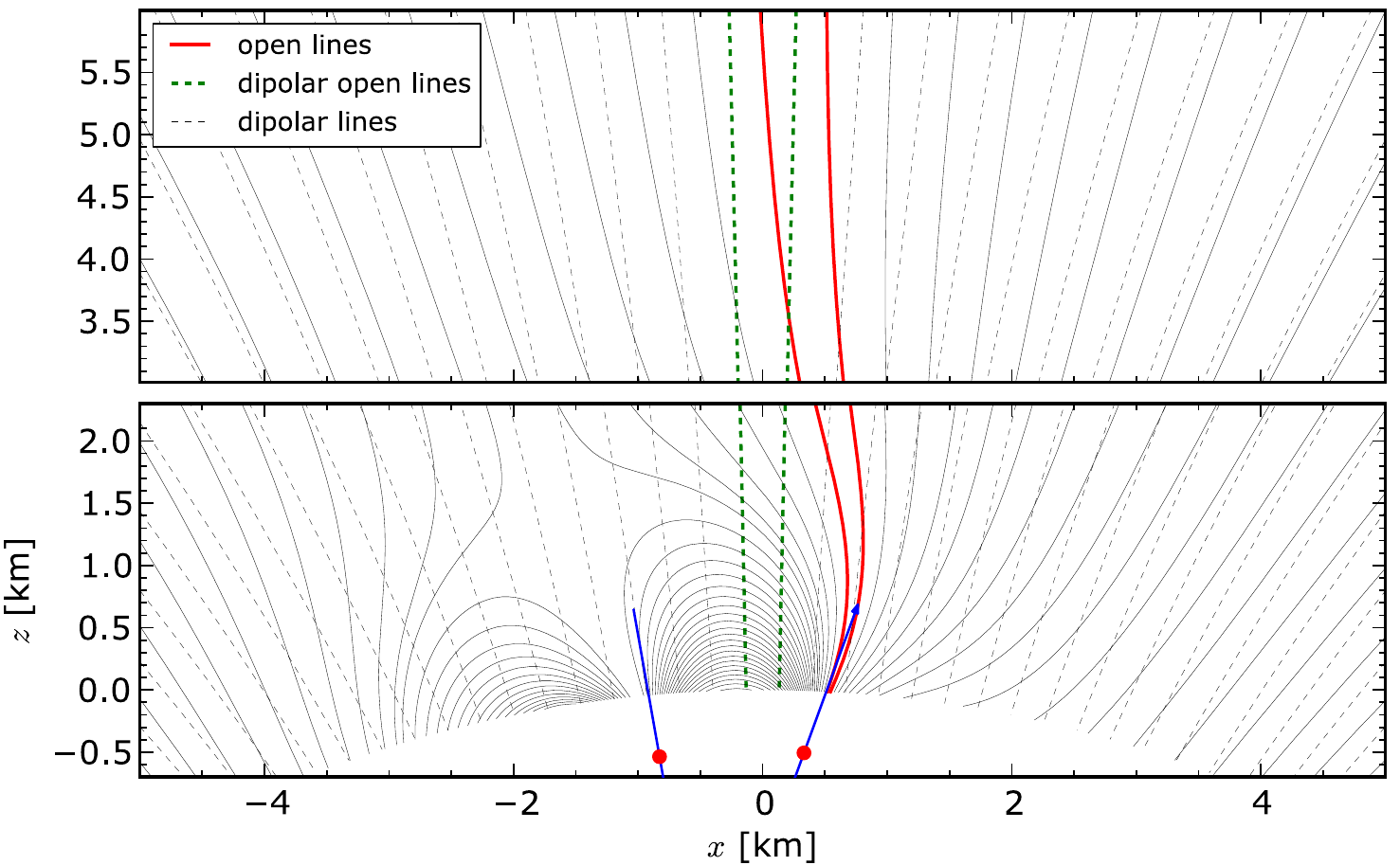}
\par\end{centering}

\caption[{Possible non-dipolar structure of the magnetic field lines {[}PSR
B1133+16{]}}]{Possible non-dipolar structure of the magnetic field lines of PSR
B1133+16.\protect \linebreak{}
The structure was obtained using two crust anchored anomalies located
at:\protect \linebreak{}
${\bf r_{1}}=\left(0.95R,\,2^{\circ},\,0^{\circ}\right)$, ${\bf r_{2}}=\left(0.95R,\,5^{\circ},\,180^{\circ}\right)$,
with the dipole moments\protect \linebreak{}
${\bf m_{1}}=\left(8\times10^{-3}d,\,20^{\circ},\,0^{\circ}\right)$,
${\bf m_{2}}=\left(8\times10^{-3}d,\,170^{\circ},\,0^{\circ}\right)$
respectively (blue arrows). The influence of the anomalies is negligible
at distances $D\gtrsim4.2R$, where $B_{{\rm m}}/B_{{\rm d}}\approx m/d=5\times10^{-2}$
(top panel). For more details on the polar cap region see Figure \ref{fig:model.1133_zoom}.\label{fig:model.1133}}
\end{figure}

\vspace*{0.5cm}

\begin{comment}
\textasciitilde{}/Programs/studies/phd/lines/lines.py (plot\_1133\_zoom),
340 data set
\end{comment}

\begin{figure}[H]
\begin{centering}
\includegraphics{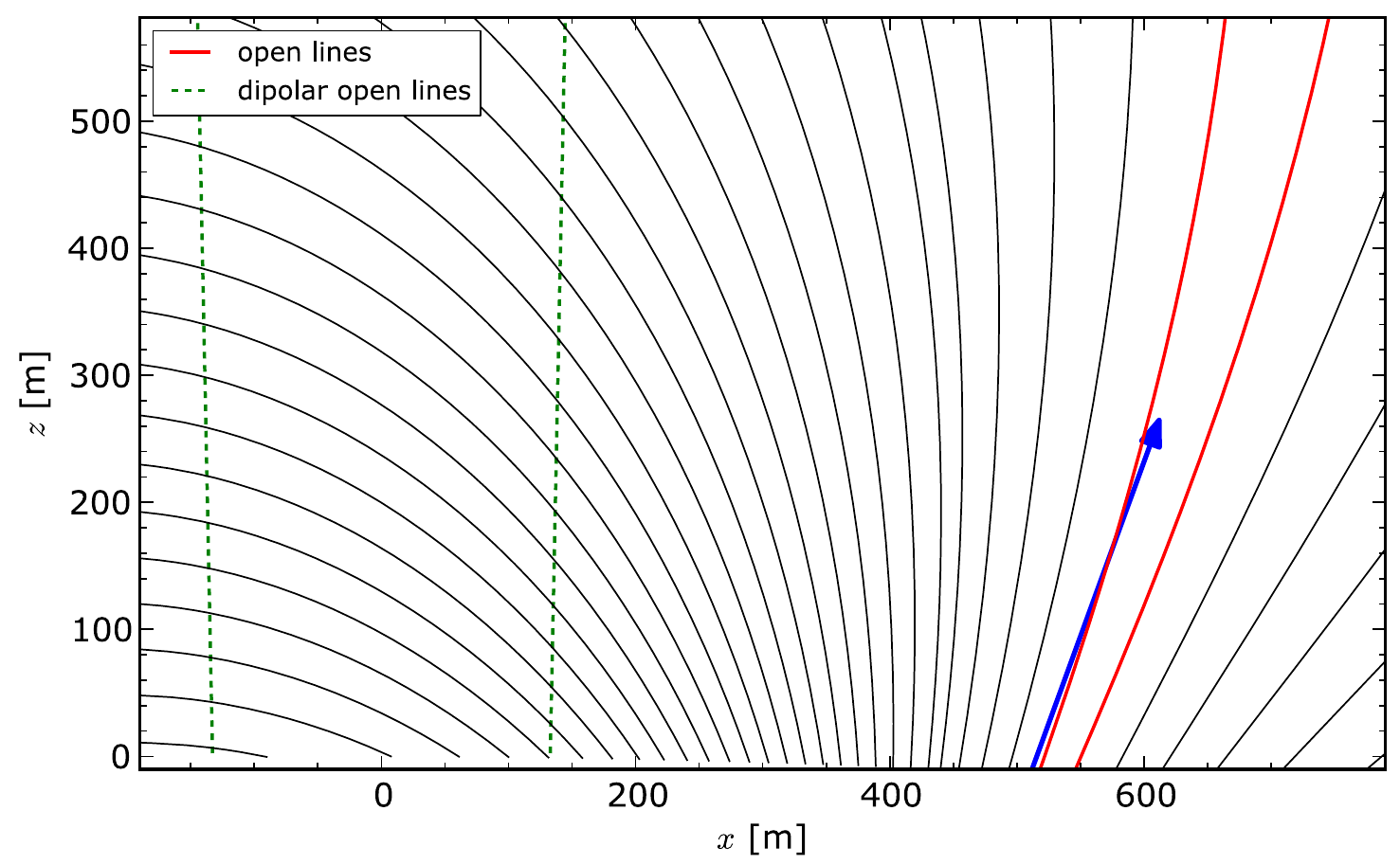}
\par\end{centering}

\caption[{Zoom of the polar cap region {[}PSR B1133+16{]}}]{Zoom of the polar cap region of PSR B1133+16. See Figure \ref{fig:model.1133}
for a description. \label{fig:model.1133_zoom}}
\end{figure}

\vspace*{0.5cm}

\begin{comment}
\textasciitilde{}/Programs/studies/phd/lines/lines.py (curvature\_1133),
343 data set
\end{comment}

\begin{figure}[H]
\begin{centering}
\includegraphics{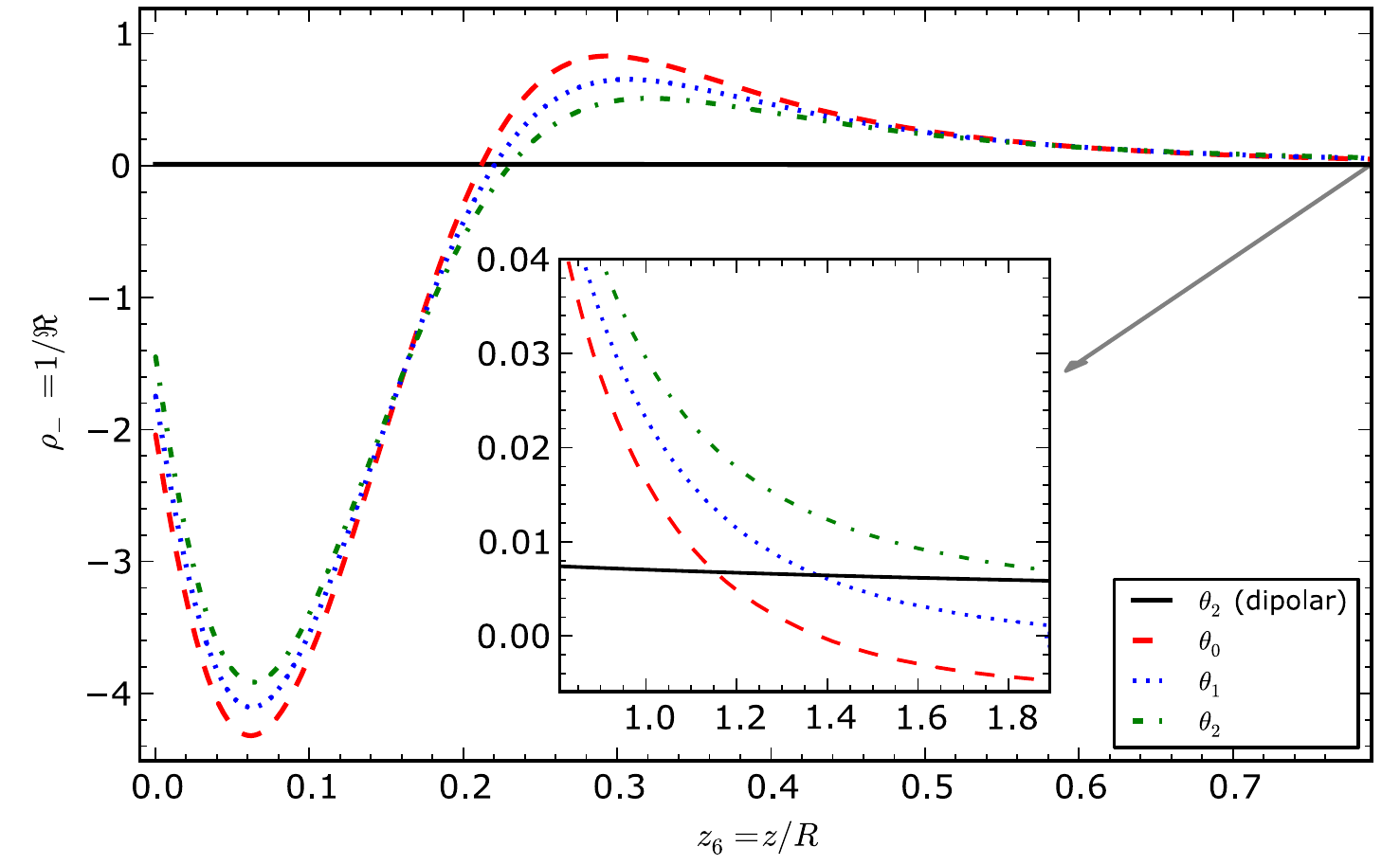}
\par\end{centering}

\caption[{Curvature of the open magnetic field lines {[}PSR B1133+16{]}}]{Dependence of a curvature of the open magnetic field lines on the
distance from the stellar surface for PSR B1133+16. The distance is
in units of stellar radius ($z_{6}=z/R$) and the curvature of the
magnetic field lines is $\rho_{-6}=1/\Re_{6}=\rho/\left(10^{-6}\,{\rm cm}^{-1}\right)$.
\label{fig:model.1133_curva}}
\end{figure}

\clearpage{}

\subsection{PSR B1929+10\label{sec:model.1929}}

With a pulse period of $P=0.23\,{\rm s}$ and a period derivative
of $\dot{P}_{-15}\approx1.2$, the pulsar's characteristic age is
determined to be $\tau_{c}=3.1\,{\rm Myr}$. These spin parameters
imply a dipolar component of the magnetic field at the neutron star
magnetic poles $B_{{\rm d}}=1.0\times10^{12}$. The distance to the
pulsar $D=0.36\,{\rm kpc}$ was estimated using the parallax.

\citet{1996_Pavlov} identified a candidate optical counterpart of
PSR B1929+10 with brightness $U\sim25.7^{^{{\rm mag}}}$, which was
later confirmed by proper motion measurements performed by \citet{2002_Mignani}. 

The X-ray pulse profile of PSR B1929+10 consists of a single, broad
peak which is in contrast with the sharp radio one of \citet{2008_Misanovic}.
The two-component spectral fit (BB+PL) suggests that both the thermal
and nonthermal luminosities are of the same order. The derived surface
temperature $T_{{\rm s}}=4.5\,{\rm MK}$ and the surface magnetic
field $B_{{\rm s}}=1.3\times10^{14}\,{\rm G}$ do not coincide with
the theoretical curve $T_{{\rm s}}-B_{{\rm s}}$ of the critical temperature
calculated by \citet{2008_Medin}. We believe that this inconsistency
can be removed by adding an additional blackbody component (the whole
surface or the warm spot radiation).

\begin{comment}
\textasciitilde{}/Programs/studies/phd/lines/lines.py (plot\_1929),
322 +324 data sets
\end{comment}

\begin{figure}[H]
\begin{centering}
\includegraphics{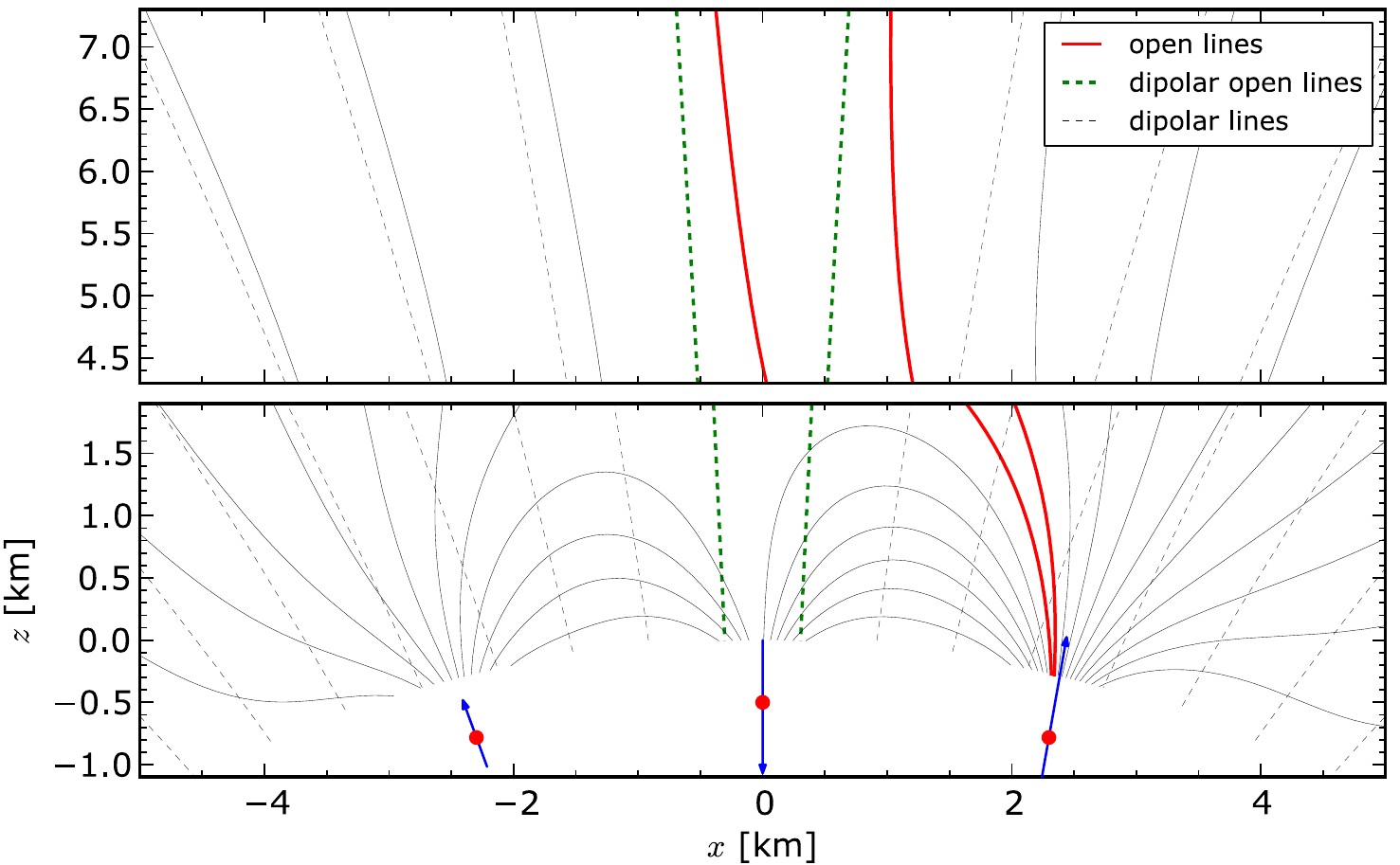}
\par\end{centering}

\caption[{Possible non-dipolar structure of the magnetic field lines {[}PSR
B1929+10{]}}]{Possible non-dipolar structure of the magnetic field lines of PSR
B1929+10.\protect \linebreak{}
The structure was obtained using two crust anchored anomalies located
at:\protect \linebreak{}
${\bf r_{1}}=\left(0.95R,\,14^{\circ},\,180^{\circ}\right)$, ${\bf r_{2}}=\left(0.95R,\,0^{\circ},\,0^{\circ}\right)$,
${\bf r_{3}}=\left(0.95R,\,14^{\circ},\,0^{\circ}\right)$, with the
dipole moments\protect \linebreak{}
${\bf m_{1}}=\left(1\times10^{-2}d,\,20^{\circ},\,180^{\circ}\right)$,
${\bf m_{2}}=\left(2\times10^{-2}d,\,180^{\circ},\,0^{\circ}\right)$,
${\bf m_{3}}=\left(3\times10^{-2}d,\,10^{\circ},\,0^{\circ}\right)$
respectively (blue arrows). The influence of the anomalies is negligible
at distances $D\gtrsim4.5R$, where $B_{{\rm m}}/B_{{\rm d}}\approx m/d=3\times10^{-2}$
(top panel). For more details on the polar cap region see Figure \ref{fig:model.b1929_zoom}.\label{fig:model.b1929}}
\end{figure}

\vspace*{0.5cm}

\begin{comment}
\textasciitilde{}/Programs/studies/phd/lines/lines.py (plot\_1929\_zoom),
315 data sets
\end{comment}

\begin{figure}[H]
\begin{centering}
\includegraphics{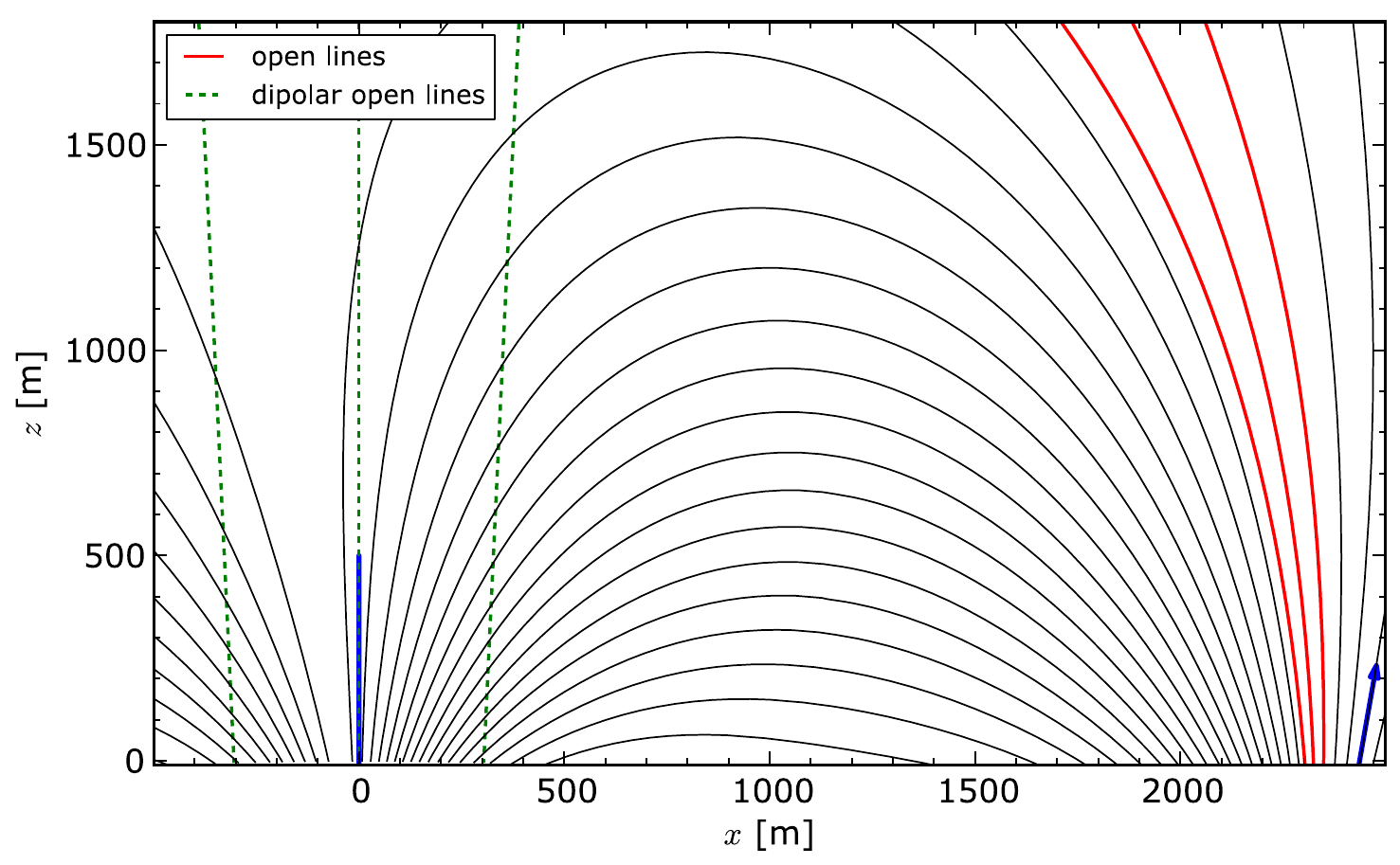}
\par\end{centering}

\caption[{Zoom of the polar cap region {[}PSR B1929+10{]}}]{Zoom of the polar cap region of PSR B1929+10. See Figure \ref{fig:model.b1929}
for a description. \label{fig:model.b1929_zoom}}
\end{figure}

\vspace*{0.5cm}

\begin{comment}
\textasciitilde{}/Programs/studies/phd/lines/lines.py (curvature\_1929),
315 data set
\end{comment}

\begin{figure}[H]
\begin{centering}
\includegraphics{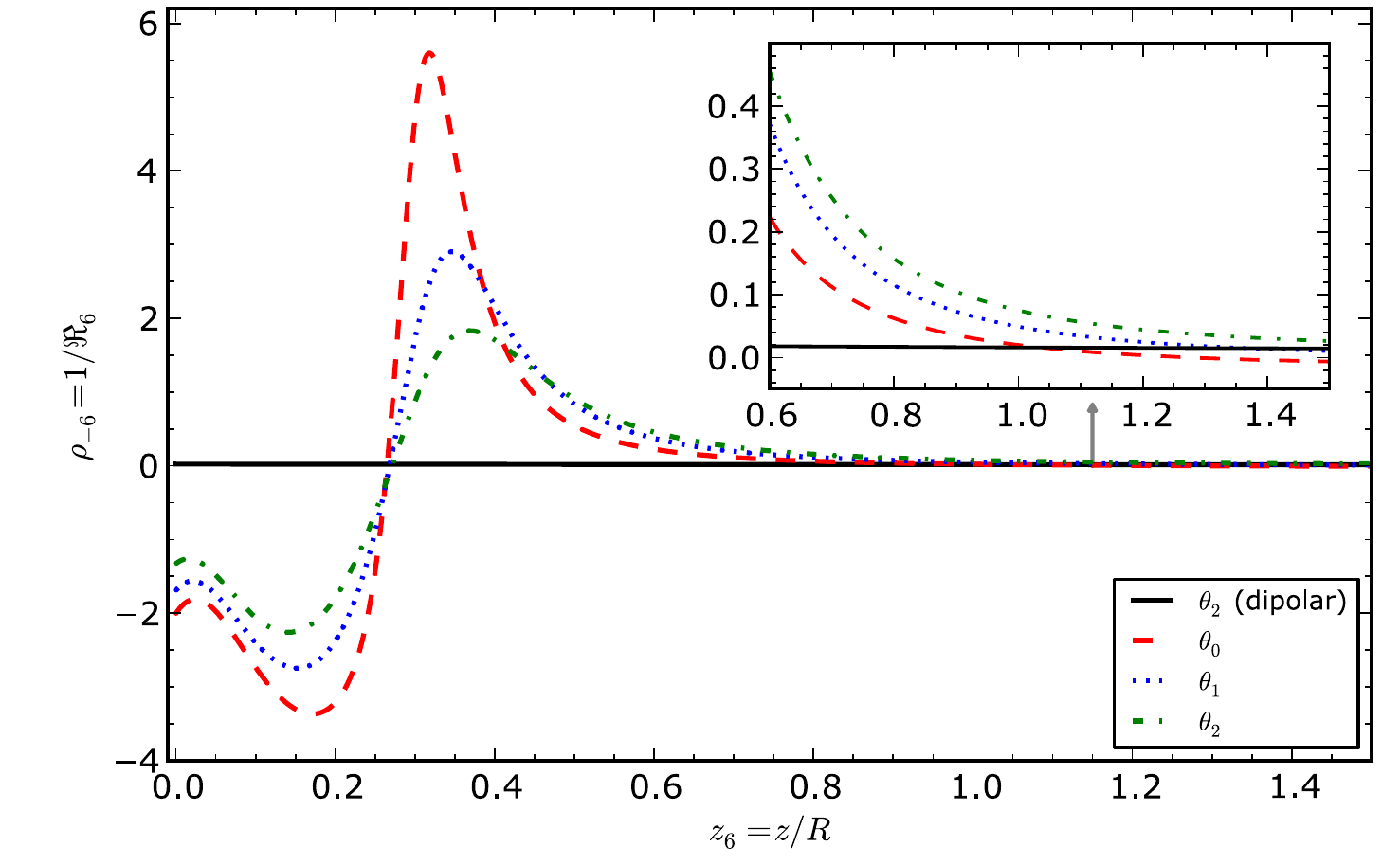}
\par\end{centering}

\caption[{Dependence of the curvature of open magnetic field lines {[}PSR B1929+10{]}}]{Dependence of a curvature of the open magnetic field lines on the
distance from the stellar surface for PSR B1929+10. The distance is
in units of stellar radius ($z_{6}=z/R$) and the curvature of the
magnetic field lines is $\rho_{-6}=1/\Re_{6}=\rho/\left(10^{-6}\,{\rm cm}^{-1}\right)$.\label{fig:model.b1929_curva}}
\end{figure}

\chapter{Partially Screened Gap\label{chap:psg}}

\thispagestyle{headings}
\renewcommand{\baselinestretch}{1.23}\normalsize

The charge-depleted inner acceleration region above the polar cap
can be formed if a local charge density differs from the co-rotational
charge density \citep{1969_Goldreich}. We assume that the crust of
the neutron stars mainly consists of iron $\left({\rm _{26}^{56}Fe}\right)$
formed at the neutron star’s birth (e.g. \citealp{2001_Lai}). Depending
on the mutual orientation of ${\bf \Omega}$ and $\boldsymbol{\mu}$,
the stellar surface at the polar caps is either positively (${\bf \Omega}\cdot\boldsymbol{\mu}<0$)
or negatively (${\bf \Omega}\cdot\boldsymbol{\mu}>0$) charged. Therefore,
the charge depletion above the polar cap depends on the binding energy
of either the positive ${\rm _{26}^{56}Fe}$ ions or electrons. In
this thesis we consider the case of positively charged polar caps
(${\bf \Omega}\cdot\boldsymbol{\mu}<0$). We assume that due to the
high cohesive energy of iron ions, the positive charges cannot be
supplied at a rate that would compensate for the inertial outflow
through the light cylinder (see \citet{2006_Medin,2007_Medin,2007_Gil_b}).
This is actually possible if the surface temperature $T_{{\rm s}}$
is below the critical value $T_{{\rm crit}}$. Since the number density
of the iron ions in the neutron star crust is many orders of magnitude
larger than the co-rotational charge density (the so-called Goldreich-Julian
density) $\rho_{{\rm GJ}}={\bf \Omega}\cdot{\bf B}/\left(2\pi c\right)$,
then a thermionic emission from the polar cap surface is not simply
described by the usual condition $\epsilon_{{\rm i}}\approx kT_{{\rm s}}$
, where $\epsilon_{i}$ is the cohesive energy and/or work function,
$T_{{\rm s}}$ is the actual surface temperature, and $k$ is the
Boltzman constant. The outflow of iron ions can be described in the
form (\citealt{2003_Gil} and references therein)

\begin{equation}
\frac{\rho_{{\rm i}}}{\rho_{{\rm GJ}}}\approx\left(C_{{\rm i}}-\frac{\epsilon_{i}}{kT_{{\rm s}}}\right),
\end{equation}
where $\rho_{{\rm i}}\leq\rho_{{\rm GJ}}$ is the charge density of
the outflowing ions. As soon as the surface temperature $T_{{\rm s}}$
reaches the critical value

\begin{equation}
T_{{\rm crit}}=\frac{\epsilon_{i}}{C_{{\rm i}}k},
\end{equation}
the ion outflow reaches the maximum value $\rho_{{\rm i}}=\rho_{{\rm GJ}}$.
The numerical coefficient\linebreak{}
 $C_{{\rm i}}=30\pm3$ is determined from the tail of the exponential
function with an accuracy of about 10\%. Thus, for a given value of
the cohesive energy, the critical temperature $T_{{\rm crit}}$ is
also estimated within an accuracy of about 10\%. The cohesive energy
is mainly defined by the strength of the magnetic field and was calculated
by \citet{2006_Medin,2007_Medin}. 

\onehalfspacing

\section{The Model}

As it follows from the X-ray observations (see Section \ref{sec:x-ray.thermal}),
the temperature of the hot spot (which is associated with the actual
polar cap) is more than $10^{6}\,{\rm K}$. As we mentioned above,
in order to sustain such a high temperature bombardment by the backstreaming
particles is required. But particle acceleration (and therefore the
surface heating) is possible only if $T_{{\rm s}}<T_{{\rm crit}}$.
\citet{2003_Gil} introduced the model of the Partially Screened Gap
to describe the polar gap sparking discharge specifically under such
circumstances.

The PSG model assumes the existence of heavy iron ions (${\rm _{26}^{56}Fe}$)
with a density near but still below the co-rotational charge density
($\rho_{{\rm GJ}}$), thus the actual charge density causes partial
screening of the potential drop just above the polar cap. The degree
of screening can be described by screening factor

\begin{equation}
\eta=1-\rho_{{\rm i}}/\rho_{{\rm GJ}}.
\end{equation}
where $\rho_{{\rm i}}$ is the charge density of the heavy ions in
the gap. The thermal ejection of ions from the surface causes partial
screening of the acceleration potential drop 
\begin{equation}
\Delta V=\eta\Delta V_{{\rm max}},\label{eq:psg.d_v1}
\end{equation}
where $\Delta V_{{\rm max}}$ is the potential drop in a vacuum gap.
We can express the dependence of the critical temperature on the pulsar
parameters by fitting to the numerical calculations of \citet{2007_Medin}
\begin{equation}
T_{{\rm crit}}=1.6\times10^{4}\left\{ \left[\left(P\dot{P}_{-15}\right)^{0.5}b\right]^{1.1}+17.7\right\} ,\label{eq:psg.t_s}
\end{equation}
or $T_{{\rm crit}}=1.1\times10^{6}\left(B_{14}^{1.1}+0.3\right)$,
where $B_{14}=B_{{\rm s}}/\left(10^{14}\,{\rm G}\right)$ , $B_{{\rm s}}=bB_{{\rm d}}$
is a surface magnetic field (applicable only if hot spot components
are observed, i.e. $b>1$).

The actual potential drop $\Delta V$ should be thermostatically regulated
and a quasi-equilibrium state should be established in which heating
due to the electron/positron bombardment is balanced by cooling due
to thermal radiation (see \citealt{2003_Gil} for more details). The
necessary condition for this quasi-equilibrium state is 
\begin{equation}
\sigma T_{{\rm s}}^{4}=\eta e\Delta Vcn_{{\rm GJ}},\label{eq:psg.heating_condition}
\end{equation}
where $\sigma$ is the Stefan-Boltzmann constant, $e$ - the electron
charge, and \linebreak{}
$n_{{\rm GJ}}=\rho_{{\rm GJ}}/e=1.4\times10^{11}b\dot{P}_{-15}^{0.5}P^{-0.5}$
is the co-rotational number density. The Goldreich-Julian co-rotational
number density can be expressed in terms of $B_{14}$ as 
\begin{equation}
n_{{\rm GJ}}=6.93\times10^{12}B_{14}P^{-1}.\label{eq:psg.n_gj}
\end{equation}
Here we assume that the density of backstreaming relativistic electrons
is $\eta n_{{\rm GJ}}$.

By using Equations \ref{eq:psg.heating_condition}, \ref{eq:psg.t_s}
and \ref{eq:psg.n_gj} we can express the acceleration potential drop
that satisfies the heating condition (Equation \ref{eq:psg.heating_condition})
as follows

\begin{equation}
\Delta V=7.3\times10^{5}\frac{\left(B_{14}^{1.1}+0.3\right)^{4}P}{\eta B_{14}}.\label{eq:psg.potential_heating}
\end{equation}

The above equation may suggest that the acceleration potential drop
is inversely proportional to the screening factor. In fact, it is
just the opposite (see Equations \ref{eq:psg.d_v1} and \ref{eq:psg.potential_drop}).

Knowing that $\Delta V=\gamma_{{\rm max}}mc^{2}/e$, where $m$ is
the mass of a particle (electron or positron), we can calculate the
maximum Lorentz factor of the primary particles in PSG as

\begin{equation}
\gamma_{{\rm max}}=450\frac{\left(B_{14}^{1.1}+0.3\right)^{4}P}{\eta B_{14}}.
\end{equation}

\subsection{Acceleration potential drop\label{sec:psg.potential_drop}}

As the actual polar cap is much smaller than the conventional polar
cap (see section \ref{sub:x-ray.thermal_observations}), we cannot
use the approximation proposed by \citet{1975_Ruderman} that the
gap height is of the same order as the gap width ($h\approx h_{\perp}$).
On the contrary, the small polar cap size and subpulse phenomenon
suggest that in the PSG model the spark half-width is considerably
smaller than the gap height ($h_{\perp}<h$). For such a regime we
need to recalculate a formula for the acceleration potential drop
$\Delta V$.

\begin{figure}[H]
\begin{centering}
\includegraphics[height=7.5cm]{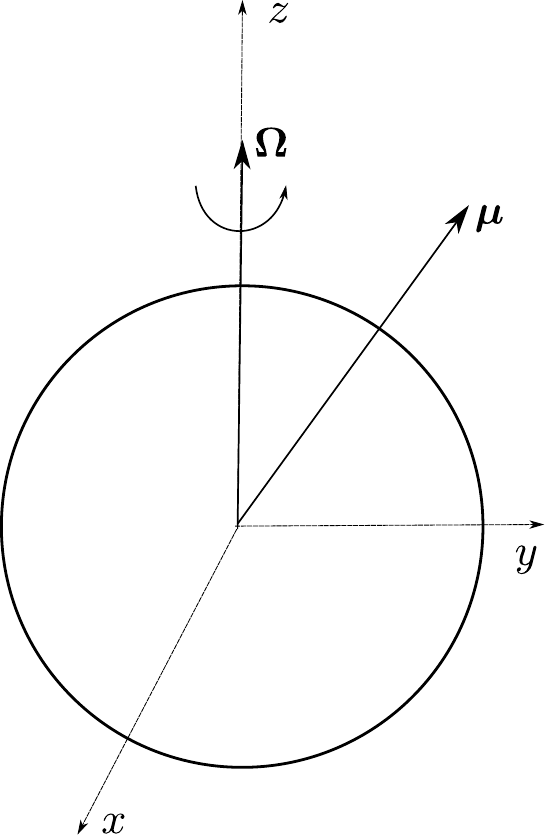}
\par\end{centering}

\caption[Co-rotating frame of reference (acceleration potential drop)]{Co-rotating frame of reference with the z-axis aligned with the angular
velocity ${\bf \Omega}$. The magnetic dipole moment $\boldsymbol{\mu}$
is constant in this frame of reference, thus $\partial{\bf B}/\partial t=0$.
\label{fig:psg.coordinate_star}}
\end{figure}

Let us consider a reference frame co-rotating with a star and with
the z-axis aligned with the star's angular velocity ${\bf \Omega}$
(see Figure \ref{fig:psg.coordinate_star}). 

Let us underline that we will neglect the effects of non-inertiality
of the co-rotating system. Thus, we assume that in any given moment
we have a system moving with a constant velocity.

In this co-rotating frame of reference we can write the spherical
components of an angular velocity as follows

\begin{equation}
{\bf \Omega}=\left(\Omega\cos\theta,\,-\Omega\sin\theta,\,0\right).
\end{equation}

Gauss's law in the co-rotating frame (after Lorentz transformations)
takes the form 

\begin{equation}
\nabla\cdot{\bf E}=4\pi\rho\left({\bf r}\right)-4\pi\left(\frac{{\bf \Omega}\cdot{\bf B}}{2\pi c}\right).\label{eq:psg.divergence}
\end{equation}

While Faraday's law of induction can be written as

\begin{equation}
\nabla\times{\bf E}=0.\label{eq:psg.curl}
\end{equation}

Note that if we consider a drift of plasma in the Inner Acceleration
Region (IAR), we should expect temporal variations of the magnetic
field ($\nabla\times{\bf E}=-\partial{\bf B}/\left(c\partial t\right)$)
\citep{1939_Schiff}, but as was shown by \citet{2012_Leeuwen}, even
if we consider fluctuations of the electric current of the order of
the Goldreich-Julian current $\rho_{{\rm GJ}}c$, the resulting variation
of the magnetic field is so small that $\nabla\times{\bf E}=0$ with
a high accuracy, and circulation of the non-co-rotational electric
field along a closed path is zero.

Equation \ref{eq:psg.divergence} in the spherical system of coordinates
has the following form

\begin{equation}
\frac{2}{r}E_{r}+\frac{\partial E_{r}}{\partial r}+\frac{\cos\theta}{r\sin\theta}E_{\theta}+\frac{1}{r}\frac{\partial E_{\theta}}{\partial\theta}+\frac{1}{r\sin\theta}\frac{\partial E_{\phi}}{\partial\phi}=4\pi\rho\left(r,\theta,\phi\right)-4\pi\left(\frac{{\bf \Omega}\cdot{\bf B}}{2\pi c}\right).
\end{equation}

The PSG model assumes the existence of ions in the IAR region that
affects the charge density. Using the screening factor, $\eta$, we
can write that

\[
\rho\left(r,\theta,\phi\right)=\left(1-\eta\right)\rho_{{\rm GJ}}\left(r,\theta,\phi\right)=\left(1-\eta\right)\frac{{\bf \Omega}\cdot{\bf B}}{2\pi c}.
\]

In general, $\eta$ depends on the curvature and strength of the magnetic
field, thus it varies across the polar cap, but we can still assume
that $\eta$ is approximately constant at least for a given spark.
Then 

\begin{equation}
\frac{2}{r}E_{r}+\frac{\partial E_{r}}{\partial r}+\frac{\cos\theta}{r\sin\theta}E_{\theta}+\frac{1}{r}\frac{\partial E_{\theta}}{\partial\theta}+\frac{1}{r\sin\theta}\frac{\partial E_{\phi}}{\partial\phi}=-4\pi\eta\left(\frac{B_{r}\Omega\cos\theta-B_{\theta}\Omega\sin\theta}{2\pi c}\right).
\end{equation}

Let us change the variables as follows: $r=R+z$ and $\theta=\alpha+\vartheta$.
Here $R$ is the stellar radius and $\alpha$ is the inclination angle
between the rotation and the magnetic axis.

\begin{multline}
\frac{2}{R+z}E_{r}+\frac{\partial E_{r}}{\partial z}+\frac{\cos\left(\alpha+\vartheta\right)}{\left(R+z\right)\sin\left(\alpha+\vartheta\right)}E_{\theta}+\frac{1}{R+z}\frac{\partial E_{\theta}}{\partial\vartheta}+\frac{1}{\left(R+z\right)\sin\left(\alpha+\vartheta\right)}\frac{\partial E_{\phi}}{\partial\phi}=\\
=-4\pi\eta\left(\frac{\left(B_{r}\Omega\cos\theta-B_{\theta}\Omega\sin\theta\right)}{2\pi c}\right).\label{eq:psg.potential_long}
\end{multline}

Assuming that $R\gg z$, which is correct as the gap height is less
than the stellar radius ($h\ll R$), $\alpha\gg\vartheta$, and $B_{r}\gg B_{\theta}$,
which is correct for the polar cap region, we can write Equation \ref{eq:psg.potential_long}
in the first approximation ($R\rightarrow\infty$) as follows

\begin{equation}
\frac{\partial E_{r}}{\partial z}+\frac{1}{R}\frac{\partial E_{\theta}}{\partial\vartheta}=-4\pi\eta\left(\frac{B_{r}\Omega\cos\theta}{2\pi c}\right).\label{eq:psg.potential_estiamte}
\end{equation}

Note that for spark widths considerably smaller than the stellar radius\linebreak{}
 $h_{\perp}\ll R$ ($\Delta\vartheta\approx h_{\perp}/R$) we can
write that $\frac{1}{R}\frac{\partial E_{\theta}}{\partial\vartheta}\gg\frac{\cot\left(\alpha+\vartheta\right)}{R}E_{\theta}$.

Let us now consider Faraday's law (Equation \ref{eq:psg.curl}). The
curl of an electric field in spherical coordinates can be written
as

\begin{equation}
\begin{split}\left({\bf \nabla}\times{\bf E}\right)_{r}= & \frac{1}{r\sin\theta}\left(\frac{\partial}{\partial\theta}\left(E_{\phi}\sin\theta\right)-\frac{\partial E_{\phi}}{\partial\phi}\right)=0,\\
\left({\bf \nabla}\times{\bf E}\right)_{\theta}= & \frac{1}{r}\left(\frac{1}{\sin\theta}\frac{\partial E_{r}}{\partial\phi}-\frac{\partial}{\partial r}\left(rE_{\phi}\right)\right)=0,\\
\left({\bf \nabla}\times{\bf E}\right)_{\phi}= & \frac{1}{r}\left(\frac{\partial}{\partial r}\left(rE_{\theta}\right)-\frac{\partial E{}_{r}}{\partial\theta}\right)=0.
\end{split}
\label{eq:psg.curl_system}
\end{equation}

Using the same change of variables we performed above ($r=R+z$ and
$\theta=\alpha+\vartheta$), the third equation of System \ref{eq:psg.curl_system}
can be written as

\begin{equation}
R\frac{\partial E_{\theta}}{\partial z}=\frac{\partial E_{r}}{\partial\vartheta}.
\end{equation}

From this equation in the zeroth approximation we can estimate the
variations of the electric field components as

\begin{equation}
R\Delta E_{\theta}\Delta\vartheta\approx\Delta E_{r}\Delta z.
\end{equation}

Since $h_{\perp}\ll R$ we can write that

\begin{equation}
\left\langle h_{\perp}E_{\theta}\right\rangle =\left\langle hE_{r}\right\rangle =\Delta V.\label{eq:psg.potential_est2}
\end{equation}

From Equation \ref{eq:psg.potential_estiamte} we can also briefly
estimate that

\begin{equation}
\frac{\Delta E_{r}}{h}+\frac{\Delta E_{\theta}}{h_{\perp}}=-4\pi\eta\left(\frac{B_{r}\Omega\cos\theta}{2\pi c}\right).
\end{equation}

Using Equations \ref{eq:psg.potential_estiamte} and \ref{eq:psg.potential_est2}
we can write that

\begin{equation}
\frac{\left\langle hE_{r}\right\rangle }{h^{2}}+\frac{\left\langle h_{\perp}E_{\theta}\right\rangle }{h_{\perp}^{2}}=\frac{\Delta V}{h^{2}}+\frac{\Delta V}{h_{\perp}^{2}}.
\end{equation}

Finally, we can estimate the potential drop in a spark region

\begin{equation}
\frac{\Delta V}{h^{2}}+\frac{\Delta V}{h_{\perp}^{2}}=\frac{2\eta B_{r}\Omega\cos\left(\alpha+\vartheta\right)}{c}.\label{eq:psg.delta_v_h_hperp}
\end{equation}

If we use the same assumption as \citet{1975_Ruderman}, i.e.: (1)
the spark half-width is of the same order as the gap height $h_{\perp}=h$,
(2) there is no ion extraction from the stellar surface ($\eta=1$),
and (3) the pulsar magnetic and rotation axes are aligned ($\alpha=0^{\circ}$),
we get: 
\begin{equation}
\Delta V_{{\rm RS}}=\frac{B_{r}\Omega}{c}h^{2}.
\end{equation}

Note that the potential drop defined by Equation \ref{eq:psg.delta_v_h_hperp}
differs from that used in the Standard Model by the screening factor
(as the presence of ions screens the gap) and by the factor of $\cos\left(\alpha+\vartheta\right)$
which also takes into account non-aligned pulsars. In our case the
polar cap size is much smaller than the conventional polar cap size.
It seems reasonable to also consider sparks with widths much smaller
than the gap height ($h_{\perp}\ll h$), in that case the potential
drop can be calculated as

\begin{equation}
\Delta V=\frac{2\eta B_{r}\Omega\cos\left(\alpha+\vartheta\right)}{c}h_{\perp}^{2}.
\end{equation}

Even for a relatively small inclination angle between the rotation
and magnetic axis, we can still write $\vartheta\ll\alpha$, thus

\begin{equation}
\Delta V=\frac{4\pi\eta B_{r}\cos\alpha}{cP}h_{\perp}^{2}.\label{eq:psg.potential_drop}
\end{equation}

\subsection{Acceleration path}

Since the exact dependence of the electric field on $z$ is unknown
we use the same linear approximation that \citet{1975_Ruderman} used.
In the frame of the PSG model as $h_{\perp}<h$ or even $h_{\perp}\ll h$,
we can use Equations \ref{eq:psg.potential_est2} and \ref{eq:psg.potential_drop}
to describe the component of the electric field along the magnetic
field line: 
\begin{equation}
E\approx\frac{8\pi\eta B_{{\rm s}}\cos\alpha}{cP}\frac{h_{\perp}^{2}}{h^{2}}\left(h-z\right),\label{eq:psg.acceleration_field}
\end{equation}
which vanishes at the top $z=h$. The Lorentz factor of particles
after passing distance $l_{{\rm acc}}$ can be calculated as follows

\begin{equation}
\gamma_{{\rm acc}}=\frac{e}{mc^{2}}\int_{z_{1}}^{z_{2}}Edz\approx\frac{8\pi\eta B_{{\rm s}}e\cos\alpha}{mc^{3}P}\frac{h_{\perp}^{2}}{h^{2}}\left(z_{2}-z_{1}\right)\left(h-\frac{z_{1}+z_{2}}{2}\right),
\end{equation}
where $m$ is the mass of a particle (electron or positron) and $z_{2}-z_{1}=l_{{\rm acc}}$.
Then we can approximate $z_{1}+z_{2}\approx h$, thus 
\begin{equation}
l_{{\rm acc,ap}}=\frac{\gamma_{{\rm acc}}mc^{3}P}{4\pi\eta B_{{\rm s}}e\cos\alpha}\frac{h}{h_{\perp}^{2}}.\label{eq:psg.acceleration}
\end{equation}

Assuming that a non-relativistic particle is accelerated from the
stellar surface ($z_{1}=0$, $\gamma_{0}=1$) we can calculate the
distance $l_{{\rm acc}}$ which it should pass to gain a Lorentz factor
$\gamma_{{\rm acc}}$:

\begin{equation}
l_{{\rm acc}}=h\left(1-\sqrt{1-\frac{2\gamma}{\ell}}\right),\label{eq:psg.acceleration-1}
\end{equation}
where $\ell=8\pi\eta B_{{\rm s}}eh_{\perp}^{2}\cos\left(\alpha\right)/\left(Pc^{3}m\right)$.
Although the approximate formula \ref{eq:psg.acceleration} is much
more readable, in the calculations we use the exact value (see Equation
\ref{eq:psg.acceleration-1}) as for Lorentz factors that are considerably
smaller than the maximum value, the discrepancy is about a factor
of two, $l_{{\rm acc,ap}}\approx2l_{{\rm acc}}$.

\subsection{Electron/positron mean free path}

The mean free path of a particle (electron and/or positron) $l_{{\rm p}}$
can be defined as the mean length that a particle passes until a $\gamma$-photon
is emitted. In the case of the CR particle, mean free path can be
estimated as a distance that a particle with a Lorentz factor $\gamma$
travels during the time which is necessary to emit a curvature photon
(see \citealt{1997_Zhang}) 
\begin{equation}
l_{{\rm CR}}\sim c\left(\frac{P_{{\rm CR}}}{E_{\gamma,{\rm CR}}}\right)^{-1}=\frac{9}{4}\frac{\hbar\Re c}{\gamma e^{2}},\label{eq:psg.le_cr}
\end{equation}
where $P_{{\rm CR}}=2\gamma^{4}e^{2}c/3\Re^{2}$ is the power of CR,
$E_{\gamma,{\rm CR}}=3\hbar\gamma^{3}c/2\Re$ is the photon characteristic
energy, and $\Re$ is the curvature radius of the magnetic field lines.

For the ICS process calculation of the particle mean free path $l_{{\rm ICS}}$
is not as simple as that of the CR process. Although we can define
$l_{{\rm ICS}}$ in the same way that we defined $l_{{\rm CR}}$,
it is difficult to estimate the characteristic frequency of emitted
photons. We have to take into account photons of various frequencies
with various incident angles. An estimation of the mean free path
of an electron (or positron) to produce a photon is in \citet{1985_Xia}
\begin{equation}
l_{{\rm ICS}}\sim\left[\int_{\mu_{0}}^{\mu_{1}}\int_{0}^{\infty}\sigma^{\prime}\left(\epsilon,\mu\right)\left(1-\beta\mu_{i}\right)n_{{\rm ph}}\left(\epsilon\right)d\epsilon d\mu\right]^{-1}.\label{eq:psg.le_ics}
\end{equation}
Here $\epsilon$ is the incident photon energy in units of $mc^{2}$,
$\mu=\cos\psi$ is the cosine of the photon incident angle, $\beta=v/c$
is the velocity in terms of speed of light, $\sigma^{\prime}$ is
the cross section of ICS in the particle rest frame, 
\begin{equation}
n_{{\rm ph}}\left(\epsilon,\, T\right)d\epsilon=\frac{4\pi}{\lambda_{c}^{3}}\frac{\epsilon^{2}}{\exp\left(\epsilon/\mho\right)-1}d\epsilon\label{eq:psg.nph}
\end{equation}
represents the photon number density distribution of semi-isotropic
blackbody radiation, $\mho=kT/mc^{2}$, $k$ is the Boltzmann constant,
and $\lambda_{c}=h/mc=2.424\times10^{-10}$ cm is the electron Compton
wavelength. A detailed description of how to calculate $\sigma^{\prime}$
can be found in Section \ref{sec:cascade.ics_cross_sec}.

We should expect two modes of ICS: resonant and thermal-peak (see
Section \ref{sec:cascade.ics_electron_path} for more details). The
Resonant ICS (RICS) takes place if the photon frequency in the particle
rest frame is equal to the electron cyclotron frequency. As shown
in Section \ref{sec:cascade.background_photons}, the particle mean
free path strongly depends on the distance from the polar cap. Both
the photon density and incident angles ($\mu_{0}$ and $\mu_{1}$)
change with increasing altitude. In our calculations we take into
account both of those effects, thus we replace $n_{{\rm ph}}\left(\epsilon,\, T\right)$,
$\mu_{0}$ and $\mu_{1}$ with $n_{{\rm sp}}\left(\epsilon,\, T,\, L\right)$,
$\mu_{{\rm min}}$$\left(L\right)$ and $\mu_{{\rm max}}\left(L\right)$,
respectively (for more details see Section \ref{sec:cascade.background_photons}).
Here, $L$ is the location of the particle, $n_{{\rm sp}}\left(\epsilon,\, T,\, L\right)$
is the photon density at location $L$, and $\mu_{{\rm min}}\left(L\right)$
and $\mu_{{\rm max}}$$\left(L\right)$ correspond to the highest
and lowest angle between the photons and particle at a given location
$L$. Thus, just above the polar cap for RICS the mean free path of
outflowing positrons is:

\begin{equation}
l_{{\rm RICS}}\approx\left[\int_{\mu_{{\rm min}}\left(L\right)}^{\mu_{{\rm max}}\left(L\right)}\int_{\epsilon_{_{{\rm res}}}^{{\rm ^{min}}}}^{\epsilon_{_{{\rm res}}}^{{\rm ^{max}}}}\left(1-\beta\mu\right)\sigma^{\prime}\left(\epsilon,\mu\right)n_{{\rm sp}}\left(\epsilon,\, T,\, L\right)d\epsilon d\mu\right]^{-1},\label{eq:cascade.ics_free_path-1}
\end{equation}
where the limits of integration over energy, $\epsilon_{{\rm _{res}}}^{{\rm ^{min}}}$
and $\epsilon_{{\rm _{res}}}^{^{{\rm max}}}$, are chosen to cover
the resonant energy (for more details see Section \ref{sec:cascade.ics_electron_path}).

The thermal-peak ICS (TICS) includes all scattering processes of photons
with frequencies around the maximum of the thermal spectrum. As an
example we adopt\linebreak{}
 $\epsilon_{_{{\rm th}}}^{{\rm ^{min}}}\approx0.05\epsilon_{{\rm _{th}}}$,
and $\epsilon_{_{{\rm th}}}^{{\rm ^{{\rm max}}}}\approx2\epsilon_{_{{\rm th}}}$
where $\epsilon_{_{{\rm th}}}=2.82kT/\left(mc^{2}\right)$ is the
energy, in units of $mc^{2}$, at which blackbody radiation with temperature
$T$ has the largest photon number density. The electron/positron
mean free path for the TICS process is 

\begin{equation}
l_{{\rm TICS}}\approx\left[\int_{\mu_{{\rm min}}\left(L\right)}^{\mu_{{\rm max}}\left(L\right)}\int_{\epsilon_{_{{\rm th}}}^{{\rm ^{min}}}}^{\epsilon_{_{{\rm th}}}^{{\rm {\rm ^{max}}}}}\left(1-\beta\mu\right)\sigma^{\prime}\left(\epsilon,\mu\right)n_{{\rm ph}}\left(\epsilon,\, T,\, L\right)d\epsilon d\mu\right]^{-1}.\label{eq:cascade.t_ics-1}
\end{equation}

\subsection{Photon mean free path}

The photons with energy $E_{\gamma}>2mc^{2}$ propagating obliquely
to the magnetic field lines can be absorbed by the field, and as a
result, an electron-positron pair is created. To describe the strength
of the magnetic field we use $\beta_{q}=B/B_{q}$, where $B_{q}=m^{2}c^{3}/e\hbar=4.413\times10^{13}\,{\rm G}$
is the critical magnetic field strength.

For strong magnetic fields ($\beta_{q}\gtrsim0.2$, see Section \ref{sec:cascade.pairs})
the photon mean free path can be calculated as (see Section \ref{sec:cascade.photon_path}
for more details) 
\begin{equation}
l_{{\rm ph}}\approx\Re\frac{2mc^{2}}{E_{\gamma}},\label{eq:psg.l_ph}
\end{equation}
 while for weaker magnetic fields ($\beta_{q}\lesssim0.2$) we can
use an asymptotic approximation derived by \citet{1966_Erber} 
\begin{equation}
l_{{\rm ph}}=\frac{4.4}{(e^{2}/\hbar c)}\frac{\hbar}{mc}\frac{B_{q}}{B\sin\Psi}\exp\left(\frac{4}{3\chi}\right),\label{eq:psg.l_ph2}
\end{equation}
 
\begin{equation}
\chi\equiv\frac{E_{\gamma}}{2mc^{2}}\frac{B\sin\Psi}{B_{q}}\hspace{1cm}(\chi\ll1),
\end{equation}
where $\Psi$ is the angle of intersection between the photon and
the local magnetic field.

\section{Gap height \label{sec:psg.finding_height}}

By knowing the acceleration potential drop in PSG $\Delta V$ we can
evaluate the gap height $h$ and the screening factor $\eta$, which
actually depends on the details of the avalanche pair production in
the gap. First, we need to determine which process, Curvature Radiation
(CR) or Inverse Compton Scattering (ICS), is responsible for the $\gamma$-photon
generation in the gap region. In order to identify the proper process
we need the following parameters: $l_{{\rm acc}}$ - the distance
which a particle should pass to gain the Lorentz factor $\gamma_{{\rm acc}}$,
$l_{{\rm p}}$ - the mean length a particle (electron and/or positron)
travels before a $\gamma$-photon is emitted, and $l_{{\rm ph}}$
- the mean free path of the $\gamma$-photon before being absorbed
by the magnetic field.

As mentioned above, PSG can exist if Equation \ref{eq:psg.heating_condition}
is satisfied. On the other hand, in order to heat the polar cap surface
to high enough temperatures the high enough flux of back-streaming
particles is required. By using Equations \ref{eq:psg.potential_heating}
and \ref{eq:psg.potential_drop} we can find the relationship between
the screening factor, the spark half-width and pulsar parameters

\begin{equation}
\eta h_{\perp}=4.17\frac{\left(B_{14}^{1.1}+0.3\right)^{2}P}{B_{14}\sqrt{\left|\cos\alpha\right|}}.\label{eq:psg.eta_hperp}
\end{equation}

Thus, for specific pulsar parameters we can define a product of the
two main parameters of PSG, namely the screening factor $\eta$ and
the spark half-width $h_{\perp}$.

\subsection{Particle mean free paths, CR vs. ICS gap\label{sec:psg.particles_mean}}

\begin{figure}[H]
\begin{centering}
\includegraphics{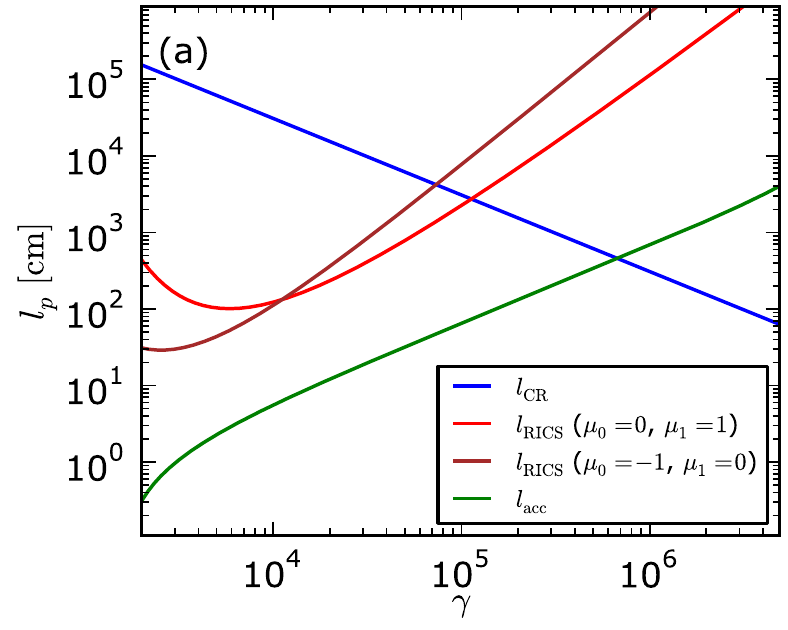}\includegraphics{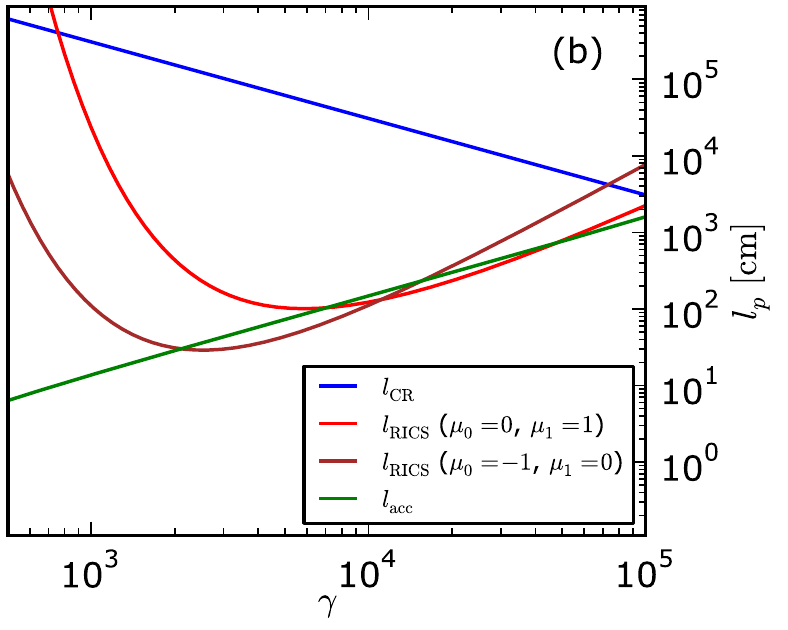}
\par\end{centering}

\centering{}\caption[{Dependence of the mean free path of the primary particle {[}CR + ICS{]}}]{Dependence of the mean free path of the primary particle on Lorentz
factor $\gamma$ for both the CR and ICS processes. Panel (a) corresponds
to calculations for a relatively higher potential drop (e.g. a wider
spark with $h_{\perp}=3\,{\rm m}$ and $\eta=1$), while panel (b)
corresponds to calculations for a relatively lower potential drop
(e.g. a narrow spark with $h_{\perp}=1\,{\rm m}$ and $\eta=0.1$).
The acceleration paths on both panels were calculated for the same
pulsar parameters ($B_{14}=3.5$, $T_{6}=4.4$, $\Re_{6}=1$, $P=1$,
$\alpha=10^{\circ}$). Note that for the RICS process the particle
mean free paths were calculated for optimal conditions (just above
the polar cap). \label{fig:psg.le_gamma} }
\end{figure}

The Figure \ref{fig:psg.le_gamma} shows the dependence of particle
mean free paths on the Lorentz factor $\gamma$ for some pulsar parameters
(the dependence on pulsar parameters will be discussed in Section
\ref{sec:psg.model_parameters}). Let us note that these free paths
do not depend on the gap height $h$ (see Equations \ref{eq:psg.le_cr},
\ref{eq:psg.le_ics} and \ref{eq:psg.l_ph}). The results presented
in the Figure do not allow to define the gap height unambiguously.
However, we can find which process is responsible for generation of
the $\gamma$-photon in PSG. For narrow sparks the acceleration potential
drop decreases, and as a result the Lorentz factor of the primary
particles is about $\gamma\sim10^{3}-10^{4}$. In this regime $l_{{\rm ICS}}\ll l_{{\rm CR}}$,
so the gap will be dominated by ICS. Thus, ICS will dominate the gap
if deceleration due to Inverse Compton Scattering prevents further
acceleration by the electric field. Let us remember that ICS is not
efficient for particles with the Lorentz factor $\gamma\gtrsim10^{5}$.
If the sparks are wider or $\eta\approx1$, the acceleration potential
drop increases and the Lorentz factors of primary particles reach
values about $\gamma\sim10^{5}-10^{6}$. In this regime the $\gamma$-photon
emission is dominated by CR. Let us note that the condition $l_{{\rm ICS}}\ll l_{{\rm CR}}$
is satisfied for particles with $\gamma\sim10^{3}-10^{4}$, as one
can see from Figure \ref{fig:psg.le_gamma} (panel a), but this does
not mean that the ICS event happens. Since $l_{{\rm acc}}\ll l_{{\rm ICS}}$,
the particles will be accelerated to higher energies ($\gamma\sim10^{5}-10^{6}$)
before they upscatter the X-ray photons. Thus, the particles start
emission of $\gamma$-photons (via CR) as soon as condition $l_{{\rm acc}}\approx l_{{\rm CR}}$
is met.

\begin{comment}
\textasciitilde{}Programs/magnetic/magnetic/src/radiation/gap.py (le\_gamma\_phd)
\end{comment}

\subsection{Possible scenarios of the gap breakdown: PSG-on and PSG-off modes\label{sec:psg.multiplicity}}

\begin{figure}[H]
\begin{centering}
\includegraphics{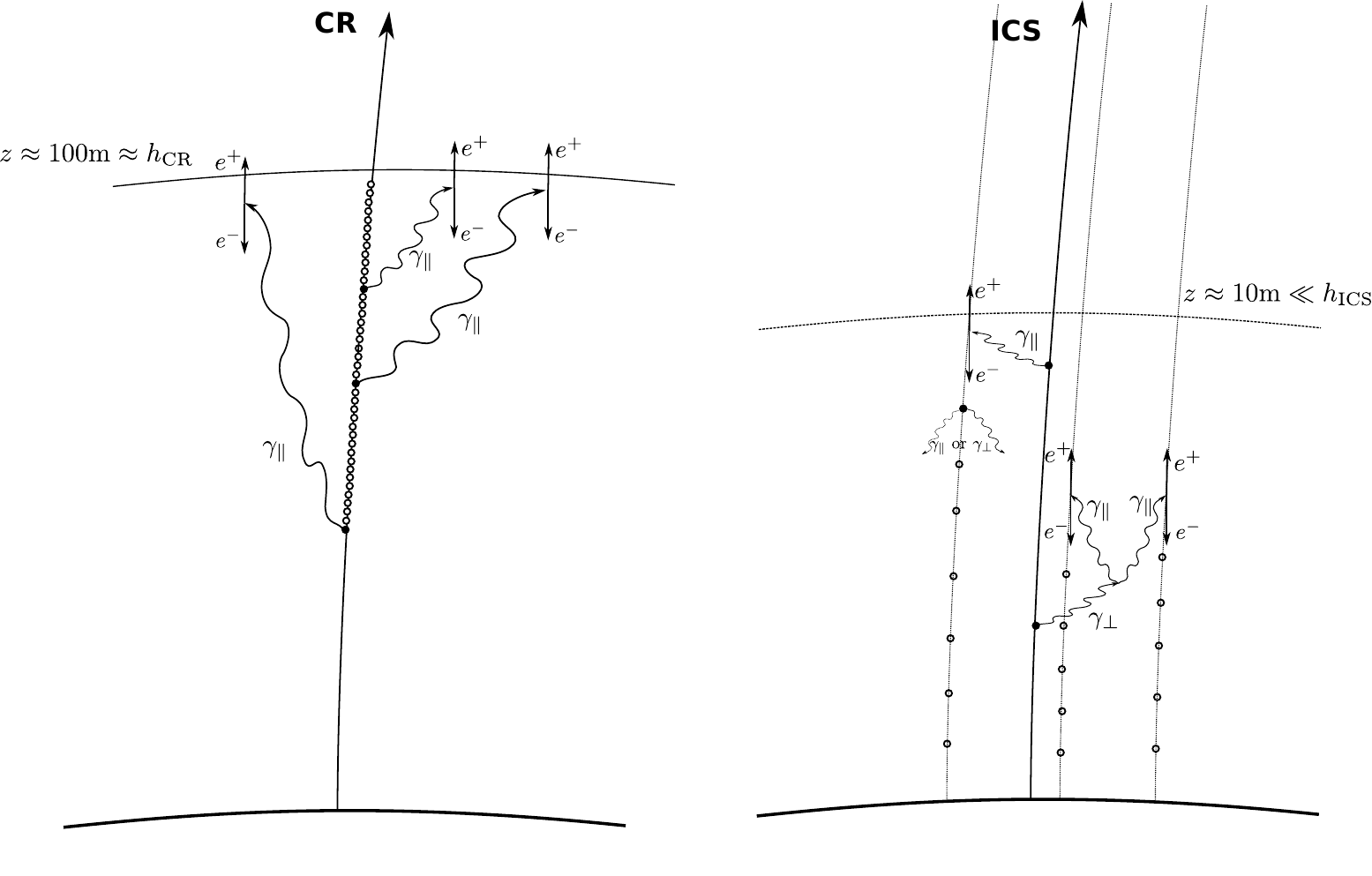}
\par\end{centering}

\caption[{Sketch of the differences in a cascade formation {[}CR- and ICS-dominated
gaps{]}}]{Sketch of differences in a cascade formation for the CR-dominated
gap (left panel) and the ICS-dominated gap (right panel). In order
to increase readability, only a few points (filled circles) are shown
which correspond to altitudes where $\gamma$-photons are emitted.
The unfilled circles correspond to places where $\gamma$-photons
are also emitted, but those photons (and their evolution) are not
included in the diagram. Note that for the ICS-dominated gap we plot
only the bottom (active) part of the gap ($z\ll h_{{\rm ICS}}$),
furthermore, points of radiation are tracked only for the first population
of newly created particles. The avalanche nature of the ICS-dominated
gap will result in a much higher multiplicity and continuous backflow
of relativistic particles. \label{fig:psg.schematic_ics_cr}}
\end{figure}

\begin{figure}[H]
\begin{centering}
\includegraphics[width=7.5cm,height=7cm]{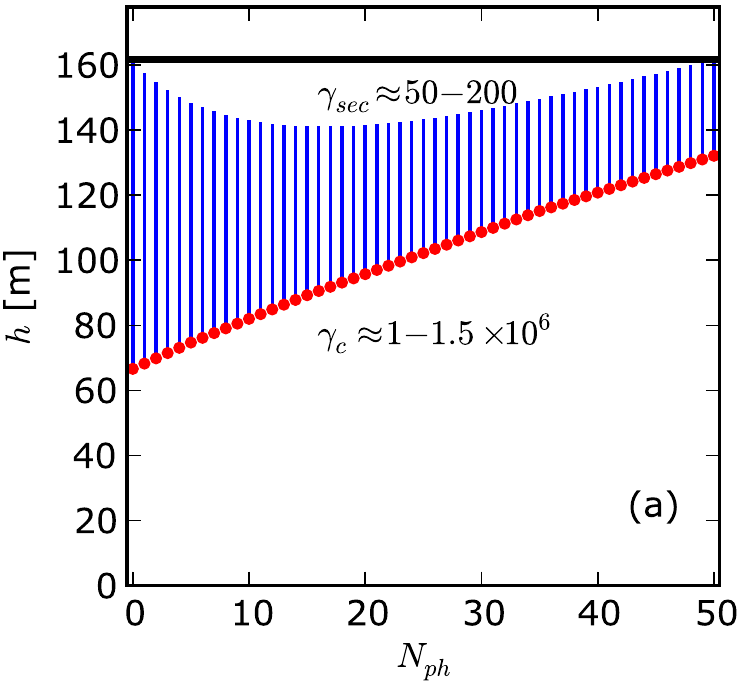}\hspace{0.5cm}\includegraphics[width=7.5cm,height=7cm]{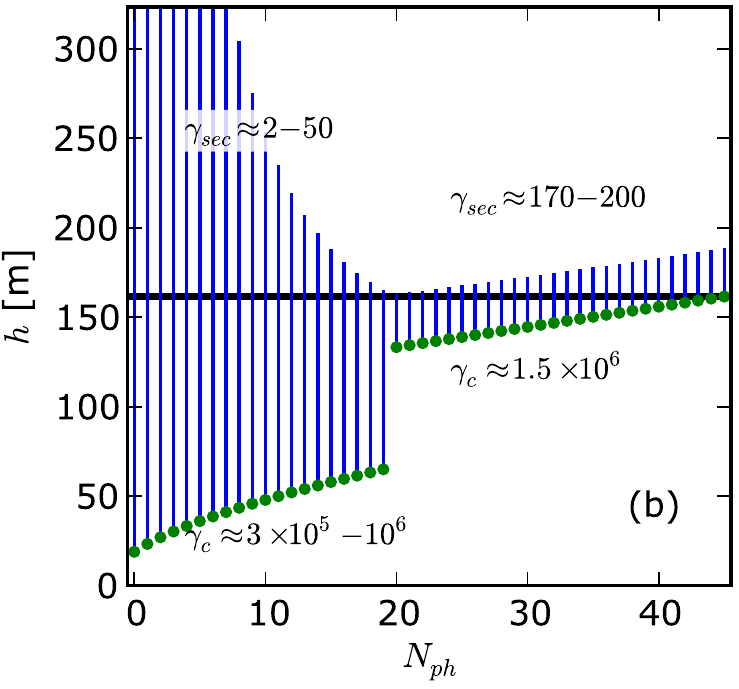}
\par\end{centering}

\caption[Cascade formation for a CR-dominated gap]{Cascade formation for a CR-dominated gap. Blue lines represent the
mean free path of $\gamma$-photons. The filled circles correspond
to places of $\gamma$-photon emission. Panel (a) includes the free
paths of $\gamma$-photons which produce pairs below ZPF (red circles)
while panel (b) includes the free paths of $\gamma$-photons which
produce pairs above the acceleration gap (blue circles). The results
were obtained using the following parameters: $N_{{\rm ph}}^{^{{\rm CR}}}=50$,
$B_{{\rm s}}=2.3\times10^{14}\,{\rm G}$, $B_{{\rm d}}=2.9\times10^{14}\,{\rm G}$,
$T_{}=3\,{\rm MK}$, $P=1.3\,{\rm s}$, $\Re_{6}=0.5$, and $\alpha=60.7^{\circ}$.
\label{fig:psg.cr_sol}}
\end{figure}

\begin{figure}[H]
\begin{centering}
\includegraphics[width=9.5cm,height=7cm]{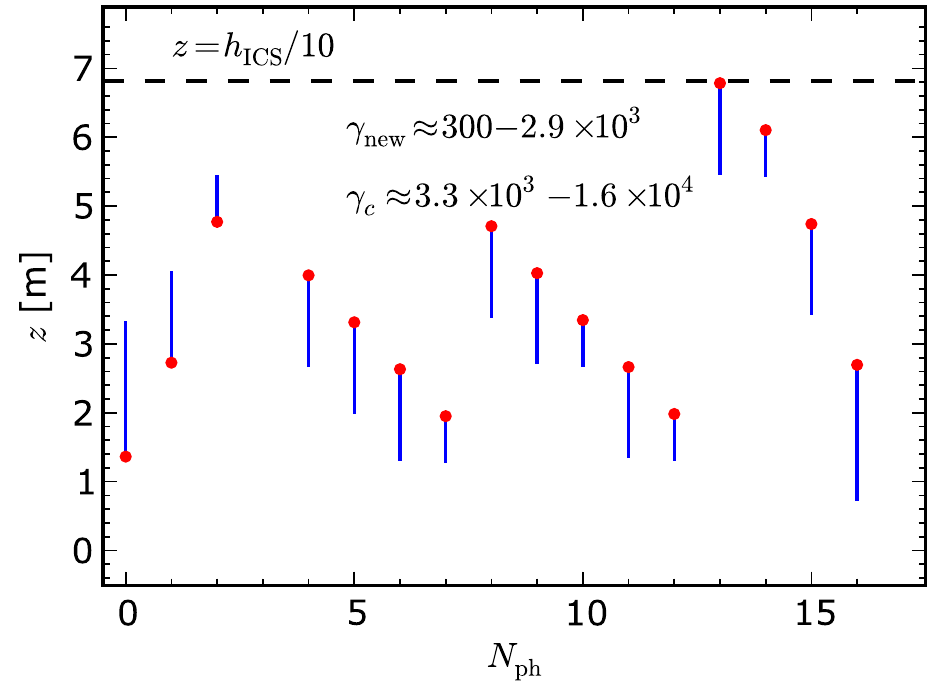}
\par\end{centering}

\caption[Cascade formation for an ICS-dominated gap]{Cascade formation for an ICS-dominated gap. Blue lines represent
the mean free path of $\gamma$-photons. The filled circles correspond
to places of $\gamma$-photon emission. The results were obtained
using the following parameters: $N_{{\rm ph}}^{^{{\rm ICS}}}=15$,
$B_{{\rm s}}=2.3\times10^{14}\,{\rm G}$, $B_{{\rm d}}=2.9\times10^{14}\,{\rm G}$,
$T_{{\rm s}}=3\,{\rm MK}$, $P=1.3\,{\rm s}$, $\Re_{6}=0.5$, and
$\alpha=60.7^{\circ}$.\label{fig:psg.ics_sol}}
\end{figure}

As is seen from Figures \ref{fig:psg.schematic_ics_cr} and \ref{fig:psg.cr_sol},
in the CR-dominated gap the primary particle should travel a distance
comparable with gap height $l_{{\rm acc}}\approx h/2$ in order to
gain an energy corresponding to the characteristic Lorentz factor
$\gamma_{{\rm c}}^{{\rm ^{CR}}}$. On the other hand, the primary
particles in the ICS-dominated gap reach a characteristic value $\gamma_{{\rm c}}^{{\rm ^{ICS}}}$
at altitudes that are considerably smaller than gap height $l_{{\rm acc}}\ll h$
(see Figures \ref{fig:psg.schematic_ics_cr} and \ref{fig:psg.ics_sol}).
Thus, $\gamma_{{\rm c}}^{{\rm ^{CR}}}\approx10^{6}\approx\gamma_{{\rm max}}$
is about three orders of magnitude higher than $\gamma_{{\rm c}}^{^{{\rm ICS}}}\approx10^{3}\ll\gamma_{{\rm max}}$,
here $\gamma_{{\rm max}}$ is the value of the Lorentz factor after
the particle travels a distance $h$. Furthermore, the characteristic
energy of CR photons is considerably smaller than the energy of emitting
(primary) particles, e.g. for $\gamma=10^{6}$, $\Re_{6}=1$, $\gamma_{{\rm sec}}\approx10^{2}$.
On the other hand, RICS photons upscattered in an ultrastrong ($B>B_{{\rm crit}}$)
magnetic field gain a significant part of the energy of the scattering
(primary) particle. Therefore, the electron/positron pair created
by the RICS photon has energies comparable with the energy of the
scattering (primary) particle. This will essentially influence the
multiplicity $M_{{\rm pr}}$ in the ICS gap, as all the newly created
particles will participate in further cascade pair-production. Additionally,
RICS in ultrastrong magnetic fields produces approximately the same
amount of photons with $\parallel$ and $\perp$ polarisation (see
Section \ref{sec:cascade.R-ICS_cross}), while most of the photons
produced by CR are $\parallel$-polarised (see Section \ref{sec:cascade.cr}).
Splitting of the $\perp$-polarised photons will increase the photon
mean free path, but it will also increase the multiplicity in the
ICS gaps.

Figure \ref{fig:psg.schematic_ics_cr} presents a sketch of a cascade
formation for CR- and ICS-dominated gaps. The CR photons are emitted
in the upper half of the gap. Most of these photons produce pairs
at about the same height, in the region where the acceleration potential
is almost equal to zero, hereinafter we will call this region the
Zero-Potential Front (ZPF). The newly created particles have much
lower Lorentz factors as compared with the primary particle, thus
they are not able to emit CR photons.

Figure \ref{fig:psg.cr_sol} presents the primary particle evolution
and photon mean free paths of\linebreak{}
$\gamma$-rays produced in the CR-dominated gap. As can be seen, the
first $\gamma$-photon produces a pair approximately at the same time
(and same place) as the primary particle reaches ZPF. Thus, the multiplicity
in a gap region (the number of particles created by a single primary
particle) in the CR scenario is strictly related to the number of
photons produced by the primary particle $M_{{\rm CR}}\approx2\times N_{{\rm ph}}^{{\rm ^{CR}}}$.

\begin{comment}
\textasciitilde{}/Programs/magnetic/magnetic/src/radiation/gap.py
(find\_solution\_cr\_psgoff\_plot ,plot\_solution\_cr, plot\_solution\_acr,
$n_{end}=50$ $B_{14}=2.3$, $B_{{\rm d}}=0.02978$, $T_{6}=3.0$,
$P=1.273768291578$, $\Re_{6}=0.5$, $\alpha=60.7^{\circ}$
\end{comment}

The energy of $\gamma$-photons produced by ICS depends on the Lorentz
factor of the primary particles and on the strength of the magnetic
field. In ultrastrong magnetic fields the energy of newly created
particles is comparable with the energy of the scattering particle
$\gamma_{{\rm new}}\approx\gamma_{{\rm c}}/2$. Figure \ref{fig:psg.ics_sol}
shows schematically the locations at which $\gamma$-photons are emitted
by ICS. The first $\gamma$-photon is produced already at altitudes
of about a few metres and then converted to an electron-positron pair
well below ZPF. Note that already at relatively low altitudes ($z\gtrsim100\,{\rm m}$)
the photon density decreases rapidly (see Section \ref{sec:cascade.photon_density}),
furthermore, the small size of the polar cap entails a rapid change
of the particle-photon incident angles (see Section \ref{sec:cascade.incident_angles}).
Those two effects make the ICS process significant only in the lower
parts of the gap ($z\lesssim20\,{\rm m}$). On the other hand, the
multiplicity in the ICS-dominated gap is enhanced by all newly created
particles which are created in the lower part of the gap. Furthermore,
the ICS is more effective for backstreaming particles (see Figure
\ref{fig:psg.le_gamma}), thus most of the $\gamma$-photons in the
gap region will be created by scatterings on electrons. For the ICS
scenario it is not possible to evaluate a simple expression for the
multiplicity produced by a single primary particle in a gap region.
Furthermore, it is not possible to determine the actual value of $N_{{\rm ph}}$
required to break the gap (both for CR and ICS) without a full cascade
simulation.

\begin{comment}
\textasciitilde{}/Programs/magnetic/magnetic/src/radiation/gap\_ics.py
(read\_data(910), find\_solution\_hperp(200.), plot\_solution()
\end{comment}

The differences between the CR and ICS gaps that we mention above
have drastic consequences on the cascade formation process. Since
the cooling time of the hot spot is very short ($\tau_{{\rm cool}}\lesssim10^{-8}\,{\rm s}$
, see \citealp{2003_Gil}), to sustain the hot spot temperature just
below the critical temperature a continuous backflow of relativistic
particles is required. An energetic enough flux of backstreaming particles
can be produced only in ICS-dominated gaps. The heating of the surface
will sustain the outflow of iron ions from the crust, maintaining
$\eta<1$, hence we call this mode the PSG-on mode. As the temperature
of the polar cap is in quasi equilibrium with the backstreaming particles
(temperature is close to the critical value) the gap can break only
due to production of a dense enough plasma $n_{p}\gg\eta n_{{\rm GJ}}$
in the gap region. The multiplicity in the PSG-on mode is much higher
than the multiplicity of CR-dominated gaps. Moreover, in the gap dominated
by CR the particles are created in a cloud-like fashion (see Figure
\ref{fig:psg.cr_sol}). The successive clouds heat up the surface
once per $\tau_{{\rm 0}}\approx2h/c$, which for a typical gap height
$h\approx100\,{\rm m}$ is much longer than the time needed for the
surface to cool down $\tau_{{\rm 0}}\approx6\times10^{-7}\gg\tau_{{\rm cool}}$.
Therefore, in the CR-dominated gaps the backstreaming particles cannot
sustain the temperature that is close to the critical value during
$\tau_{{\rm 1}}\gg\tau_{0}\gg\tau_{{\rm cool}}$, thus for most of
the time the screening factor is $\eta\approx1$ and we call this
mode the PSG-off mode. The low multiplicity of a cascade in the PSG-off
mode can cause that the gap to breakdown only due to overheating of
the surface, but not due to production of a dense enough plasma. The
growth of particle density will continue to the point when the backstreaming
particles heat up the surface to a temperature equal to or higher
than the critical temperature, $\tau_{{\rm heat}}\gg\tau_{0}$. Let
us note that the primary particles in the PSG-off mode are very energetic
$\gamma\approx10^{6}$, and hence the density of particles required
to close gap $\rho_{c}$ is much lower than the Goldreich-Julian density.
To describe this difference we use the overheating parameter $\kappa=\rho_{c}/\rho_{{\rm GJ}}$.
Knowing that in the PSG-off mode $\eta\approx1$, we use Equation
\ref{eq:psg.heating_condition} and the relation $\Delta V=\gamma_{{\rm acc}}mc^{2}/e$
to calculate the overheating parameter:

\begin{equation}
\kappa=\frac{\sigma\, T^{4}}{n_{{\rm GJ}}\,\gamma_{{\rm max}}\, mc^{3}}.\label{eq:psg.overheating_parameter}
\end{equation}

\subsection{PSG-off mode\label{sec:psg.off_gap}}

\begin{figure}[H]
\begin{centering}
\includegraphics{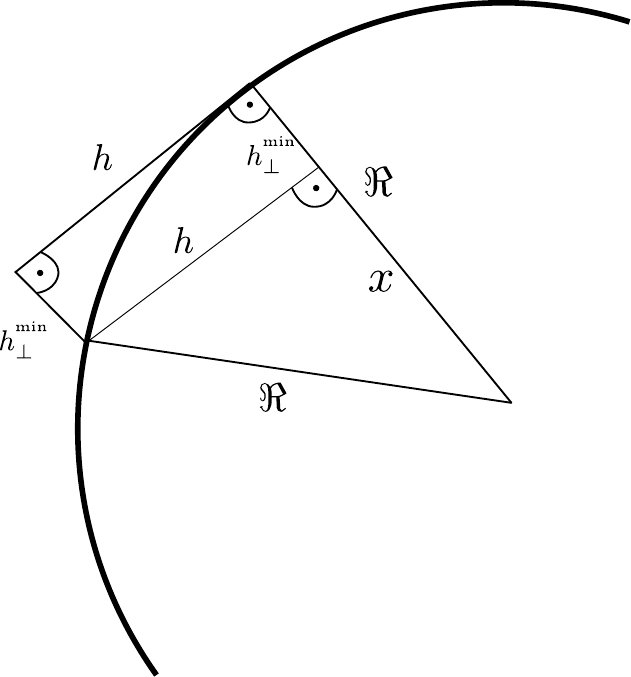}
\par\end{centering}

\centering{}\caption[Diagram of the minimum spark half-width]{Diagram of the minimum spark half-width $h_{\perp}^{{\rm ^{min}}}$
for a given gap height $h$ and a radius of curvature $\Re$. \label{fig:psg.hperp_min_theory} }
\end{figure}

Curvature emission by a primary particle is effective for Lorentz
factors $\gamma\gtrsim10^{5}$ (when $l_{{\rm CR}}\leq l_{{\rm acc}}$).
An equilibrium between acceleration and deceleration (by reaction
force) would be established if the CR power were equal to the ''electric
power''. In our case ($\Re_{6}\approx1$, $\gamma_{{\rm c}}\approx10^{6}$),
the reaction force is not high enough to stop acceleration by the
electric field. In the PSG-off mode the spark region is free from
ions ($\eta\approx1$), thus the heating condition (Equations \ref{eq:psg.heating_condition}
and \ref{eq:psg.eta_hperp}) is no longer satisfied. Taking into account
the curvature of magnetic field lines just above the stellar surface,
we can estimate the dependence of the minimum spark half-width on
the gap height (see Figure \ref{fig:psg.hperp_min_theory}):

\begin{equation}
h_{\perp}^{{\rm ^{min}}}=\Re-\sqrt{\Re^{2}-h^{2}}.\label{eq:psg.hperp_min}
\end{equation}

\begin{figure}[H]
\begin{centering}
\includegraphics{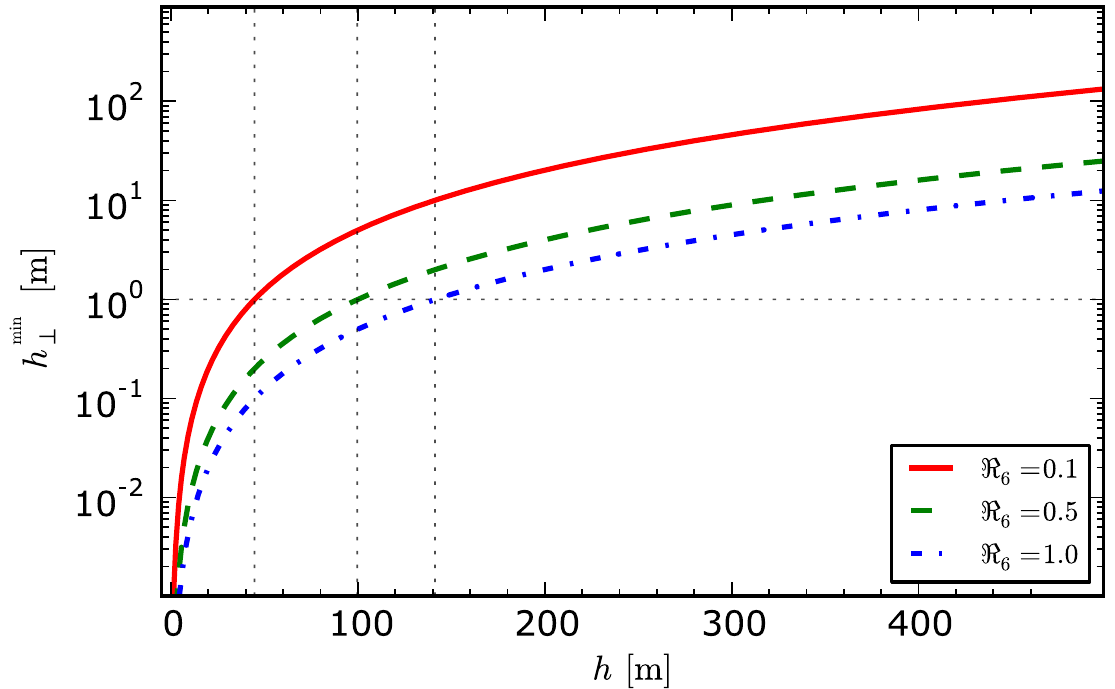}
\par\end{centering}

\centering{}\caption[Minimum spark half-width vs. the gap height]{Minimum spark half-width vs. gap height calculated for three different
radii of curvature: $\Re_{6}=0.1$ - red solid line, $\Re_{6}=0.5$
- green dashed line, and $\Re_{6}=1$ - blue dotted line. \label{fig:psg.hperp_min_res} }
\end{figure}

Figure \ref{fig:psg.hperp_min_res} presents the minimum spark half-width
for three different radii of curvature: $\Re_{6}=0.1$, $\Re_{6}=0.5$,
$\Re_{6}=1$. Note that as long as the gap height does not exceed
some specific value ($h\approx40\,{\rm m}$, $h\approx100\,{\rm m}$,
$h\approx140\,{\rm m}$, respectively for the given curvature radii)
the minimum spark half-width is well below $1\,{\rm m}$.

\begin{comment}
\textasciitilde{}/Programs/studies/phd/spark\_width/spark\_width.py
(plot\_hperp)
\end{comment}

\begin{figure}[H]
\begin{centering}
\includegraphics{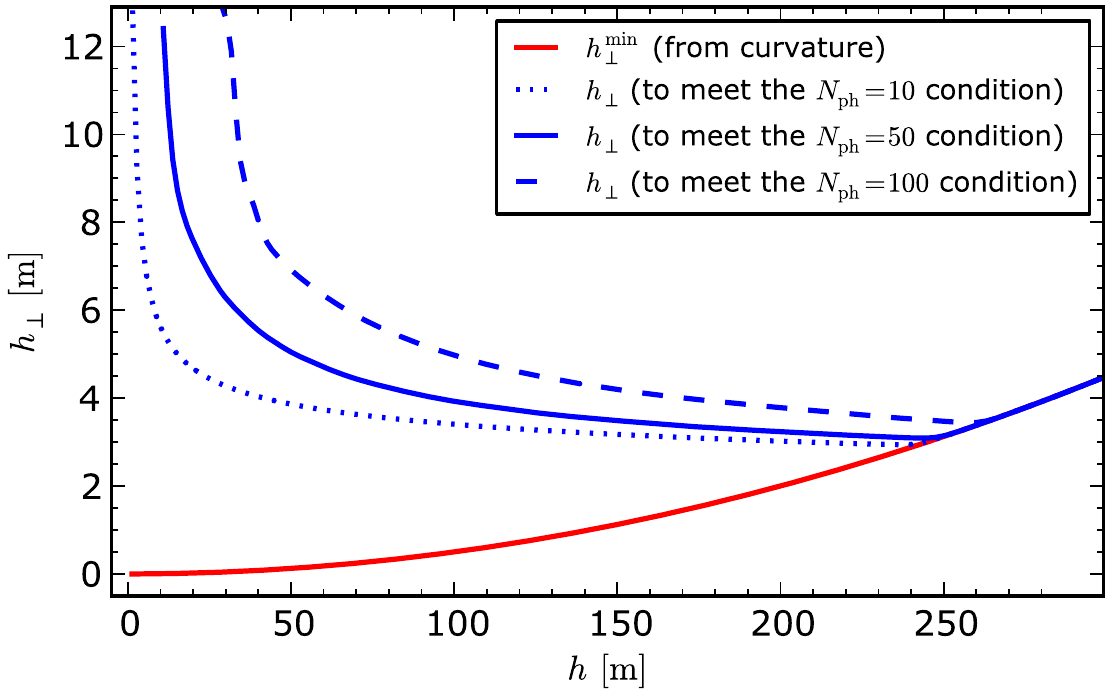}
\par\end{centering}

\centering{}\caption[{Dependence of a spark half-width on the gap height {[}PSG-off mode{]}}]{Dependence of a spark half-width on the gap height for the PSG-off
mode. The results were obtained using the following pulsar parameters:
$B_{14}=2.3$, $T_{6}=3.0$, $P=1.3\,{\rm s}$, $\Re_{6}=1.0$, $\alpha=60.7^{\circ}$.\label{fig:psg.hperp_h_nph} }
\end{figure}

On the other hand, we can estimate the acceleration potential $\Delta V$
(and thus the spark half-width $h_{\perp}^{^{{\rm N_{{\rm ph}}}}}$)
required to produce a specified number of photons $N_{{\rm ph}}^{^{{\rm CR}}}$
within a gap. Figure \ref{fig:psg.hperp_h_nph} presents the dependence
of both $h_{\perp}^{^{{\rm min}}}$ and $h_{\perp}^{{\rm ^{N_{{\rm ph}}}}}$
on the gap height. As results from the Figure, the gap height in PSG-off
does not change drastically with $N_{{\rm ph}}^{^{{\rm CR}}}$, and
for these specific parameters of a pulsar it is $h\approx240\,{\rm m}$.
For historical reasons, hereafter unless stated otherwise, we will
use $N_{{\rm ph}}^{^{{\rm CR}}}=50$ to calculate the gap parameters
of the PSG-off mode. Note that in order to find the gap height, we
assume $h_{\perp}^{^{{\rm min}}}=h_{\perp}^{^{N_{{\rm ph}}}}$, which
results in a gap that allows both overheating of the entire spark
surface by backstreaming particles and the creation of the required
number of photons $N_{{\rm ph}}^{^{{\rm CR}}}$.

\begin{comment}
\textasciitilde{}/Programs/magnetic/magnetic/src/radiation/gap.py
(show\_solution\_cr\_psgoff, plot\_psgoff\_cr (or just this to use
files) $B_{14}=2.3$, $B_{{\rm d}}=0.02978$, $T_{6}=3.0$, $P=1.291578$,
$\Re_{6}=1.0$, $\alpha=60.7^{\circ}$)
\end{comment}

\begin{figure}[H]
\begin{centering}
\includegraphics[height=13cm]{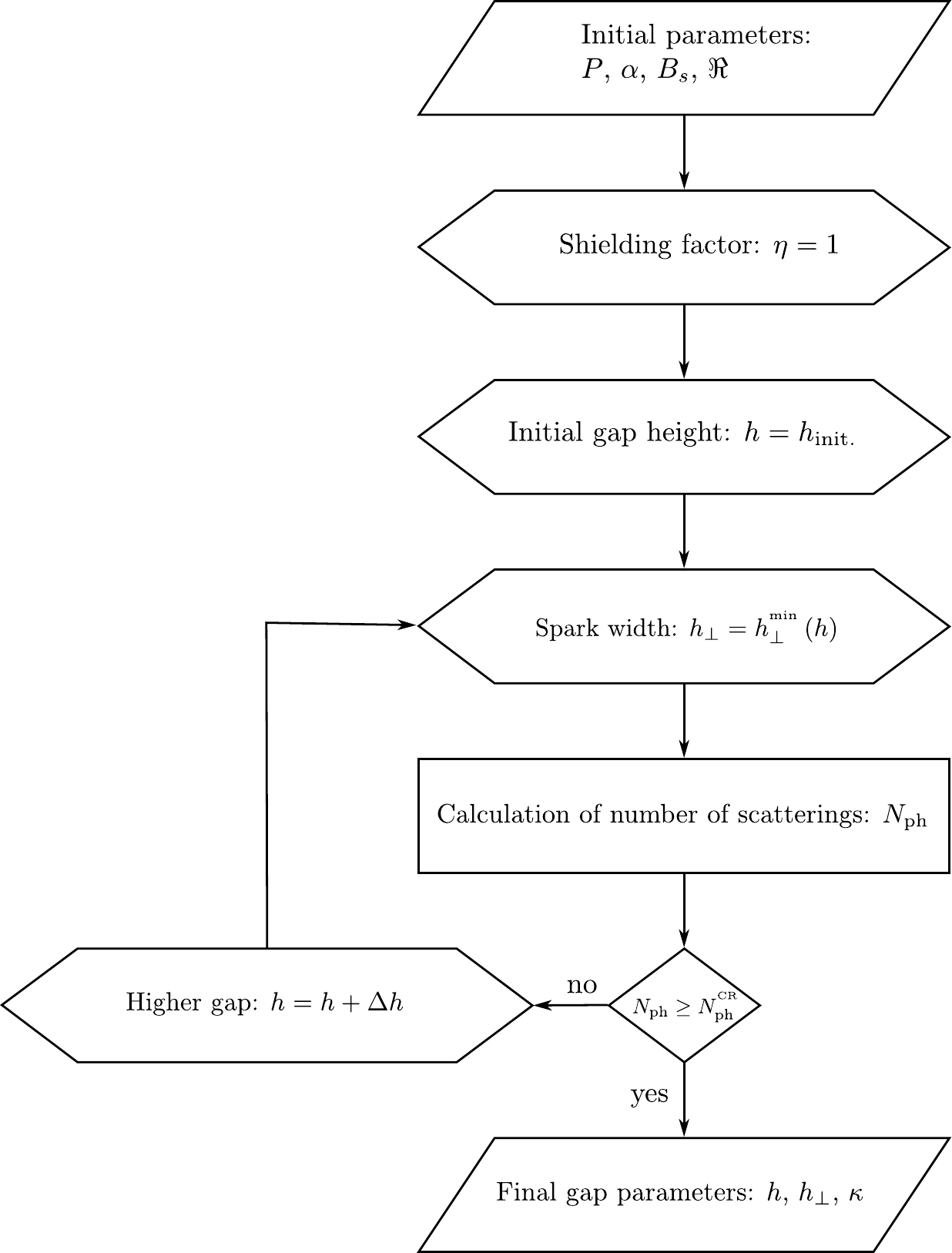}
\par\end{centering}

\centering{}\caption[{Flowchart of the algorithm used to estimate the gap height {[}PSG-off
mode{]}}]{Flowchart of the algorithm used to estimate the gap height in the
PSG-off mode. The initial gap height from which we begin our calculations
is an arbitrary set to $h_{{\rm init.}}=10\,{\rm m}$, while the step
$\Delta h$ depends on the required accuracy. The number of $\gamma$-ray
photons created in a spark by a single primary particle is set to
$N_{{\rm ph}}^{{\rm ^{CR}}}=50$ (see text for more details).\label{fig:psg.flowchart_cr} }
\end{figure}

In our calculations we use the algorithm presented in Figure \ref{fig:psg.flowchart_cr}
to find the gap height in the PSG-off mode for given pulsar parameters:
a pulsar period $P$, a pulsar inclination angle $\alpha$, a surface
magnetic field strength $B_{{\rm s}}$, and a curvature radius of
field lines $\Re$.

\begin{figure}[H]
\begin{centering}
\includegraphics[height=7cm]{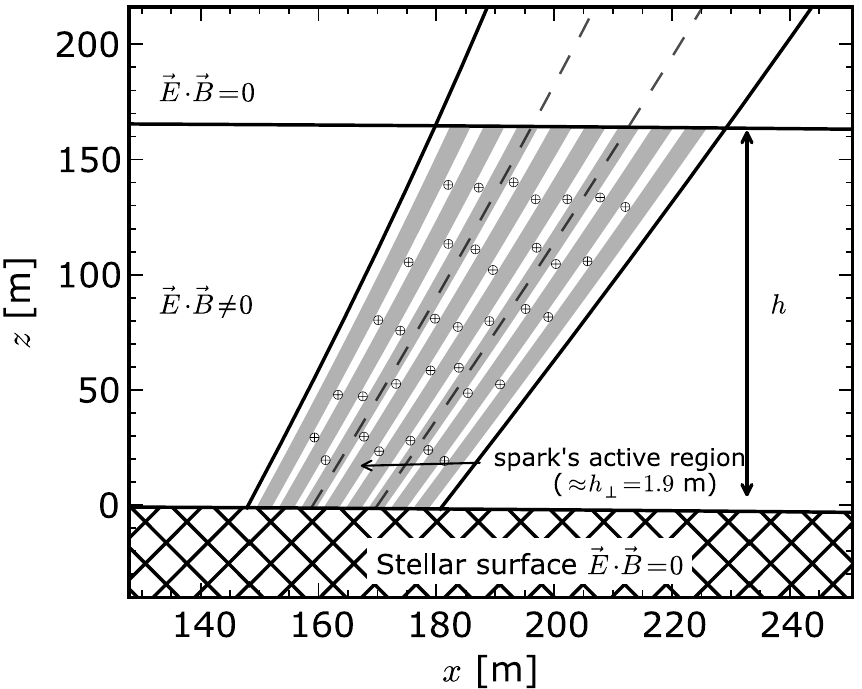}
\par\end{centering}

\caption[{Gap structure in the PSG-off mode {[}PSR B0943+10{]}}]{Gap structure in the PSG-off mode for PSR B0943+10. Filled columns
represent the locations and sizes of the active regions of sparks.
Here we assumed that the active region of a spark (the place where
acceleration is high enough to produce a cascade) has a size comparable
with the spark half-width. The iron ions extracted from the surface
(due to a high surface temperature) are represented by circle-plus
symbols. \label{fig:psg.cr_breakdown}}
\end{figure}

Figure \ref{fig:psg.cr_breakdown} presents the result of finding
the gap height in the PSG-off mode for PSR B0943+10. The presented
solution corresponds to the magnetic field structure presented in
Section \ref{sec:model.b0943}. The average radius of curvature in
the gap region is relatively high, $\Re_{6}=0.7$, hence the inclination
of the gap region. The polar gap conditions, the strength of magnetic
field $B_{14}=2.4$ ($R_{{\rm bb}}=17\,{\rm m}$) and the polar cap
temperature $T_{6}=3.0$ were restrained to follow the observed values
(see Table \ref{tab:x-ray_thermal}). The presented solution corresponds
to the following PSG parameters: gap height $h=166\,{\rm m}$, spark
half-width $h_{\perp}=1.9\,{\rm m}$, $\eta=1$ (fixed), $\kappa=7\times10^{-3}$,
$\gamma_{{\rm c}}=1.4\times10^{6}$. Note that the primary particles
will gain $\gamma_{{\rm max}}=1.9\times10^{6}$ as the CR efficiency
is not high enough to stop the acceleration.

\begin{comment}
\textasciitilde{}/Programs/studies/phd/gap\_breakdown/gap\_breakdown.py
(plot\_cr), set\_=318, ds=1e2
\end{comment}

\subsection{PSG-on mode\label{sec:psg.on_gap}}

In the PSG-on mode, radiation of the surface just below the spark
is in quasi-equilibrium with the flux of backstreaming particles.
When the surface temperature rises, the density of iron ions increases,
thus resulting in a decrease in the potential drop, which in turn,
reduces the flux of backstreaming particles. On the other hand, when
the surface temperature decreases it entails the drop of iron ion
density and, consequently, an increase in the flux of backstreaming
particles. Thus the polar cap temperature is maintained slightly below
the critical value. This quasi-equilibrium state prevents the gap
breakdown due to surface overheating. However, a high multiplicity
in the PSG-on mode leads to a production of dense plasma. When the
density of the plasma $n_{p}\gg\eta n_{{\rm GJ}}$, the acceleration
potential drop will be completely screened due to charge separation. 

\begin{figure}[H]
\begin{centering}
\includegraphics[height=18cm]{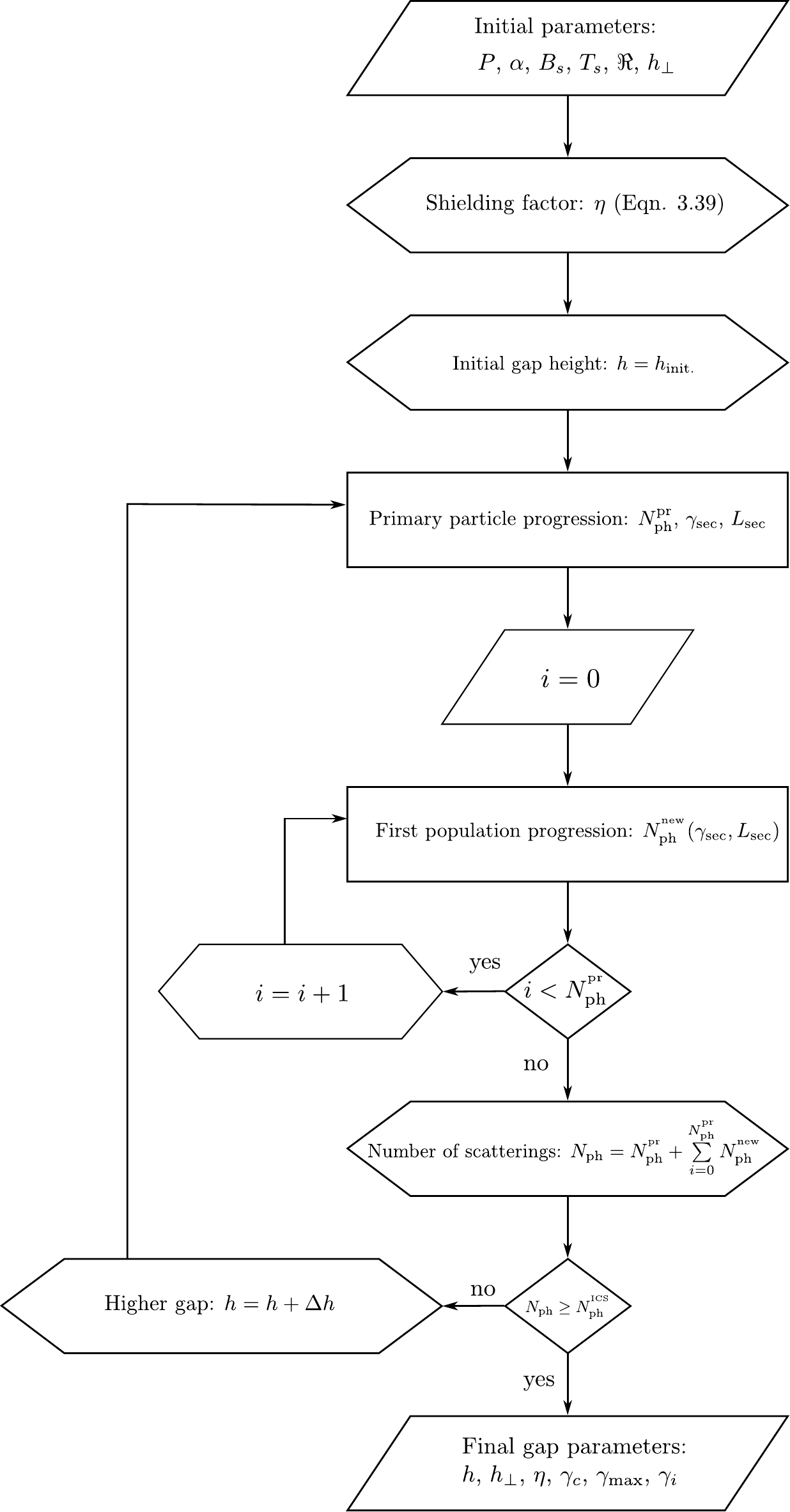}
\par\end{centering}

\centering{}\caption[{Flowchart of the algorithm used to estimate the gap height {[}PSG-on
mode{]}}]{Flowchart of algorithm used to estimate the gap height in PSG-on
mode for a given spark half-width (see text for more details).\label{fig:psg.flowchart_ics_hperp} }
\end{figure}

Alongside the pulsar parameters the gap height in the PSG-on mode
also depends on the spark half-width $h_{\perp}$ and on the number
of scatterings by the first population of newly created particles
$N_{{\rm ph}}^{{\rm ^{ICS}}}$. For a sample of pulsars we can use
drift information to put constraints on the spark half-width (see
Section \ref{sec:psg.drift}). Figure \ref{fig:psg.flowchart_ics_hperp}
presents the procedure of finding the gap height in the PSG-on mode
for the following pulsar parameters: a pulsar period $P$, a pulsar
inclination angle $\alpha$, a surface magnetic field strength $B_{{\rm s}}$,
a surface temperature $T_{{\rm s}}$, a curvature radius of magnetic
field lines $\Re$, and a spark half-width $h_{\perp}$. First we
use Equation \ref{eq:psg.heating_condition} to estimate the screening
factor $\eta$ which defines the electric field, and thus the particle
acceleration. Then we estimate the number of scatterings for a single
outflowing particle $N_{{\rm ph}}^{{\rm ^{pr}}}$ for the initial
gap height. The initial gap height from which we begin our calculations
is an arbitrary set to $h_{{\rm init.}}=10\,{\rm m}$. We track the
propagation of $\gamma$-photons produced by ICS on a primary particle
to find the location $L_{{\rm new}}$ where pairs are created. Then
we calculate their propagation through the acceleration region and
we estimate the number of scatterings by every newly created particle
of the first population $N_{{\rm ph}}^{{\rm ^{new}}}$. If the total
number of scatterings by the first population (including the primary
particle) is $N_{{\rm ph}}<N_{{\rm ph}}^{{\rm ^{ICS}}}$ , we resume
our calculations assuming a higher gap until the $N_{{\rm ph}}\geq N_{{\rm ph}}^{{\rm ^{ICS}}}$
is met.

As a result we obtain the gap parameters: the gap height $h$, the
screening factor $\eta$, the characteristic Lorentz factor of a particle
at the moment of ICS photon emission $\gamma_{{\rm c}}$, the maximum
value of the Lorentz factor $\gamma_{{\rm max}}$, and the characteristic
Lorentz factor of iron ions $\gamma_{{\rm i}}$. In our calculations,
if not stated otherwise, we use $N_{{\rm ph}}^{^{{\rm ICS}}}=25$
to calculate the gap parameters of the PSG-on mode. Note that in this
approximation we take into account only the first population of newly
created particles. In fact, the avalanche nature of the ICS-dominated
gap will result in a much higher multiplicity than in the PSG-off
mode $M_{{\rm ICS}}\gg M_{{\rm CR}}$. For details of particle/photon
propagation, see Chapter \ref{chap:cascade}. Subpulse drift observations
are available only for a few X-ray pulsars with the hot spot component.
Thus, to find the approximate gap parameters for pulsars without the
predicted spark half-width we use $h_{\perp}=2\,{\rm {\rm m}}$.

\begin{figure}[H]
\begin{centering}
\includegraphics[height=7cm]{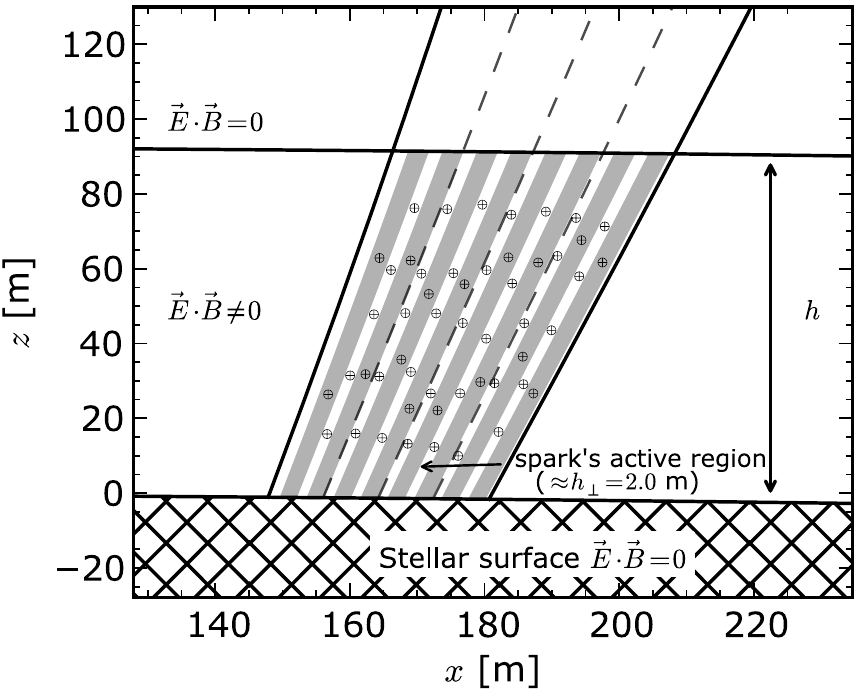}
\par\end{centering}

\centering{}\caption[{Gap structure in the PSG-on mode {[}PSR B0943+10{]}}]{Gap structure in the PSG-on mode for PSR B0943+10. Filled columns
represent the locations and sizes of the active regions of sparks.
Here we assumed that the active region of a spark (the place where
acceleration is high enough to produce a cascade) has a size comparable
with the spark half-width. The iron ions extracted from the surface
(due to a high surface temperature) are represented by circle-plus
symbols. Note that iron ions are present in both the non-active and
active regions. The density of ions in the non-active regions is so
high that it prevents cascade formation of pairs.\label{fig:psg.ics_breakdown}}
\end{figure}

Figure \ref{fig:psg.ics_breakdown} presents the result of finding
the gap height in the PSG-on mode for PSR B0943+10. In this model
the gap parameters, such as the magnetic field strength $B_{{\rm s}}$
and the surface temperature $T_{{\rm s}}$, were restrained to follow
the observed values (see Table \ref{tab:x-ray_thermal}). The result
was obtained for the non-dipolar structure of a surface magnetic field
presented in Section \ref{sec:model.b0943} and for the predicted
value of a spark half-width $h_{\perp}\approx2\,{\rm m}$ (see Table
\ref{tab:psg.drift}). The height required to produce $N_{{\rm ph}}^{{\rm ^{ICS}}}=25$
photons by the first population of particles was estimated as $h\approx92\,{\rm m}$.
Other gap parameters for this solution can be found in Table \ref{tab:psg.psg_top}.

\begin{comment}
\textasciitilde{}/Programs/studies/phd/gap\_breakdown/gap\_breakdown.py
(plot\_ics, 2, old in 0)
\end{comment}

\subsection{Results}

In Table \ref{tab:psg.psg_top} we present the results of finding
the gap height for the sample of pulsars. For the PSG-on mode we show
the estimated PSG parameters found using the predicted spark half-width
and the spark half-width $h_{\perp}=2\,{\rm m}$. The only exception
is Geminga (PSR J0633+1746), for which drift information is not available
and we can only present calculations for $h_{\perp}=2\,{\rm m}$.
For PSR B0628-28 the predicted spark half-width is large ($h_{\perp}=3.9\,{\rm m}$),
which entails a high acceleration potential. For such wide sparks
it is not possible to find the PSG-on solution with the required number
of scatterings $N_{{\rm ph}}^{^{{\rm ICS}}}$. We believe that for
this specific pulsar the predicted spark half-width is overestimated.
Actually, if a spark is narrower ($h_{\perp}=2\,{\rm m}$), it can
operate in the PSG-on mode (see Table \ref{tab:psg.psg_top}). This
result may suggest that for this specific pulsar the parameters of
the subpulse phenomenon could be overestimated (e.g. due to aliasing).
On the other hand, X-ray observations of Geminga suggest a relatively
low temperature of the hot spot \linebreak{}
($T_{{\rm s}}\approx1.9\,{\rm MK}$). The low density of the background
photons requires the formation of narrow sparks ($h_{{\rm \perp}}=1\,{\rm m}$)
to allow the gap to operate in the PSG-on mode. We believe that the
relatively large hot spot ($R_{{\rm pc}}=44.5\,{\rm m}$) of Geminga
causes the width of the sparks to grow so fast that it can operate
only in the PSG-off mode. We believe that this can explain the very
weak radio luminosity of the Geminga pulsar.

\begin{comment}
Results from results.log. files in: 403 (\textbf{PSR B0628-28}), 373
(\textbf{PSR J0633+1746}), 383 (\textbf{PSR B0834+06}), 315 (\textbf{PSR
B0943+10}), 350 (\textbf{PSR B0950+08}), 341 (\textbf{PSR B1133+16}),
322 (\textbf{PSR B1929+10})
\end{comment}

\begin{table}[H]
\caption[Estimated parameters of PSG for the sample of pulsars]{Estimated parameters of PSG for the sample of pulsars. The conditions
in the polar cap region: surface temperature, magnetic field strength,
polar cap radius, and curvature radius of the field lines are given
the in headers next to the pulsar name. The individual columns are
as follows: (1) PSG mode (see Section \ref{sec:psg.multiplicity}),
(2) Gap height, (3) Spark half-width, (4) Screening factor, (5) Overheating
parameter, (6) Characteristic Lorentz factor of scattering particles
, (7) Maximum Lorentz factor of primary particles, (8) Lorentz factor
of iron ions (if they are relativistic), (9) Particle mean free path,
and (10) Photon mean free path. The results are presented for two
different gap breakdown scenarios: the PSG-off and PSG-on modes (see
Section \ref{sec:psg.finding_height} for more details).\protect \\
 $^{a}$ The modes correspond to calculations using the predicted
spark half-width (see Table \ref{tab:psg.drift})\protect \\
$^{b}$ The modes correspond to calculations with a spark half-width
$h_{\perp}=2\,{\rm m}$\label{tab:psg.psg_top}}

\centering{}%
\begin{tabular}{|c|c|c|c|c|c|c|c|c|c|}
\hline 
 &  &  &  &  &  &  &  &  & \tabularnewline
mode & $h$ & $h_{\perp}$ & $\eta$ & $\kappa$ & $\gamma_{{\rm c}}$ & $\gamma_{{\rm max}}$ & $\gamma_{{\rm i}}$ & $l_{{\rm p}}$ & $l_{{\rm ph}}$\tabularnewline
{\footnotesize $\left(N_{{\rm ph}}\right)$} & {\footnotesize $\left({\rm m}\right)$} & {\footnotesize $\left({\rm m}\right)$} &  &  &  &  &  & {\footnotesize $\left({\rm m}\right)$} & {\footnotesize $\left({\rm m}\right)$}\tabularnewline
\hline 
\hline 
\multicolumn{10}{|c|}{}\tabularnewline
\multicolumn{10}{|c|}{\textbf{PSR B0628-28}\hspace{1cm}$T_{6}=2.8$\hspace{1cm} $B_{14}=2.2$\hspace{1cm}
$R_{{\rm pc}}=21.3\,{\rm m}$\hspace{1cm} $\Re_{6}=0.6$}\tabularnewline
\multicolumn{1}{|c}{} & \multicolumn{1}{c}{} & \multicolumn{1}{c}{} & \multicolumn{1}{c}{} & \multicolumn{1}{c}{} & \multicolumn{1}{c}{} & \multicolumn{1}{c}{} & \multicolumn{1}{c}{} & \multicolumn{1}{c}{} & \tabularnewline
\hline 
 &  &  &  &  &  &  &  &  & \tabularnewline
off  & $198.3$  & $3.2$ & -- & $0.007$ & $1.3\times10^{6}$ & $1.6\times10^{6}$ & -- & $1.4$ & $58.9$\tabularnewline
on$^{a}$ & -- & $3.6$ & -- & -- & -- & -- & -- & -- & --\tabularnewline
on$^{b}$ & $78.6$ & $2.0$ & $0.15$ & -- & $6.1\times10^{3}$ & $8.9\times10^{4}$ & $23$ & $1.9$ & $1.3$\tabularnewline
\hline 
\multicolumn{10}{|r|}{\emph{Continued on next page}}\tabularnewline
\hline 
\end{tabular}
\end{table}

\begin{table}[H]
\begin{centering}
Table \ref{tab:psg.psg_top} - continued from previous page
\par\end{centering}

\vspace{0.3cm}

\centering{}%
\begin{tabular}{|c|c|c|c|c|c|c|c|c|c|}
\hline 
 &  &  &  &  &  &  &  &  & \tabularnewline
mode & $h$ & $h_{\perp}$ & $\eta$ & $\kappa$ & $\gamma_{{\rm c}}$ & $\gamma_{{\rm max}}$ & $\gamma_{{\rm i}}$ & $l_{{\rm p}}$ & $l_{{\rm ph}}$\tabularnewline
 & {\footnotesize $\left({\rm m}\right)$} & {\footnotesize $\left({\rm m}\right)$} &  &  &  &  &  & {\footnotesize $\left({\rm m}\right)$} & {\footnotesize $\left({\rm m}\right)$}\tabularnewline
\hline 
\hline 
\multicolumn{10}{|c|}{}\tabularnewline
\multicolumn{10}{|c|}{\textbf{PSR B0628-28}\hspace{1cm}$T_{6}=2.8$\hspace{1cm} $B_{14}=2.2$\hspace{1cm}
$R_{{\rm pc}}=21.3\,{\rm m}$\hspace{1cm} $\Re_{6}=0.6$}\tabularnewline
\multicolumn{1}{|c}{} & \multicolumn{1}{c}{} & \multicolumn{1}{c}{} & \multicolumn{1}{c}{} & \multicolumn{1}{c}{} & \multicolumn{1}{c}{} & \multicolumn{1}{c}{} & \multicolumn{1}{c}{} & \multicolumn{1}{c}{} & \tabularnewline
\hline 
 &  &  &  &  &  &  &  &  & \tabularnewline
off  & $198.3$  & $3.2$ & -- & $0.007$ & $1.3\times10^{6}$ & $1.6\times10^{6}$ & -- & $1.4$ & $58.9$\tabularnewline
on$^{a}$ & -- & $3.6$ & -- & -- & -- & -- & -- & -- & --\tabularnewline
on$^{b}$ & $78.6$ & $2.0$ & $0.15$ & -- & $6.1\times10^{3}$ & $8.9\times10^{4}$ & $23$ & $1.9$ & $1.3$\tabularnewline
 &  &  &  &  &  &  &  &  & \tabularnewline
\hline 
\hline 
\multicolumn{1}{|c}{} & \multicolumn{1}{c}{} & \multicolumn{1}{c}{} & \multicolumn{1}{c}{} & \multicolumn{1}{c}{} & \multicolumn{1}{c}{} & \multicolumn{1}{c}{} & \multicolumn{1}{c}{} & \multicolumn{1}{c}{} & \tabularnewline
\multicolumn{10}{|c|}{\textbf{PSR J0633+1746}\hspace{1cm}$T_{6}=1.9$\hspace{1cm} $B_{14}=1.5$\hspace{1cm}
$R_{{\rm pc}}=44.5\,{\rm m}$\hspace{1cm} $\Re_{6}=2.1$}\tabularnewline
\multicolumn{1}{|c}{} & \multicolumn{1}{c}{} & \multicolumn{1}{c}{} & \multicolumn{1}{c}{} & \multicolumn{1}{c}{} & \multicolumn{1}{c}{} & \multicolumn{1}{c}{} & \multicolumn{1}{c}{} & \multicolumn{1}{c}{} & \tabularnewline
\hline 
 &  &  &  &  &  &  &  &  & \tabularnewline
off & $252.1$ & $1.5$ & -- & $0.0002$ & $2.9\times10^{6}$ & $3.5\times10^{6}$ & -- & $2.2$ & $66.6$\tabularnewline
on$^{b}$ & -- & $2.0$ & -- & -- & -- & -- & -- & -- & --\tabularnewline
 &  &  &  &  &  &  &  &  & \tabularnewline
\hline 
\hline 
\multicolumn{1}{|c}{} & \multicolumn{1}{c}{} & \multicolumn{1}{c}{} & \multicolumn{1}{c}{} & \multicolumn{1}{c}{} & \multicolumn{1}{c}{} & \multicolumn{1}{c}{} & \multicolumn{1}{c}{} & \multicolumn{1}{c}{} & \tabularnewline
\multicolumn{10}{|c|}{\textbf{PSR B0834+06}\hspace{1cm}$T_{6}=2.4$\hspace{1cm} $B_{14}=1.9$\hspace{1cm}
$R_{{\rm pc}}=22.7\,{\rm m}$\hspace{1cm} $\Re_{6}=0.6$}\tabularnewline
\multicolumn{1}{|c}{} & \multicolumn{1}{c}{} & \multicolumn{1}{c}{} & \multicolumn{1}{c}{} & \multicolumn{1}{c}{} & \multicolumn{1}{c}{} & \multicolumn{1}{c}{} & \multicolumn{1}{c}{} & \multicolumn{1}{c}{} & \tabularnewline
\hline 
 &  &  &  &  &  &  &  &  & \tabularnewline
off  & $172.5$ & $3.2$ & -- & $0.0027$ & $1.2\times10^{6}$ & $1.4\times10^{6}$ & -- & $1.3$ & $52.1$\tabularnewline
on$^{a}$ & $82.0$ & $1.8$ & $0.12$ & -- & $5.3\times10^{3}$ & $7.0\times10^{4}$ & $18$ & $2.3$ & $1.5$\tabularnewline
on$^{b}$ & $102.9$ & $2.0$ & $0.11$ & -- & $4.9\times10^{3}$ & $7.8\times10^{4}$ & $20$ & $2.4$ & $1.8$\tabularnewline
 &  &  &  &  &  &  &  &  & \tabularnewline
\hline 
\hline 
\multicolumn{1}{|c}{} & \multicolumn{1}{c}{} & \multicolumn{1}{c}{} & \multicolumn{1}{c}{} & \multicolumn{1}{c}{} & \multicolumn{1}{c}{} & \multicolumn{1}{c}{} & \multicolumn{1}{c}{} & \multicolumn{1}{c}{} & \tabularnewline
\multicolumn{10}{|c|}{\textbf{PSR B0943+10}\hspace{1cm}$T_{6}=3.1$\hspace{1cm} $B_{14}=2.4$\hspace{1cm}
$R_{{\rm pc}}=17.6\,{\rm m}$\hspace{1cm} $\Re_{6}=0.7$}\tabularnewline
\multicolumn{1}{|c}{} & \multicolumn{1}{c}{} & \multicolumn{1}{c}{} & \multicolumn{1}{c}{} & \multicolumn{1}{c}{} & \multicolumn{1}{c}{} & \multicolumn{1}{c}{} & \multicolumn{1}{c}{} & \multicolumn{1}{c}{} & \tabularnewline
\hline 
 &  &  &  &  &  &  &  &  & \tabularnewline
off  & $168.3$ & $1.9$ & -- & $0.0068$ & $1.6\times10^{6}$ & $2.0\times10^{6}$ & -- & $1.4$ & $46.5$\tabularnewline
on$^{a,b}$  & $71.6$ & $2.0$ & $0.1$ & -- & $8.8\times10^{3}$ & $2.5\times10^{5}$ & $63$ & $1.1$ & $1.1$\tabularnewline
 &  &  &  &  &  &  &  &  & \tabularnewline
\hline 
\multicolumn{10}{|r|}{\emph{Continued on next page}}\tabularnewline
\hline 
\end{tabular}
\end{table}

\begin{table}[H]
\begin{centering}
Table \ref{tab:psg.psg_top} - continued from previous page
\par\end{centering}

\vspace{0.3cm}

\centering{}%
\begin{tabular}{|c|c|c|c|c|c|c|c|c|c|}
\hline 
 &  &  &  &  &  &  &  &  & \tabularnewline
mode & $h$ & $h_{\perp}$ & $\eta$ & $\kappa$ & $\gamma_{{\rm c}}$ & $\gamma_{{\rm max}}$ & $\gamma_{{\rm i}}$ & $l_{{\rm p}}$ & $l_{{\rm ph}}$\tabularnewline
 & {\footnotesize $\left({\rm m}\right)$} & {\footnotesize $\left({\rm m}\right)$} &  &  &  &  &  & {\footnotesize $\left({\rm m}\right)$} & {\footnotesize $\left({\rm m}\right)$}\tabularnewline
\hline 
\hline 
\multicolumn{1}{|c}{} & \multicolumn{1}{c}{} & \multicolumn{1}{c}{} & \multicolumn{1}{c}{} & \multicolumn{1}{c}{} & \multicolumn{1}{c}{} & \multicolumn{1}{c}{} & \multicolumn{1}{c}{} & \multicolumn{1}{c}{} & \tabularnewline
\multicolumn{10}{|c|}{\textbf{PSR B0950+08}\hspace{1cm} $T_{6}=2.6$\hspace{1cm} $B_{14}=2.0$\hspace{1cm}
$R_{{\rm pc}}=14.0\,{\rm m}$\hspace{1cm} $\Re_{6}=0.8$}\tabularnewline
\multicolumn{1}{|c}{} & \multicolumn{1}{c}{} & \multicolumn{1}{c}{} & \multicolumn{1}{c}{} & \multicolumn{1}{c}{} & \multicolumn{1}{c}{} & \multicolumn{1}{c}{} & \multicolumn{1}{c}{} & \multicolumn{1}{c}{} & \tabularnewline
\hline 
 &  &  &  &  &  &  &  &  & \tabularnewline
off  & $172.8$ & $1.9$ & -- & $0.0009$ & $1.7\times10^{6}$ & $2.0\times10^{6}$ & -- & $1.6$ & $47.9$\tabularnewline
on$^{a}$ & $16.9$ & $0.7$ & $0.09$ & -- & $3.9\times10^{3}$ & $2.3\times10^{4}$ & $6$ & $1.4$ & $0.6$\tabularnewline
on$^{b}$ & $61.7$ & $2.0$ & $0.03$ & -- & $5.1\times10^{3}$ & $6.6\times10^{4}$ & $17$ & $1.8$ & $1.6$\tabularnewline
 &  &  &  &  &  &  &  &  & \tabularnewline
\hline 
\hline 
\multicolumn{1}{|c}{} & \multicolumn{1}{c}{} & \multicolumn{1}{c}{} & \multicolumn{1}{c}{} & \multicolumn{1}{c}{} & \multicolumn{1}{c}{} & \multicolumn{1}{c}{} & \multicolumn{1}{c}{} & \multicolumn{1}{c}{} & \tabularnewline
\multicolumn{10}{|c|}{\textbf{PSR B1133+16}\hspace{1cm}$T_{6}=2.9$\hspace{1cm} $B_{14}=2.3$\hspace{1cm}
$R_{{\rm pc}}=17.9\,{\rm m}$\hspace{1cm} $\Re_{6}=0.6$}\tabularnewline
\multicolumn{1}{|c}{} & \multicolumn{1}{c}{} & \multicolumn{1}{c}{} & \multicolumn{1}{c}{} & \multicolumn{1}{c}{} & \multicolumn{1}{c}{} & \multicolumn{1}{c}{} & \multicolumn{1}{c}{} & \multicolumn{1}{c}{} & \tabularnewline
\hline 
 &  &  &  &  &  &  &  &  & \tabularnewline
off & $167.4$ & $2.4$ & -- & $0.0076$ & $1.4\times10^{6}$ & $1.7\times10^{6}$ & -- & $1.3$ & $48.4$\tabularnewline
on$^{a}$ & $95.9$ & $2.9$ & $0.08$ & -- & $7.0\times10^{3}$ & $1.9\times10^{5}$ & $47$  & $1.2$ & $1.3$\tabularnewline
on$^{b}$ & $54.3$ & $2.0$ & $0.11$ & -- & $7.9\times10^{3}$ & $1.3\times10^{5}$ & $33$ & $1.4$ & $1.1$\tabularnewline
 &  &  &  &  &  &  &  &  & \tabularnewline
\hline 
\hline 
\multicolumn{1}{|c}{} & \multicolumn{1}{c}{} & \multicolumn{1}{c}{} & \multicolumn{1}{c}{} & \multicolumn{1}{c}{} & \multicolumn{1}{c}{} & \multicolumn{1}{c}{} & \multicolumn{1}{c}{} & \multicolumn{1}{c}{} & \tabularnewline
\multicolumn{10}{|c|}{\textbf{PSR B1929+10}\hspace{1cm}$T_{6}=3.0$\hspace{1cm} $B_{14}=2.4$\hspace{1cm}
$R_{{\rm pc}}=20\,{\rm m}$\hspace{1cm} $\Re_{6}=0.6$}\tabularnewline
\multicolumn{1}{|c}{} & \multicolumn{1}{c}{} & \multicolumn{1}{c}{} & \multicolumn{1}{c}{} & \multicolumn{1}{c}{} & \multicolumn{1}{c}{} & \multicolumn{1}{c}{} & \multicolumn{1}{c}{} & \multicolumn{1}{c}{} & \tabularnewline
\hline 
 &  &  &  &  &  &  &  &  & \tabularnewline
off & $112.7$ & $1.0$ & -- & $0.0012$ & $1.8\times10^{6}$ & $2.1\times10^{6}$ & -- & $1.1$ & $28.0$\tabularnewline
on$^{a}$ & $50.5$ & $1.6$ & $0.02$ & -- & $9.8\times10^{3}$ & $1.2\times10^{5}$ & $31$  & $1.6$ & $1.0$\tabularnewline
on$^{b}$ & $75.1$ & $2.0$ & $0.02$ & -- & $8.4\times10^{3}$ & $1.5\times10^{5}$ & $39$ & $1.5$ & $1.5$\tabularnewline
 &  &  &  &  &  &  &  &  & \tabularnewline
\hline 
\end{tabular}
\end{table}

\clearpage{}

\section{PSG model parameters \label{sec:psg.model_parameters}}

We can distinguish two types of PSG parameters: observed and derived.
As we have mentioned above, in some cases when X-ray observations
are available we can directly estimate the surface magnetic field
$B_{{\rm s}}$. On the one hand, $B_{{\rm s}}$ can be calculated
using the size of the hot spot $A_{{\rm bb}}$, and on the other hand
we can find $B_{{\rm s}}$ by using the estimation of the critical
temperature and the assumption that $T_{{\rm s}}=T_{{\rm crit}}$.
One of the most important requirements for the PSG model is that these
two estimations should coincide with each other. As is clear from
Figure \ref{fig:x-ray.medin_lai}, in most cases when the hot spot
parameters are available this requirement is fulfilled. Thus, we can
assume that the characteristic values of $B_{{\rm s}}$ vary in the
range of $(1-4)\times10^{14}\,{\rm G}$, which corresponds to the
critical surface temperature in the range of $(1.3-5)\times10^{6}\,{\rm K}$
(see Table \ref{tab:x-ray_thermal}). By using these values we can
estimate the derived parameters of PSG, such as the gap height $h$,
the screening factor $\eta$ (or the overheating parameter $\kappa$
in the PSG-off mode) and the characteristic Lorentz factor of primary
particles $\gamma_{{\rm c}}$. Let us note that these parameters also
depend on the curvature radius of the magnetic field lines $\Re$.
The curvature can be neither observed nor derived, but modelling of
the surface magnetic field (see Chapter \ref{chap:model}) indicates
that the curvature radius varies in the range of $(0.1-10)\times10^{6}$
cm. Below we will discuss the influence of pulsar parameters, such
as the magnetic field, the curvature of field lines and the period
on derived PSG parameters.

\subsection{Influence of the magnetic field}

\begin{figure}[H]
\centering{} \includegraphics{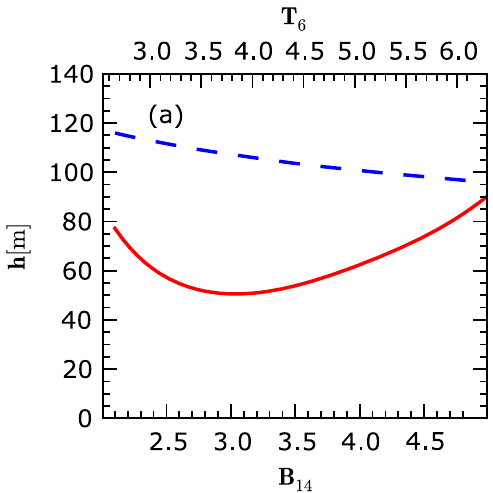} \includegraphics{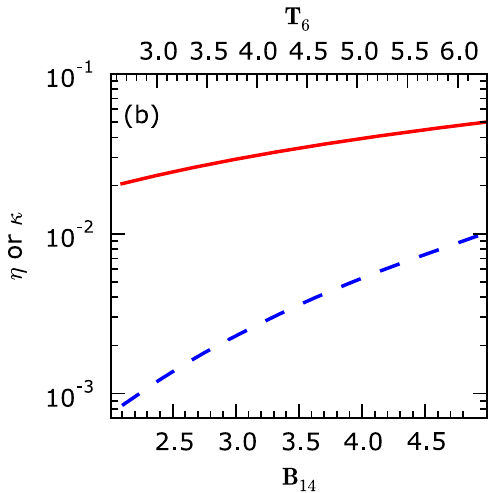}
\includegraphics{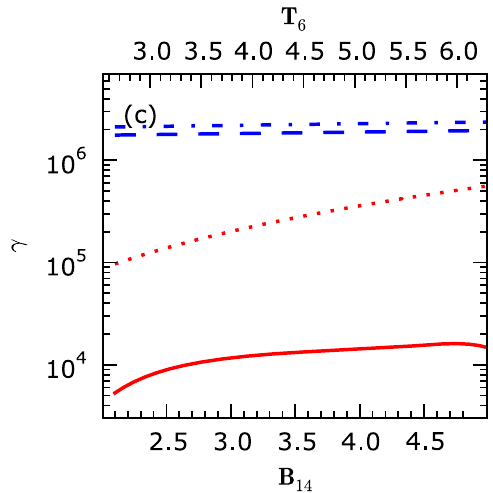} \caption[Dependence of the PSG model parameters on the surface magnetic field.]{Dependence of the gap height (panel a), the screening factor or the
overheating parameter (panel b), and the particle Lorentz factor (panel
c) on the surface magnetic field. Solid red lines correspond to the
PSG-on mode (ICS-dominated gaps) while dashed blue lines correspond
to the PSG-off mode (CR-dominated gaps). Calculations were performed
using the following parameters: $P=0.23$, $\Re_{6}=0.6$, $B_{{\rm d}}=1.2\times10^{12}\,{\rm G}$,
and $\alpha=36^{\circ}$. The actual polar cap radius was calculated
separately for a given surface magnetic field as $R_{{\rm pc}}=R_{{\rm dp}}\sqrt{B_{{\rm d}}/B_{{\rm s}}}$.
In panel (c) the red solid and dotted lines correspond to characteristic
and maximum Lorentz factors ($\gamma_{{\rm c}}$, $\gamma_{{\rm max}}$)
in the PSG-on mode while blue dashed and dashed-dotted lines correspond
to $\gamma_{{\rm c}}$ and $\gamma_{{\rm max}}$ in the PSG-off mode.
Corresponding critical temperature is shown on top axis of the figures.
\label{fig:psg.h_eta_b14} }
\end{figure}

The conditions in PSG are mainly defined by the surface magnetic field.
In Figure \ref{fig:psg.h_eta_b14}, panel (a) we present the dependence
of the gap height on the surface magnetic field calculated according
to the approach described in Section \ref{sec:psg.finding_height}.
It is clear that in the PSG-off mode the gap height decreases as the
surface magnetic field increases. In the PSG-on mode, on the other
hand, the gap height shows a minimum at a specific value of the magnetic
field strength (for a given pulsar's parameters it is $B_{14}\approx3$).
This behaviour is the result of an increasing potential acceleration
drop with an increasing surface magnetic field. When the magnetic
field strength exceeds the optimal value, which corresponds to acceleration
when ICS is more effective, the increase in the acceleration potential
results in less effective scattering. Panel (b) shows the dependence
of the screening factor (or the overheating parameter $\kappa$ in
the PSG-off mode) on the surface magnetic field. We can see that for
stronger magnetic fields both $\eta$ and $\kappa$ increase, which
means that: (1) the density of heavy ions above the polar cap in the
PSG-on mode decreases, (2) the density of particles required to overheat
(and thus to close) the polar cap increases. Let us note that the
surface temperature $T_{{\rm s}}$ stays very near to the critical
temperature $T_{{\rm crit}}$, which is shown on the top axis of the
Figures. In panel (c) the red, solid and dotted lines correspond to
characteristic and maximum Lorentz factors ($\gamma_{{\rm c}}$, $\gamma_{{\rm max}}$)
in the PSG-on mode, while the blue, dashed and dashed-dotted lines
correspond to $\gamma_{{\rm c}}$ and $\gamma_{{\rm max}}$ in the
PSG-off mode. We see that especially for the PSG-off mode $\gamma_{{\rm c}}$
does not depend on the magnetic field strength. Note also that in
the PSG-off mode (CR-dominated gap), the characteristic Lorentz factor
(the Lorentz factor for which most of the gamma photons are produced)
slightly differs from the maximum value, $\gamma_{{\rm c}}\approx\gamma_{{\rm max}}$.
On the other hand, in the PSG-on mode $\gamma_{{\rm c}}\ll\gamma_{{\rm max}}$,
which reflects the fact that most of the scatterings take place in
the bottom part of the gap.

\begin{comment}
\textasciitilde{}/Programs/magnetic/magnetic/src/radiation/gap.py
(show\_b14, show\_b14\_cr)

\textasciitilde{}/Programs/studies/phd/plot\_psg\_params/plot\_psg\_params.py
(plot\_b14)
\end{comment}

\subsection{Influence of the curvature radius \label{sec:psg.curvature_radius}}

\begin{figure}[H]
\centering{} \includegraphics{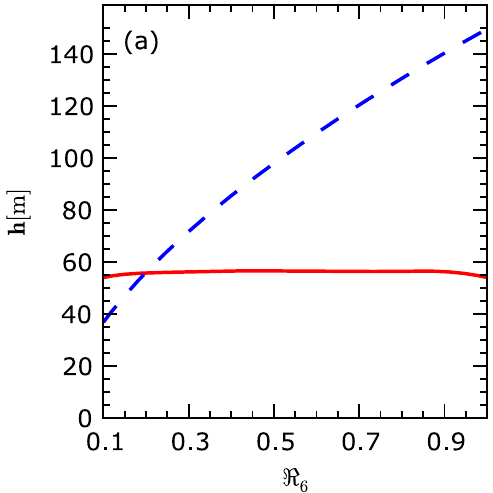} \includegraphics{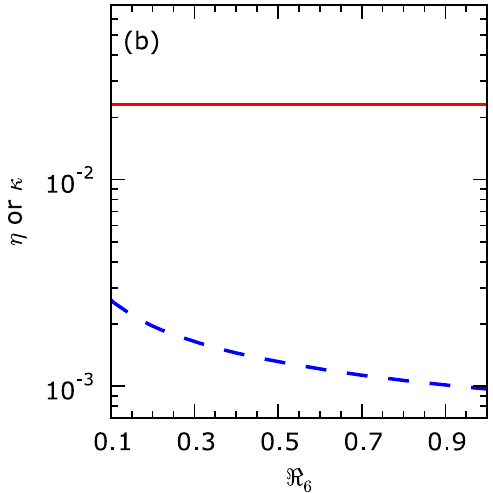}
\includegraphics{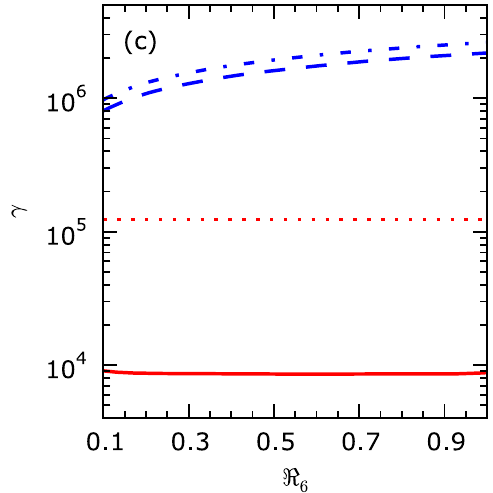} \caption[Dependence of the PSG model parameters on the curvature radius of
magnetic field lines.]{Dependence of the gap height (panel a), the screening factor or the
overheating parameter (panel b), and the particle Lorentz factor (panel
c) on the curvature radius of magnetic field lines. Calculations were
performed using the following parameters: $P=0.23$, $B_{{\rm d}}=1.2\times10^{12}\,{\rm G}$,
$B_{{\rm s}}=2.4\times10^{14}\,{\rm G}$, $\alpha=36^{\circ}$. For
a more detailed description see Figure \ref{fig:psg.h_eta_b14}. \label{fig:psg.h_re6} }
\end{figure}

The curvature of the magnetic field lines significantly affects the
gap height in the PSG-off mode (see Figure \ref{fig:psg.h_re6}, panel
a). In the case of the CR-dominated gap, the curvature of the magnetic
field lines affects not only the photons' mean free path (for higher
curvature the magnetic field will absorb photons faster), but also
the particle mean free path and, more importantly, the energy of photons
generated in the gap region. The higher energy of photons further
reduces the photon mean free path, thus resulting in lower heights
of the PSG. In contrast, the gap height in the PSG-on mode is only
slightly affected by changes in the curvature of the magnetic field
lines. In this case the most important parameter which determines
the cascade properties is the primary particle mean free path which
does not depend on the curvature of the magnetic field lines. 

The overheating parameter in the PSG-off mode inversely depends on
the radius of curvature of the magnetic field lines (see Figure \ref{fig:psg.h_re6},
panel b). The higher the curvature, the higher the overheating parameter,
which means that the sparks are narrower. This is consistent with
the expectation that for a higher curvature of the magnetic field
lines, the gap breakdown is easier to develop and takes place before
the sparks manage to grow in width. On the other hand, the screening
factor in the PSG-off mode does not depend on the curvature of the
magnetic field lines.

With an increasing radius of curvature the Lorentz factor of primary
particles (both $\gamma_{{\rm c}}$ and $\gamma_{{\rm max}}$) required
to close the gap in the PSG-off mode also increases. This reflects
the fact that in order to produce a sufficient number of photons in
the gap region, the primary particles should be accelerated to higher
energies (if the curvature is lower). Higher energies of the primary
particles will increase the emitted $\gamma$-photon energy, thereby
they will partly inhibit the growth of the photon mean free path due
to the lower curvature. As mentioned above, the gap height in the
PSG-on mode very weakly depends on the photon mean free path, thus
both $\gamma_{{\rm c}}$ and $\gamma_{{\rm max}}$ are not affected
by the increase in the radius of curvature.

\begin{comment}
\textasciitilde{}/Programs/magnetic/magnetic/src/radiation/gap.py
(show\_re6)

\textasciitilde{}/Programs/studies/phd/plot\_psg\_params/plot\_psg\_params.py
(plot\_re6)
\end{comment}

\subsection{Influence of the pulsar period}

\begin{figure}[H]
\centering{}\includegraphics{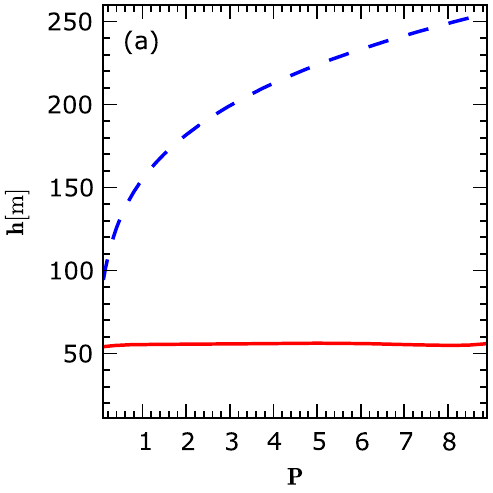} \includegraphics{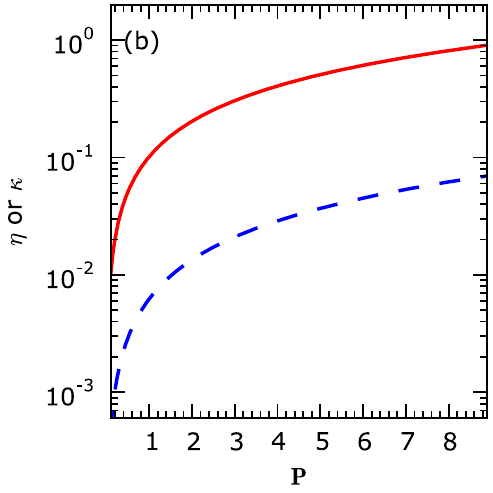}
\includegraphics{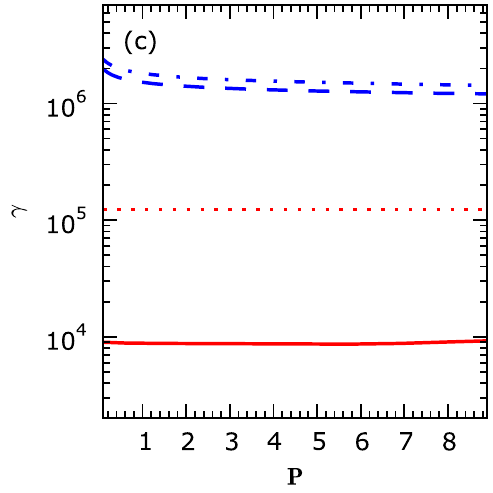} \caption[Dependence of the PSG model parameters on the pulsar period.]{Dependence of the gap height (panel a), the screening factor or a
overheating parameter (panel b), and the particle Lorentz factor (panel
c) on the pulsar period. Calculations were performed using the following
parameters: $P=0.23$, $B_{{\rm d}}=1.2\times10^{12}\,{\rm G}$, $B_{{\rm s}}=2.4\times10^{14}\,{\rm G}$,
$\alpha=36^{\circ}$, $\Re_{6}=0.6$. The actual polar cap radius
was calculated separately for a given pulsar period as $R_{{\rm pc}}=R_{{\rm dp}}\sqrt{B_{{\rm d}}/B_{{\rm s}}}$,
where $R_{{\rm dp}}=\sqrt{2\pi R^{3}/\left(cP\right)}$. For a more
detailed description see Figure \ref{fig:psg.h_eta_b14}.\label{fig:psg.h_p} }
\end{figure}

As we can see from Figure \ref{fig:psg.h_p}, panel (a) and panel
(c), in the PSG-on mode the gap height and the Lorentz factor of primary
particles do not depend on the pulsar period. The increase in the
screening factor (see Figure \ref{fig:psg.h_p}b) compensates the
increase in the acceleration potential drop (see Equation \ref{eq:psg.potential_heating}).
Thus the particles in the gap region are accelerated in the same way
independently of the pulsar period. On the other hand, the gap height
in the PSG-off mode increases with the increasing pulsar period. This
reflects the fact that in the PSG-off mode the acceleration potential,
and hence $\gamma_{{\rm c}}$ and $\gamma_{{\rm max}}$, decreases
with longer periods (see Equation \ref{eq:psg.potential_drop}). Longer
pulsar periods entail an increase in the screening factor (in the
PSG-on mode, see Equation \ref{eq:psg.eta_hperp}) and in the overheating
parameter (in the PSG-off mode). Note that for periods longer than
some specific value (for a given pulsar's parameters it is $P_{{\rm max}}\approx9\,{\rm s}$),
the screening factor in the PSG-on mode would exceed unity. This means
that the PSG-on mode cannot be responsible for the gap breakdown for
pulsars with such long periods.

\section{Drift model\label{sec:psg.drift}}

The existence of IAR in general causes a rotation of the plasma relative
to the NS, as the charge density differs from the Goldreich-Julian
co-rotational density. The power spectrum of radio emission must have
a feature due to this plasma rotation. This feature is indeed observed
and is called the drifting subpulse phenomenon.

\subsection{Aligned pulsars}

An explanation for drifting subpulses was offered by \citet{1975_Ruderman}
as being due to a rotating carousel of sub-beams within a hollow emission
cone. According to this model a pair cascades may not occur simultaneously
across the whole polar cap but is localised in the form of discharges
of small regions in the polar gap. Such sparks may produce plasma
columns that stream into the magnetosphere to produce the observed
radio emission. The location of the discharges on the polar cap determines
the geometrical pattern of instantaneous subpulses within a pulsar's
integrated pulse profile.

In the PSG model the stable pattern of subpulses is due to heating
of the inactive part of the spark (the place where no cascade forms
due to a low acceleration potential) by all the neighbouring discharges.
The lifetime of a single spark is very short. On the other hand, an
inactive region is continuously heated by all the neighbouring sparks.
Even when one of them dies, the temperature is still high enough (high
ion density) to prevent spark formation in this region. As the discharges
do not exchange information (they are not synchronised) and their
lifetime is very small, the geometrical pattern of sparks on the polar
cap should be stable. 

For pulsars with an aligned magnetic and rotation axis the sparks
circulate around the rotation axis. Note that this circulation is
not related with the magnetic axis but with the direction of the co-rotational
velocity. Namely, the drift velocity is opposed to the co-rotational
velocity.

\begin{figure}[H]
\begin{centering}
\includegraphics[height=6.7cm]{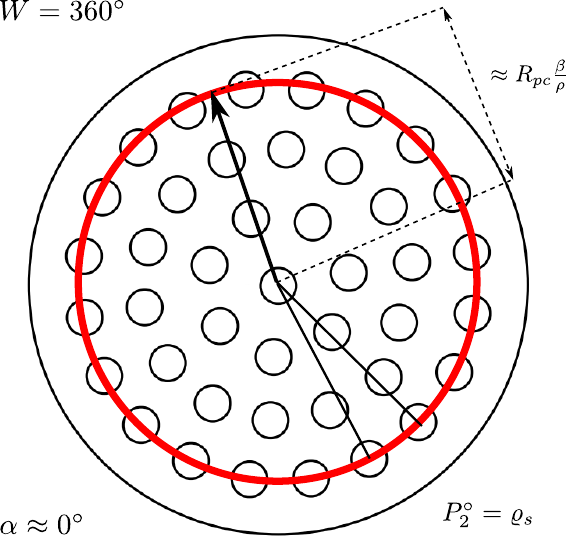}
\par\end{centering}

\caption[Top view of a polar cap region of an aligned pulsar]{Top view of a polar cap region of an aligned pulsar. Small circles
represent sparks, while the red line corresponds to the line of sight.
If we neglect the transition from a non-dipolar structure of the magnetic
field on the stellar surface to a dipolar structure in the region
where radio emission is produced, we can assume that the observed
subpulse separation $P_{2}^{\circ}$ also describes spark separation
$\varrho_{s}$ (angular separation between the adjacent sparks on
the polar cap). \label{fig:psg.top_aligned}}
\end{figure}

We can calculate the drift velocity of an aligned pulsar using the
following approximation (see Figure \ref{fig:psg.top_aligned})

\begin{equation}
v_{{\rm dr}}\approx\frac{2\pi R_{{\rm pc}}}{PP_{3}}\frac{\beta}{\rho}\frac{P_{2}^{\circ}}{360^{\circ}},
\end{equation}
where $R_{{\rm pc}}$ is the actual polar cap size, $P_{2}^{\circ}$
is the characteristic spacing between subpulses in the pulse longitude,
$P_{3}$ is the period at which a pattern of subpulses crosses the
pulse window (in units of the pulsar period), $\beta$ is the impact
angle, and $\rho$ is the opening angle.

In such an approximation the assumption that only half of the spark
is active can be written as

\begin{equation}
\frac{P_{2}^{\circ}}{360^{\circ}}\approx\frac{2h_{\perp}}{2\pi R_{{\rm pc}}\frac{\beta}{\rho}}.
\end{equation}

Finally, we can define the observed drift velocity of aligned pulsars
as

\begin{equation}
v_{dr}\approx\frac{2h_{\perp}}{PP_{3}}.
\end{equation}

\subsection{Non-aligned pulsars}

Most observed pulsars are non-aligned rotators. It is very common
to apply the carousel model to interpret observations of the drifting
subpulses of non-aligned pulsars \citep{1975_Ruderman}. Despite the
fact that the carousel model can explain some properties of subpulses
(for example the change in intensity), we believe that this model
is not suitable for describing the spark's behaviour on the polar
cap. There is no physical reason for a spark to circulate around the
magnetic axis. The circulation in aligned pulsars is caused by a lack
of coronation with respect to the rotation axis. For non-aligned pulsars,
the co-rotation velocity in the polar cap region has more or less
the same direction: what is more, if we assume circulation around
the magnetic axis we will get plasma with a velocity that is higher
than the co-rotational velocity, which is difficult to explain in
a region where the charge density is lower than the co-rotational
density.

As in our model, the drift is caused by a lack of charge in IAR, thus
the plasma should drift in approximately the same direction, i.e.
in the direction opposite to the co-rotation velocity. We believe
that the change in subpulse intensity is caused by the observation
of a different part of a spark and/or different conditions across
the polar cap at which the spark is formed (e.g. magnetic field strength,
curvature of the magnetic field lines, background photon flux).

\begin{figure}[H]
\begin{centering}
\includegraphics[height=7.5cm]{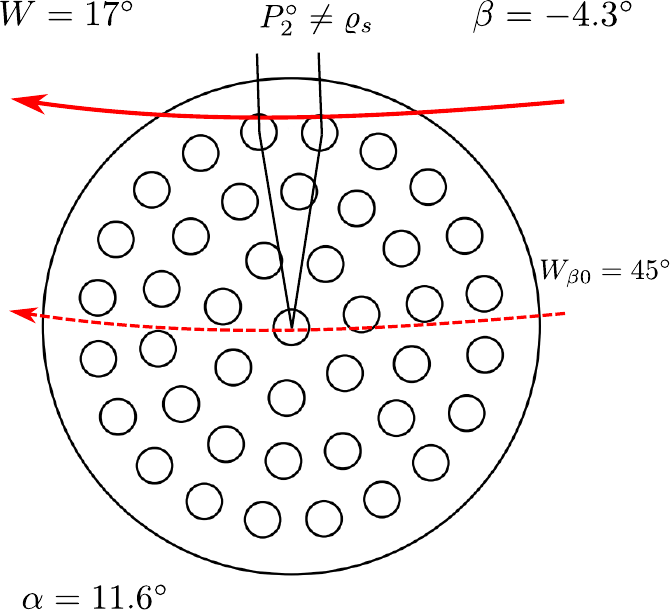}
\par\end{centering}

\caption[Top view of the polar cap region in the case of a non-aligned pulsar]{Top view of the polar cap region in the case of a non-aligned pulsar.
Small circles represent sparks, the red line corresponds to the line
of sight. In general, the observed subpulse separation $P_{2}^{\circ}$
does not describe the actual spark separation $\varrho_{s}$ (the
angular separation between the adjacent sparks on the polar cap).
In order to calculate the distance between the sparks we use an approximation
from Equation \ref{eq:psg.vdr_nona_approx}. \label{fig:psg.top_nonaligined}}
\end{figure}

For pulsars with a relatively high inclination angle $\alpha$ we
can calculate the drift velocity using the following approximation

\begin{equation}
v_{dr}\approx\frac{2R_{{\rm pc}}\frac{W}{W_{\beta0}}}{PP_{3}}\frac{P_{2}^{\circ}}{W}\approx\frac{2R_{{\rm pc}}}{PP_{3}}\frac{P_{2}^{\circ}}{W_{\beta0}},\label{eq:psg.vdr_nona_approx}
\end{equation}
where $W$ is the profile width and $W_{\beta0}\approx W/\sqrt{1-\left(\frac{\beta}{\rho}\right)^{2}}$
is the profile width calculated assuming $\beta=0$ (see Figure \ref{fig:psg.top_nonaligined}).
Using the assumption that only half of the spark is active, we can
write that

\begin{equation}
\frac{P_{2}^{\circ}}{W}\approx\frac{2h_{\perp}}{2R_{{\rm pc}}\frac{W}{W_{\beta0}}}\longrightarrow\frac{P_{2}^{\circ}}{W_{\beta0}}\approx\frac{h_{\perp}}{R_{{\rm pc}}},\label{eq:psg.hperp_wb0}
\end{equation}
and the drift velocity

\begin{equation}
v_{dr}=\frac{2h_{\perp}}{PP_{3}}.\label{eq:psg.vdr_same_order}
\end{equation}

The spark half-width can be calculated using Equation \ref{eq:psg.hperp_wb0}
as follows

\begin{equation}
h_{\perp}=R_{{\rm pc}}\frac{P_{2}^{\circ}}{W_{\beta0}}.\label{eq:psg.hperp_p2}
\end{equation}

\subsection{Screening factor}

In our model the drift is caused by a lack of charge in IAR, thus
we can write the equation for the drift velocity as follows 
\begin{equation}
{\bf v_{\perp}={\bf v}}_{{\rm dr}}=\frac{c{\bf \Delta E}\times{\bf B}}{B^{2}},
\end{equation}
where $\Delta{\bf E}$ is the electric field caused by the difference
of an actual charge density from the Goldreich-Julian co-rotational
density. We can a use calculation of the circulation of an electric
field, Equations \ref{eq:psg.potential_est2} and \ref{eq:psg.potential_drop},
to find the dependence of the drift velocity on the screening factor:

\begin{equation}
v_{{\rm dr}}=c\frac{E_{\theta}B_{r}}{B_{r}^{2}}=\frac{4\pi\eta h_{\perp}\cos\alpha}{P}.\label{eq:psg.vdr_shielding}
\end{equation}
Finally, by using Equations \ref{eq:psg.vdr_same_order} and \ref{eq:psg.vdr_shielding}
we can find the dependence of the screening factor on the observed
drift parameters

\begin{equation}
\eta=\frac{1}{2\pi P_{3}\cos\alpha}.\label{eq:psg.eta_p3}
\end{equation}

\subsection{Profile width and subpulse separation \label{sec:psg.profile_width}}

The key parameters in the above calculations are the pulse width $W$
(or $W_{\beta0}$), the characteristic spacing between subpulses $P_{2}^{\circ}$,
and the period at which a pattern of subpulses crosses the pulse window
$P_{3}$. Of these three only $P_{3}$ is easy to apply, both $W$
and $P_{2}^{\circ}$ need serious study before they can be used.

In general, the profile width depends on the frequency at which we
observe the pulsar, and most normal pulsars show a systematic increase
in pulse width and the separation of profile components when observed
at lower frequencies. The model known as radius-to-frequency mapping
explains this effect as a direct consequence of the emission at higher
frequencies being produced closer to the neutron star surface than
at lower frequencies. For this reason both the pulse width and the
spacing between subpulses should be measured at the same frequency.
Note that $P_{3}$ is not affected by this effect since its determination
involves analyses of many pulses and does not depend on the pulse
width. The observed pulse width $W$, measured in longitude of rotation,
can be calculated by applying simple spherical geometry \citep{1984_Gil}:

\begin{equation}
\sin^{2}\frac{W}{4}=\frac{\sin^{2}\left(\rho/2\right)-\sin^{2}\left(\beta/2\right)}{\sin\alpha\sin\left(\alpha+\beta\right)}.
\end{equation}

In the above calculations we are using the $W_{\beta0}\approx W/\sqrt{1-\left(\frac{\beta}{\rho}\right)^{2}}$approximation,
where $W_{\beta0}$ is the pulse width calculated assuming $\beta=0$.
In the first approximation we can assume that $W_{\beta0}$ corresponds
to the distance $2R_{{\rm pc}}$ at the polar cap which, is valid
for non-aligned pulsars with a relatively high inclination angle.
A more accurate value can be calculated using formulas presented in
\citet{1984_Gil}. The running polar coordinates along the line of
sight trajectory can be expressed in the form

\begin{equation}
\rho\left(\varphi\right)=2\arcsin\left(\sqrt{\sin^{2}\frac{\varphi}{2}\sin\alpha\sin\left(\alpha+\beta\right)+\sin^{2}\frac{\beta}{2}}\right),
\end{equation}

\begin{equation}
\sigma\left(\varphi\right)=\arctan\left(\frac{\sin\varphi\sin\alpha\sin\left(\alpha+\beta\right)}{\cos\left(\alpha+\beta\right)-\cos\alpha\cos\rho\left(\varphi\right)}\right).
\end{equation}

\begin{figure}[H]
\begin{centering}
\includegraphics[height=6.5cm]{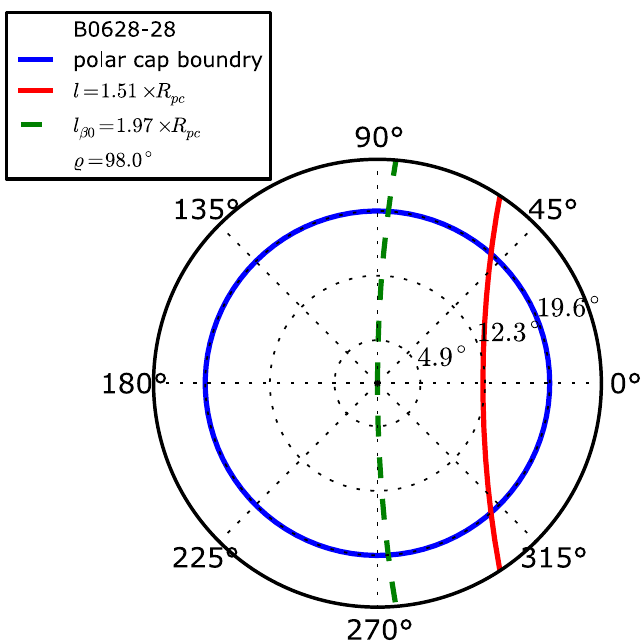}\hspace{0.5cm}\includegraphics[height=6.5cm]{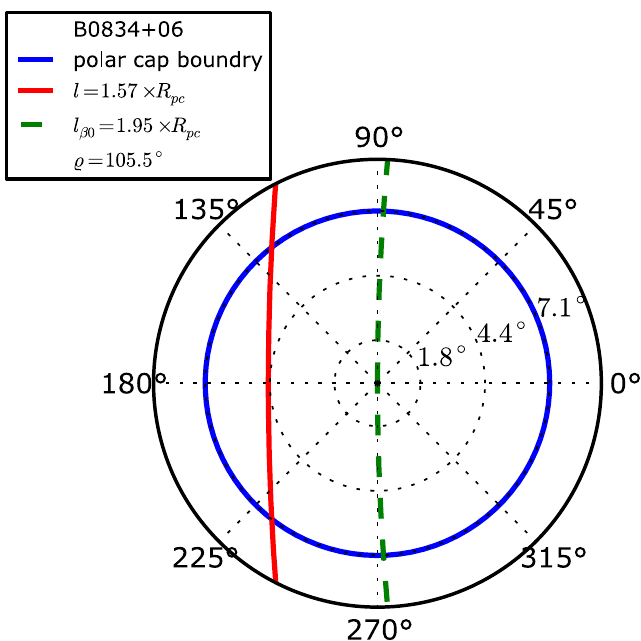}
\par\end{centering}

\begin{centering}
\includegraphics[height=6.5cm]{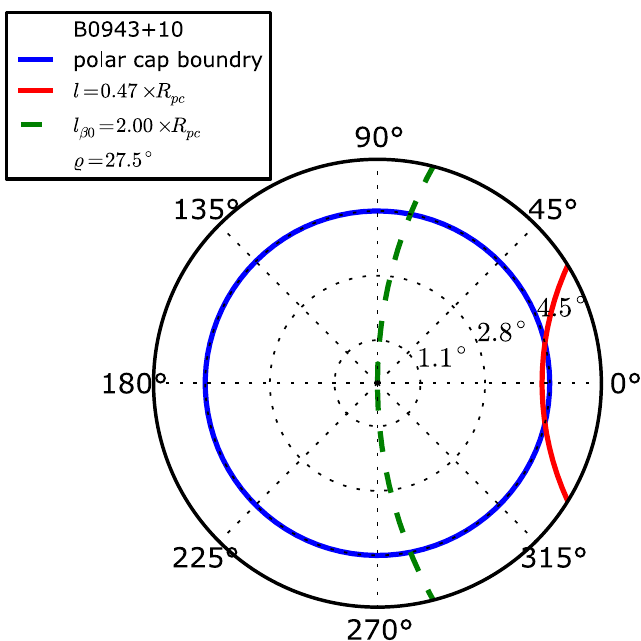}\hspace{0.5cm}\includegraphics[height=6.5cm]{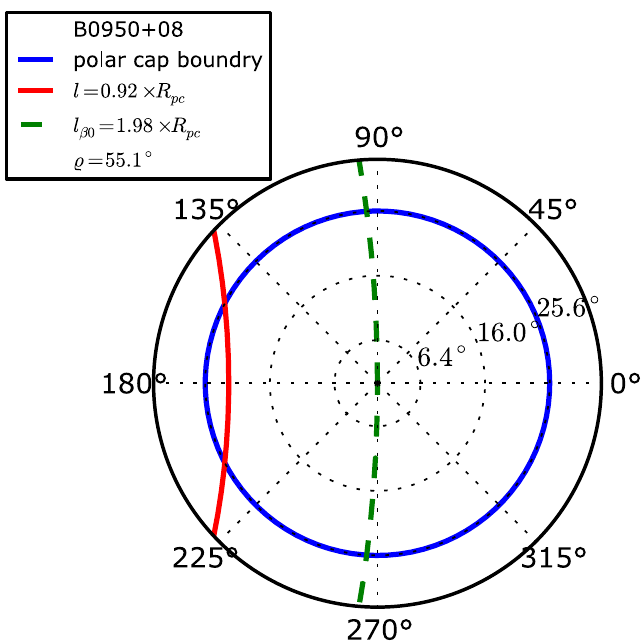}
\par\end{centering}

\begin{centering}
\includegraphics[height=6.5cm]{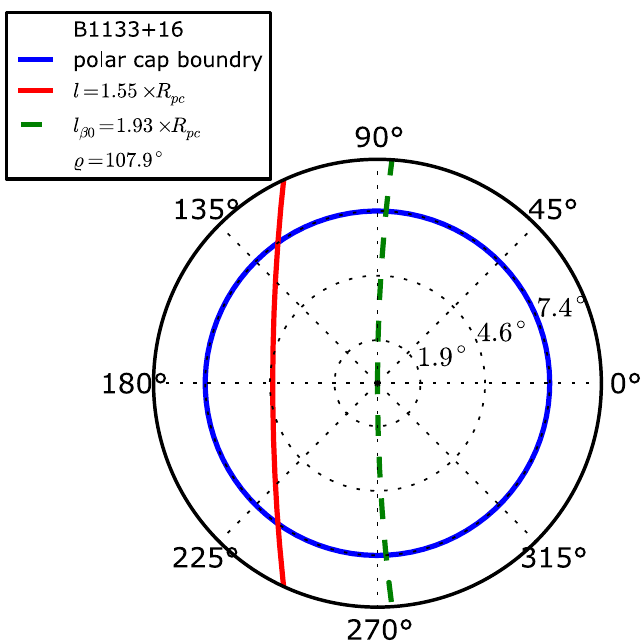}\hspace{0.5cm}\includegraphics[height=6.5cm]{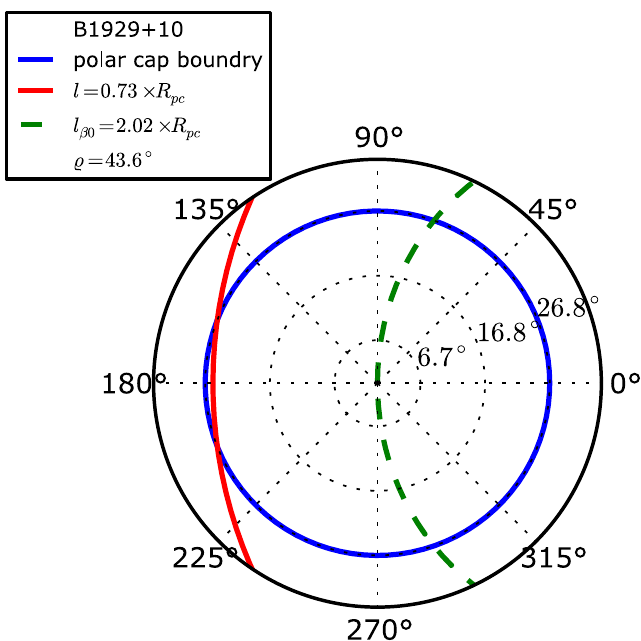}
\par\end{centering}

\caption[Top view of the polar cap region of pulsars with radio and X-ray observations]{Top view of the polar cap region of pulsars with radio drift observations
and X-ray hot spot radiation. Red lines correspond to the line of
sight while green dashed lines correspond to the theoretical lines
of sight calculated with an assumption that $\beta=0^{\circ}$. The
geometry of pulsars can be found in Table \ref{tab:x-ray_inclination}.
\label{fig:psg.polar_caps}}
\end{figure}

In numerical calculations of $\sigma\left(\varphi\right)$ it is convenient
to use the ``${\rm atan2}$'' function which takes into account
the signs of both components and places the angle in the correct quadrant
(see the footnote on page \pageref{fn:model.atan2}). Figure \ref{fig:psg.polar_caps}
presents the geometry of the emission region for pulsars with available
radio observations of the subpulse drift and X-ray observations of
the hot spot. By knowing the actual polar cap radius $R_{{\rm pc}}$
we can determine the transverse size of the region responsible for
the generation of plasma clouds in IAR (the spark half-width). 

\begin{comment}
\textasciitilde{}/Programs/studies/phd/line\_of\_sight/line\_of\_sight.py
\end{comment}

In our model the motion of sparks and the progressively different
positions of the associated plasma columns are responsible for the
observed drift of subpulses. For some pulsars it is possible to measure
directly the subpulse separation using a single pulse. In most calculations
it is assumed that the observed subpulses correspond to the adjacent
sparks. In general, this is not necessarily true. The distribution
of sparks on the polar cap is unknown and it is very likely that the
line of sight does not cross the adjacent sparks but it omits some
sparks in between. Therefore, the observed value of $P_{2}^{\circ}$
should be considered rather as an upper limit for spark separation.
Furthermore, for many pulsars the observed value $P_{2}^{\circ}>W$,
which means that it is not related to any structure at the polar cap
but that it corresponds to some other periodicity. We can use Equations
\ref{eq:psg.eta_hperp} and \ref{eq:psg.eta_p3} to calculate the
spark half-width as follows

\begin{equation}
h_{\perp}=26.2\frac{\left(B_{14}^{1.1}+0.3\right)^{2}PP_{3}\sqrt{\left|\cos\alpha\right|}}{B_{14}}.
\end{equation}

Finally, using Equation \ref{eq:psg.hperp_wb0} we can determine the
predicted value of the subpulse separation

\begin{equation}
\tilde{P}_{2}^{\circ}\approx\frac{26.2\left(B_{14}^{1.1}+0.3\right)^{2}PP_{3}\sqrt{\left|\cos\alpha\right|}}{B_{14}R_{{\rm pc}}}W_{\beta0}.
\end{equation}

\subsection{Heating efficiency}

The spin-down energy loss is

\begin{equation}
L_{{\rm SD}}=3.9\times10^{31}\frac{\dot{P}_{-15}}{P^{3}}.
\end{equation}

We can use Equations \ref{eq:psg.potential_drop}, \ref{eq:psg.hperp_p2}
and \ref{eq:psg.eta_p3} to calculate the dependence of the acceleration
potential drop on the parameters of drifting subpulses:

\begin{equation}
\Delta V\approx2.824\times10^{10}\left(\frac{\dot{P}_{-15}}{P^{3}}\right)^{0.5}\frac{1}{P_{3}}\left(\frac{P_{2}^{\circ}}{W_{\beta0}}\right)^{2}.\label{eq:psg.potential_drop_radio}
\end{equation}

The power of heating by backstreaming particles can be calculated
as follows

\begin{equation}
L_{{\rm heat}}=\eta n_{{\rm GJ}}c\left(\Delta Ve\right)\pi R_{{\rm pc}}^{2}.\label{eq:psg.l_heat}
\end{equation}

The number density of the Goldreich-Julian co-rotational charge can
be calculated using

\begin{equation}
n_{{\rm GJ}}=\frac{{\bf \Omega}\cdot{\bf B_{{\rm s}}}}{2\pi ce},
\end{equation}
where $\Omega=2\pi/P$ is an angular velocity, $B_{{\rm s}}=bB_{{\rm d}}$
is the surface magnetic field,\linebreak{}
 $b=R_{dp}^{2}/R_{pc}^{2}$, $B_{{\rm d}}=2.02\times10^{12}\sqrt{P\dot{P}_{-15}}\,{\rm G}$,
and $R_{{\rm dp}}=\sqrt{2\pi R^{3}/\left(cP\right)}\approx1.45\times10^{4}P^{-0.5}$.
The Goldreich-Julian density in terms of observed parameters can be
written as

\begin{equation}
n_{{\rm GJ}}=2.9\times10^{19}\left(P^{-3}\dot{P}_{-15}\right)^{1/2}\frac{\cos\alpha}{R_{{\rm pc}}^{2}}.\label{eq:psg.ngj_radio}
\end{equation}

Finally, using Equations \ref{eq:psg.eta_p3}, \ref{eq:psg.potential_drop_radio}
and \ref{eq:psg.ngj_radio} we can estimate the dependence of the
heating power and thus the X-ray luminosity of the hot spot radiation
on the parameters of radio observations as follows

\begin{equation}
L_{{\rm heat}}=L_{{\rm X}}=6\times10^{30}\left(\frac{\dot{P}_{-15}}{P^{3}}\right)\left(\frac{1}{P_{3}}\frac{P_{2}^{\circ}}{W_{\beta0}}\right)^{2}.
\end{equation}

The heating efficiency by the backstreaming particles can be calculated
as

\begin{equation}
\xi_{_{{\rm heat}}}=\frac{L_{{\rm heat}}}{L_{{\rm SD}}}=0.15\left(\frac{1}{P_{3}}\frac{P_{2}^{\circ}}{W_{\beta0}}\right)^{2}.
\end{equation}

\subsection{Ion luminosity}

In the PSG-on mode the bulk of energy is transferred to the iron ions
which shield the acceleration potential drop. Similar as for the backstreaming
particles, we can estimate the power of ion acceleration as

\begin{equation}
L_{{\rm ion}}=\left(1-\eta\right)n_{{\rm GJ}}c\left(\Delta Vq_{{\rm ion}}\right)\pi R_{{\rm pc}}^{2},\label{eq:psg.l_ion}
\end{equation}
where $q_{{\rm ion}}=26e=1.25\times10^{-8}\,{\rm erg^{0.5}cm^{0.5}}$
is the ion charge. Using the same approach as for electron, we can
calculate the dependence of energy transformed to the ions per second
on the parameters of the radio observations as follows

\begin{equation}
L_{{\rm ion}}=9.75\times10^{32}\left(1-\eta\right)\frac{\dot{P}_{-15}}{P^{3}P_{3}}\left(\frac{P_{2}^{\circ}}{W_{\beta0}}\right)^{2}\cos\alpha.
\end{equation}

It is clearly visible that if the screening factor is low $\eta\ll1$,
most of the energy in IAR is transferred to the iron ions. Using Equations
\ref{eq:psg.l_heat} and \ref{eq:psg.l_ion} we can show that

\begin{equation}
\frac{L_{{\rm ion}}}{L_{{\rm heat}}}=\frac{26\left(1-\eta\right)}{\eta}\approx\frac{26}{\eta}.
\end{equation}

Finally, the ion acceleration efficiency can be calculated as

\begin{equation}
\xi_{{\rm _{ion}}}=\frac{L_{{\rm ion}}}{L_{{\rm SD}}}\approx25\frac{1}{P_{3}}\left(\frac{P_{2}^{\circ}}{W_{\beta0}}\right)^{2}\cos\alpha.
\end{equation}

It may seem that ion luminosity exceeds the spin-down luminosity,
but note that both $P_{3}>1$ and $P_{2}^{\circ}<W_{\beta0}$. The
predicted values of heating efficiency $\xi_{{\rm heat}}$ and ion
acceleration efficiency $\xi_{{\rm ion}}$ are presented in the next
section.

\subsection{Observations}

In this section we confront the values of the subpulse drift and X-ray
radiation as estimated by other authors with predicted values estimated
using the approach presented above. In Table \ref{tab:psg.radio_drift},
alongside our predicted value of $\tilde{P}_{2}^{\circ}$ we present
two other estimates: (1) the subpulse separation estimated using the
carousel model developed by \citet{1975_Ruderman}, $P_{2,{\rm RS}}^{\circ}$;
(2) the subpulse separation found using the analysis of the Longitude-Resolved
Fluctuation Spectrum \citep{1970_Backer} and the integrated Two-Dimensional
Fluctuation Spectrum \citep{2003_Edwards}, performed by \citet{2006_Weltevrede},
$P_{2,{\rm W}}^{\circ}$. We have found that the subpulse separation
estimated using the fluctuations spectra is overestimated (in most
cases $P_{2,{\rm W}}^{\circ}>W$). By definition $P_{2}^{\circ}$
should correspond to the structure within a single pulse, thus if
the geometry is not extreme ($\varrho>1^{\circ}$) it should comply
with $P_{2}^{\circ}\leq W$. For this specific sample of pulsars $P_{2,{\rm W}}^{\circ}$
should not be interpreted as the actual subpulse separation. On the
other hand, $\tilde{P}_{2}^{\circ}$ is in good agreement with $P_{2,{\rm RS}}^{\circ}$.
The predicted values $\tilde{P}_{2}^{\circ}$ for B0834+06 and B0943+10
suggest that $P_{2,{\rm RS}}^{\circ}$ for those pulsars could be
overestimated due to the aliasing phenomenon ($P_{2,{\rm RS}}^{\circ}\approx2\tilde{P}_{2}^{\circ}$).
For B1929+10 we do not list $P_{2,{\rm RS}}^{\circ}$ as its value
presented in \citet{2008_Gil} does not comply with the $P_{2,{\rm RS}}^{\circ}\leq W$
condition. We believe that the overestimated value of $P_{2,{\rm RS}}^{\circ}$
for B1929+10 is a result of using the fluctuations spectra presented
in \citet{2006_Weltevrede} to calculate the number of sparks in the
carousel model. Note that the coincidence of $\tilde{P}_{2}^{\circ}$
and $P_{2,{\rm RS}}^{\circ}$ is yet to be clarified, as in our model
there is no physical reason for sparks to circulate around the magnetic
axis. In fact, the PSG model assumes the non-dipolar structure of
the magnetic field lines in the gap region and the actual position
of the polar cap is not necessarily coincident with the global dipole
(e.g.  see Figures \ref{fig:model.b0628}, \ref{fig:model.b0950},
\ref{fig:model.b1929}).

\begin{table}[H]
\caption[Details of a subpulse drift for pulsars with X-ray hot spot radiation]{Details of a subpulse drift for pulsars with X-ray hot spot radiation.
The individual columns are as follows: (1) Pulsar name, (2) Predicted
characteristic spacing between subpulses in the pulse longitude, $\tilde{P}_{2}^{\circ}$;
(3) Spacing between subpulses, found in the literature, estimated
using the carousel model, $P_{2,{\rm RS}}^{\circ}$; (4) Spacing between
subpulses estimated using fluctuations spectra, $P_{2,{\rm W}}^{\circ}$;
(5) Period at which a pattern of subpulses crosses the pulse window
(in units of the pulsar period), $P_{3}$; (6) Number of sparks estimated
using the carousel model, $N$; (7) Profile width at 10\%, $W$; (8)
Profile width calculated assuming $\beta=0$, $W_{\beta0}$; (9) Angular
width of the observed region on the polar cap $\varrho$ (see Figure
\ref{fig:psg.polar_caps}); (10) References; (11) Number of the pulsar.
\label{tab:psg.radio_drift}}

\centering{}%
\begin{tabular}{|l|c|c|c|c|c|c|c|c|c|c|}
\hline 
 &  &  &  &  &  &  &  &  &  & \tabularnewline
Name  & $\tilde{P}_{2}^{\circ}$  & $P_{2,{\rm RS}}^{\circ}$  & $P_{2,{\rm W}}^{\circ}$  & $P_{3}$  & $N$ & $W$ & $W_{\beta0}$ & $\varrho$ & Ref. & No.\tabularnewline
 & {\scriptsize $\left({\rm deg}\right)$}  & {\scriptsize $\left({\rm deg}\right)$}  & {\scriptsize $\left({\rm deg}\right)$}  & {\scriptsize $\left(P\right)$}  &  & {\scriptsize $\left({\rm deg}\right)$}  & {\scriptsize $\left({\rm deg}\right)$}  & {\scriptsize $\left({\rm deg}\right)$} &  & \tabularnewline
\hline 
\hline 
 &  &  &  &  &  &  &  &  &  & \tabularnewline
B0628--28  & $8.4$  & $6$ & $30$  & $7.0$  & $24$ & $38.1$ & $48.2$ & $98.0$ & \citetalias{2007_Gil_b}, \citetalias{2006_Weltevrede} & $8$\tabularnewline
B0834+06  & $1.4$ & $3$ & $20$  & $2.2$  & $14$ & $12.2$ & $15.8$ & $105.5$ & \citetalias{2007_Rankin}, \citetalias{2006_Weltevrede} & $14$\tabularnewline
B0943+10  & $9.8$ & $17$ & -- & $1.8$  & $20$ & $25.6$ & $84.7$ & $27.5$ & \citetalias{1999_Deshpande}, \citetalias{2001_Deshpande}, \citetalias{2001_Asgekar} & $15$\tabularnewline
B0950+08  & $4.0$ & -- & -- & $6.5$  & -- & $32.2$ & $63.7$ & $55.1$ & \citetalias{1980_Wolszczan}  & $16$\tabularnewline
B1133+16  & $3.2$ & $4$ & $130$ & $3.0$  & $11$ & $14.4$ & $18.1$ & $107.9$ & \citetalias{2006_Gil}, \citetalias{2007_Herfindal}, \citetalias{2006_Weltevrede} & $22$\tabularnewline
B1929+10  & $5.8$ & $?$ & $90$  & $9.8$  & $?$ & $24.5$ & $80.8$ & $43.6$ & \citetalias{2008_Gil}, \citetalias{2006_Weltevrede} & $42$\tabularnewline
 &  &  &  &  &  &  &  &  &  & \tabularnewline
\hline 
\end{tabular}
\end{table}

Table \ref{tab:psg.drift} presents the observed and derived parameters
of PSG for pulsars with available radio and X-ray observations. In
the calculations we used the predicted value of subpulse separation
$\tilde{P_{2}^{\circ}}$. Note that we consider only pulsars with
a visible hot spot component since only for these pulsars we can estimate
the size of the polar cap. The low value of the estimated screening
factor ($\eta\ll1$) suggests that when the drift is visible, the
pulsar operates in the PSG-on mode. If the pulsar operates in the
PSG-off mode, $\eta\approx1$, the drift velocity is much higher (see
Equation \ref{eq:psg.vdr_shielding}) and the drift phenomenon should
be more chaotic and much difficult to identify. Observations of PSR
0943+10 show a strong, regular subpulse drifting in the radio-bright
mode, with only a hint of modulation in the radio-quiescent mode.
Based on this fact we believe that the two different scenarios of
the gap breakdown (PSG-on and PSG-off modes) can explain the mode
changing. 

\begin{table}[H]
\caption[Derived parameters of PSG for pulsars with available radio observations
of the subpulse drift and X-ray hot spot radiation]{Derived parameters of PSG for pulsars with available radio observations
of the subpulse drift and X-ray hot spot radiation. The individual
columns are as follows: (1) Pulsar name, (2) Screening factor, $\eta$;
(3) Predicted heating efficiency, $\xi_{{\rm heat}}$; (4) Observed
bolometric efficiency, $\xi_{_{{\rm BB}}}$; (5) Predicted ion acceleration
efficiency, $\xi_{{\rm _{ion}}}$; (6) Surface temperature, $T_{{\rm s}}$;
(7) Strength of the surface magnetic field, $B_{{\rm s}}$; (8) Observed
polar cap radius, $R_{{\rm pc}}$; (9) Estimated spark half-width,
$h_{\perp}$; (10) Number of the pulsar. $T_{{\rm s}}$, $R_{{\rm pc}}$,
$b$ were chosen to fit $1\sigma$ uncertainty. Note that in the calculations
$\tilde{P}_{2}^{\circ}$ was used. \label{tab:psg.drift}}

\vspace{0.3cm}

\centering{}%
\begin{tabular}{|l|c|c|c|c|c|c|c|c|c|}
\hline 
 &  &  &  &  &  &  &  &  & \tabularnewline
Name  & $\eta$  & $\log\xi_{_{{\rm heat}}}$  & $\log\xi_{_{{\rm BB}}}$  & $\log\xi_{_{{\rm ion}}}$  & $T_{{\rm s}}$  & $B_{{\rm s}}$ & $R_{{\rm pc}}$  & $h_{\perp}$  & No.\tabularnewline
 &  & {\scriptsize $\left({\rm radio}\right)$}  & {\scriptsize $\left({\rm x-ray}\right)$}  & {\scriptsize $\left({\rm ions}\right)$}  & {\scriptsize $\left(10^{6}{\rm K}\right)$}  & {\scriptsize $\left(10^{14}{\rm G}\right)$} & {\scriptsize $\left({\rm m}\right)$}  & {\scriptsize $\left({\rm m}\right)$} & \tabularnewline
\hline 
\hline 
 &  &  &  &  &  &  &  &  & \tabularnewline
B0628--28  & $0.07$  & $-4.03$  & $-3.61$  & $-1.43$ & $2.5$  & $2.0$  & $23$  & $3.9$  & $8$\tabularnewline
B0834+06  & $0.15$  & $-3.60$  & $-3.34$  & $-1.35$ & $3.0$  & $2.4$  & $20$  & $1.8$  & $14$\tabularnewline
B0943+10  & $0.09$  & $-3.24$  & $-3.27$  & $-0.75$ & $3.3$  & $2.5$  & $17$  & $2.0$  & $15$\tabularnewline
B0950+08  & $0.09$  & $-5.08$  & $-4.54$  & $-2.62$  & $2.6$  & $2.1$  & $14$  & $0.7$ & $16$\tabularnewline
B1133+16  & $0.09$  & $-3.29$  & $-3.13$  & $-0.81$  & $3.4$  & $2.7$  & $17$  & $2.9$ & $22$\tabularnewline
B1929+10  & $0.02$  & $-5.10$  & $-4.17$  & $-1.98$  & $4.2$  & $2.0$  & $22$  & $1.6$ & $42$\tabularnewline
 &  &  &  &  &  &  &  &  & \tabularnewline
\hline 
\end{tabular}
\end{table}

\chapter{Cascade simulation\label{chap:cascade}}

\thispagestyle{headings}

\textit{In this chapter we present the approach of calculating the
pair cascades developed by \citet{2010_Medin} which has been applied
to cases with non-dipolar structure of magnetic field. The original
approach was adapted to perform full three-dimensional calculations
and extended with effects that may have a greater importance for non-dipolar
configuration of surface magnetic fields (e.g. aberration). Additionally,
to perform a thorough analysis of the Inverse Compton Scattering we
present the detailed description of calculating the ICS cross section
originally developed by \citet{2000_Gonthier}.}

\vspace{1cm}

Following the approach presented by \textit{\citet{2010_Medin}} we
can divide the cascade simulation into three parts:
\begin{itemize}
\item propagation of the primary particle (including photon emission),
\item photon propagation in strong magnetic field (pair production, photon
splitting), 
\item propagation and photon emission of the secondary%
\footnote{In this thesis the term ''secondary'' refers to any newly created
particle except for the primary particles accelerated in IAR, e.g.
the third generation of electrons and positrons are all considered
as ''secondary'' particles.%
}particles.
\end{itemize}
We use the ''co-rotating'' frame of reference (the frame which rotates
with the star) to track both photons and particles. In calculations
we consider regions far inside the light cylinder. Thus, following
\citet{2010_Medin}, we ignore any bending of the photon path due
to rotation of the star. Furthermore, we also ignore effects of general
relativity on trajectories of photons and particles. 

\begin{figure}[H]
\begin{centering}
\includegraphics[width=16.5cm]{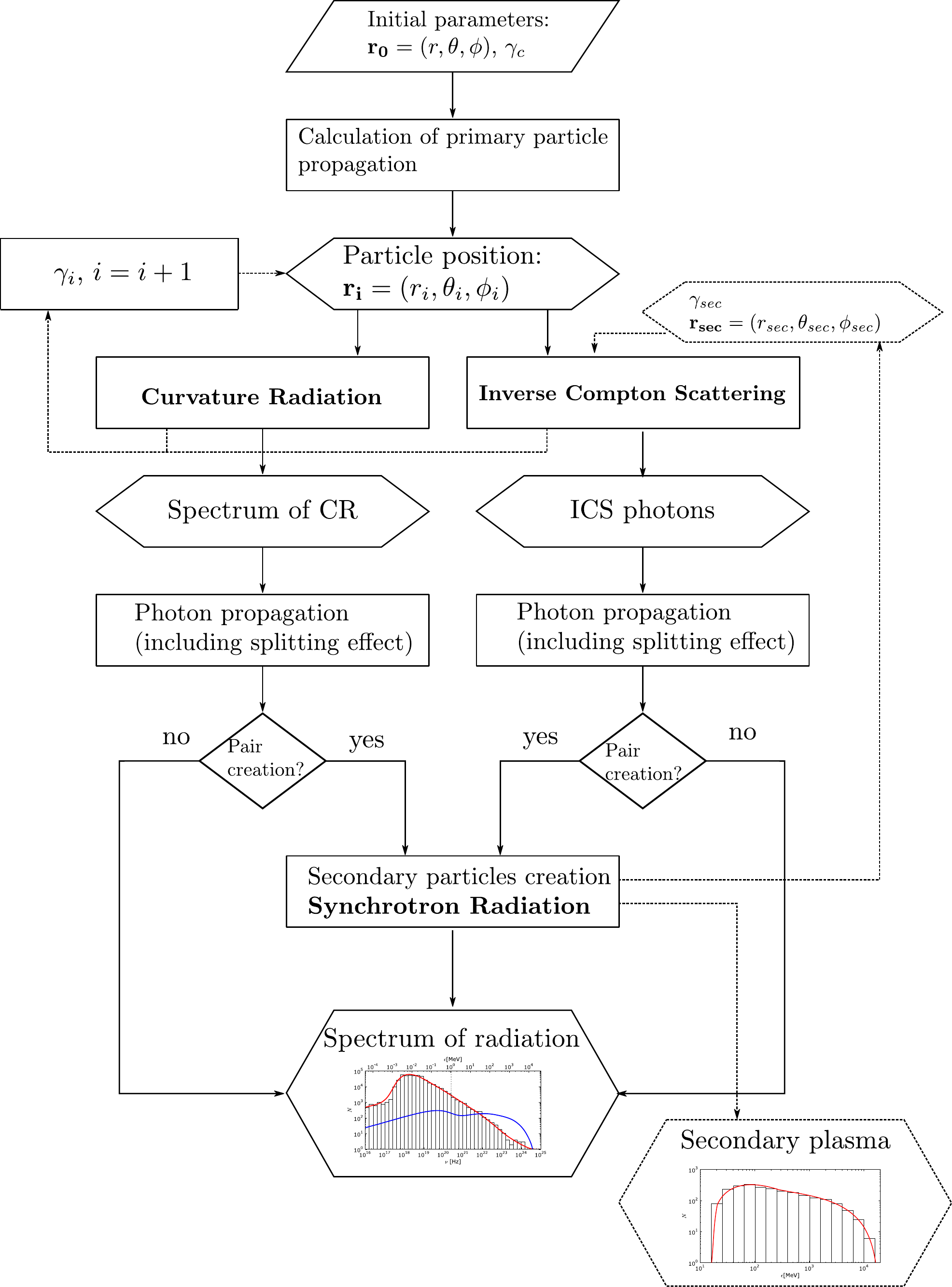}
\par\end{centering}

\centering{}\caption[Flowchart of the algorithm used to calculate a cascade simulation]{Flowchart of algorithm used to calculate a cascade simulation. \label{fig:cascade.flowchart_cascade} }
\end{figure}

Figure \ref{fig:cascade.flowchart_cascade} presents a summary flowchart
of the algorithm used to calculate the properties of secondary plasma
and the spectrum of radiation for a given structure of a neutron's
star magnetic field and gap parameters.

\section{Curvature Radiation \label{sec:cascade.cr}}

As we have shown in Chapter \ref{chap:model}, an ultrastrong surface
magnetic field ($B_{{\rm s}}>10^{14}\,{\rm G}$) is accompanied by
high curvature (curvature radius $\Re_{6}\approx0.1-10$). This suggests
that one of the important processes of radiation which should be considered
is Curvature Radiation (CR).

CR is quite similar to ordinary synchrotron radiation (radiation of
ultrarelativistic particles in the magnetic field), the only difference
being that the radius of circular motion (the gyroradius) is in fact
the curvature radius of magnetic field lines. Due to beaming effects
the radiation appears to be concentrated in a narrow set of directions
about the velocity of the particle. The angular width of the cone
of emission is of the order $\sim1/\gamma$, where $\gamma$ is a
Lorentz factor of an emitting particle (for more details see \citealp{1979_Rybicki}).

We track the primary particle above the acceleration zone (the gap
region) as it moves along the magnetic field line. The length of the
step $\Delta s$ is chosen so as to achieve sufficient accuracy even
for large curvature of the magnetic field line, $\Delta s\approx0.01\Re_{{\rm min}}$,
where $\Re_{{\rm min}}$ is the minimum radius of curvature. The distribution
of CR photon energy can be written as (see Equation 14.93 in \citealp{1998_Jackson})
\begin{equation}
\frac{{\rm d}N}{{\rm d}\epsilon}=\frac{E}{\epsilon_{{\rm _{CR}}}}\frac{9\sqrt{3}}{8\pi}\int_{\epsilon/\epsilon_{_{{\rm CR}}}}^{\infty}K_{5/3}(t){\rm d}t,\label{eq:cascade.dn_deps}
\end{equation}
where $E=4\pi e^{2}\gamma^{4}/3\Re$ is the total energy radiated
per revolution, $\epsilon_{_{{\rm CR}}}=3\gamma^{3}\hbar c/(2\Re)$
is the characteristic energy of curvature photons, and $K_{5/3}$
is the $n=5/3$ Bessel function of the second kind. The total energy
radiated by a particle after it passes the length $\Delta s$, $E_{_{\Delta s}}$,
can be written as 
\begin{equation}
E_{_{\Delta s}}=E\frac{\Delta s}{2\pi\Re}.\label{eq:cascade.i_ds}
\end{equation}

Thus, by using Equations \ref{eq:cascade.dn_deps} and \ref{eq:cascade.i_ds}
we can write the formula for the distribution on CR photon energy
after a particle passes length $\Delta s$ 
\begin{equation}
\frac{{\rm d}N}{{\rm d}\epsilon}=\frac{\Delta s}{2\pi\Re}\frac{\sqrt{3}e^{2}\gamma}{\hbar c}\int_{\epsilon/\epsilon_{{\rm _{CR}}}}^{\infty}K_{5/3}(t){\rm d}t.
\end{equation}

It is convenient to divide the spectrum of photon energy into discrete
bins. Then, the number of photons in each energy bin can be calculated
as 
\begin{equation}
N_{\epsilon}=\int_{\epsilon_{_{i}}}^{\epsilon_{_{i}}+\Delta\epsilon}\frac{dN}{d\epsilon}{\rm d}\epsilon,
\end{equation}
where $\epsilon_{_{i}}$ is the lowest energy for the $i$-th bin
and $\Delta\epsilon$ is the energy bin width. Our simulation uses
$50$ bins with an energy range of $\epsilon_{_{0}}=4\times10^{-2}\,{\rm keV}$
(soft X-ray) to\linebreak{}
 $\epsilon_{_{49}}=4\times10^{5}\,{\rm MeV}$ (hard $\gamma$-rays).

Depending on the photon frequency the polarisation fraction of CR
photons is between $50\%$ and $100\%$ polarised parallel to the
magnetic field (see \citealp{1998_Jackson}, \citealp{1979_Rybicki}).
Therefore, using similar approach as \citet{2010_Medin} we randomly
assign the polarisation in the ratio of one photon $\perp$-polarised
to every seven $\parallel$-polarised photons, which corresponds to
$75\%$ parallel polarisation.

\section{Photon propagation \label{sec:cascade.photons}}

To explain some of the properties of pulsars and their surroundings
(e.g. nebulae radiation), large magnetospheric plasma densities exceeding
the Goldreich-Julian density (see Equation \ref{eq:psg.n_gj}) by
many orders of magnitude are required. In order to simulate the process
of generation of such a dense plasma it is necessary to check the
conditions of photon decay into electron-positron pairs.

A photon with energy $E_{\gamma}>2mc^{2}$ and propagating with a
nonzero angle $\Psi$ with respect to an external magnetic field can
be absorbed by the field and, as a result an electron-positron pair
is created. The concurrent process is photon splitting $\gamma\rightarrow\gamma\gamma$,
which may occur even if the photon energy is below the pair creation
threshold ($E_{\gamma}<2mc^{2}$).

In the cascade simulation the photon is emitted (or scattered in the
case of ICS) from point $P_{{\rm ph}}$ in a direction tangent to
the magnetic field line $\Delta\mathbf{s_{\parallel}}$. The direction
vector is calculated as the value of the magnetic field at the point
of photon creation (see Equations \ref{eq:model.field}, \ref{eq:model.b_d}
and \ref{eq:model.b_m}) normalised so that its length is equal to
the desired step $\Delta\mathbf{s_{\parallel}}=\mathbf{B}\Delta s/B$.
However, the direction of the magnetic field at the point of photon
emission does not take into account the randomness of the emission
direction due to the relativistic beaming effect. In Section \ref{sec:cascade.beaming}
we describe a procedure to include the beaming effect in the emission
process which alters $\Delta\mathbf{s_{\parallel}}\rightarrow{\bf \Delta s_{ph}}$.
Finally, we can write that at the point of curvature emission photons
are created with energy $\epsilon_{{\rm ph}}$, polarisation $\parallel$
or $\perp$, weighting factor $N_{\epsilon}$ (number of photons),
and with both optical depths (for pair production $\tau$ and for
photon splitting $\tau_{{\rm sp}}$) set to zero. Since we neglect
any banding of the photon path we assume that from the point of emission
it travels in a straight line. In each following step the photon travels
a distance ${\bf \Delta s_{ph}}$. In the co-rotating frame of reference
in every step we need to take into the account aberration due to pulsar
rotation. In order to do so, in every step we alter the photon position
according to the procedure described in Section \ref{sec:cascade.aberration}.
As stated by \citet{2010_Medin} we can calculate the change in the
pair production optical depth , $\Delta\tau$, and in the photon splitting
optical depth $\Delta\tau_{{\rm sp}}$, at the new position as: 
\begin{equation}
\Delta\tau\simeq\Delta s_{{\rm ph}}R_{\|,\perp},
\end{equation}
 
\begin{equation}
\Delta\tau_{{\rm sp}}\simeq\Delta s_{{\rm ph}}R_{\|,\perp}^{{\rm sp}},
\end{equation}
where $R_{\|,\perp}$ and $R_{\|,\perp}^{{\rm sp}}$ are the attenuation
coefficients for $\|$ or $\perp$ polarised photons for pair production
and photon splitting, receptively.

\subsection{Relativistic beaming (emission direction) \label{sec:cascade.beaming}}

\begin{figure}[H]
\begin{centering}
\includegraphics[height=9cm]{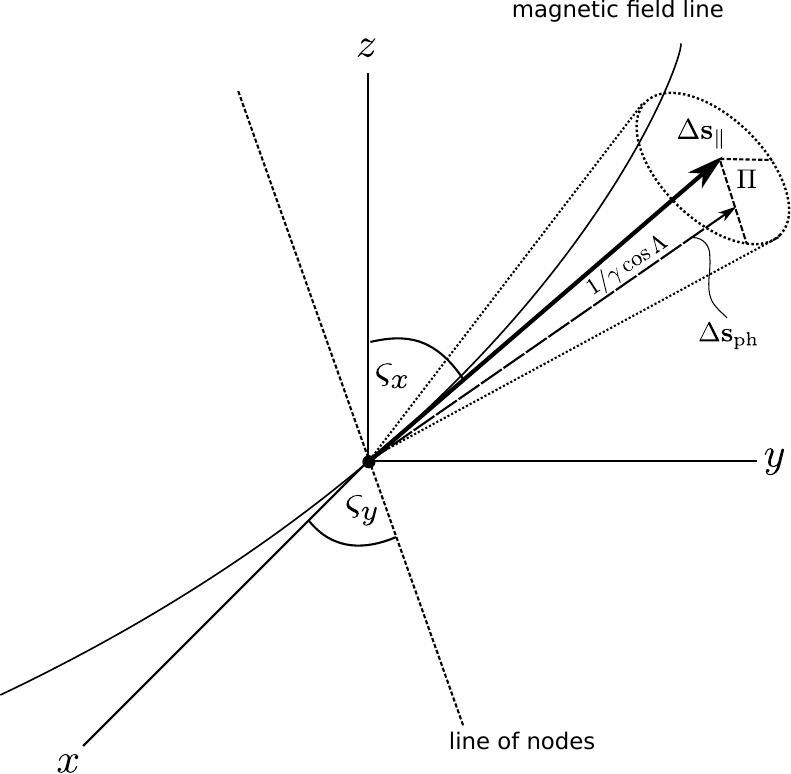} 
\par\end{centering}

\centering{}\caption[Relativistic beaming effect of photon emission]{Relativistic beaming effect of photon emission (for both CR and ICS).
In the simulation we include the beaming effect by performing three
steps: (I) rotation of the $xyz$ frame of reference in order to align
the $z$-axis with $\mathbf{\Delta s_{\parallel}}$, (II) transformation
of the step vector from a Cartesian to a spherical system of coordinates
and alteration of the $\theta$ and $\phi$ components with random
values $1/\gamma\cos\Lambda$ and $\Pi$, respectively, (III) transformation
of the step vector from a spherical to a Cartesian system of coordinates
and rotation back to the original system of reference. Note that after
these steps we get a new vector ${\bf \Delta s_{ph}}$ inclined to
the primary one, ${\bf \Delta s_{ph}}$, at an angle ranging from
$0$ to $1/\gamma$. \label{fig:cascade.beaming}}
\end{figure}

Due to relativistic beaming the emission direction should be modified
by an additional emission angle of order $\sim1/\gamma$. We use the
following steps to include the beaming effect in our simulation (see
Figure \ref{fig:cascade.beaming}).

(I) The first step is rotation of the $xyz$ frame of reference in
order to align the $z$-axis with $\mathbf{\Delta s_{\parallel}}$.
In our calculations we used rotation by angle $\varsigma_{y}$ around
the $y$-axis, $R_{y}\left(\varsigma_{y}\right)$, and rotation by
angle $\varsigma_{x}$ around the $x$-axis, $R_{x}\left(\varsigma_{x}\right)$.
The final rotation matrix can be written as

\begin{equation}
R_{yx}=R_{y}\left(\varsigma_{y}\right)R_{x}\left(\varsigma_{x}\right)=\left[\begin{array}{ccc}
\cos\varsigma_{y} & \sin\varsigma_{x}\sin\varsigma_{y} & \sin\varsigma_{y}\cos\varsigma_{x}\\
0 & \cos\alpha & -\sin\alpha\\
-\sin\varsigma_{y} & \cos\varsigma_{y}\sin\varsigma_{x} & \cos\varsigma_{y}\cos\varsigma_{x}
\end{array}\right].
\end{equation}

The Euler angles for rotations can be calculated as

\begin{equation}
\begin{array}{c}
\varsigma_{x}={\rm atan2}\left(s_{y},s_{z}\right),\\
\varsigma_{y}=\begin{cases}
\arctan\left(-\frac{s_{x}}{s_{z}}\cos\varsigma_{x}\right) & {\rm if\ }s_{z}\neq0\\
\arctan\left(-\frac{s_{x}}{s_{y}}\sin\varsigma_{x}\right) & {\rm if\ }s_{y}\neq0\\
\frac{\pi}{2} & {\rm if\ }s_{x}=0\ {\rm and}\ s_{y}=0.
\end{cases}
\end{array}
\end{equation}

Note that in order to increase readability, the $\Delta$ symbol and
${\rm \parallel}$ index were discarded (e.g. $s_{x}=\Delta s_{{\rm \parallel},x}$). 

(II) The second step is the transformation of the step vector's coordinates
in the double rotated frame of reference ${\bf \Delta s_{{\rm ph}}^{\prime\prime}=}\left(s_{x}^{\prime\prime},\ s_{y}^{\prime\prime},\ s_{z}^{\prime\prime}\right)$
to spherical system of coordinates and alteration of the $\theta$
and $\phi$ components as follows

\begin{eqnarray}
s_{r}^{\prime\prime} & = & \sqrt{s_{x}^{\prime\prime2}+s_{y}^{\prime\prime2}+s_{z}^{\prime\prime2}},\nonumber \\
s_{\theta}^{\prime\prime} & = & \arccos\left(\frac{s_{z}^{\prime\prime}}{\sqrt{s_{x}^{\prime\prime2}+s_{y}^{\prime\prime2}+s_{z}^{\prime\prime2}}}\right)+\frac{1}{\gamma}\cos\Lambda,\nonumber \\
s_{\phi}^{\prime\prime} & = & \arctan\left(\frac{s_{y}^{\prime\prime}}{s_{x}^{\prime\prime}}\right)+\Pi,
\end{eqnarray}
where $\Lambda$ and $\Pi$ are random angles between $0$ and $2\pi$.
The inverse tangent denoted in the $\phi$-coordinate must be suitably
defined by taking into account the correct quadrant (see the ``${\rm atan2}$''
description in the footnote on page \pageref{fn:model.atan2}).

(III) The last step is the transformation of vector components to
the Cartesian system of coordinates, ${\bf s_{ph}^{\prime\prime}}=\left[s_{r}^{\prime\prime}\sin\left(s_{\theta}^{\prime\prime}\right)\cos\left(s_{\phi}^{\prime\prime}\right),\, s_{r}^{\prime\prime}\sin\left(s_{\theta}^{\prime\prime}\right)\sin\left(s_{\phi}^{\prime\prime}\right),\, s_{r}^{\prime\prime}\cos\left(s_{\theta}^{\prime\prime}\right)\right]$
and rotation back to the original coordinate system ${\bf \Delta s_{ph}}=\left(R_{yx}\right)^{-1}{\bf {\bf s_{ph}^{\prime\prime}}}$.

The rotation matrix of this transformation can be written as

\begin{equation}
\left(R_{yx}\right)^{-1}=\left(R_{yx}\right)^{T}=\left[\begin{array}{ccc}
\cos\varsigma_{y} & 0 & -\sin\varsigma_{y}\\
\sin\varsigma_{x}\sin\varsigma_{y} & \cos\varsigma_{x} & \sin\varsigma_{x}\cos\varsigma_{y}\\
\sin\varsigma_{y}\cos\varsigma_{x} & -\sin\varsigma_{x} & \cos\varsigma_{x}\cos\varsigma_{y}
\end{array}\right].
\end{equation}

\subsection{Aberration due to pulsar rotation \label{sec:cascade.aberration}}

Note that in our frame of reference (co-rotating with a star) the
path of the photon should be curved (see \citealt{1978_Harding}).
In the dipolar case the angular deviation increases approximately
as $s_{{\rm ph}}\Omega/c=s_{{\rm ph}}/R_{LC}$. When the configuration
of magnetic field in non-dipolar inclusion of an aberration is even
more important. Therefore, the location of photon decay should be
modified to include the growth of the photon-magnetic field intersection
angle.

\begin{figure}[H]
\begin{centering}
\includegraphics[height=11cm]{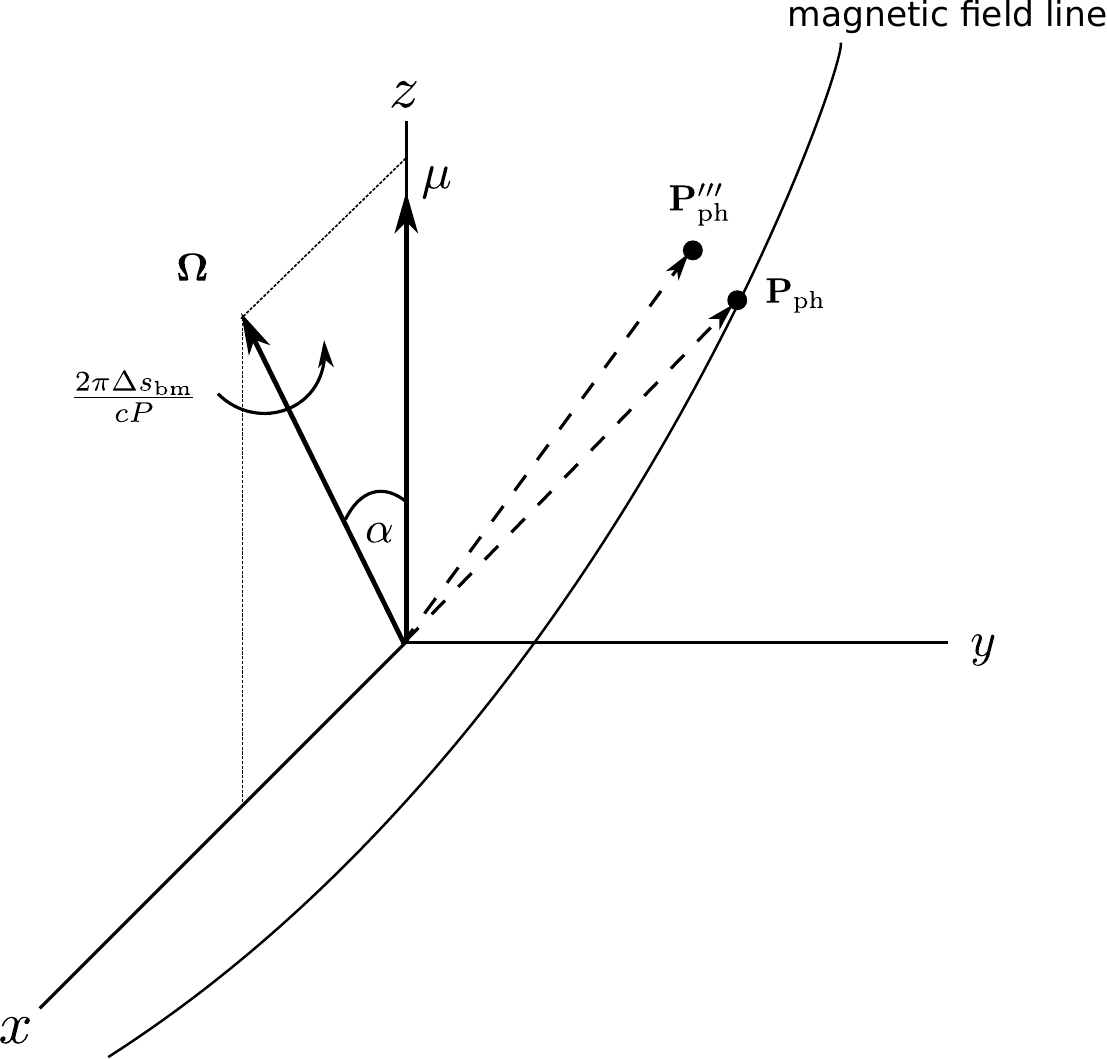} 
\par\end{centering}

\centering{}\caption[Aberration due to pulsar rotation]{ Aberration due to pulsar rotation. We use the following procedure
to include the aberration effect: (I) rotation around the $y$-axis
to align $\Omega$ with $\mu$, (II) rotation by angle\protect \linebreak{}
 $\omega=2\pi\Delta s_{{\rm bm}}/\left(cP\right)$ around the $z$-axis
(which reflects the pulsar rotation), (III) rotation back to the original
frame of reference (in which $\mu$ is aligned with the $z$-axis).
\label{fig:cascade.aberration}}
\end{figure}

In our simulation we include the aberration effect by alteration of
photon position $P_{{\rm ph}}$ in every step ${\bf \Delta s_{ph}}$
(see Figure \ref{fig:cascade.aberration}).

We use the three-step procedure to alter the photon position.

(I) Rotation of the $xyz$ frame of reference around the $y$-axis
by angle $\alpha$,\linebreak{}
${\bf P_{ph}^{\prime}}=R_{y}\left(\alpha\right){\bf P_{ph}}$. Note
that here $\alpha$ refers to the inclination of the magnetic axis
with respect to the rotation axis and we assume that the pulsar's
angular velocity vector ${\bf \Omega}$ lies in the $xz$-plane. The
rotation matrix of this transformation can be written as

\begin{equation}
R_{y}\left(\alpha\right)=\left[\begin{array}{ccc}
\cos\alpha & 0 & \sin\alpha\\
0 & 1 & 0\\
-\sin\alpha & 0 & \cos\alpha
\end{array}\right].
\end{equation}

(II) After step (I) the $z$-axis is aligned to $\Omega$, and in
order to include the rotation of the pulsar we need to again rotate
the frame of reverence by angle $\omega=2\pi\Delta s_{{\rm bm}}/\left(cP\right)$
around the $z$-axis, ${\bf P_{ph}^{\prime\prime}}=R_{z}\left(\omega\right){\bf P_{ph}^{\prime}}$.
We use the following rotation matrix
\begin{equation}
R_{z}\left(\omega\right)=\left[\begin{array}{ccc}
\cos\omega & -\sin\omega & 0\\
\sin\omega & \cos\omega & 0\\
0 & 0 & 1
\end{array}\right].
\end{equation}

(III) The final step is a rotation back to the original frame of reference,\linebreak{}
 ${\bf P_{ph}^{\prime\prime\prime}=}\left(R_{y}\left(\alpha\right)\right)^{-1}{\bf P_{ph}^{\prime\prime}}$,
using the following rotation matrix
\begin{equation}
\left(R_{y}\left(\alpha\right)\right)^{-1}=\left(R_{y}\left(\alpha\right)\right)^{T}=\left[\begin{array}{ccc}
\cos\alpha & 0 & -\sin\alpha\\
0 & 1 & 0\\
\sin\alpha & 0 & \cos\alpha
\end{array}\right].
\end{equation}

\subsection{Pair production attenuation coefficient \label{sec:cascade.pairs}}

The pair production attenuation coefficient can be written as \citep{2010_Medin}

\begin{equation}
R_{\|,\perp}=R^{\prime}\sin\Psi,
\end{equation}
where $R^{\prime}$ is the attenuation coefficient in the frame where
the photon propagates perpendicular to the local magnetic field (the
so-called ''perpendicular'' frame), $\Psi$ is the intersection
angle between the propagation direction of the photon and the local
magnetic field. To increase readability we suppress the subscripts
$\parallel$ and $\perp$, but \foreignlanguage{english}{$R^{\prime}$}
has to be calculated for both polarisations separately.

As stated by \citet{2010_Medin} the total attenuation coefficient
for pair production can be calculated as $R^{\prime}=\sum_{jk}R^{\prime}{}_{j,k}$,
where $R{}_{j,k}^{\prime}$ is the attenuation coefficient for the
process producing an electron in Landau level $j$ and a positron
in Landau level $k$. For the electron-positron pair the sum is taken
over all possible states ($j$ and $k$). Note that production of
electron-positron pairs is symmetric $R{}_{jk}^{\prime}=R{}_{kj}^{\prime}$.
Thus, to represent the pair creation probability in either the $\left(jk\right)$
or $\left(kj\right)$ state we will use $R{}_{jk}^{\prime}$. For
a given Landau levels $j$ and $k$, the pair production threshold
condition is \citep{2010_Medin} 
\begin{equation}
E_{\gamma}^{\prime}>E_{j}^{\prime}+E_{k}^{\prime},\label{eq:cascade.energy}
\end{equation}
where $E_{\gamma}^{\prime}=E_{\gamma}\sin\Psi$ is the photon energy
in the perpendicular frame and\linebreak{}
 $E_{n}^{\prime}=mc^{2}\sqrt{1+2\epsilon_{_{B}}n}$ is the minimum
energy of a particle (electron or positron) in Landau Level $n$.
This condition can be written in a dimensionless form as 
\begin{equation}
x=\frac{E_{\gamma}^{\prime}}{2mc^{2}}=\frac{E_{\gamma}}{2mc^{2}}\sin\Psi>\frac{1}{2}\left[\sqrt{1+2\epsilon_{_{B}}j}+\sqrt{1+2\epsilon_{_{B}}k}\right],
\end{equation}
where $\epsilon_{{\rm _{B}}}=\hbar eB/\left(mc\right)$ is the cyclotron
energy of a particle (electron or positron) in magnetic field $B$
in units of $mc^{2}$.

The first nonzero pair production attenuation coefficients for both
polarisations ($\perp$ and $\parallel$) are \citep{1983_Daugherty,2010_Medin}
\begin{equation}
R{}_{\|,00}^{\prime}=\frac{1}{2a_{0}}\frac{\epsilon_{_{B}}}{x^{2}\sqrt{x^{2}-1}}e^{-2x^{2}/\epsilon_{_{B}}},\hspace{1cm}x>\left(x_{0}=1\right),
\end{equation}

\begin{equation}
R{}_{\perp,01}^{\prime}=2\times\frac{1}{2a_{0}}\frac{\epsilon_{_{B}}}{2x^{2}}\frac{2x^{2}-\epsilon_{_{B}}}{\sqrt{x^{2}-1-\epsilon_{_{B}}+\frac{\epsilon_{_{B}}^{2}}{4x^{2}}}}e^{-2x^{2}/\epsilon_{_{B}}},\hspace{1cm}x>\left(x_{1}=\left(1+\sqrt{1+2\epsilon_{_{B}}}\right)/2\right),
\end{equation}

\begin{equation}
R{}_{\|,01}^{\prime}=2\times\frac{1}{2a_{0}}\frac{2+\epsilon_{_{B}}-\frac{\epsilon_{_{B}}^{2}}{4x^{2}}}{\sqrt{x^{2}-1-\epsilon_{_{B}}+\frac{\epsilon_{_{B}}^{2}}{4x^{2}}}}e^{-2x^{2}/\epsilon_{_{B}}},\hspace{1cm}x>x_{1},
\end{equation}

\begin{equation}
R{}_{\|,02}^{\prime}=2\times\frac{1}{2a_{0}}\frac{2x^{2}}{\epsilon_{_{B}}}\frac{1+\epsilon_{_{B}}-\frac{\epsilon_{_{B}}^{2}}{4x^{2}}}{\sqrt{x^{2}-1-2\epsilon_{_{B}}+\frac{\epsilon_{_{B}}^{2}}{4x^{2}}}}e^{-2x^{2}/\epsilon_{_{B}}},\hspace{1cm}x>\left(x_{2}=\left(1+\sqrt{1+4\epsilon_{_{B}}}\right)/2\right),
\end{equation}

\begin{equation}
R{}_{\perp,02}^{\prime}=2\times\frac{1}{2a_{0}}\frac{x^{2}-\epsilon_{_{B}}}{\sqrt{x^{2}-1-2\epsilon_{_{B}}+\frac{\epsilon_{_{B}}^{2}}{4x^{2}}}}e^{-2x^{2}/\epsilon_{_{B}}},\hspace{1cm}x>x_{2},
\end{equation}
where $a_{0}$ is the Bohr radius (let us note that $R'_{\perp,00}=0$).
In the above equations for all channels except $00$ the pair production
attenuation coefficients are multiplied by a factor of two (see the
text above Equation \ref{eq:cascade.energy}).

The pair production optical depth is defined as \citep{2010_Medin}:
\begin{equation}
\tau=\int_{0}^{s_{{\rm ph}}}R(s){\rm d}s=\int_{0}^{s_{{\rm ph}}}R^{\prime}(s)\sin\Psi{\rm d}s.\label{eq:cascade.tau_in}
\end{equation}

We can assume $\Psi\ll1$, because all high-energy photons ($x>1$)
will produce pairs much earlier than $\Psi$ reaches a value near
unity. In this limit $\sin\Psi\simeq s_{{\rm ph}}/\Re$, so the relation
between $x$ and $s_{{\rm ph}}$ can be expressed by 
\begin{equation}
x\simeq\frac{s_{{\rm ph}}}{\Re}\frac{E_{\gamma}}{2mc^{2}}.
\end{equation}
Equation \ref{eq:cascade.tau_in} can be rewritten as

\[
\tau=\tau_{1}+\tau_{\|,2}+\tau_{\perp,2}+...;
\]
 
\begin{equation}
\tau_{1}=\int_{s_{0}}^{s_{1}}R_{\|,00}{\rm d}s,\hspace{0.5cm}\tau_{\|,2}=\int_{s_{1}}^{s_{2}}\left(R_{\|,00}+R_{\|,01}\right){\rm d}s,\hspace{0.5cm}\tau_{\perp,2}=\int_{s_{1}}^{s_{2}}R_{\perp,01}{\rm d}s,
\end{equation}
where $s_{0}$ and $s_{1}$ are distances which the photon should
pass in order to have energy $x_{0}$ and $x_{1}$, respectively (in
the perpendicular frame of reference). Let us note that $s_{0}$,
$s_{1}$ and $s_{2}$ are of the same order, and if $s<s_{0}$ the
attenuation coefficient is zero.

The pair production optical depth to reach the second threshold is
\begin{equation}
\int_{s_{0}}^{s_{1}}{\rm d}sR_{\|,00}(s)=\frac{\epsilon_{{\rm _{B}}}}{2a_{0}}\left(\frac{2mc^{2}}{E_{\gamma}}\right)^{2}\Re\int_{x_{1}}^{x_{2}}\frac{{\rm d}x}{x\sqrt{x^{2}-1}}e^{-2x^{2}/\epsilon_{{\rm _{B}}}},
\end{equation}
where $s_{0}$ is the distance travelled by the photon to reach the
threshold $x_{0}\equiv1$, and $s_{1}$ is the distance travelled
by the photon to reach the second threshold $x_{1}\equiv\left(1+\sqrt{1+2\epsilon_{{\rm _{B}}}}\right)/2$.

\begin{comment}
\textasciitilde{}/Programs/magnetic/magnetic/src/radiation/photon.py
(a - show\_tau, 500 MeV, Re=1e6, size=1e2, b - show\_tau\_psi, 500
MeV, b=B\_crit, Re=1e6, size=1e4, )
\end{comment}

\begin{figure}[H]
\begin{centering}
\includegraphics[height=9.5cm]{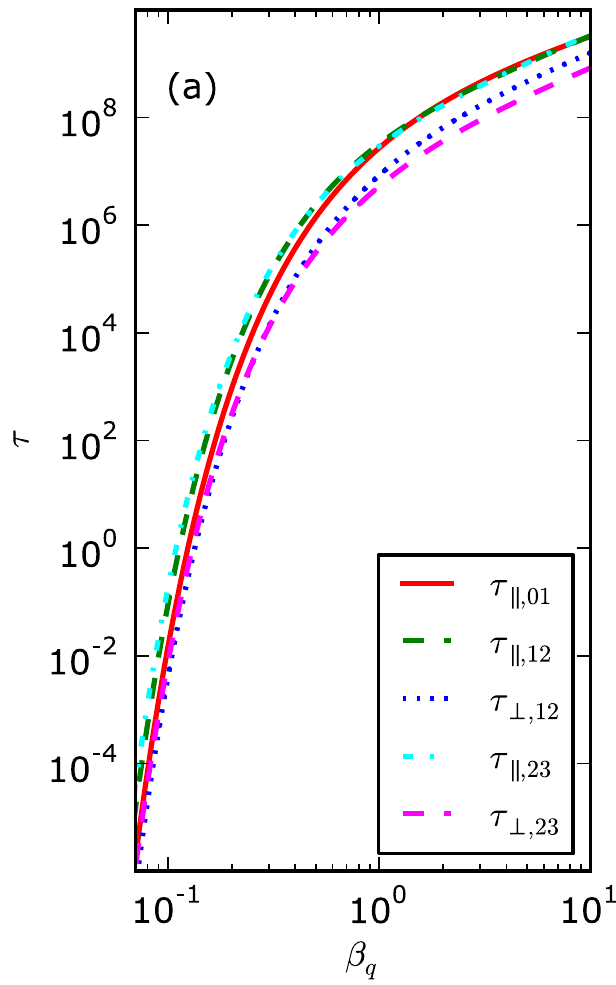}\hspace{1cm}\includegraphics[height=9.5cm]{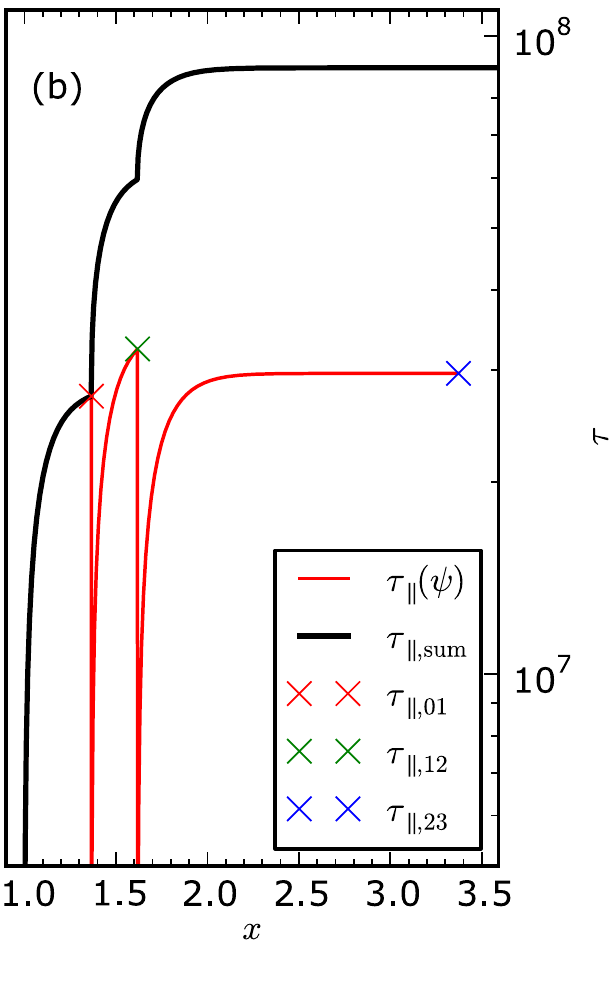}
\par\end{centering}

\centering{}\caption[Pair production optical depth ]{Panel (a) presents the dependence of the pair production optical
depth on the magnetic field strength ($\beta_{q}=B/B_{q}$). Panel
(b) presents the dependence of the optical depth on photon energy
in the perpendicular frame of reference ($x=\epsilon\sin\Psi/\left(2mc^{2}\right)$).
On both panels the photon energy is $\epsilon=500\,{\rm MeV}$, while
panel (b) was obtained for magnetic field strength $\beta_{q}=1$.
\label{fig:cascade.pair_creation}}
\end{figure}

\subsection{Photon mean free path \label{sec:cascade.photon_path}}

As was shown in the previous section (see Figure \ref{fig:cascade.pair_creation})
for strong magnetic fields (e.g. $\beta_{q}\gtrsim0.2$), $\tau_{1}$,
$\tau_{\|,2}$, and $\tau_{\perp,2}$ are much larger than one. Therefore,
the pair production process takes place according to two scenarios
(see also \citealp{2010_Medin}). If $\beta_{q}\gtrsim0.2$ pairs
are produced by photons almost immediately upon reaching the first
threshold, the created pairs will be in the low Landau levels ($n\lesssim2$).
If $\beta_{q}\lesssim0.2$, the photons will travel longer distances
to be absorbed and the created pairs will be in the higher Landau
levels.

Thus, for strong magnetic fields ($\beta_{q}\gtrsim0.2$) the photon
mean free path can be approximated as 
\begin{equation}
l_{{\rm ph}}\approx s_{0}=\Re\frac{2mc^{2}}{E_{\gamma}},\label{eq:cascade.l_ph}
\end{equation}
while for relatively weak magnetic fields ($\beta_{q}\lesssim0.2$)
we can use the asymptotic approximation derived by \citet{1966_Erber}:
\begin{equation}
l_{{\rm ph}}\approx\frac{4.4}{(e^{2}/\hbar c)}\frac{\hbar}{mc}\frac{B_{q}}{B\sin\Psi}\exp\left(\frac{4}{3\chi}\right),\label{eq:cascade.l_ph2}
\end{equation}
 
\begin{equation}
\chi\equiv\frac{E_{\gamma}}{2mc^{2}}\frac{B\sin\Psi}{B_{q}}\hspace{1cm}(\chi\ll1).
\end{equation}

\subsection{Photon-splitting attenuation coefficient \label{sec:cascade.splitting}}

\begin{figure}[H]
\begin{centering}
\includegraphics[height=8cm]{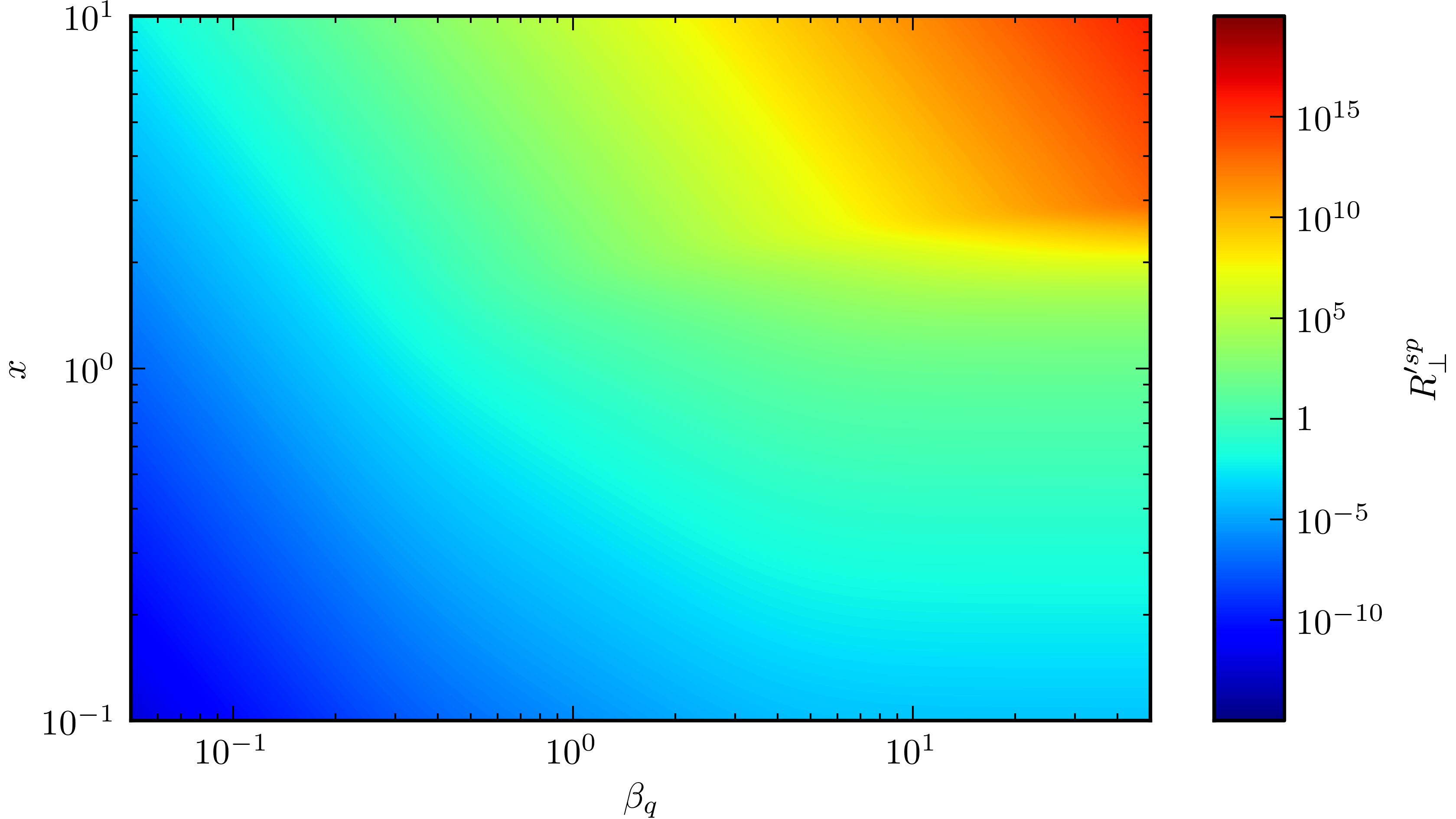}
\par\end{centering}

\centering{}\caption[Photon-splitting attenuation coefficient]{Dependence of the photon-splitting attenuation coefficient on the
energy of the photon in the perpendicular frame ($x=\epsilon\sin\Psi/\left(2mc^{2}\right)$,
vertical axis) and on the strength of the magnetic field ($\beta_{q}=B/B_{q}$,
horizontal axis).\label{fig:cascade.r_spl_fig} }
\end{figure}

In our calculations we include photon splitting by following the approach
presented by \citet{2010_Medin}. Since only the $\perp\rightarrow\parallel\parallel$
process is allowed, for $\parallel$-polarised photons the photon
splitting attenuation coefficient is zero $R_{\parallel}^{sp}=0$
(\citealp{1971_Adler}, \citealp{2002_Usov}, \citealp{2001_Baring}).
To calculate the splitting attenuation coefficient in the perpendicular
frame for $\perp$-polarised photons we use the formula adopted from
the numerical calculation of \citet{1997_Baring} : 
\begin{equation}
R{}_{\perp}^{\prime{\rm sp}}\simeq\frac{\frac{\alpha_{f}^{2}}{60\pi^{2}}\left(\frac{26}{315}\right)^{2}\left(2x\right)^{5}\epsilon_{{\rm _{B}}}^{6}}{\left[\epsilon_{_{B}}^{3}\exp\left(-0.6x^{3}\right)+0.05\right]\left[0.25\epsilon_{{\rm _{B}}}^{3}\exp\left(-0.6x^{3}\right)+20\right]}.
\end{equation}
For photon energies $x\leq1$ this expression underestimates the results
of \citet{1997_Baring} at $\beta_{q}=1$ by less than $30\%$, while
at both $\beta_{q}\le0.5$ and $\beta_{q}\gg1$ the discrepancy is
less than $10\%$. As can be seen from Figure \ref{fig:cascade.r_spl_fig},
the attenuation coefficient $R{}_{\perp}^{\prime{\rm sp}}$ drops
rapidly with the magnetic field strength for $\beta_{q}<1$, thus
photon splitting is unimportant for $\beta_{q}\lesssim0.5$ (e.g.
\citealp{2001_Baring,2010_Medin}). 

\begin{comment}
\textasciitilde{}/Programs/magnetic/magnetic/src/cascade/photon\_evolution.py
(show\_splitting3d)
\end{comment}

\subsection{Pair creation vs photon splitting\label{sec:cascade.pair_cr_spl}}

\begin{figure}[H]
\begin{centering}
\includegraphics[width=8cm,height=6.5cm]{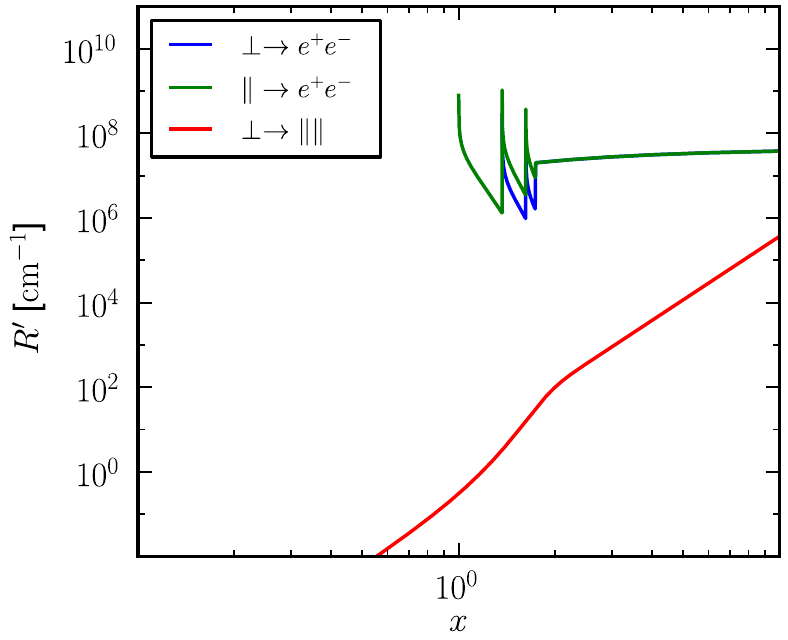}\includegraphics[width=8cm,height=6.5cm]{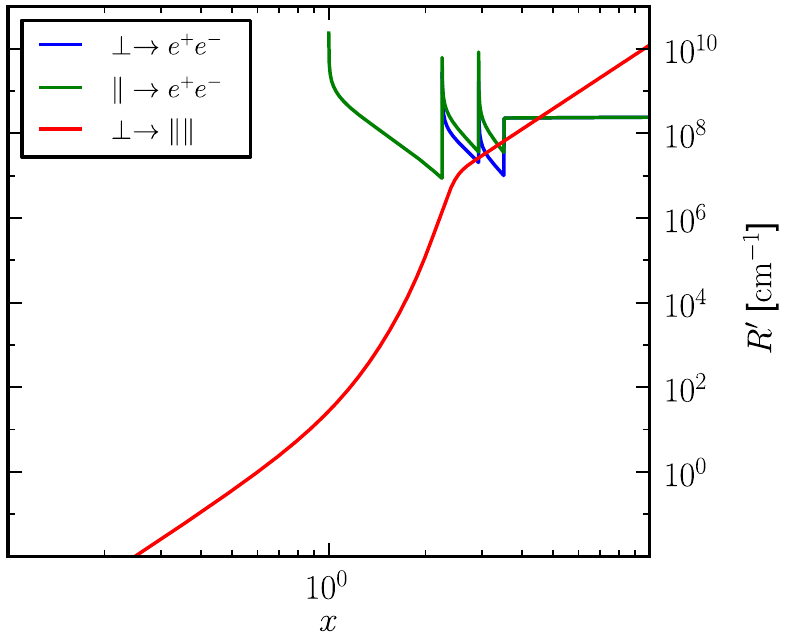}
\par\end{centering}

\caption[Attenuation coefficients of pair production and photon splitting]{Attenuation coefficients of pair production and photon splitting
in the perpendicular frame of reference. Panel (a) was obtained using
photon energy $E_{\gamma}=10^{3}\,{\rm MeV}$ and magnetic field strength
$B=B_{q}=4.414\times10^{13}\,{\rm G}$ ($\beta_{q}=1$). Panel (b)
presents calculations for photon energy $E_{\gamma}=10^{3}\,{\rm MeV}$
and magnetic field strength $B=2.5\times10^{14}\,{\rm G}$ ($\beta_{q}=5.7$).\label{fig:cascade.pair_split_0}}
\end{figure}

\begin{figure}[H]
\begin{centering}
\includegraphics[width=8cm,height=6.5cm]{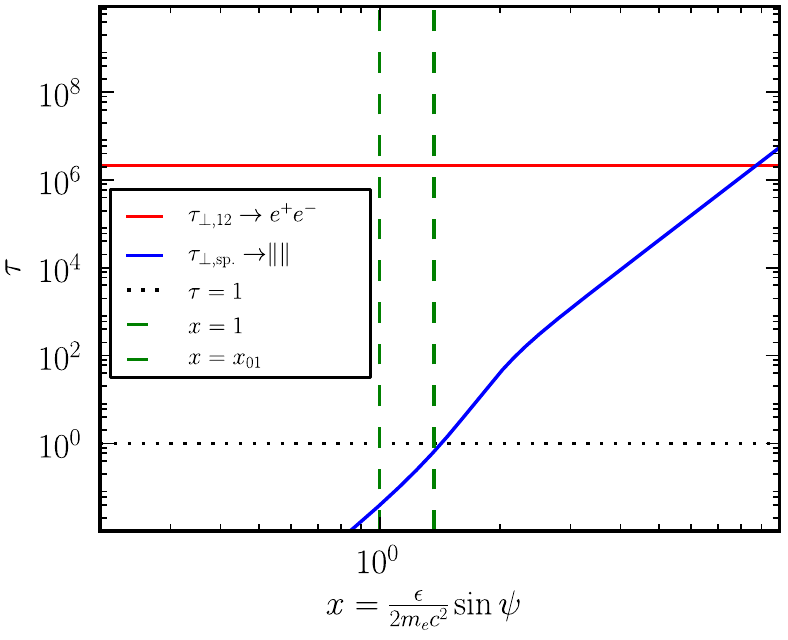}\includegraphics[width=8cm,height=6.5cm]{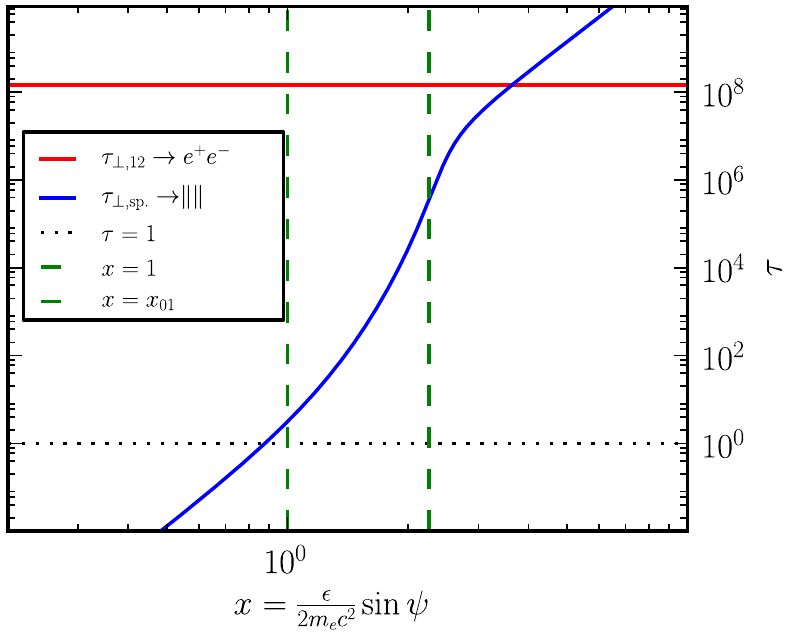}
\par\end{centering}

\caption[Optical depth for pair production and photon splitting]{Optical depth for pair production and photon splitting for $\perp$-polarised
photons. Panel (a) presents results for $E_{\gamma}=10^{3}\,{\rm MeV}$
and $B=B_{q}=4.4\times10^{13}\,{\rm G}$ ($\beta_{q}=1$), while panel
(b) was obtained using the same photon energy but a stronger magnetic
field $B=2.5\times10^{14}\,{\rm G}$ ($\beta_{q}=5.7$). If $\beta_{q}=1$
the photon creates an electron-positron pair, while in an ultrastrong
magnetic field ($\beta_{q}=5.7$) the photon splits before it reaches
the first threshold, $x=x_{0}$. \label{fig:cascade.pair_split}}
\end{figure}

As noted by \citet{2010_Medin}, even though the photon splitting
attenuation coefficient above the first threshold ($x>x_{0}$) is
much smaller than for pair production (see Figure \ref{fig:cascade.pair_split_0}),
in ultrastrong magnetic fields ($\beta_{q}\gtrsim0.5$) the \textbf{$\perp$}-polarised
photons split before reaching the first threshold (see Figure \ref{fig:cascade.pair_split}).
On the other hand, the\textbf{ $\parallel$}-polarised photons produce
pairs in the zeroth Landau level.

\begin{comment}
\textasciitilde{}/Programs/magnetic/magnetic/src/radiation/photon.py
(show\_r)
\end{comment}

\begin{comment}
\textasciitilde{}/Programs/magnetic/magnetic/src/radiation/photon.py
(show\_tau\_split)
\end{comment}

\subsection{Secondary plasma}

Following the approach presented by \citet{2010_Medin} whenever $\tau\geq1$
and the threshold for pair production is reached ($x=x_{0}$ for $\parallel$-polarised
photons and $x=x_{1}$ for $\perp$-polarised photons), the photon
is turned into an electron-positron pair. Whereas if $\tau_{{\rm sp}}\geq1$
the photon is turned into two photons. Following the results of \citealp{1997_Baring}
we assume that the energy of parent photon is equally distributed
between both newly created photons. A new $\parallel$-polarised photon
is created with an energy $0.5\epsilon_{{\rm ph}}$ and weighting
factor $2\Delta N_{\epsilon}$. We assume that the newly created photon
travels in the same direction as the parent photon, ${\bf \Delta s_{ph}}$.
Note that the photon should split with probability $1-e^{-\tau}$,
but as shown by \citet{2010_Medin} for cascade results this effect
is negligible.

For $\beta_{q}\lesssim0.1$, the particles are produced in high Landau
levels with energy equal to half of the photon energy each (see \citealp{1983_Daugherty}).
In our calculations we assume that the newly created particles (electron-positron
pairs) travel in the same direction as the photon. When $\beta_{q}\gtrsim0.1$,
on the other hand, we choose the maximum allowed values of $j$ and
$k$ for the newly created electron and positron. Note that for $\beta_{q}\gtrsim0.1$
the particles are created in low Landau levels. 

\begin{figure}[H]
\begin{centering}
\includegraphics[height=8.5cm]{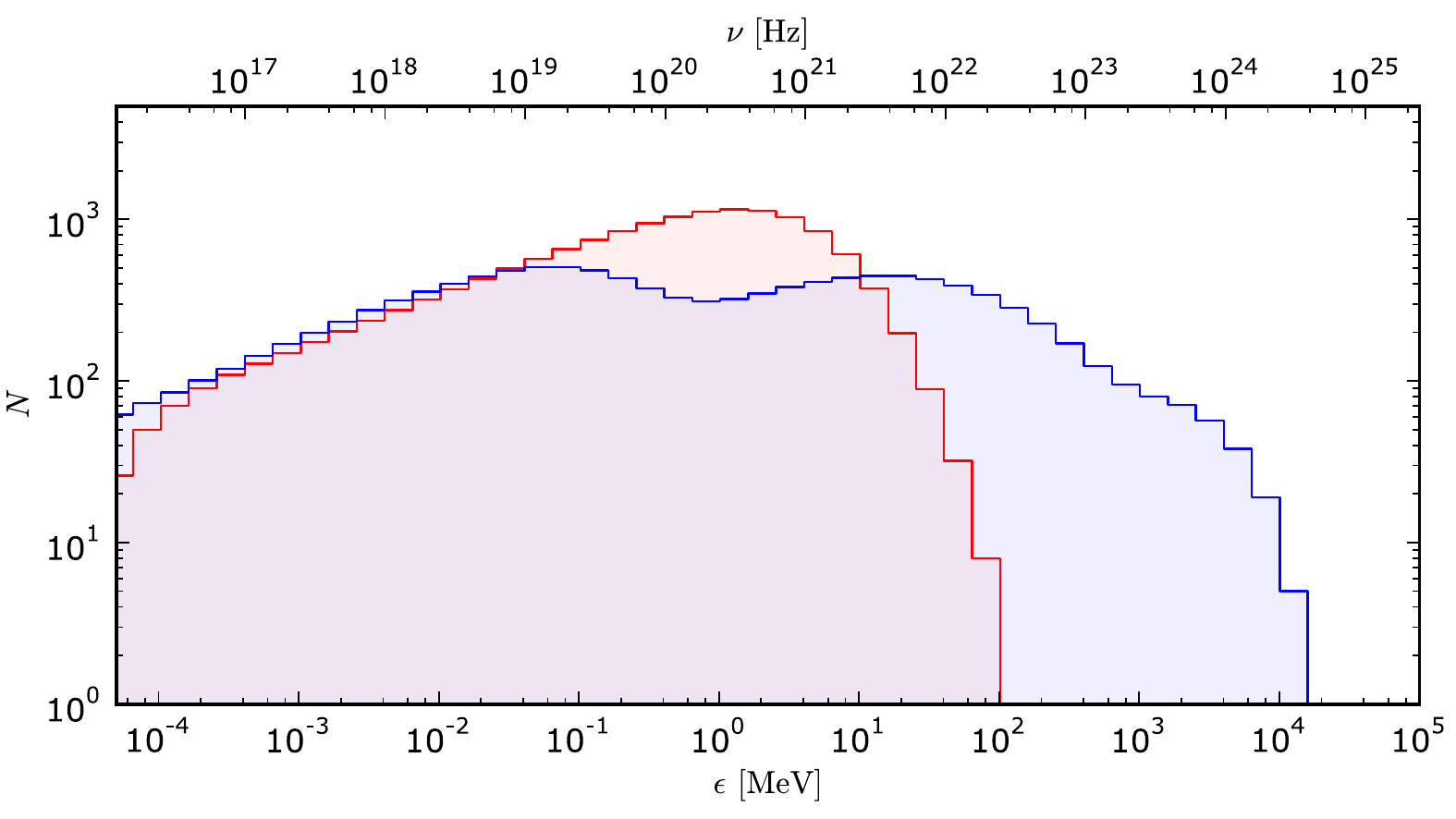}
\par\end{centering}

\centering{}\caption[Distribution of CR-photons produced by a single primary particle]{ Distribution of CR photons produced by a single primary particle
for a dipolar (blue line) and non-dipolar (red line) structure of
the magnetic field. The minimum radius of curvature in thee dipolar
case is $\Re_{6}^{^{{\rm min}}}\approx50$, while in the non-dipolar
case $\Re_{6}^{^{{\rm min}}}\approx2$. In both cases the radiation
was calculated up to a distance of $D=100R$, and with an initial
Lorentz factor of the primary particle $\gamma_{{\rm c}}=3.5\times10^{6}$.
\label{fig:cascade.cr_spec}}
\end{figure}

Figure \ref{fig:cascade.cr_spec} presents the spectrum of Curvature
Radiation for a dipolar and non-dipolar structure of magnetic field
lines. Note the characteristic three peaks in the CR distribution
for the non-dipolar structure.

\begin{comment}
\textasciitilde{}/Programs/magnetic/magnetic/src/new\_cascade/new\_cascade.py
(379 - gamma=3.5e6, h=252e2 - manual photon\_evolution=0, 373 photon\_evolution=0)

\textasciitilde{}/Programs/studies/phd/cascade\_plot/cascade\_plot.py
(plot\_spectrum2 or t2)
\end{comment}

Formation of the peaks is caused by the fact that the particle passes
regions with three different values of curvature: (I) just above the
stellar surface, $z\approx1\,{\rm km}$, where curvature is the highest;
(II) at altitudes where the influence of anomalies is comparable with
the global dipole, $z\approx2.5\,{\rm km}$, also with strong curvature;
(III) and at altitudes where the influence of anomalies is negligible,
$z\gtrsim3.1R$, with approximately dipolar curvature (see Figure
\ref{fig:model.j0633_curva}). Hence, the spectrum is a sum of radiation
generated in a highly non-dipolar magnetic field (high energetic and
soft $\gamma$-rays) and with radiation at higher altitudes where
the magnetic field is dipolar (X-rays). The primary particle loses
about $63\%$ and $1\%$ of its initial energy in the non-dipolar
and dipolar case, respectively. As can be seen from the Figure, to
get high emission of CR photons and, thus, a significant density of
secondary plasma, a non-dipolar structure of the magnetic field is
required.

\begin{figure}[H]
\begin{centering}
\includegraphics[height=7.3cm]{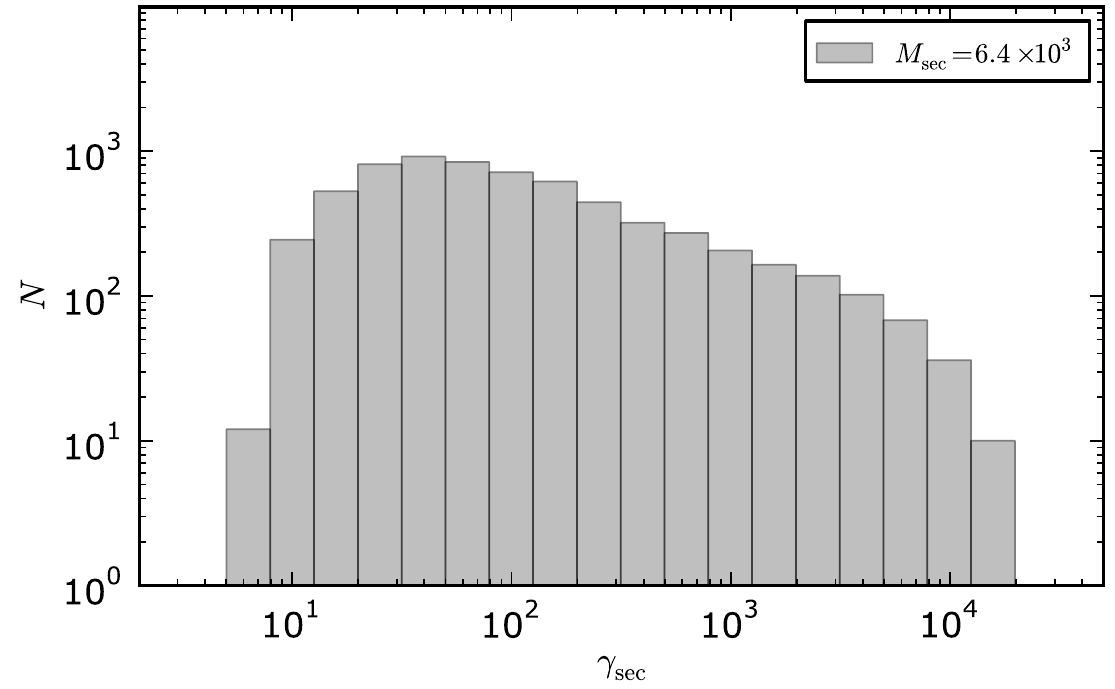}
\par\end{centering}

\centering{}\caption[Distribution of particle energy created by CR photons ]{ Distribution of particle energy created by CR photons calculated
for a non-dipolar structure of the magnetic field. For this specific
magnetic field configuration and initial parameters (see the caption
of Figure \ref{fig:cascade.cr_spec}) the secondary plasma multiplicity
is $M_{{\rm sec}}\approx6\times10^{3}$. Note that this result does
not include Synchrotron Radiation and the actual energies of the created
pairs are lower as they lose their transverse momenta (see Section
\ref{sec:cascade.synchrotron}). \label{fig:cascade.cr_pairs}}
\end{figure}

The high energetic photons produced in a strongly non-dipolar magnetic
field will either split or create electron-positron pairs. Figure
\ref{fig:cascade.cr_pairs} presents the distribution of particle
energy created by CR photons. Note that for $\beta_{q}\lesssim0.1$
the pairs are created in high Landau levels and in order to get the
final distribution of secondary plasma energy we should consider the
loss of particle energy due to Synchrotron Radiation (see Section
\ref{sec:cascade.synchrotron}).

\begin{comment}
\textasciitilde{}/Programs/studies/phd/cascade\_plot/cascade\_plot.py

(plot\_pairs, data/373\_cr\_1e04\_100m\_CR\_noSR\_PH\_e5\_leftline/
)
\end{comment}

\section{Synchrotron Radiation \label{sec:cascade.synchrotron}}

When pairs (electrons and positrons) are created in high Landau Levels
they radiate away their transverse momentum through Synchrotron Radiation
(SR). The secondary positron (or electron) is created with energy
$\gamma mc^{2}$ and pitch angle $\Psi$, which corresponds to a specific
value of Landau Level $n$. Following \citet{2010_Medin} we choose
the frame in which the particle has no momentum along the direction
of external magnetic field. In such a frame of reference the particle
propagates in a circular motion transverse to the magnetic field (the
so-called ''circular'' frame). The relation of the energy of the
newly created particle in the circular frame of reference ($E_{\perp}=\gamma_{\perp}mc^{2}$)
with the particle energy in the co-rotating frame can be written as
\citep{2010_Medin} 
\begin{equation}
\gamma_{\perp}=\sqrt{\gamma^{2}\sin^{2}\Psi+\cos^{2}\Psi}=\sqrt{1+2\epsilon_{_{B}}n}.\label{eq:cascade.gamma_perp}
\end{equation}

The power of synchrotron emission, $P_{{\rm SR}}$, can calculated
as follows 
\begin{equation}
P_{{\rm SR}}=\frac{2e^{2}}{3c^{3}}\left(\gamma_{\perp}^{2}-1\right)c^{2}\epsilon_{_{B}}^{2},
\end{equation}
In the circular frame $E_{\perp}$, is radiated away through synchrotron
emission after a particle travels a distance 
\begin{equation}
l_{{\rm p}}^{{\rm SR}}\approx\left|\frac{E_{\perp}}{P_{{\rm SR}}}c\right|=\frac{\gamma_{\perp}mc^{3}}{\frac{2e^{2}}{3c^{3}}\left(\gamma_{\perp}^{2}-1\right)c^{2}\epsilon_{{\rm _{B}}}^{2}}.
\end{equation}

The particle (electron or positron) mean free path for SR is much
shorter than for other relevant cascade processes (see Section \ref{sec:cascade.cr}
for Curvature Radiation, and Section \ref{sec:cascade.ics} for ICS).
In fact, it is so short that in our calculations we assume that before
moving from its initial position the particle loses all of its perpendicular
momentum $p_{\perp}$ due to SR (see \citealp{1982_Daugherty,2010_Medin}).
Once the particle reaches the ground Landau level ($n=0$, $p_{\perp}=0$)
its final energy can be calculate as 
\begin{equation}
\gamma_{\parallel}=\left(1-\beta^{2}\cos^{2}\Psi\right)^{-1/2}=\gamma/\gamma_{\perp},\label{eq:cascade.gamma_par}
\end{equation}
here $\beta=v/c=\sqrt{1-1/\gamma^{2}}$ is the particle velocity in
units of speed of light.

Following the approach presented by \citet{2010_Medin}, to simplify
the simulation we assume that in the circular frame synchrotron photons
are emitted isotropically in the plane of motion such that there is
no perpendicular velocity change of the particle (the Lorentz factors
$\gamma$ and $\gamma_{\perp}$ decrease but $\gamma_{\parallel}$
is constant). Thus, the Equation \ref{eq:cascade.gamma_par} remains
valid until the particle reaches the ground state. In order to simulate
the full SR process the following procedure was adopted:  the particle
Lorentz factor in the circular frame $\gamma_{\perp}$ drops from
its initial value to $\gamma_{\perp}=1$ (i.e., $n=0$) in a series
of steps. Each step entails emission of one synchrotron photon, with
energy $\epsilon_{_{\perp}}$ depending on the current value of $\gamma_{\perp}$.
After the photon emission the energy of the particle is reduced by
$\epsilon_{_{\perp}}$, $\Delta\gamma_{\perp}=\epsilon_{_{\perp}}/mc^{2}$.
Subsequently, the particle with reduced energy emits a photon with
a new value of $\epsilon_{_{\perp}}$. This process continues until
the particle is at $n=0$ Landau level. Depending on the particle's
Landau level $n$, the SR photon energy $\epsilon_{_{\perp}}$ is
chosen in one of three ways.

\clearpage{}

(I) When the particle is created in a high Landau Level ($n\geq3$),
we choose the energy of the photon randomly but according to a probability
based on the asymptotic synchrotron spectrum (e.g. \citet{1968_Sokolov},
\citet{1987_Harding}): 
\begin{equation}
\frac{{\rm d}^{2}N}{{\rm d}t{\rm d}\epsilon_{\perp}}=\frac{\sqrt{3}}{2\pi}\frac{\alpha_{f}\epsilon_{{\rm _{B}}}}{\epsilon_{\perp}}\times\left[f\cdot F\left(\frac{\epsilon_{\perp}}{f\epsilon_{{\rm _{SR}}}}\right)+\left(\frac{\epsilon_{\perp}}{\gamma_{\perp}mc^{2}}\right)^{2}G\left(\frac{\epsilon_{_{\perp}}}{f\epsilon_{{\rm _{SR}}}}\right)\right],\label{eq:cascade.synchrotron_spectrum}
\end{equation}
where 
\begin{equation}
\epsilon_{_{{\rm SR}}}=\frac{3}{2}\gamma_{\perp}^{2}\hbar\epsilon_{_{B}}
\end{equation}
 is the characteristic energy of the synchrotron photons, $f=1-\epsilon_{_{\perp}}/\left(\gamma_{\perp}mc^{2}\right)$
is the fraction of the electron's energy after photon emission, $F\left(x\right)=x\int_{x}^{\infty}K_{5/3}\left(t\right){\rm d}t$,
and $G\left(x\right)=xK_{2/3}\left(x\right)$. The functions $K_{5/3}$
and $K_{2/3}$ correspond to modified Bessel functions of the second
kind. The expression in Equation \ref{eq:cascade.synchrotron_spectrum}
differs from the classical synchrotron spectrum (e.g. \citealp{1979_Rybicki})
by a factor of $f=1-\epsilon_{_{\perp}}/\left(\gamma_{\perp}mc^{2}\right)$
which appears in several places in Equation \ref{eq:cascade.synchrotron_spectrum}
and by a term with the function $G\left(x\right)$. Note that in the
classical expressions for the total radiation spectra these terms
cancel out. However, as noted by \citet{2010_Medin} when the quantum
effects are considered there is asymmetry between the perpendicular
and parallel polarisations such that term $G\left(x\right)$ remain.

(II) If $n=2$, the photon's energy is either that required to lower
the particle energy to its first excited state ($n=1$) or to the
ground state ($n=0$). The probability of each process depends on
the local magnetic field strength. We use the simplified prescription
based on the results of \citet{1982_Herold} to calculate the transition
rates (see also \citealp{1987_Harding}). If $\beta_{q}<1$ the energy
of the photon is set to lower the particle energy to the first excited
state, $\epsilon_{_{\perp}}=mc^{2}\left(\sqrt{1+4\beta_{q}}-\sqrt{1+2\beta_{q}}\right)$.
If $\beta_{q}\gtrsim1$ the photon's energy is randomly chosen to
be that which is required to lower the particle energy to either the
first excited state, or the ground state ($\epsilon_{\perp}=mc^{2}\left(\sqrt{1+4\beta_{q}}-1\right)$),
with probability $50\%$ each.

(III) When $n=1$, the photon's energy is chosen to lower the particle's
energy to its ground state, $\epsilon_{_{\perp}}=mc^{2}\left(\sqrt{1+2\beta_{q}}-1\right)$.
If after emission of SR photon the particle is not in the ground state,
$\gamma_{\perp}$ is recalculated and a new energy of photon is chosen.

The photon energy in the co-rotating frame can be calculated as 
\begin{equation}
\epsilon=\gamma_{\parallel}\epsilon_{_{\perp}}.
\end{equation}
The weighting factor of the emitted photon is the same as the secondary
particle that emitted it ($\Delta N_{\epsilon}$ ). In the circular
frame the photon is emitted in a random direction perpendicular to
the magnetic field. Hence, in the co-rotating frame the emission angle
can be calculated using Equations \ref{eq:cascade.gamma_perp} and
\ref{eq:cascade.gamma_par} as follows 
\begin{equation}
\Psi=\arcsin\sqrt{\frac{\gamma_{\perp}^{2}-1}{\gamma_{\perp}^{2}\gamma_{\parallel}^{2}-1}}\cos\Pi,
\end{equation}
where $\Pi$ is a random number from $0$ to $2\pi$. In our simulation
we include this emission angle by using the same approach as presented
in Section \ref{sec:cascade.beaming}, but as the maximum value we
use $\Psi$ instead of $1/\gamma$.

The polarisation fraction of SR photons is the exact opposite of the
CR case and it ranges from $50\%$ to $100\%$ polarised perpendicular
to the magnetic field. Following the approach presented by \citet{2010_Medin}
in our calculations the photon polarisation is randomly assign in
the ratio of one $\parallel$ to every seven $\perp$ photons, which
corresponds to a $75\%$ perpendicular polarisation.

\begin{figure}[H]
\begin{centering}
\includegraphics[height=9.3cm]{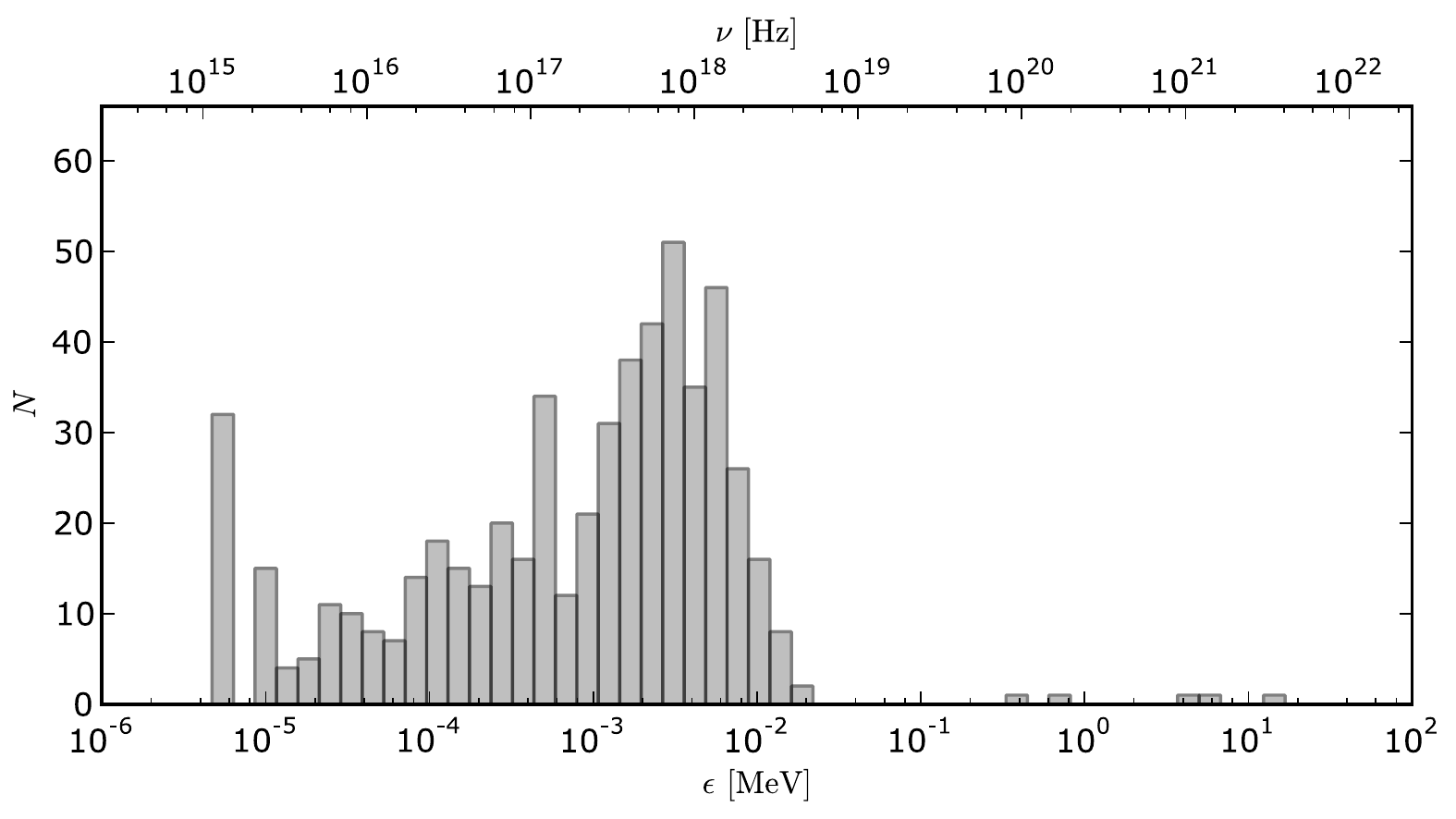}
\par\end{centering}

\centering{}\caption[Distribution of SR produced by a single secondary particle]{ Distribution of SR produced by a single secondary particle with
Lorentz factor $\gamma=10^{2}$. We have assumed that the particle
was created in a region where the magnetic field strength was $B=4.14\times10^{8}$
($\beta_{q}=10^{-5}$) and with a pitch angle $\Psi=7^{\circ}$. For
such a relatively high pitch angle the particle loses most of its
energy ending with Lorentz factor $\gamma_{{\rm end}}\approx6$.\label{fig:cascade.sr_spec}}
\end{figure}

Figure \ref{fig:cascade.sr_spec} presents the distribution of SR
produced by a single secondary particle. To show the nature of the
distribution, a relatively high pitch angle was used. Note that when
a particle is created at a distance where the magnetic field is relatively
weak (e.g. $\beta_{q}=10^{-5}$ for $\gamma=10^{2}$) then most of
the energy is radiated in the range of $1-10\,{\rm keV}$. Thus, we
believe that if a strong enough instability forms (that increases
the particle's pitch angle), the SR process could be responsible for
the production of a non-thermal component of the X-ray spectrum.

\begin{comment}
\textasciitilde{}/Programs/magnetic/magnetic/src/cascade/sr.py (show\_spectrum\_new)
\end{comment}

\begin{figure}[H]
\begin{centering}
\includegraphics[height=7.5cm]{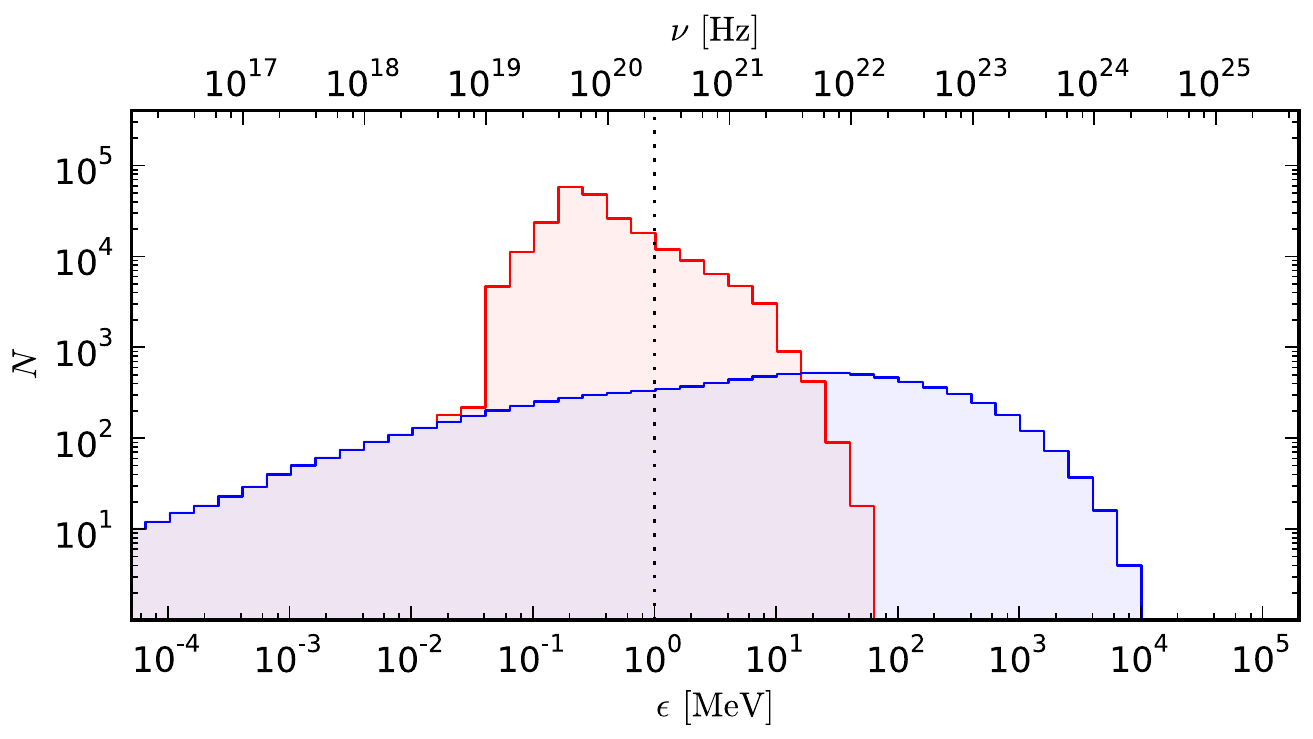}
\par\end{centering}

\centering{}\caption[{Final photon distribution produced by a single primary particle {[}CR{]}}]{Final photon distribution produced by a single primary particle.
The blue line corresponds to the initial CR photons distribution for
a non-dipolar structure of the magnetic field, while the red line
presents the final distribution with the inclusion of photon splitting,
pair production and SR. \label{fig:cascade.final_spec}}
\end{figure}

Figure \ref{fig:cascade.final_spec} presents the final spectrum produced
by a single primary particle with an initial Lorentz factor of $\gamma_{{\rm c}}=3.5\times10^{6}$
for a non-dipolar configuration of the surface magnetic field of PSR
J0633+1746 (see Section \ref{sec:model.0633}). Due to CR the particle
loses about $68\%$ of its initial energy ($\Delta\epsilon=2.2\times10^{6}mc^{2}$),
which is radiated mainly in close vicinity of a neutron star, where
curvature of the magnetic field is the highest. As the $\gamma$-photons
propagate they will split (only if the magnetic field is strong enough)
and eventually most photons will be absorbed by the magnetic field
- as a result electron-positron pairs emerge. These pairs radiate
away their transverse momenta through SR, producing mainly X-ray photons
(at larger distances) and only a few $\gamma$-photons (in a strong
magnetic field just above the stellar surface). Note that at the end
(after pair production) only $14\%$ of the primary particle's energy
($\Delta\epsilon_{{\rm ph}}=4\times10^{5}mc^{2}$) is converted into
photons and the bulk of its energy, $54\%$ ($\Delta\epsilon_{{\rm pairs}}=1.8\times10^{6}mc^{2}$),
is allocated into secondary plasma. The multiplicity for this specific
simulation is of the order $M_{{\rm sec}}=10^{4}$. Note that we use
$M_{{\rm sec}}$ to describe the multiplicity of secondary plasma
in contrast to $M_{{\rm pr}}$ which describes particle multiplicity
in the IAR. 

\begin{comment}
\textasciitilde{}/Programs/studies/phd/cascade\_plot/cascade\_plot.py
t4(plot\_spectrum\_final,373\_cr\_1e05\_10m\_CR\_SR\_e4\_leftline/)
\end{comment}

\begin{figure}[H]
\begin{centering}
\includegraphics[height=6.5cm]{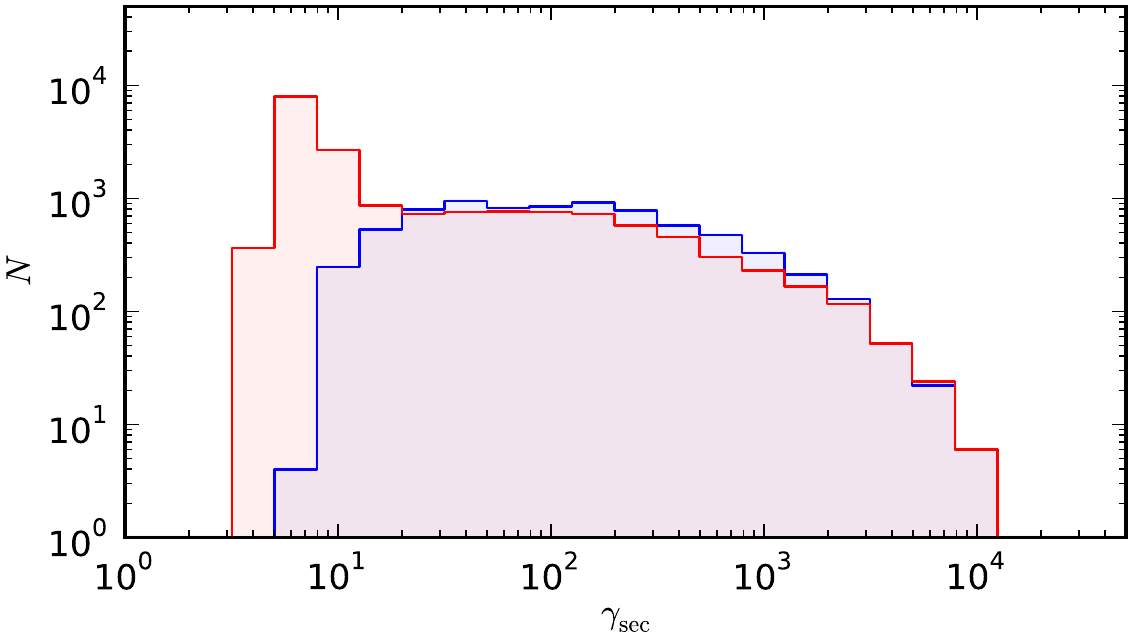}
\par\end{centering}

\centering{}\caption[Distribution of particle energy created by CR photons]{ Distribution of particle energy created by CR photons calculated
for a non-dipolar structure of the magnetic field. For this specific
magnetic field configuration and initial parameters (see the caption
of Figure \ref{fig:cascade.cr_spec}) the secondary plasma multiplicity
is $M_{{\rm sec}}\approx10^{4}$. Note that this result does not include
Synchrotron Radiation and the actual energies of the created pairs
are lower as they lose their transverse momenta (see Section \ref{sec:cascade.synchrotron}).
\label{fig:cascade.cr_sr_pairs}}
\end{figure}

Figure \ref{fig:cascade.cr_sr_pairs} presents the distribution of
particle energy created by CR photons but with the inclusion of SR
emission (red line). Note that synchrotron emission both lowers the
particle energy (after SR maximum at $\gamma\approx5-8$, while without
SR at $\gamma\approx15-20$) and increases the multiplicity of secondary
plasma $M_{{\rm sec}}\approx10^{4}$.

\begin{comment}
\textasciitilde{}/Programs/studies/phd/cascade\_plot/cascade\_plot.py
(plot\_pairs2;373\_cr\_1e05\_10m\_CR\_noSR\_PH\_e4\_leftline, 373\_cr\_1e05\_10m\_CR\_SR\_e4\_leftline)
\end{comment}

\section{Inverse Compton Scattering \label{sec:cascade.ics}}

The Inverse Compton Scattering (hereafter ICS) process in the neutron
star vicinity has been studied extensively by \citet{1982_Xia,1984_Kardashev,1985_Xia,1989_Daugherty,1989_Dermer,1990_Dermer,1992_Bednarek,1995_Chang,1995_Sturner,1996_Zhang,1997_Zhang,2000_Zhang,2002_Harding},
etc. According to these studies, the ICS process may play a significant
role in the physics of a neutron star's magnetosphere. Relativistic
particles (positrons and electrons) can Compton-scatter thermal radiation
from the neutron star surface. As a particle with a certain relativistic
velocity scatters the thermal photons with a blackbody distribution,
it will produce radiation in quite a wide energy range. However, we
can distinguish two characteristic frequencies of upscattered photons:
one is the frequency due to resonant scattering, another is the range
of frequencies contributed by the scattering of photons with frequencies
around the ''thermal-peak''. The Resonant Inverse Compton Scattering
(RICS) corresponds to a scenario when the scattering cross section
is largest. On the other hand, Thermal-peak Inverse Compton Scattering
(TICS) corresponds to interactions with photons with the maximum number
density. These two modes are very different when it comes to the nature
of the process. The photons' energy in RICS depends on the strength
of the magnetic field, thus at low altitudes (where the field is very
strong), it can power pair cascades, while TICS can be responsible
for magnetospheric radiation at much higher altitudes. Note that for
some specific combinations of magnetic strength and distribution of
background photons, RICS and TICS are indistinguishable as the resonance
frequency falls into the thermal peak range.

\subsection{The cross section of ICS \label{sec:cascade.ics_cross_sec}}

\begin{figure}[!ht]
\begin{centering}
\includegraphics{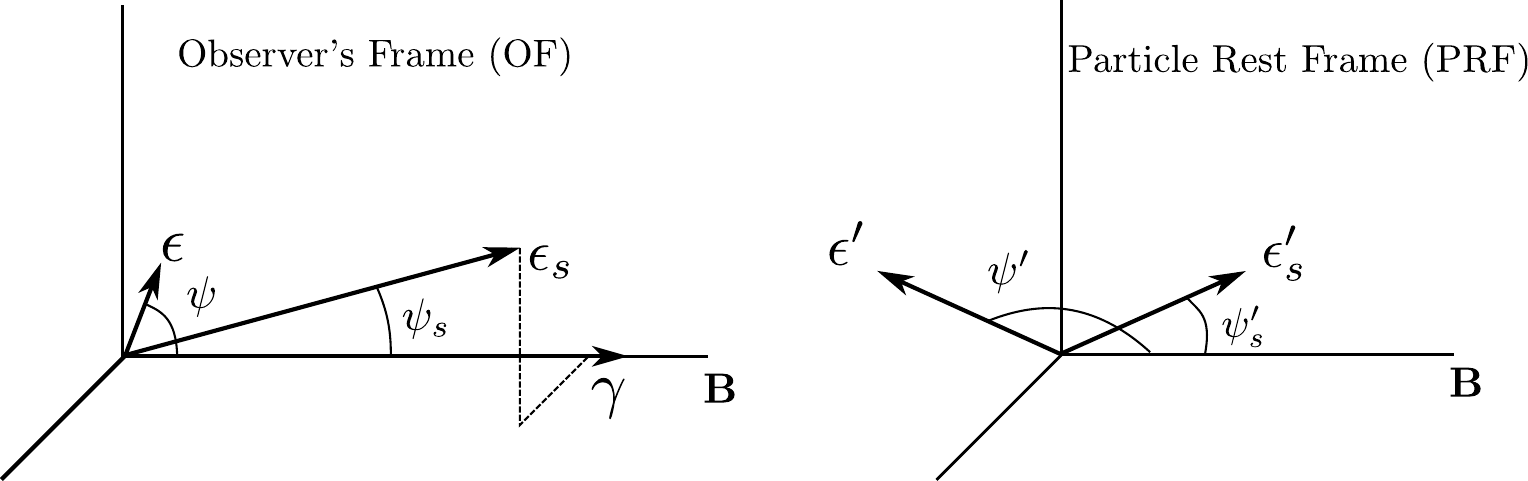}
\par\end{centering}

\centering{}\caption[Geometry of Inverse Compton Scattering]{ Reproduction of the Figure from \citet{1990_Dermer}. Geometry of
the ICS event in the observer's frame (left) and the particle rest
frame (right). A particle with Lorentz factor $\gamma$, beamed along
the direction of the magnetic field, scatters a photon with energy
$\epsilon$ directed at angle $\psi$ with respect to the magnetic
field line. After scattering, the energy and angle of the photon are
denoted by $\epsilon_{s}$ and $\psi_{s}$, respectively. Quantities
in the particle rest frame are denoted by a prime.\label{fig:cascade.ics_fig}}
\end{figure}

Due to the rapid time scale for synchrotron emission (see section
\ref{sec:cascade.synchrotron}), a particle in an excited Landau level
almost instantaneously de-excites to the ground level. The particle
motion is therefore strongly confined to the magnetic field direction.
In our calculations we consider the geometry illustrated in Figure
\ref{fig:cascade.ics_fig}. In the observer's frame of reference (OF),
a particle with Lorentz factor $\gamma$ travelling along the magnetic
field line scatters a photon. Let $\psi=\arccos\mu$ be the angle
between the magnetic field line (particle propagation) and the direction
of photon propagation in OF and $\psi^{\prime}=\arccos\mu^{\prime}$
in the particle rest frame (PRF). The energy of the photon in PRF
is given by 
\begin{equation}
\epsilon^{\prime}=\gamma\epsilon\left(1-\beta\mu\right).\label{cascade.erf_phot_eng}
\end{equation}
 After scattering, the photon energy is denoted by $\epsilon{}_{s}^{\prime}$
in PRF and $\epsilon_{s}$ in OF. The angle between the direction
of propagation of the scattered photon and ${\bf B}$ (which describes
the direction of particle propagation) is denoted by $\psi_{s}=\arccos\mu_{s}$
in OF and $\psi{}_{s}^{\prime}=\arccos\mu{}_{s}^{\prime}$, where
$\mu{}_{s}^{\prime}=\left(\mu_{s}-\beta\right)/\left(1-\beta\mu_{s}\right)$
in PRF \citep{1990_Dermer}.

\subsubsection{ICS cross section in the Thomson regime}

Restriction to the Thomson regime requires that $\gamma\epsilon\left(1-\mu\right)\ll1$.
In the particle rest frame, the angle $\psi^{\prime}=\arcsin\left\{ \gamma^{-1}\left[\sin\psi/\left(1-\beta\cos\psi\right)\right]\right\} $,
and when $\gamma\gg1$, $\left|\mu^{\prime}\right|\to1$. In the Thomson
regime the only important Compton scattering process involves transitions
between ground-state Landau levels. \citet{1989_Daugherty} and \citet{1989_Dermer}
calculated the differential cross section (after summing over polarisation
modes and integrating over azimuth) for a photon scattered from $\psi^{\prime}=0^{\circ}$
into angle $\psi{}_{s}^{\prime}=\arccos\mu{}_{s}^{\prime}$ as follows

\begin{equation}
\frac{{\rm d}\sigma^{\prime}}{{\rm d}\mu{}_{s}^{\prime}}=\frac{3\sigma_{_{{\rm T}}}}{16}\left(1+\mu{}_{s}^{\prime2}\right)\left[\frac{\epsilon{}^{\prime2}}{\left(\epsilon^{\prime}+\epsilon_{_{B}}\right){}^{2}}+\frac{\epsilon{}^{\prime2}}{\left(\epsilon^{\prime}-\epsilon_{_{B}}\right){}^{2}+\left(\Gamma/2\right){}^{2}}\right],\label{cascade.ics_simple_cross}
\end{equation}
where $\Gamma=4\alpha_{f}\epsilon_{_{B}}^{2}/3$ is the resonant width
\citep{1978_Daugherty,1985_Xia}, $\sigma_{_{{\rm T}}}$ is the Thomson
cross section, and $\alpha_{f}=e^{2}/\hbar c$ is the fine-structure
constant. In the Thomson limit $\epsilon^{\prime}\ll1$, and thus
the scattered photon energy in PRF can be approximated as 
\begin{equation}
\epsilon{}_{s}^{\prime}\simeq\epsilon^{\prime}+\epsilon{}^{\prime2}(\mu^{\prime}-\mu{}_{s}^{\prime})^{2}/2\approx\epsilon^{\prime}.\label{cascade.ics_simple_scateng}
\end{equation}

Equations \ref{cascade.ics_simple_cross} and \ref{cascade.ics_simple_scateng}
show that a differential magnetic Compton scattering cross section
when $\gamma\gg1$ is similar in form to a nonmagnetic Thomson cross
section. The important difference is that the magnitude of the cross
section is enhanced when $\epsilon^{\prime}$ approaches $\epsilon_{_{B}}$
and is depressed at energies $\epsilon^{\prime}<\epsilon_{_{B}}$.
The total cross section for magnetic Compton scattering, obtained
by integrating Equation \ref{cascade.ics_simple_cross} over $\mu{}_{s}^{\prime}$,
was calculated by \citet{1989_Dermer,1997_Zhang} and is given by 

\begin{equation}
\sigma^{\prime}=\frac{\sigma_{_{{\rm IC}}}}{2}\left[\frac{u^{2}}{\left(u+1\right){}^{2}}+\frac{u^{2}}{\left(u-1\right){}^{2}+a^{2}}\right],\label{eq:cascade.ics_cross}
\end{equation}
where $\sigma_{_{{\rm IC}}}=\sigma_{{\rm _{T}}}$, $\sigma_{{\rm _{T}}}$
is the Thomson cross section, $u=\epsilon^{\prime}/\epsilon_{_{B}}$,
$a=\frac{2}{3}\alpha_{f}\epsilon_{_{B}}$.

\subsubsection{ICS cross section in the Klein-Nishina regime}

The Klein-Nishina regime includes quantum effects due to the relativistic
nature of scattering, and it requires that $\gamma\epsilon\left(1-\mu\right)\gtrsim1$.
The principal effect is to reduce the cross section from its classical
value as the photon energy in PRF becomes large. In the Klein-Nishina
regime instead of $\sigma_{{\rm _{IC}}}=\sigma_{_{{\rm T}}}$ we can
use the following relationship

\begin{equation}
\sigma_{_{{\rm IC}}}=\sigma_{_{{\rm KN}}}=\frac{3}{4}\sigma_{_{{\rm T}}}\left\{ \frac{1+\epsilon^{\prime}}{\epsilon^{\prime3}}\left[\frac{2\epsilon^{\prime}\left(1+\epsilon^{\prime}\right)}{1+2\epsilon^{\prime}}-\ln\left(1+2\epsilon^{\prime}\right)\right]+\frac{1}{2\epsilon^{\prime}}\ln\left(1+2\epsilon^{\prime}\right)-\frac{1+3\epsilon^{\prime}}{\left(1+2\epsilon^{\prime}\right)^{2}}\right\} .
\end{equation}

In an extreme relativistic regime $\epsilon^{\prime}\gg1$ the Klein-Nishina
formula can be simplified to
\begin{equation}
\sigma_{{\rm _{KN}}}\approx\frac{3}{8}\sigma_{_{{\rm T}}}\epsilon^{\prime-1}\left[\ln\left(2\epsilon^{\prime}\right)+\frac{1}{2}\right].\label{eq:cascade.kn_cross}
\end{equation}

The above formula clearly shows that Inverse Compton Scattering is
less efficient for photons with energy in PRF significantly exceeding
particle rest energy.

\subsubsection{QED Compton Scattering cross section}

Previous studies on upscattering and energy loss by relativistic particles
have used the non-relativistic, magnetic Thomson cross section for
resonant scattering or the Klein-Nishina cross section for thermal-peak
scattering. As noted by \citet{2000_Gonthier}, this approach does
not account for the relativistic quantum effects of strong magnetic
fields ($B>10^{12}\,{\rm G}$). When the photon energy exceeds $mc^{2}$
in the particle rest frame, the strong magnetic field significantly
lowers the Compton scattering cross section below and at the resonance.
\citet{2000_Gonthier} developed expressions for the scattering of
ultrarelativistic electrons with $\gamma\gg1$ moving parallel to
the magnetic field. Because of the large Lorentz Factor of particle
$\gamma$, the photon incident angle $\psi$ gets Lorentz concentrated
to $\psi^{\prime}\approx\psi/2\gamma\approx0^{\circ}$ in the PRF.
The differential cross section in the rest frame of the particle can
be written as

\begin{equation}
\frac{{\rm d}\sigma{}_{\|,\perp}^{\prime}}{{\rm d}\cos\psi{}_{s}^{\prime}}=\frac{3\sigma_{_{{\rm T}}}}{16\pi}\frac{\epsilon{}_{s}^{\prime2}e^{-\epsilon{}_{s}^{\prime2}\sin^{2}\left(\psi_{s}^{\prime}/2\epsilon_{_{B}}\right)}}{\epsilon^{\prime}\left(2+\epsilon^{\prime}-\epsilon{}_{s}^{\prime}\right)\left[\epsilon{}_{s}^{\prime}+\epsilon^{\prime}\epsilon{}_{s}^{\prime}\left(1-\cos\psi{}_{s}^{\prime}\right)-\epsilon{}_{s}^{\prime2}\sin^{2}\psi{}_{s}^{\prime}\right]}\frac{1}{l!}\left(\frac{\epsilon{}_{s}^{\prime2}\sin^{2}\psi{}_{s}^{\prime}}{2\epsilon_{_{B}}}\right)G_{\|,\perp},\label{eq:cascade.cross_gonthier2}
\end{equation}
 where 
\begin{equation}
G_{\|}=\hat{G}_{\mathrm{no-flip}}^{\|}+\hat{G}_{\mathrm{flip}}^{\|},\hspace{1cm}G_{\perp}=\hat{G}_{\mathrm{no-flip}}^{\perp}+\hat{G}_{\mathrm{flip}}^{\perp}
\end{equation}
 and 
\begin{equation}
\begin{split}\hat{G}_{\mathrm{no-flip}}^{\|}= & \int_{0}^{2\pi}\left|G_{\mathrm{no-flip}}^{\|,\|}\right|^{2}{\rm d}\phi^{\prime}=\int_{0}^{2\pi}\left|G_{\mathrm{no-flip}}^{\perp,\|}\right|^{2}{\rm d}\phi^{\prime}=\\
= & 2\pi\left\{ \left[\left(B_{1}+B_{3}+B_{7}\right)\cos\psi{}_{s}^{\prime}-\left(B_{2}+B_{6}\right)\sin\psi{}_{s}^{\prime}\right]^{2}+\left(B_{4}\cos\psi{}_{s}^{\prime}-B_{5}\sin\psi{}_{s}^{\prime}\right)^{2}\right\} ,\\
\hat{G}_{\mathrm{no-flip}}^{\perp}= & \int_{0}^{2\pi}\left|G_{\mathrm{no-flip}}^{\|,\perp}\right|^{2}{\rm d}\phi^{\prime}=\int_{0}^{2\pi}\left|G_{\mathrm{no-flip}}^{\perp,\perp}\right|^{2}{\rm d}\phi^{\prime}=\\
= & 2\pi\left[\left(B_{1}-B_{3}-B_{7}\right)^{2}+B_{4}^{2}\right],\\
\hat{G}_{\mathrm{flip}}^{\|}= & \int_{0}^{2\pi}\left|G_{\mathrm{flip}}^{\|,\|}\right|^{2}{\rm d}\phi^{\prime}=\int_{0}^{2\pi}\left|G_{\mathrm{flip}}^{\perp,\|}\right|^{2}{\rm d}\phi^{\prime}=\\
= & 2\pi\left\{ \left[\left(C_{1}+C_{3}+C_{7}\right)\cos\psi{}_{s}^{\prime}-\left(C_{2}+C_{6}\right)\sin\psi{}_{s}^{\prime}\right]^{2}+\left(C_{4}\cos\psi{}_{s}^{\prime}-C_{5}\sin\psi{}_{s}^{\prime}\right)^{2}\right\} ,\\
\hat{G}_{\mathrm{flip}}^{\perp}= & \int_{0}^{2\pi}\left|G_{\mathrm{flip}}^{\|,\perp}\right|^{2}{\rm d}\phi^{\prime}=\int_{0}^{2\pi}\left|G_{\mathrm{flip}}^{\perp,\perp}\right|^{2}{\rm d}\phi^{\prime}=\\
= & 2\pi\left[\left(C_{1}-C_{3}-C_{7}\right)^{2}+C_{4}^{2}\right].
\end{split}
\end{equation}
The imaginary terms and the $\phi^{\prime}$ dependence are isolated
in the polarisation components and in the phase exponentials, leading
to elementary integrations over the azimuthal angle, $\phi^{\prime}$
\citep{2000_Gonthier}. 

The differential cross section depends on the final Landau state $l$,
thus a sum must be calculated over all the contributing Landau states.
The energy of the scattered photon is given by \citep{2000_Gonthier}
\begin{equation}
\epsilon{}_{s}^{\prime}=\frac{2\left(\epsilon^{\prime}-l\epsilon_{_{B}}\right)}{1+\epsilon^{\prime}\left(1-\cos\psi{}_{s}^{\prime}\right)+\left\{ \left[1+\epsilon^{\prime}\left(1-\cos\psi{}_{s}^{\prime}\right)\right]^{2}-2\left(\epsilon^{\prime}-l\epsilon_{_{B}}\right)\sin^{2}\psi{}_{s}^{\prime}\right\} ^{\frac{1}{2}}},
\end{equation}
 where $l$ is the final Landau level of the scattered particle. Each
final state has an energy threshold of $l\epsilon_{_{B}}$, thus the
maximum contributing Landau state $l_{{\rm max}}$ can be expressed
as: $\epsilon^{\prime}/\epsilon_{_{B}}-1<l_{{\rm max}}<\epsilon^{\prime}/\epsilon_{_{B}}$.
To obtain the energy-dependent cross section, the Romberg's method
can be used to numerically integrate the differential cross section
over $\psi{}_{s}^{\prime}$. For this particular case (scattering
of relativistic particles) there is only one resonance appearing at
the fundamental cyclotron frequency $\epsilon_{_{B}}=\beta_{q}=eB/\left(mc\right)$.

The values of $B$ and $C$ can be expressed as:

\begin{equation}
\begin{split}B_{1}= & \frac{2\epsilon^{\prime}-\epsilon^{\prime}\epsilon{}_{s}^{\prime}\left(1-\cos\psi{}_{s}^{\prime}\right)}{2\left(\epsilon^{\prime}-\epsilon_{_{B}}\right)},\\
B_{2}= & -\frac{\left(\epsilon^{\prime}-\epsilon{}_{s}^{\prime}\cos\psi{}_{s}^{\prime}\right)\left(2l\epsilon_{_{B}}-\epsilon{}_{s}^{\prime2}\sin^{2}\psi{}_{s}^{\prime}\right)+2l\epsilon_{_{B}}\epsilon^{\prime}}{2\epsilon{}_{s}^{\prime}\sin\psi{}_{s}^{\prime}\left(\epsilon^{\prime}-\epsilon_{_{B}}\right)},\\
B_{3}= & \frac{l\epsilon_{_{B}}\left(2l\epsilon_{_{B}}-2\epsilon_{_{B}}-\epsilon{}_{s}^{\prime2}\sin^{2}\psi{}_{s}^{\prime}\right)}{\epsilon{}_{s}^{\prime2}\sin^{2}\left[\psi{}_{s}^{\prime}\left(\epsilon^{\prime}-\epsilon_{_{B}}\right)\right]},\\
B_{4}= & -\frac{2\epsilon{}_{s}^{\prime}+\epsilon^{\prime}\epsilon{}_{s}^{\prime}\left(1-\cos\psi{}_{s}^{\prime}\right)-\epsilon{}_{s}^{\prime2}\sin^{2}\psi{}_{s}^{\prime}}{2\left[\epsilon^{\prime}\epsilon{}_{s}^{\prime}\left(1-\cos\psi{}_{s}^{\prime}\right)-\epsilon^{\prime}-\epsilon_{_{B}}\right]},\\
B_{5}= & -\frac{\left(\epsilon^{\prime}-\epsilon{}_{s}^{\prime}\cos\psi{}_{s}^{\prime}\right)\epsilon{}_{s}^{\prime}\sin\psi{}_{s}^{\prime}}{2\left[\epsilon^{\prime}\epsilon{}_{s}^{\prime}\left(1-\cos\psi{}_{s}^{\prime}\right)-\epsilon^{\prime}-\epsilon_{_{B}}\right]},\\
B_{6}= & \frac{l\epsilon_{_{B}}\cos\psi{}_{s}^{\prime}}{\sin\psi{}_{s}^{\prime}\left[\epsilon^{\prime}\epsilon{}_{s}^{\prime}\left(1-\cos\psi_{s}^{\prime}\right)-\epsilon^{\prime}+\epsilon_{_{B}}\right]},\\
B_{7}= & \frac{2l\left(l-1\right)\epsilon_{_{B}}^{2}}{\epsilon{}_{s}^{\prime2}\sin^{2}\psi{}_{s}^{\prime}\left[\epsilon^{\prime}\epsilon{}_{s}^{\prime}\left(1-\cos\psi{}_{s}^{\prime}\right)-\epsilon^{\prime}+\epsilon_{_{B}}\right]},\\
C_{1}= & \sqrt{2l\epsilon_{_{B}}}\frac{\epsilon^{\prime}}{2\left(\epsilon^{\prime}-\epsilon_{_{B}}\right)},\\
C_{2}= & -\sqrt{2l\epsilon_{_{B}}}\frac{2\epsilon^{\prime}+2\epsilon^{\prime}{}^{2}-\epsilon^{\prime}\epsilon{}_{s}^{\prime}\left(1-\cos\psi{}_{s}^{\prime}\right)-2l\epsilon_{_{B}}+\epsilon{}_{s}^{\prime2}\sin^{2}\psi{}_{s}^{\prime}}{2\epsilon^{\prime}{}_{s}\sin\psi{}_{s}^{\prime}\left(\epsilon^{\prime}-\epsilon_{_{B}}\right)},\\
C_{3}= & \sqrt{2l\epsilon_{_{B}}}\frac{\left(\epsilon^{\prime}-\epsilon_{s}^{\prime}\cos\psi{}_{s}^{\prime}\right)\left(2l\epsilon_{_{B}}-2\epsilon_{_{B}}-\epsilon{}_{s}^{\prime2}\sin^{2}\psi{}_{s}^{\prime}\right)}{2\epsilon{}_{s}^{\prime2}\sin^{2}\psi{}_{s}^{\prime}\left(\epsilon^{\prime}-\epsilon_{_{B}}\right)},\\
C_{4}= & -\sqrt{2l\epsilon_{_{B}}}\frac{\epsilon{}_{s}^{\prime}\cos\psi{}_{s}^{\prime}}{2\left[\epsilon{}_{s}^{\prime}\epsilon^{\prime}\left(1-\cos\psi{}_{s}^{\prime}\right)-\epsilon^{\prime}-\epsilon_{_{B}}\right]},\\
C_{5}= & \sqrt{2l\epsilon_{_{B}}}\frac{\epsilon{}_{s}^{\prime}\sin\psi{}_{s}^{\prime}}{2\left[\epsilon{}_{s}^{\prime}\epsilon^{\prime}\left(1-\cos\psi{}_{s}^{\prime}\right)-\epsilon^{\prime}-\epsilon_{_{B}}\right]},\\
C_{6}= & -\sqrt{2l\epsilon_{_{B}}}\frac{2\epsilon{}_{s}^{\prime}+\epsilon^{\prime}\epsilon{}_{s}^{\prime}\left(1-\cos\psi{}_{s}^{\prime}\right)-\epsilon{}_{s}^{\prime2}\sin^{2}\psi{}_{s}^{\prime}}{2\epsilon{}_{s}^{\prime}\sin\psi{}_{s}^{\prime}\left[\epsilon{}_{s}^{\prime}\epsilon^{\prime}\left(1-\cos\psi{}_{s}^{\prime}\right)-\epsilon^{\prime}+\epsilon_{_{B}}\right]},\\
C_{7}= & \sqrt{2l\epsilon_{_{B}}}\frac{\left(l-1\right)\epsilon_{_{B}}\left(\epsilon^{\prime}-\epsilon{}_{s}^{\prime}\cos\psi{}_{s}^{\prime}\right)}{\epsilon{}_{s}^{\prime2}\sin^{2}\psi{}_{s}^{\prime}\left[\epsilon{}_{s}^{\prime}\epsilon^{\prime}\left(1-\cos\psi{}_{s}^{\prime}\right)-\epsilon^{\prime}+\epsilon_{_{B}}\right]}.
\end{split}
\end{equation}

Although the expressions presented above describe the exact cross
section for ICS in strong magnetic fields, due to their complexity
their usage in cascade simulation is limited.

\subsubsection{Approximate cross section (final states l=0)}

\begin{figure}[H]
\begin{centering}
\includegraphics[height=7cm]{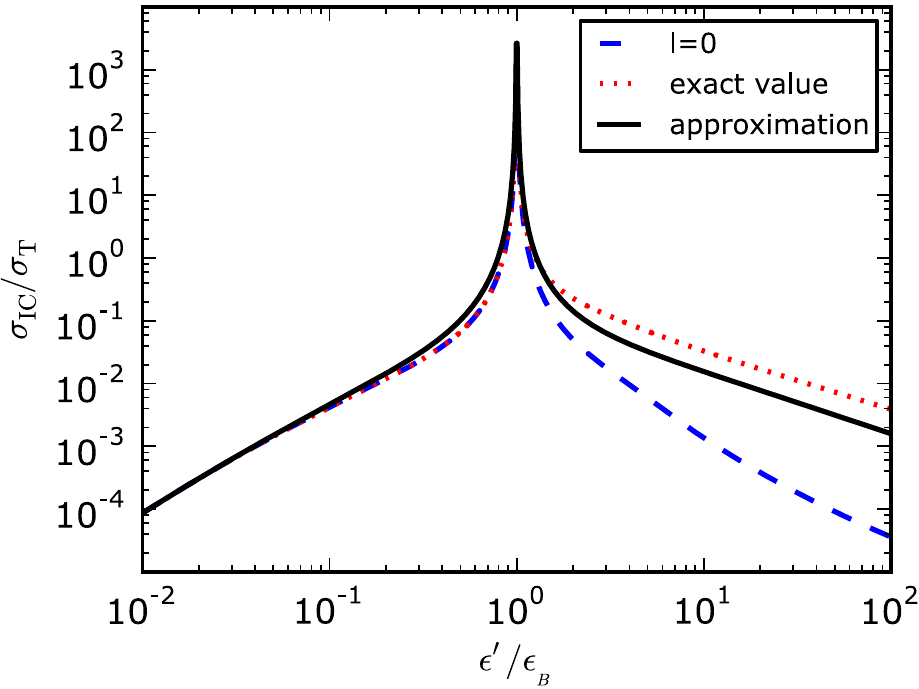}
\par\end{centering}

\centering{}\caption[Total cross section of ICS as a function of an incident photon energy]{ Total cross section of Compton scattering (in Thomson units) as
a function of an incident photon energy in PRF (in units of the cyclotron
energy) calculated for a magnetic field strength $B_{14}=3.5$. The
exact QED scattering cross section, summed over all contributing final
electron/positron Landau states, is indicated as the red dotted curve.
The cross section for final Landau states $l=0$ is plotted as a blue
dashed line. \label{fig:cascade.ics_sigma}}
\end{figure}

An approximation to the exact $l=0$ differential cross section can
be given by assuming that the scattering is significantly below the
resonance, where $\epsilon^{\prime}<\epsilon_{_{B}}$ and also $\epsilon^{\prime}<1$.
\citet{2000_Gonthier} showed that by keeping only terms to first
order in $\epsilon^{\prime}$ and $\epsilon{}_{s}^{\prime}$ in the
region of validity, it agrees very well with the exact $l=0$ cross
section. The approximation overestimates the exact $l=0$ cross section
above the region of validity $\epsilon^{\prime}>\epsilon_{_{B}}$.
However, the approximation is close to the total cross section for
both energy regions ($\epsilon^{\prime}<\epsilon_{_{B}}$ and $\epsilon^{\prime}>\epsilon_{_{B}}$
), even for high magnetic field strengths (see Figure \ref{fig:cascade.ics_sigma}).

According to \citet{2000_Gonthier}, the polarisation-dependent and
averaged, approximate cross section can be calculated as:

\begin{equation}
\sigma{}^{\prime\|\rightarrow\|}=\sigma{}^{\prime\perp\rightarrow\|}=\frac{3\sigma_{_{{\rm T}}}}{16}\left[g\left(\epsilon^{\prime}\right)-h\left(\epsilon^{\prime}\right)\right]\left[\frac{1}{\left(\epsilon^{\prime}-\epsilon_{_{B}}\right)^{2}}+\frac{1}{\left(\epsilon^{\prime}+\epsilon_{_{B}}\right)^{2}}\right],
\end{equation}

\begin{equation}
\sigma{}^{\prime\|\rightarrow\perp}=\sigma{}^{\prime\perp\rightarrow\perp}=\frac{3\sigma_{_{{\rm T}}}}{16}\left[f\left(\epsilon^{\prime}\right)-2\epsilon^{\prime}h\left(\epsilon^{\prime}\right)\right]\left[\frac{1}{\left(\epsilon^{\prime}-\epsilon_{_{B}}\right)^{2}}+\frac{1}{\left(\epsilon^{\prime}+\epsilon_{_{B}}\right)^{2}}\right],
\end{equation}

\begin{equation}
\sigma{}_{{\rm avg}}^{\prime}=\frac{3\sigma_{_{{\rm T}}}}{16}\left[g\left(\epsilon^{\prime}\right)+f\left(\epsilon^{\prime}\right)-\left(1+2\epsilon^{\prime}\right)h\left(\epsilon^{\prime}\right)\right]\left[\frac{1}{\left(\epsilon^{\prime}-\epsilon_{_{B}}\right)^{2}}+\frac{1}{\left(\epsilon^{\prime}+\epsilon_{_{B}}\right)^{2}}\right],\label{cascade.cross_gonthier}
\end{equation}

\begin{equation}
\begin{array}{c}
g\left(\epsilon^{\prime}\right)=\frac{\epsilon^{\prime2}\left(3+2\epsilon^{\prime}\right)+2\epsilon^{\prime}}{\sqrt{\epsilon^{\prime}\left(2+\epsilon^{\prime}\right)}}\ln\left[1+\epsilon^{\prime}-\sqrt{\epsilon^{\prime}\left(2+\epsilon^{\prime}\right)}\right]+\frac{\epsilon^{\prime}}{2}\ln\left(1+4\epsilon^{\prime}\right)+\\
+\epsilon^{\prime}\left(1+2\epsilon^{\prime}\right)\ln\left(1+2\epsilon^{\prime}\right)+2\epsilon^{\prime},
\end{array}
\end{equation}

\begin{equation}
f\left(\epsilon^{\prime}\right)=-\epsilon{}^{\prime2}\ln\left(1+4\epsilon^{\prime}\right)+\epsilon^{\prime}\left(1+2\epsilon^{\prime}\right)\ln\left(1+2\epsilon^{\prime}\right),
\end{equation}

\begin{equation}
h\left(\epsilon^{\prime}\right)=\left\{ \begin{array}{ll}
\frac{\epsilon{}^{\prime2}}{\sqrt{\epsilon^{\prime}\left(2-\epsilon^{\prime}\right)}}\arctan\left[\frac{\sqrt{\epsilon^{\prime}\left(2-\epsilon^{\prime}\right)}}{1+\epsilon^{\prime}}\right] & \mbox{ for \ensuremath{\epsilon^{\prime}<2}},\\
\frac{\epsilon^{\prime}{}^{2}}{2\sqrt{\epsilon^{\prime}\left(\epsilon^{\prime}-2\right)}}\ln\left[\frac{\left(1+\epsilon^{\prime}+\sqrt{\epsilon^{\prime}(\epsilon^{\prime}-2)}\right)^{2}}{1+4\epsilon^{\prime}}\right] & \mbox{ for \ensuremath{\epsilon^{\prime}>2}}.
\end{array}\right.
\end{equation}

Figure \ref{fig:cascade.ics_sigma} presents the total approximate
cross section of Compton scattering, the exact QED cross section (summed
over all contributing final electron/positron Landau states) and the
exact cross section for final Landau state $l=0$ as a function of
energy of the incident photon in PRF (in units of cyclotron energy,
$\epsilon^{\prime}/\epsilon_{_{B}}$). As mentioned above, the approximation
is valid in the region below the resonance, $\epsilon^{\prime}<\epsilon_{_{B}}$.
Although the approximation overestimates the cross section for $l=0$
final Landau state in the regime of high energetic photons ($\epsilon^{\prime}>\epsilon_{_{B}}$),
it can be used in this regime as the approximation of the total cross
section. In our simulation we use this approach to calculate the total
ICS cross section in both regimes, $\epsilon^{\prime}<\epsilon_{_{B}}$
and $\epsilon^{\prime}>\epsilon_{_{B}}$. Calculation of the cross
section for the resonance frequency ($\epsilon^{\prime}=\epsilon_{{\rm _{B}}}$)
is presented in the next section.

\begin{comment}
\textasciitilde{}Programs/magnetic/magnetic/src/radiation/ics\_modules/old/gonthier.py
(show\_sigma) 
\end{comment}

\subsection{Resonant Compton Scattering\label{sec:cascade.R-ICS_cross}}

This section describes an approach used to calculate the RICS cross
section for ultrastrong magnetic fields ($B>10^{12}\,{\rm G}$). For
weaker fields the calculations are much simpler and resonance is already
included in Equation \ref{eq:cascade.ics_cross}.

The trend as $\beta_{q}$ increases is for the magnitude of the cross
section to drop at all energies. For weaker magnetic fields ($\beta_{q}<1$)
the width of the resonance increases with increasing $\beta_{q}$,
but for $\beta_{q}\ge1$ this width actually declines. Since the resonance
is formally divergent, the common practice (see \citealp{1985_Xia,1986_Latal,1989_Daugherty,1990_Dermer,1991_Harding,2005_Baring,2006_Harding,2007_Baring,2008_Baring})
is to truncate it at $\epsilon^{\prime}=\epsilon_{_{B}}$ by introducing
a finite width $\Gamma$. The procedure is to replace the resonant
$\left(\epsilon^{\prime}-\epsilon_{_{B}}\right)^{2}$ denominator
(see Equations \ref{eq:cascade.cross_gonthier2} and \ref{cascade.cross_gonthier})
by $\left[\left(\epsilon^{\prime}-\epsilon_{_{B}}\right)^{2}+\Gamma^{2}/4\right]$.
In the $\beta_{q}\ll1$ regime, the cyclotron decay width assumes
the well-known result $\Gamma\approx4\alpha_{f}\epsilon_{_{B}}^{2}/3$
in dimensionless units. When $\beta_{q}\gg1$, quantum and recoil
effects generate $\Gamma\approx\alpha_{f}\epsilon_{_{B}}\left(1-1/\tilde{e}\right)$
where $\tilde{e}$ is Euler's number (e.g. see \citealp{2011_Baring}).
These widths lead to areas under the resonance being independent of
$\epsilon_{_{B}}$ in the magnetic Thomson regime of $\beta_{q}\ll1$
and scaling as $\epsilon_{_{B}}^{1/2}$ when $\beta_{q}\gg1$. These
results can be deduced using the $l=0$ approximation derived in Equation
\ref{cascade.cross_gonthier}. By using this approach the averaged,
approximate cross section can be written as 
\begin{equation}
\sigma{}_{{\rm avg}}^{\prime}=\frac{3\sigma_{_{{\rm T}}}}{16}\left[g\left(\epsilon^{\prime}\right)+f\left(\epsilon^{\prime}\right)-\left(1+2\epsilon^{\prime}\right)h\left(\epsilon^{\prime}\right)\right]\left[\frac{1}{\left(\epsilon^{\prime}-\epsilon_{_{B}}\right)^{2}+\Gamma{}^{2}/4}+\frac{1}{\left(\epsilon^{\prime}+\epsilon_{_{B}}\right)^{2}}\right].\label{eq:cascade.sigma_gont}
\end{equation}

The common practice to calculate a resonant cross section in an ultrastrong
magnetic fields is to use the Dirac delta function as follows (e.g.
\citealp{2010_Medin}) 
\begin{equation}
\sigma_{{\rm res}}^{\prime}\simeq2\pi^{2}\frac{e^{2}\hbar}{mc}\delta\left(\epsilon{}_{s}^{\prime}-\epsilon_{_{B}}\right)\label{eq:cascade.res_cross_a}
\end{equation}

This simplified approach, however, does not include scatterings of
photons whose energy in a particle rest frame is not equal but very
close to the resonance frequency. The relativistic quantum effects
of strong magnetic fields that are included in the approximate solution
increase the cross section, and thus the efficiency of the ICS process
in previous estimates could be underestimated.

According to \citet{2010_Medin} in ultrastrong magnetic fields the
ICS polarisation fraction is about $50\%$ (approximately $50\%$
of the photons are slightly above resonance and $50\%$ are slightly
below). Therefore, the polarisation of ICS photons is randomly assigned
in the ratio of one $\perp$ (perpendicular to the field) to every
$\parallel$ photon.

\subsection{Particle mean free path \label{sec:cascade.ics_electron_path}}

For the ICS process the calculation of the particle mean free path
$l_{{\rm ICS}}$ is not as simple as that of the CR process. Although
we can define $l_{{\rm ICS}}$ in the same way as we defined $l_{{\rm CR}}$,
it is difficult to estimate a characteristic frequency of emitted
photons. We have to take into account photons of various frequencies
with various incident angles. An estimation of the mean free path
of a positron (or electron) to produce a photon is \citep{1985_Xia}
\begin{equation}
l_{{\rm ICS}}\approx\left[\int_{\mu_{0}}^{\mu_{1}}\int_{0}^{\infty}\left(1-\beta\mu\right)\sigma^{\prime}\left(\epsilon,\mu\right)n_{{\rm ph}}\left(\epsilon\right){\rm d}\epsilon{\rm d}\mu\right]^{-1},
\end{equation}
where (as before) $\beta=v/c$ is the velocity in terms of speed of
light, $n_{{\rm ph}}$ represents the photon number density distribution
of semi-isotropic blackbody radiation (see Equation \ref{eq:psg.nph}).
Here $\sigma^{\prime}$ is the average cross section of scattering
in the particle rest frame (see Equation \ref{eq:cascade.sigma_gont}).
We should expect two modes of the ICS process, i.e. Resonant ICS and
Thermal-peak ICS.

\subsubsection{Resonant ICS}

The RICS takes place if the photon frequency in the particle rest
frame is equal to the cyclotron electron frequency. Using Equation
\ref{cascade.erf_phot_eng} we can write that the incident photon
energy is $\epsilon=\epsilon_{_{B}}/\left[\gamma\left(1-\beta\mu\right)\right]$.
For altitudes of the same order as the polar cap size we use $\mu_{0}=1$,
$\mu_{1}=0$ as incident angle limits for outflowing particles, and
$\mu_{0}=0$, $\mu_{1}=-1$ as incident angle limits for backflowing
particles. Thus, for outflowing particles the electron/positron mean
free path above a polar cap for the RICS process is

\begin{equation}
l_{{\rm RICS}}\approx\left[\int_{0}^{1}\int_{\epsilon_{{\rm _{res}}}^{^{{\rm min}}}}^{\epsilon_{{\rm _{res}}}^{{\rm ^{max}}}}\left(1-\beta\mu\right)\sigma^{\prime}\left(\epsilon,\mu\right)n_{{\rm ph}}\left(\epsilon\right){\rm d}\epsilon{\rm d}\mu\right]^{-1},\label{eq:cascade.ics_free_path}
\end{equation}
where limits of integration, $\epsilon_{{\rm _{res}}}^{^{{\rm min}}}$
and $\epsilon_{_{{\rm res}}}^{{\rm ^{max}}}$, are chosen to cover
the resonance. In our simulation we use such limits to include the
region where the integrated function decreases up to about two orders
of magnitude from its maximum:

\begin{equation}
\epsilon_{_{{\rm res}}}^{^{{\rm min/max}}}=\frac{\epsilon_{_{B}}\pm\frac{3}{2}\sqrt{11}\Gamma}{\gamma\left(1-\beta\mu\right)}.
\end{equation}

Here $\Gamma$ is the finite width introduced in Section \ref{sec:cascade.R-ICS_cross}
to describe the decay of an excited intermediate particle state.

\begin{figure}[H]
\begin{centering}
\includegraphics[height=7.7cm]{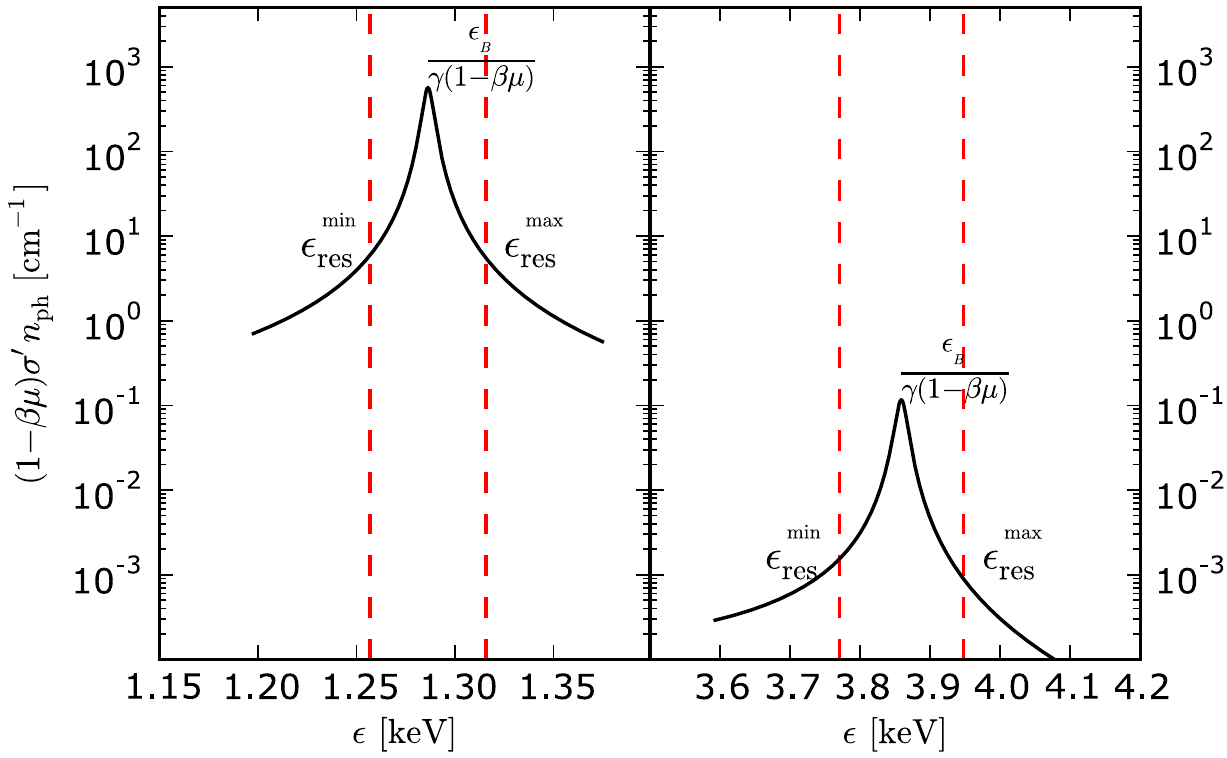}
\par\end{centering}

\caption[Dependence of the integrand from Equation \ref{eq:cascade.ics_free_path}
on the energy of the incident photon]{Dependence of the integrand from Equation \ref{eq:cascade.ics_free_path}
on the energy of the incident photon. Both panels were calculated
for surface temperature $T=3\times10^{6}\,{\rm K}$, cosine of the
incident angle $\mu=0.1$ and Lorentz factor of particle $\gamma=10^{3}$.
The left panel corresponds to resonance in magnetic field $B=10^{14}\,{\rm G}$,
while the right panel was obtained using $B=3\times10^{14}\,{\rm G}$.\label{fig:cascade.resonance_eps}}
\end{figure}

Figure \ref{fig:cascade.resonance_eps} presents the dependence of
the integrand from Equation \ref{eq:cascade.ics_free_path} on the
incident photon energy for a given incident angle. The maximum of
the integrand shows a significant decline for stronger magnetic fields.
This is due to both the drop of the cross section at all energies
with an increasing magnetic field (see Section \ref{sec:cascade.R-ICS_cross})
and due to the fact that for this specific incident angle resonance
is in a different range of photon energy. In stronger magnetic fields
resonance occurs not only for higher energetic photons but also the
width of the resonance is wider (see the right panel of Figure \ref{fig:cascade.resonance_eps}). 

\begin{comment}
\textasciitilde{}/Programs/magnetic/magnetic/src/radiation/ics\_modules/gonthier.py
(show\_le\_fun)
\end{comment}

Note that both plots do not include the dependence of the photon density
on distance from the stellar surface. Depending on whether the radiation
originates from the whole stellar surface or from the polar cap only,
the dependence of the photon number density on the height above the
surface can differ significantly (see Section \ref{sec:cascade.background_photons}).

\subsubsection{Thermal-peak ICS}

TICS includes all scattering processes of photons with frequencies
around the maximum of the thermal spectrum. In our simulation we adopt
$\epsilon_{{\rm _{th}}}^{{\rm ^{min}}}\approx0.05\epsilon_{{\rm _{th}}}$,
and $\epsilon_{{\rm _{th}}}^{^{{\rm max}}}\approx2\epsilon_{{\rm _{th}}}$
where $\epsilon_{{\rm _{th}}}=2.82kT/\left(mc^{2}\right)$ is the
energy, in units of $mc^{2}$, at which blackbody radiation with temperature
$T$ has the largest photon number density. The electron/positron
mean free path for the TICS process can be calculated as 
\begin{equation}
l_{{\rm TICS}}\approx\left[\int_{\mu_{0}}^{\mu_{1}}\int_{\epsilon_{_{{\rm th}}}^{{\rm ^{min}}}}^{\epsilon_{{\rm _{th}}}^{{\rm ^{max}}}}\left(1-\beta\mu\right)\sigma^{\prime}\left(\epsilon,\mu\right)n_{{\rm ph}}\left(\epsilon\right){\rm d}\epsilon{\rm d}\mu\right]^{-1}.\label{eq:cascade.t_ics}
\end{equation}

\begin{figure}[H]

\begin{centering}
\includegraphics[height=9cm]{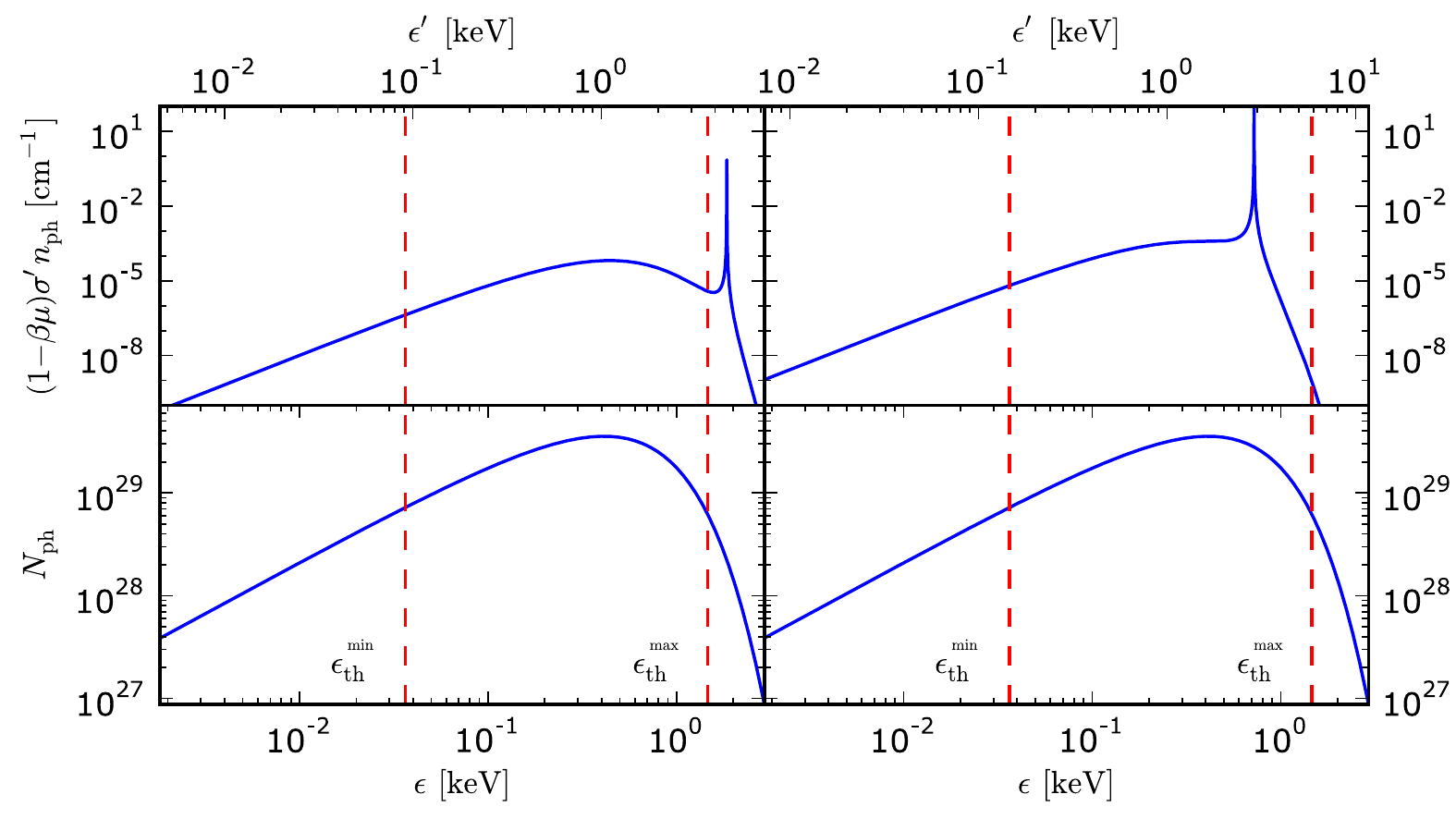}
\par\end{centering}

\caption[The integrand from Equation \ref{eq:cascade.t_ics} vs. photon number
density]{Comparison of the integrand from Equation \ref{eq:cascade.t_ics}
with photon number density. The bottom panels present the dependence
of the photon number density on photon energy in OF. The red dashed
lines correspond to limits used to calculate the particle mean free
path for TICS. The top panels present the dependence of the integrand
on photon energy in PRF. Both panels were obtained using surface temperature
$T=3\times10^{6}\,{\rm K}$, Lorentz factor of the particle $\gamma=10^{2}$
and magnetic field strength $B=10^{12}\,{\rm G}$. The cosine of the
incident angle, $\mu=0.975$ and $\mu=0.96$, was used for the left
and right panel, respectively.\label{fig:cascade.thermal_eps}}

\end{figure}

Figure \ref{fig:cascade.thermal_eps} presents the dependence of the
integrand from Equation \ref{eq:cascade.t_ics} on photon energy for
two different incident angles of background photons. As the number
density depends exponentially on the photon energy, TICS is important
only for small incident angles ($\mu\approx1$). Note that for some
specific combination of magnetic field strength, the Lorentz factor
of the primary particle and the incident angle of background photons
the resonance is in region of thermal peak. In such a case the resonant
component is dominating (a much higher cross section) and the particle
mean free path should be calculated using the approach described in
the previous section. 

\begin{comment}
\textasciitilde{}/Programs/magnetic/magnetic/src/radiation/ics\_modules/gonthier.py
(show\_le\_fun\_th)
\end{comment}

\subsubsection{Calculation results}

\begin{figure}[H]
\begin{centering}
\includegraphics[height=7cm]{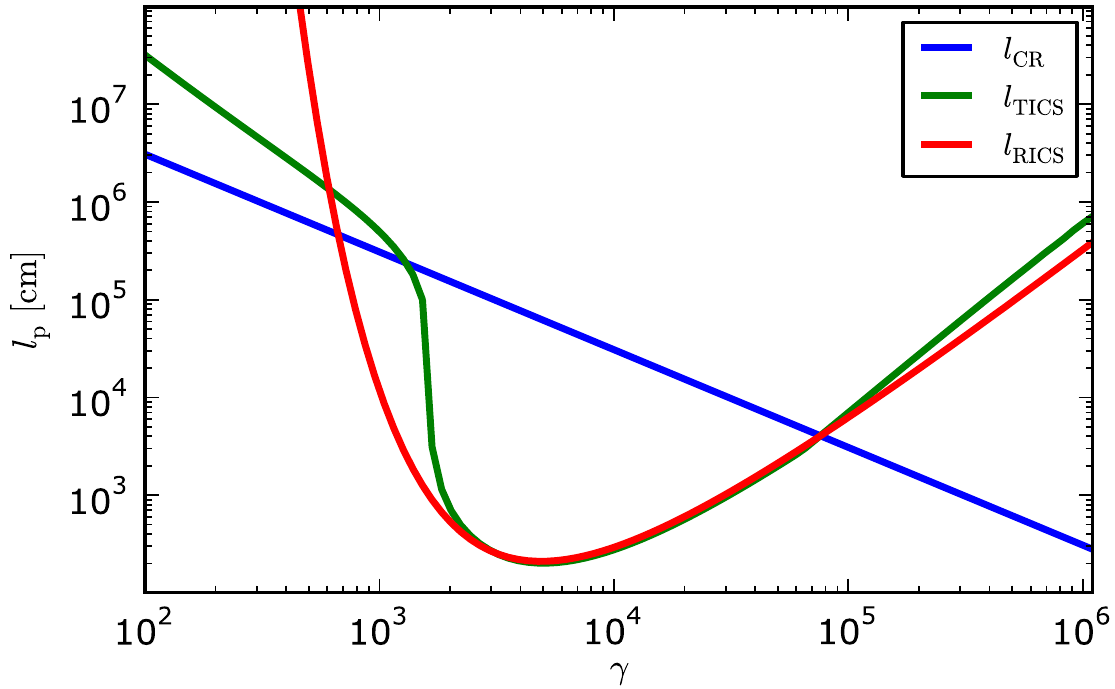}
\par\end{centering}

\centering{}\caption[Dependence of a particle mean free path on its Lorentz factor]{Dependence of a particle mean free path on its Lorentz factor for
three different processes: CR, RICS and TICS. The calculations were
performed for magnetic field strength $B_{14}=2$, radius of curvature
of magnetic field lines $\Re_{6}=1$ (for the CR process) and hot
spot temperature $T_{6}=3$ (for RICS and TICS). Both RICS and TICS
were calculated for a full range of incident angles ($\mu_{0}=0$,
$\mu_{1}=1$). Note that for a Lorentz factor in the range of $\gamma\approx2\times10^{3}-10^{5}$
the particle mean free paths of RICS and TICS are equal as the resonance
falls into the peak of the background photons.\label{fig:cascade.lp_gamma}}
\end{figure}

For ultrastrong magnetic fields quite a wide range of the particle
Lorentz factor falls into the peak of background photons (see Figure
\ref{fig:cascade.lp_gamma}). In such a case RICS is enhanced by the
fact that it involves photons with very high density. Furthermore,
the RICS process for such particles is indistinguishable from the
TICS (see Figure \ref{fig:cascade.thermal_eps}). For particles with
Lorentz Factor $\gamma\gtrsim10^{5}$, the dominant process of radiation
is CR. The exact value of this limit depends on conditions such as:
density of background photons, incident angles between particles and
photons, and curvature of magnetic field lines ($1/\Re$).

\begin{comment}
\textasciitilde{}/Programs/magnetic/magnetic/src/radiation/ics\_modules/gonthier.py
(show\_le\_gamma - Graphs, $B_{14}=2.$)
\end{comment}

\begin{figure}[H]
\begin{centering}
\includegraphics[height=9cm]{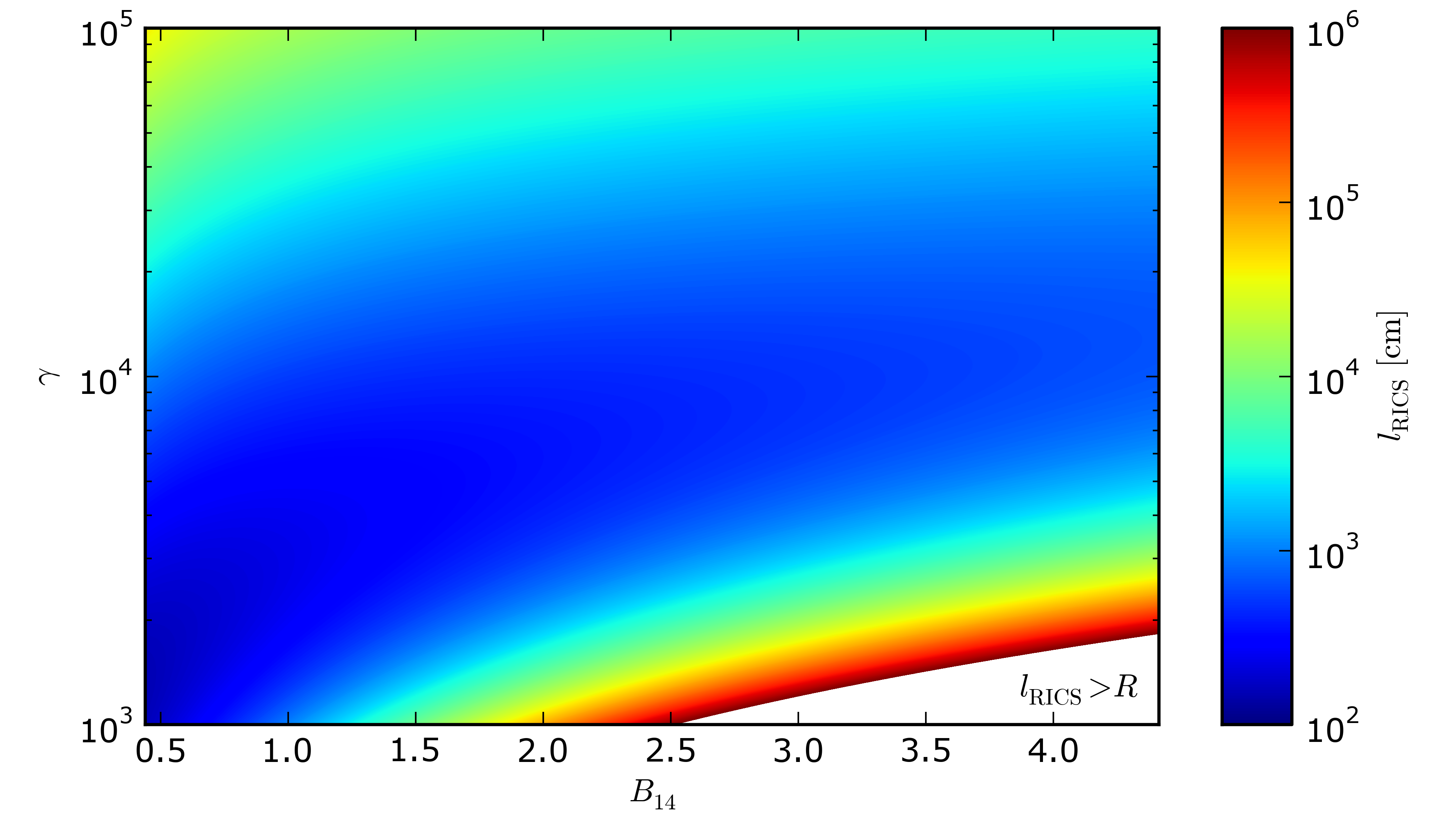}
\par\end{centering}

\centering{}\caption[{Dependence of a particle mean free path on magnetic field strength
and the Lorentz factor of a particle {[}RICS{]} }]{Dependence of a particle mean free path on magnetic field strength
($B_{14}$) and the Lorentz factor of a particle ($\gamma$) for the
RICS process. The particle mean free path was calculated for semi-isotropic
blackbody radiation ($\mu_{0}=0$, $\mu_{1}=1$) with temperature
$T_{6}=2.5$.\label{fig:cascade.lp_b14_gamma}}
\end{figure}

Figure \ref{fig:cascade.lp_b14_gamma} presents the dependence of
a particle mean free path on the magnetic field strength and the particle
Lorentz factor for RICS. The minimum of the mean free path for relatively
weak magnetic fields ($B_{14}=0.5$) is for particles with Lorentz
factor $\gamma\approx2\times10^{3}$, while for relatively stronger
magnetic fields ($B_{14}=3.5$) the RICS is most efficient for particles
with energy an order of magnitude larger ($\gamma\approx2\times10^{4}$).
This is a natural consequence of the fact that resonance takes place
when the photon energy in PRF is equal to the electron cyclotron energy,
which in stronger fields is higher. As can be seen from the Figure,
the particle mean free paths for RICS in stronger magnetic fields
increase. This is due to the decreasing resonant cross section with
increasing magnetic field strength (see Figure \ref{fig:cascade.resonance_eps}).
Note, however, that this behaviour does not include the fact that
photon density in regions with weaker magnetic fields is considerably
smaller. In fact, the results of the cascade simulation presented
in Chapter \ref{chap:physics} show that RICS is efficient only in
the immediate vicinity of a neutron star since photon density at higher
altitudes drops rapidly.

\begin{comment}
\textasciitilde{}/Programs/magnetic/magnetic/src/radiation/ics.py
(show\_le3d\_gonthier, $T_{6}=2$)
\end{comment}

\subsection{Background photons\label{sec:cascade.background_photons}}

\subsubsection{Photon density\label{sec:cascade.photon_density}}

\begin{figure}[H]
\begin{centering}
\includegraphics[height=7cm]{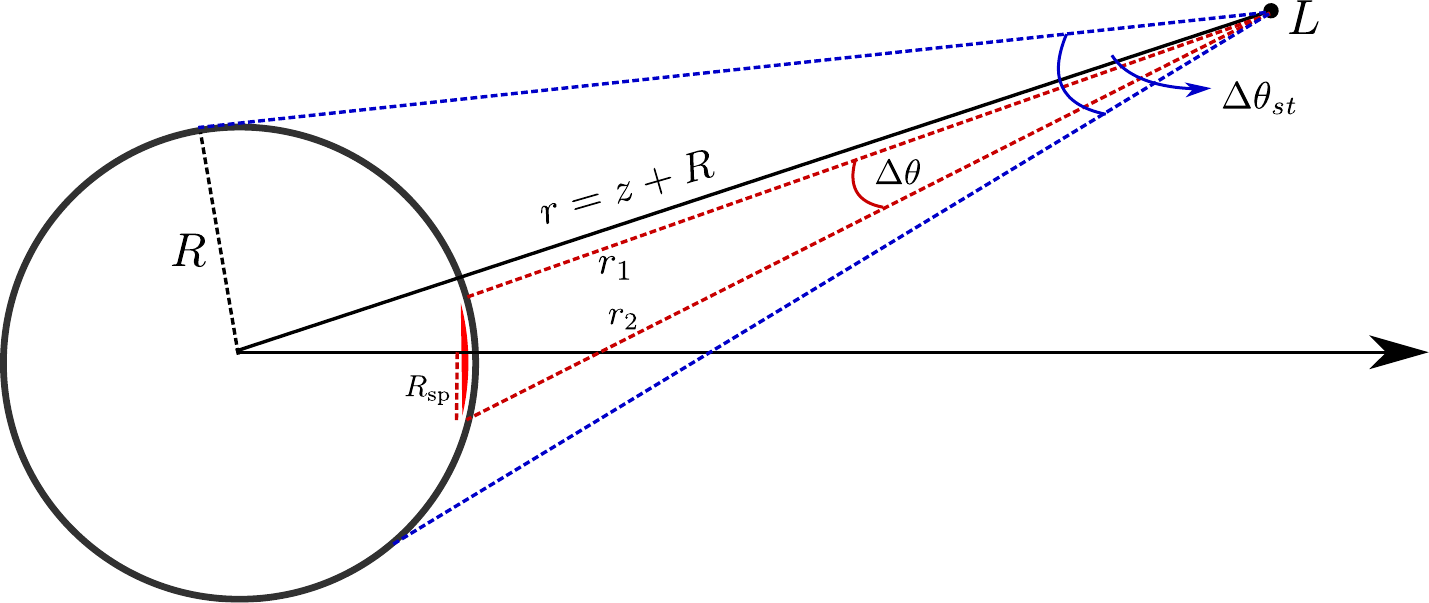}
\par\end{centering}

\caption[Simplified method used for calculating the background photon density]{Simplified method used for calculation of a photon density originating
from an entire stellar surface (blue lines) and from a hot/warm spot
(red lines). Here $R_{{\rm sp}}$ is a spot radius (either hot or
warm). Let us note that the simplified method is valid for the entire
surface component regardless of the $\phi$ component of location
$L$, while for the spot component it can be used only for small values
of $\phi$. In a more general case the spot should be projected on
the surface perpendicular to the radius vector ${\bf r}$ and passing
through point $L$.\label{fig:cascade.photon_density_dist}}
\end{figure}

One of the main parameters affecting ICS above the stellar surface
is photon density. The initial photon density (at altitude $z=0$)
highly depends on the temperature of the radiating surface. As shown
in Chapter \ref{chap:x-ray_emission} (e.g. see Table \ref{tab:x-ray_thermal}),
the entire surface has the lowest temperature ($T_{6}\lesssim0.8$),
thus the initial photon density is up to about two orders of magnitude
lower than warm spot radiation ($T_{6}\lesssim3$) and up to about
three orders magnitude lower than hot spot radiation ($T_{6}\lesssim5$).
However, the density of the photons strongly depends on the distance
from the source of radiation (especially for the hot spot). Therefore,
we used the simplified method presented in Figure \ref{fig:cascade.photon_density_dist}
to calculate photon density at a given point $L=\left(r,\,\theta,\,\phi\right)$.
Then the relative density of photons originating from the entire surface
can be calculated as

\begin{equation}
\frac{n_{{\rm st}}\left(\epsilon,\, T_{{\rm st}},\, L\right)}{n_{0}\left(\epsilon,\, T_{{\rm st}}\right)}=\sin^{2}\left(\frac{\Delta\theta_{{\rm st}}}{2}\right)=\left(\frac{R}{r}\right)^{2},\label{eq:cascade.n_ph_dist}
\end{equation}
where $n_{{\rm st},0}\left(\epsilon,\, T_{{\rm st}}\right)$ is the
density of photons with energy $\epsilon$ at the stellar surface
with temperature $T_{{\rm st}}$, and $\Delta\theta_{{\rm st}}$ is
the angular diameter of the star at a distance from the star centre
$r$.

Likewise, we can write a formula for the relative density of photons
originating from a spot (warm or hot) as

\begin{equation}
\frac{n_{{\rm sp}}\left(\epsilon,\, T_{{\rm sp}},\, L\right)}{n_{{\rm sp},0}\left(\epsilon,\, T_{{\rm sp}}\right)}=\sin^{2}\left(\frac{\Delta\theta}{2}\right),
\end{equation}
where $n_{{\rm sp},0}\left(\epsilon,\, T_{{\rm sp}}\right)$ is the
density of photons with energy $\epsilon$ at the spot surface (either
hot or warm) with temperature $T_{{\rm sp}}$. The angular diameter
of the spot can be calculated as

\begin{equation}
\Delta\theta=\arccos\left(\frac{r_{1}^{2}+r_{2}^{2}-4R_{{\rm sp}}}{2r_{1}r_{2}}\right),
\end{equation}
here $R_{{\rm sp}}$ is the spot radius and $r_{1}$, $r_{2}$ are
the distances to the outer edges of the spot (see Figure \ref{fig:cascade.photon_density_dist}).

\begin{figure}[H]
\begin{centering}
\includegraphics[height=7cm]{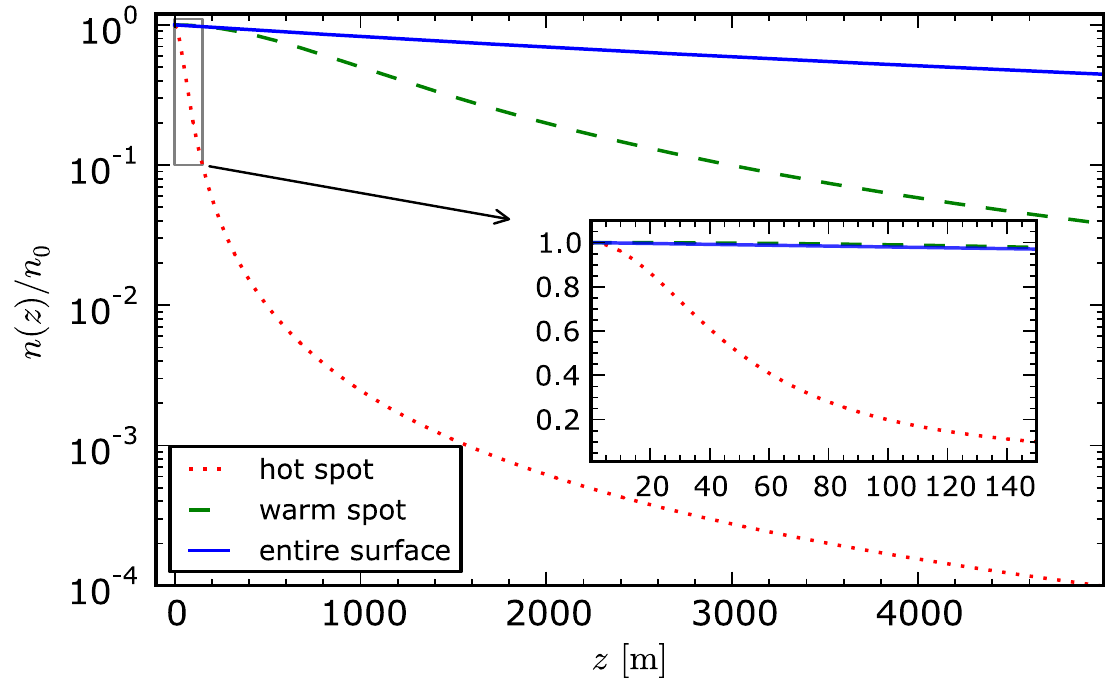}
\par\end{centering}

\caption[Dependence of the relative photon density on the distance from the
stellar surface]{Dependence of the relative photon density on the distance from the
stellar surface for three different thermal components (the entire
stellar surface, the warm spot and the hot spot). The following parameters
were used for the calculations: star radius $R=10\,{\rm km}$, warm
spot radius $R_{{\rm ws}}=1\,{\rm km}$ and hot spot radius $R_{{\rm hs}}=50\,{\rm m}$.
\label{fig:cascade.photon_fraction}}
\end{figure}

Figure \ref{fig:cascade.photon_fraction} presents the dependence
of the relative photon density ($n\left(z\right)/n_{0}$) on the distance
from the stellar surface. Due to the small size of a polar cap (hot
spot,\linebreak{}
$R_{{\rm hs}}=50\,{\rm m}$) the density of the photons drops rapidly
and already at a distance of about $z=150\,{\rm m}$ it is one order
of magnitude lower than at the polar cap surface. On the other hand,
for a larger size of the warm spot ($R_{{\rm hs}}=1\,{\rm km}$) the
photon density is reduced by an order of magnitude at a distance of
about $z=3\,{\rm {\rm km}}$. From Equation \ref{eq:cascade.n_ph_dist}
it can easily be seen that the photon density of radiation from the
entire stellar surface decreases by an order of magnitude at a distance
of about $z\approx3R\approx30\,{\rm km}$.

\begin{comment}
\textasciitilde{}/Programs/studies/phd/photon\_density/photon\_density.py
(show\_three)
\end{comment}

The very small size of the polar cap also has an additional implication
to the background photons' density. Namely, the density of the background
photons just above the polar cap highly depends not only on the distance
from the surface, but also on the position relative to the cap centre. 

\begin{figure}[H]
\begin{centering}
\includegraphics[height=7cm]{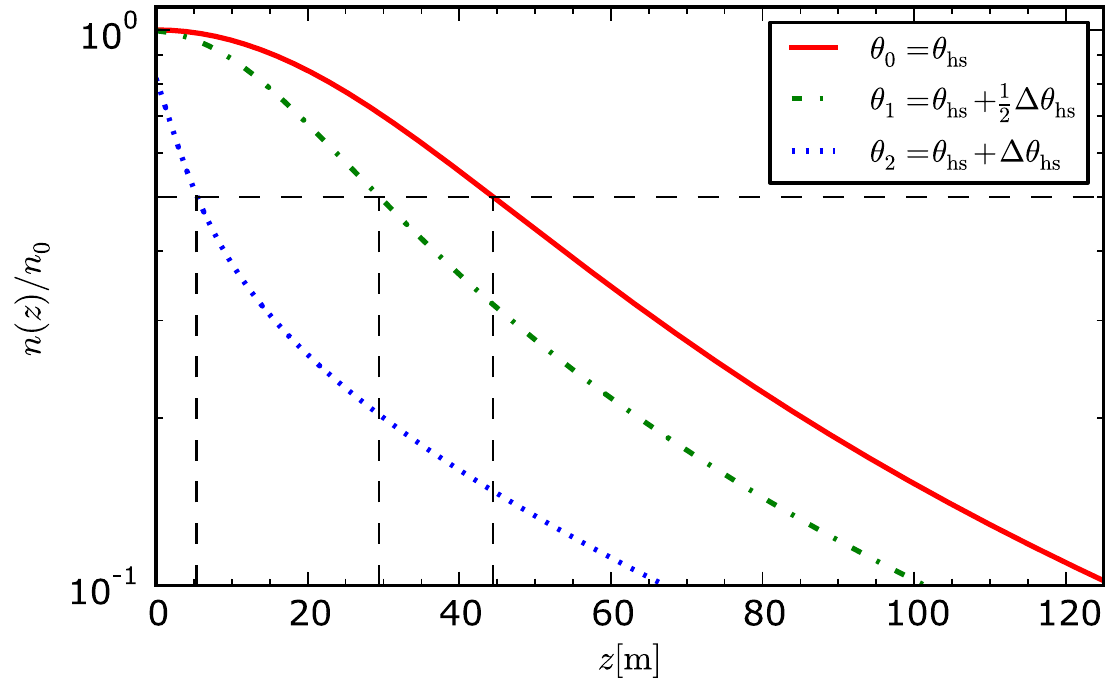}
\par\end{centering}

\caption[{Dependence of the relative photon density on the distance from the
stellar surface for a hot spot component {[}PSR B0656+14{]}}]{Dependence of the relative photon density on the distance from the
stellar surface for a hot spot component of PSR B0656+14. The relative
photon density was calculated for three different starting positions:
$\theta_{0}$ (central), $\theta_{1}$ (at the half distance to the
edge), and $\theta_{2}$ (the cap edge). The altitude ($z$) was calculated
for points which follow the magnetic field structure of \protect \linebreak{}
PSR B0656+14. \label{fig:cascade.photon_density_hot_spot}}
\end{figure}

Figure \ref{fig:cascade.photon_density_hot_spot} presents the dependence
of the relative photon density originating from a polar cap (the hot
spot) on the distance from the stellar surface for three different
starting points on the polar cap. The distance was calculated for
points which follow the magnetic field structure of PSR B0656+14.
Note that for the extreme magnetic line (which starts at the cap edge)
already at a distance of about $z_{2}\approx5\,{\rm m}$ the photon
density decreases twice, while for the central ($\theta_{0}$) and
middle line ($\theta_{1}$) the distances are respectively $z_{0}\approx45\,{\rm m}$
and $z_{1}\approx30\,{\rm m}$. This result is important as the background
photon density directly translates to the particle mean free path
in ICS (see Section \ref{sec:cascade.ics_electron_path}). This means
that for ICS-dominated gaps the sparks' height will vary depending
on their location. The breakdown of the gap (spark) in the central
region of a polar cap is easier to develop as the particle mean free
path is lower, and eventually it will result in lower heights of the
central sparks. This will influence the properties of plasma produced
in the central region of open magnetic field lines, and depending
on the conditions may result in the formation of plasma either suitable
to produce radio emission (core emission) or unsuitable to produce
radio emission (conal emission but with the line of sight crossing
the centre of the beam).

\begin{figure}[H]
\begin{centering}
\includegraphics[height=7cm]{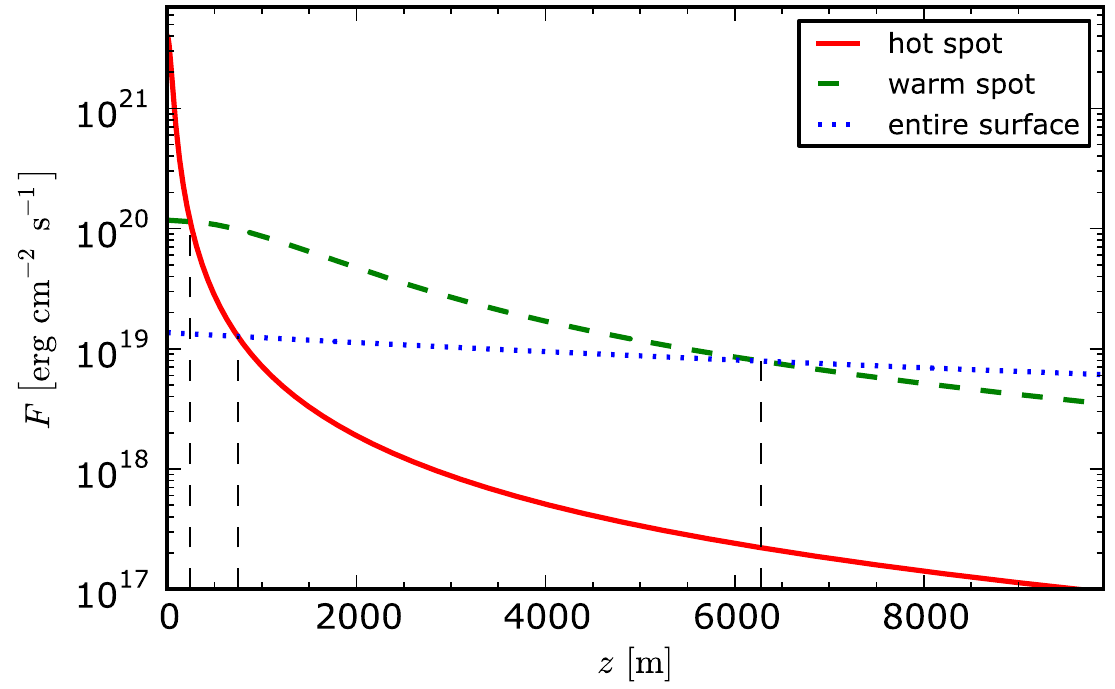}
\par\end{centering}

\caption[{Dependence of the radiation flux on the distance from the stellar
surface {[}PSR B0656+14{]}}]{Dependence of the radiation flux for three different components (the
entire stellar surface, the warm spot and the hot spot) on the distance
from the stellar surface for PSR B0656+14. The following parameters
were used for the calculations: entire stellar surface radiation,
\protect \linebreak{}
$T_{{\rm st}}=0.7\,{\rm MK}$, $R_{{\rm st}}=20\,{\rm km}$; warm
spot, $T_{{\rm ws}}=1.2\,{\rm MK}$, $R_{{\rm ws}}=1.8\,{\rm km}$;
and hot spot, $T_{{\rm hs}}=2.9\,{\rm MK}$, $R_{{\rm hs}}=50\,{\rm m}$.
\label{fig:cascade.radiation_flux_b0656}}
\end{figure}

To find the dominant component of thermal radiation at a given altitude
we need to take into account the initial flux of radiation and how
it changes with the distance. Below we present the calculations of
a radiation flux (Figure \ref{fig:cascade.radiation_flux_b0656})
for PSR B0656+14. The parameters of an entire surface and warm spot
components are in agreement with the observations (see Table \ref{tab:x-ray_thermal}),
while the hot spot component was calculated using parameters derived
from the modelling of a non-dipolar structure of the magnetic field
(see Chapter \ref{chap:model}). 

\begin{comment}
\textasciitilde{}/Programs/studies/phd/photon\_density/photon\_density.py
(show\_hotspot, 430)
\end{comment}

\begin{comment}
\textasciitilde{}/Programs/studies/phd/photon\_density/photon\_density.py
show(430 r\_surf=2e6, r\_ws=1.8e5, r\_hs=5e3, t\_surf=0.7e6,t\_ws=1.2e6,
t\_hs=2.9e6), plot\_b0656
\end{comment}

Already at a distance of $240\,{\rm m}$ the flux of the warm spot
radiation becomes higher than the flux of the hot spot radiation.
Furthermore, already at a height of $750\,{\rm m}$ flux the radiation
originating from the polar cap (hot spot) becomes lower than the flux
of radiation from the entire stellar surface. With an increasing distance
the flux of the warm spot decreases faster than the flux of the entire
surface radiation and at a distance of $6.3\,{\rm km}$ the thermal
radiation from the entire stellar surface becomes the dominant component
of the background photons. 

The results may suggest that up to a height of about $240\,{\rm m}$
(for PSR B0656+14) the hot spot radiation should be the main source
of the background photons involved in ICS. However, the actual height
is smaller as the results do not include the efficiency of ICS, which
also depends on the incident angle between the photons and the particles
(see the next Section).

\subsubsection{Photon incident angles\label{sec:cascade.incident_angles}}

\begin{figure}[H]
\begin{centering}
\includegraphics[height=7cm]{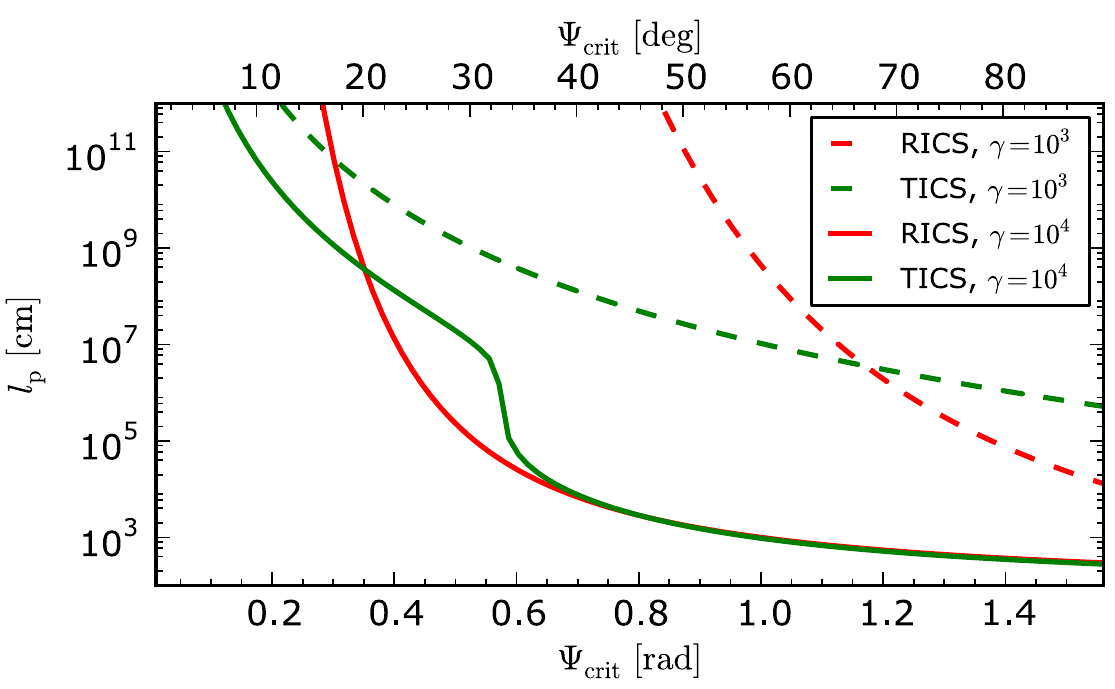}
\par\end{centering}

\caption[Dependence of the particle mean free path on the maximum value of
the incident angle]{ Dependence of the particle mean free path on the maximum value of
the incident angle $\psi_{{\rm crit}}$. The particle mean free path
$l_{{\rm p}}$ was calculated for magnetic field strength $B=10^{14}\,{\rm G}$
assuming background blackbody radiation with a temperature $T=3\,{\rm MK}$.
Two different particle Lorentz factors were used for the calculations:
$\gamma=10^{3}$ (dashed lines) and $\gamma=10^{4}$ (solid lines).
The red lines correspond to Resonant Inverse Compton Scattering, while
the blue lines correspond to Thermal-peak Inverse Compton Scattering.
\label{fig:cascade.lp_incident}}
\end{figure}

Another parameter that significantly affects the ICS is the incident
angle between the background photons and the relativistic particles.
Especially for Resonant Inverse Compton Scattering is the incident
angle of great importance. Figure \ref{fig:cascade.lp_incident} presents
the dependence of a particle mean free path for ICS on a maximum value
of the incident angle $\psi_{{\rm crit}}$. If incident angles are
low, the resonance is outside of the photon spectrum and results in
very high values of particle mean free paths. The lower the energy
of the particle (lower Lorentz factor), the incident angles should
be larger to ensure that the resonance falls into an energy range
with high photon density. 

\begin{comment}
\textasciitilde{}/Programs/magnetic/magnetic/src/radiation/ics.py
show\_psi\_phd
\end{comment}

TICS for a given magnetic field strength and the Lorentz factor of
particles is not significant (high particle mean free paths) unless
the angles of the incident photons are high enough. Note the characteristic
drop of the particle mean free path for TICS at $\psi_{{\rm crit}}\approx20^{\circ}$
(for $\gamma=10^{4}$) and $\psi_{{\rm crit}}\approx75^{\circ}$ (for
$\gamma=10^{3}$). For such high incident angles the resonance takes
place at the thermal peak of the background photons. Therefore, TICS
and RICS are indistinguishable, which results in an almost equal particle
mean free path (see the text above Figure \ref{fig:cascade.thermal_eps}
for more details). 

\begin{figure}[H]
\begin{centering}
\includegraphics[height=7cm]{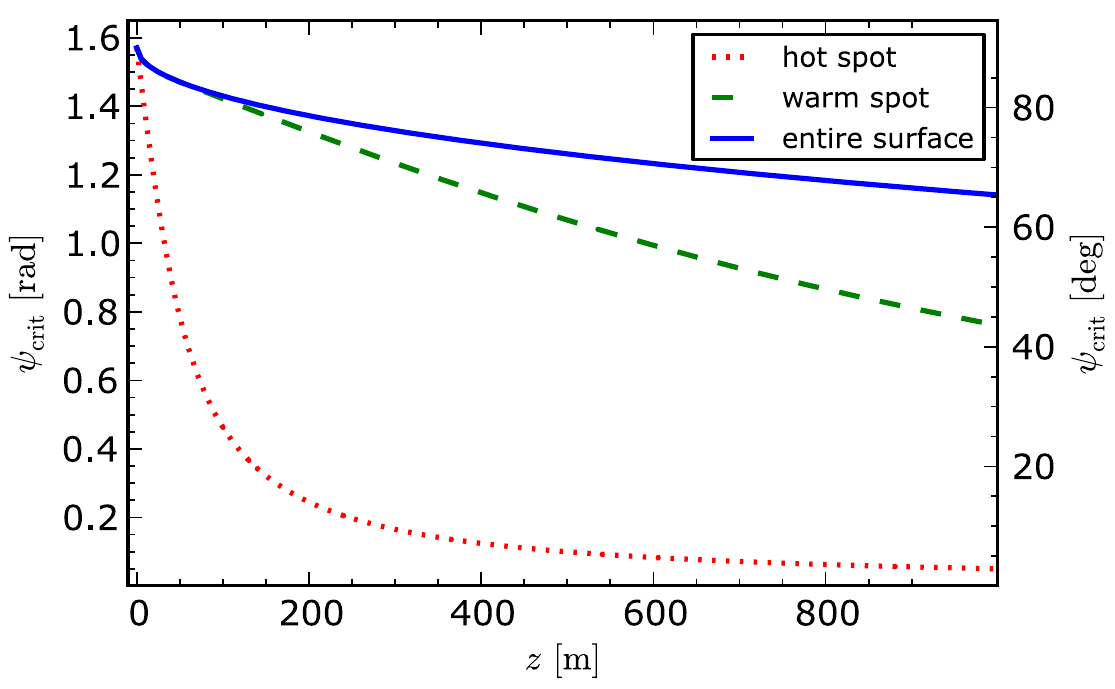}
\par\end{centering}

\caption[Dependence of the maximum incident angle on the distance from the
stellar surface]{ Dependence of the maximum incident angle on the altitude above the
stellar surface for three thermal components (the entire surface,
the warm spot and the hot spot radiation). \label{fig:cascade.incident_altitude}}
\end{figure}

Due to the very small size of the polar cap the influence of the hot
spot component will by lower not only because of the change of photon
density, but also because of the rapid change of the incident angle
between the photons and particles. Figure \ref{fig:cascade.incident_altitude}
presents the dependence of the maximum incident angle on the altitude
above the stellar surface for three thermal components (the entire
surface, the warm spot and the hot spot). As follows from the Figure,
already at an altitude of $z\approx90\,{\rm m}$ does the maximum
value of the incident angle between the photons from the hot spot
and the particles drop to $\psi_{{\rm crit}}=30^{\circ}$, which significantly
lowers the efficiency of ICS for this source of background photons
(see Figure \ref{fig:cascade.lp_incident}). Since the size of the
warm spot component is larger, the warm spot radiation will be significant
for up to higher altitudes, but already at a distance of $z\approx1.5\,{\rm {\rm km}}$
the maximum value of the incident angle also drops to $\psi_{{\rm crit}}=30^{\circ}$. 

\begin{comment}
\textasciitilde{}/Programs/studies/phd/photon\_density/photon\_density.py
show\_three, plot\_three\_psi
\end{comment}

\begin{figure}[H]
\begin{centering}
\includegraphics[height=7cm]{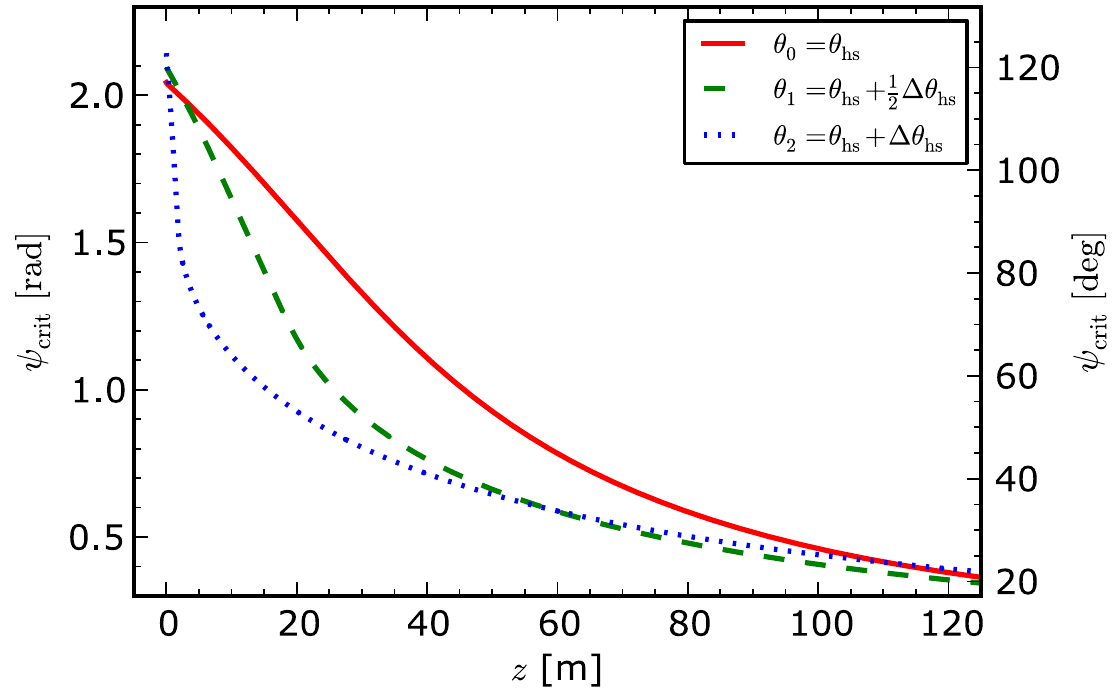}
\par\end{centering}

\caption[{Dependence of the maximum incident angle on the distance from the
stellar surface {[}PSR B0656+14{]}}]{Dependence of the maximum incident angle on the altitude above the
stellar surface for the hot spot component of PSR B0656+14. The maximum
incident angle was calculated for three different starting positions:
$\theta_{0}$ (central), $\theta_{1}$ (at the half distance to the
edge), and $\theta_{2}$ (the cap edge). \label{fig:cascade.incident_altitude_b0656}}
\end{figure}

Note that in the Figure we have calculated the maximum value of the
intersection angle at altitudes which correspond to radial progression
from the stellar surface. In fact, the actual maximum value of the
incident angle also depends on the structure of the magnetic field.
Figure \ref{fig:cascade.incident_altitude_b0656} presents the actual
maximum value of the incident angle of photons originating from the
hot spot for three different magnetic field lines calculated for PSR
B0656+14. The actual values of the maximum incident angle just above
the surface exceed $90^{\circ}$, but its rapid decline (especially
for extreme lines) causes the radiation of the hot spot component
to become insignificant for ICS at relatively low altitudes $z\approx20\,{\rm m}$.

\begin{comment}
\textasciitilde{}/Programs/studies/phd/photon\_density/photon\_density.py
show\_hotspot, plot\_hotspot\_psi
\end{comment}

Both the decrease of photon density and the decrease of the maximum
inclination angle cause the parameters of plasma produced by RICS
to highly depend on the properties (size and temperature) of the background
photons source. The hot spot component will be the dominant source
of background photons for ICS in the gap region ($z\lesssim20\,{\rm m}$),
while the radiation of the warm spot and the entire surface will be
the main source of the background photons for ICS at higher altitudes.

\chapter{Physics of pulsar radiation\label{chap:physics}}

\thispagestyle{headings}

\section{Inner Acceleration Region}

\subsection{Gamma-ray emission\label{sec:radiation.iar_gamma_ray}}

In our model most of the $\gamma$-photons are produced in the Inner
Acceleration Region or in close vicinity of a neutron star. Due to
an ultrastrong surface magnetic field, the most energetic $\gamma$-photons
are produced by Inverse Compton Scattering in the PSG-on mode. If
a pulsar is in the PSG-off mode, Curvature Radiation produces fewer
energetic photons than ICS in the PSG-on mode. Photons produced in
IAR (both the ICS and CR) are absorbed by strong magnetic fields creating
positron-electron plasma in the gap region, thereby enhancing a cascade,
or just above the gap enhancing a secondary plasma population. The
absorption of $\gamma$-photons in close vicinity of NS makes it impossible
to directly observe the radiation produced in IAR. However, a characteristic
of this emission defines the parameters of the gap (e.g. multiplicity
in the gap region, gap height, etc.), and thus the parameters of secondary
plasma.

\subsubsection{PSG-off mode}

\begin{figure}[H]

\begin{centering}
\includegraphics[height=7cm]{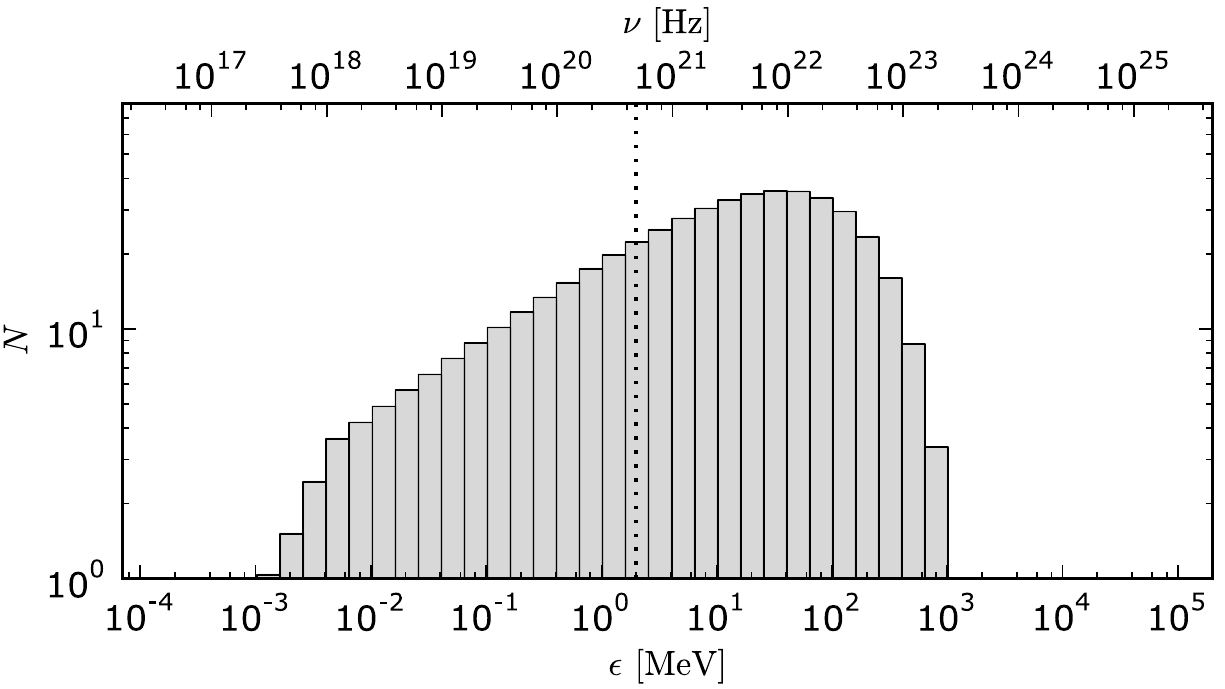}
\par\end{centering}

\caption[{Distribution of photons produced in IAR in the PSG-off mode {[}PSR
B0628-28{]}}]{Distribution of photons produced in IAR by a single particle for
PSR B0628-28. In the calculations we used parameters of the gap in
the PSG-off mode as presented in Table \ref{tab:psg.psg_top}. We
also assumed a linear change in the acceleration electric field (see
Equation \ref{eq:psg.acceleration_field}).\label{fig:radiation.cr_iar_spec_0628}}

\end{figure}

\begin{figure}[H]
\begin{centering}
\includegraphics[height=7cm]{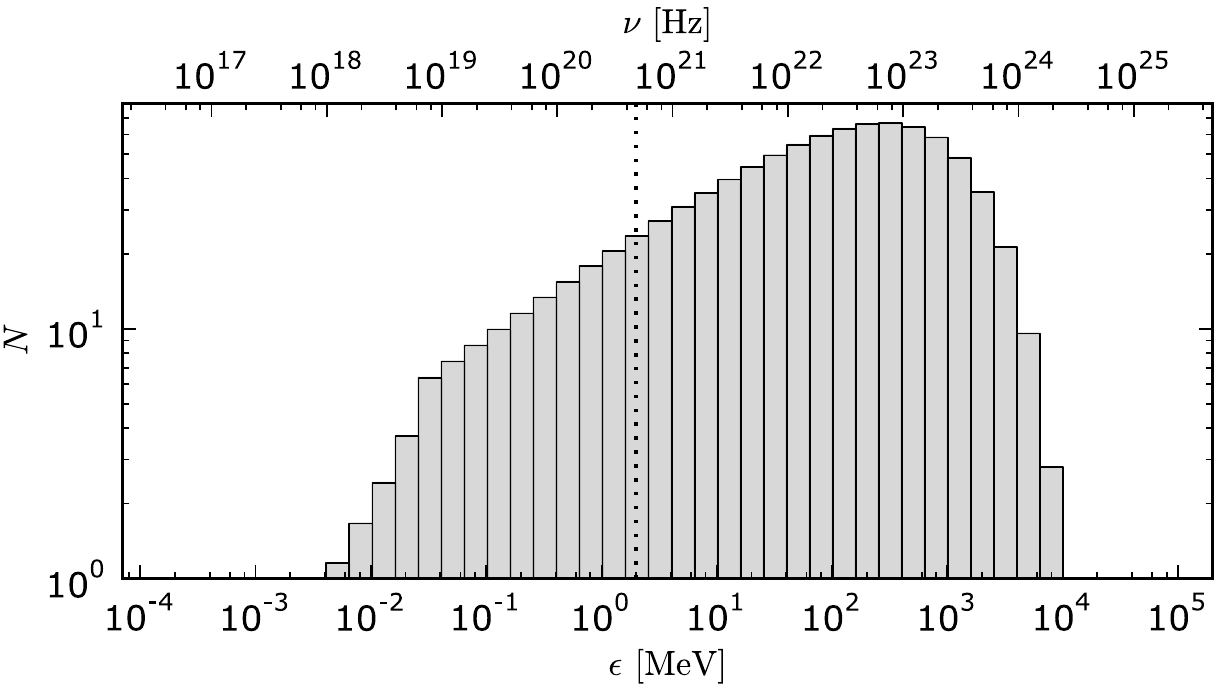}
\par\end{centering}

\caption[{Distribution of photons produced in IAR in the PSG-off mode {[}PSR
J0633-1746{]}}]{Distribution of photons produced in IAR by a single particle for
PSR J0633-1746. In the calculations we used parameters of the gap
in the PSG-off mode as presented in Table \ref{tab:psg.psg_top}.\label{fig:radiation.cr_iar_spec_0633}}
\end{figure}

In general, the existence of high potential in IAR (e.g. wide sparks
or $\eta\approx1$) results in solutions for which CR is responsible
for the emission of $\gamma$-photons. The energy of such radiation
depends on the Lorentz factor of primary particles and curvature of
the magnetic field lines. Figures \ref{fig:radiation.cr_iar_spec_0628}
and \ref{fig:radiation.cr_iar_spec_0633} present the histogram of
photons produced in IAR by CR for PSR B0628-28 and Geminga, respectively.
The curvature in IAR of Geminga is lower ($\Re_{6}\approx2.1$, see
Section \ref{sec:model.0633}), thus the primary particle should be
accelerated to higher energies in order to produce the required number
of photons in the gap region. Eventually the higher Lorentz factor
of primary particles will result in the emission of $\gamma$-photons
with energy up to $10\,{\rm GeV}$ for Geminga. On the other hand,
the curvature magnetic lines for PSR B0628-28 ($\Re_{6}=0.6$, see
Section \ref{sec:model.0628}) is higher, which reduces the photon
mean free path and it is possible to produce the required number of
photons in the gap region $N_{{\rm ph}}^{{\rm CR}}$ for lower the
Lorentz factor of primary particles. 

In CR-dominated gaps we can distinguish three types of photons: (I)
radiation with energy below $1\,{\rm MeV}$ which is unaffected by
the magnetic field (except the splitting) and can be detected by a
distant observer, (II) soft $\gamma$-ray photons which create pairs
above ZPF, (III) and high energetic $\gamma$-photons responsible
for pair production below ZPF. In an ultrastrong magnetic field the
photons from the third group will produce particles just after reaching
the first threshold. Due to the fact that most CR photons are $\parallel$-polarised,
photon splitting is insignificant in cascade pair production in the
PSG-off mode.

\begin{comment}
\textasciitilde{}/Programs/magnetic/magnetic/src/radiation/gap\_ics.py
t3 for 404 and 373
\end{comment}

\subsubsection{PSG-on mode \label{sec:radiation.ics_iar}}

\begin{figure}[H]
\begin{centering}
\includegraphics[height=7cm]{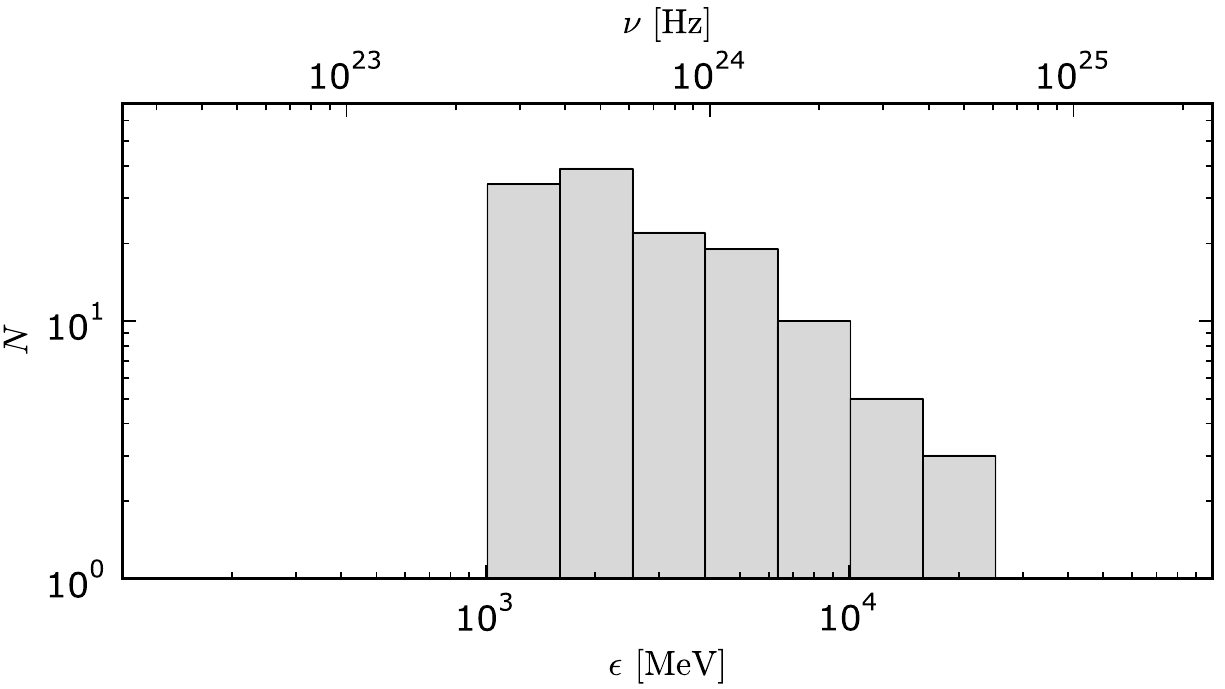}
\par\end{centering}

\caption[{Distribution of photons produced in IAR in the PSG-on mode {[}PSR
B0950+08{]}}]{Distribution of photons produced in IAR by the first population of
newly created particles for PSR B0950+08. In the calculations we used
the parameters of the gap in the PSG-on mode as presented in Table
\ref{tab:psg.psg_top}.\label{fig:radiation.ics_iar_spec_0950}}
\end{figure}

\begin{figure}[H]
\begin{centering}
\includegraphics[height=7cm]{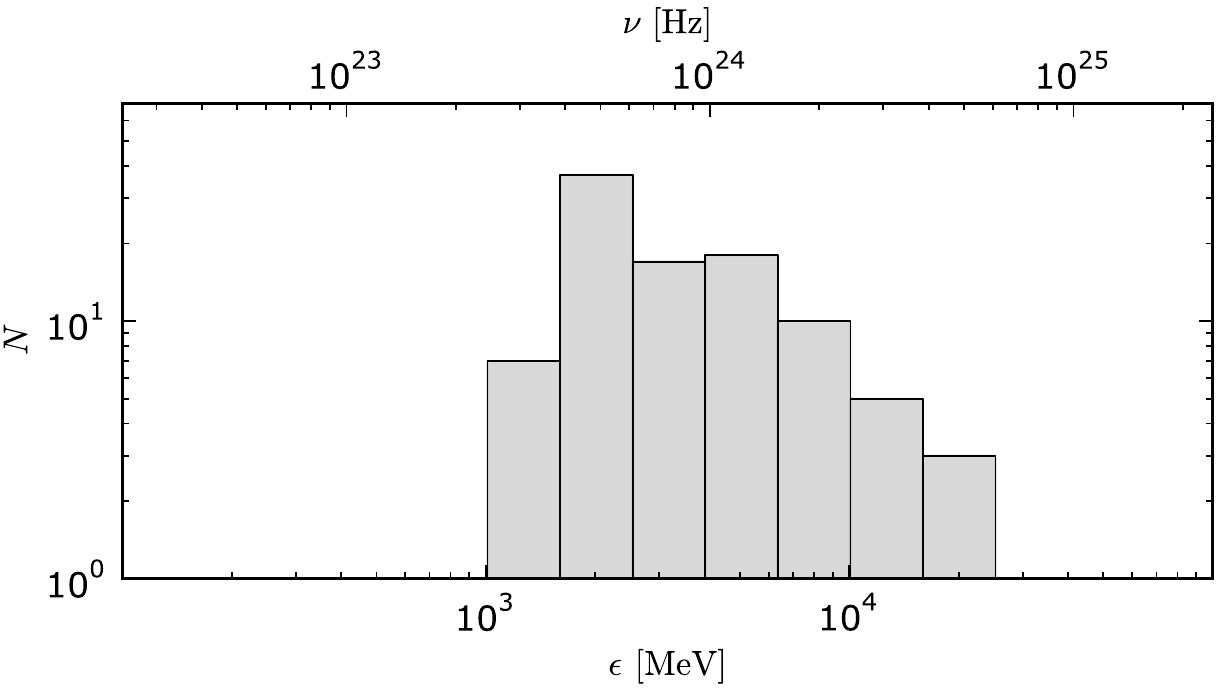}
\par\end{centering}

\caption[{Distribution of photons produced in IAR in the PSG-on mode {[}PSR
B1929+10{]}}]{Distribution of photons produced in IAR by the first population of
newly created particles for PSR B1929+10. In the calculations we used
the parameters of the gap in the PSG-on mode as presented in Table
\ref{tab:psg.psg_top}.\label{fig:radiation.ics_iar_spec_1929}}
\end{figure}

When the acceleration potential is low enough (narrow sparks with
$\eta<1$) to satisfy the condition for effective ICS ($l_{{\rm ICS}}\lesssim l_{{\rm acc}}$),
the gap will operate in the PSG-on mode. The energy of ICS radiation
in the gap region (RICS) depends on the Lorentz factor of primary
particles and the strength of magnetic field. In an ultrastrong magnetic
field of IAR implied by the PSG model, the primary particle loses
most of its energy during the scattering of background photons. Such
extremely energetic photons produce pairs on the zero-th Landau level
($\parallel$-polarised photons) or split to less energetic photons
before reaching the first threshold (see Section \ref{sec:cascade.pair_cr_spl}).
After the photons split the resulting photons are still very energetic
and create an electron-positron pair enhancing the avalanche production
of particles. In contrast to the PSG-off, most of the electron-positron
pairs in the PSG-on mode are created well below ZPF. Furthermore,
there is no additional radiation at lower energies ($\epsilon<1\,{\rm MeV}$)
which could be detected by a distant observer. Figures \ref{fig:radiation.ics_iar_spec_0950}
and \ref{fig:radiation.ics_iar_spec_1929} present the distribution
of photons produced by the first population of newly created particles
for PSR B0950+08 and PSR B1929+10, respectively. In both cases the
energy of the $\gamma$-photons ranges from $1\,{\rm GeV}$ to $\approx20\,{\rm GeV}$.
The narrow predicted spark half-width of PSR B0950+08 results in a
lower potential in IAR, thus increasing the efficiency of ICS (more
photons produced by the first population of particles). The particle
mean free path for ICS is smaller for backstreaming particles (see
Section \ref{sec:psg.particles_mean} for more details), thus most
photons in the PSG-on mode are produced in the direction towards the
stellar surface. Note that not all photons will produce electron-positron
pairs since some\linebreak{}
 $\gamma$-photons are produced so close to the stellar surface that
they reach its surface before they manage to reach the first threshold
for pair production. 

\begin{comment}
\textasciitilde{}/Programs/magnetic/magnetic/src/radiation/gap\_ics.py
t + plot\_ics for 322 and 355
\end{comment}

\subsection{X-ray and less energetic emission}

An negligible fraction of energy radiated by a primary particle in
the PSG-off mode falls in the X-ray band. What is more, in the PSG-on
mode all photons produced by ICS have energy which exceeds an electron's
rest energy by many orders of magnitude. Thus, IAR may be responsible
only for generating the thermal component of the X-ray spectrum in
the process of heating the stellar surface.

\subsubsection{Thermal emission \label{sec:radiation.thermal_radiation}}

\begin{figure}[H]
\begin{centering}
\includegraphics[height=7cm]{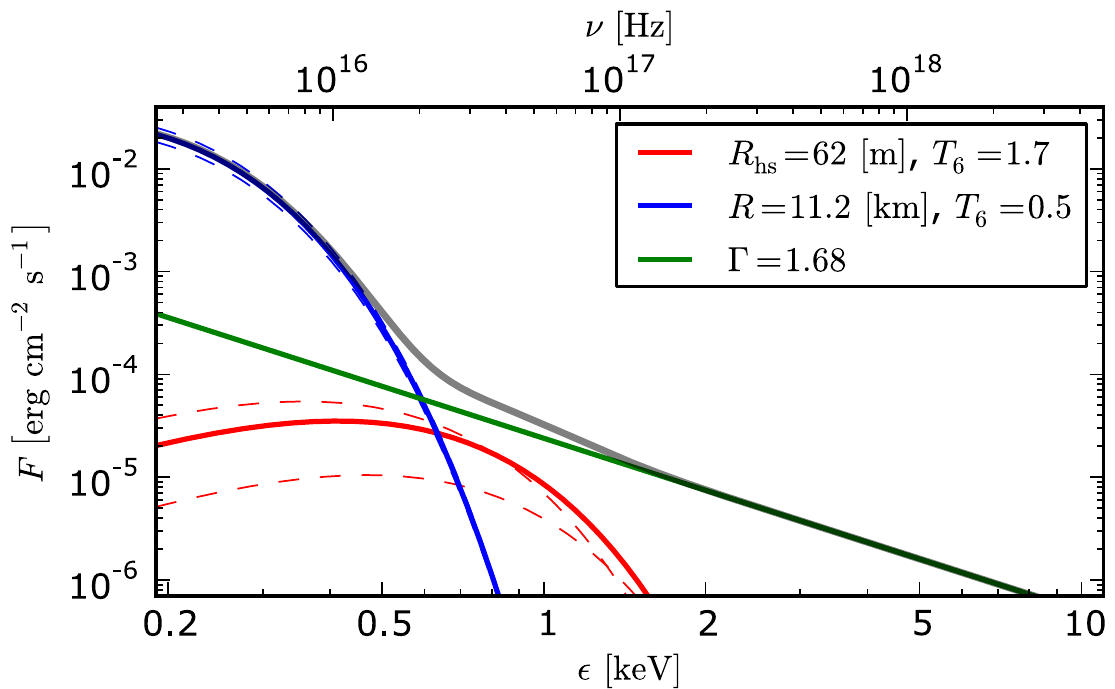}
\par\end{centering}

\caption[{Observed flux of radiation {[}PSR J0633+1746{]}}]{Observed flux of radiation for PSR J0633+1746. In the figure we present
three components of radiation: the nonthermal one (green line), the
entire surface radiation (blue line), and the hot spot component (red
line). The dashed lines correspond to uncertainties in observations
(see Table \ref{tab:x-ray_thermal}). \label{fig:radiation.bb_spectrum_j0633}}
\end{figure}

As shown in Section \ref{sec:x-ray.thermal}, thermal emission is
a common feature of neutron stars. Due to the large uncertainties
in X-ray observations, it is not possible to distinguish all three
thermal components (entire surface radiation, warm spot component
and hot spot radiation) for one specific pulsar. Furthermore, only
for a few pulsars (e.g. Geminga, PSR B0656+14) was it possible to
distinguish two thermal components alongside the nonthermal one. In
this thesis we focus on an analysis of pulsars with a visible hot
spot component ($b>1$), since only for these pulsars is it possible
to estimate the size of the actual polar cap. Most of these pulsars
are old neutron stars and only for one of them (Geminga) was the whole
surface radiation found in the X-ray spectrum. Figure \ref{fig:radiation.bb_spectrum_j0633}
presents the observed X-ray components of the Geminga pulsar: the
whole surface radiation, the polar cap (hot spot) and the nonthermal
component. The maximum of energy for the whole surface radiation is
in extreme ultraviolet and in soft X-rays for the hot spot component.
Taking into account the very small area of the polar cap, radiation
from the hot spot is unlikely to be observed in wavelengths off the
maximum. 

\begin{comment}
\textasciitilde{}/Programs/studies/phd/spectrum/spectrum.py (uncomment
tg)
\end{comment}

\begin{figure}[H]
\begin{centering}
\includegraphics[height=7cm]{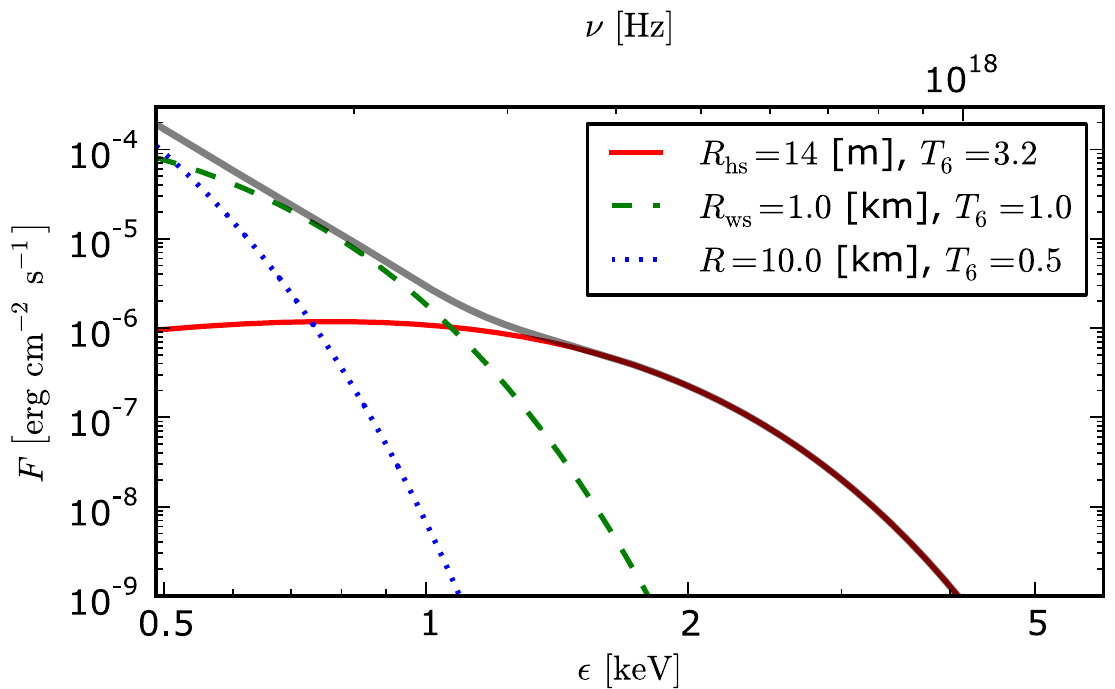}
\par\end{centering}

\caption[{X-ray spectrum {[}PSR B1133+16{]}}]{X-ray spectrum of PSR B1133+16. In addition to the observed thermal
radiation (red solid line), two other thermal components are presented:
the warm spot radiation (green dashed line) and the entire surface
radiation (blue dotted line). \label{fig:radiation.bb_spectrum_b1133}}
\end{figure}

Figure \ref{fig:radiation.bb_spectrum_b1133} presents the X-ray spectrum
of PSR B1133+16. The small number of counts detected resulted in the
fact that only separate fits for the BB and PL components were performed.
Both the BB and PL fits describe the observed spectrum with similar
accuracy. In the Figure we present additional thermal components (the
entire surface radiation and the warm spot) which have not been determined
by the observations. The Figure shows that the overlapping thermal
components can mimic the power-law dependence of the spectrum at frequencies
below $2\,{\rm keV}$. 

\begin{comment}
\textasciitilde{}/Programs/studies/phd/spectrum/spectrum.py (uncomment
t1133)
\end{comment}

Although this specific combination of thermal components for PSR B1133+16
would result in a photon index greater than the observed one $\Gamma=2.51$,
the spectral fits for all pulsars should be extended to include more
BB components in order to examine the effect of thermal components
overlapping at lower frequencies. The results of our calculations
suggest that the nonthermal X-ray radiation should dominate the spectrum
at higher frequencies $\approx3-10\,{\rm keV}$, but the power-law-like
behaviour at lower frequencies could be the result of the overlapping
of thermal components anticipated in the PSG scenario (see Section
\ref{sec:radiation.thermal_reheating}).

\subsubsection{Nonthermal emission}

The polarisation of ICS radiation in an ultrastrong magnetic field
is $50\%$ (one $\parallel$ to every $\perp$-polarised photon).
Synchrotron Radiation of secondary particles created by $\perp$-polarised
photons would generate hard X-ray photons, however, as was mentioned
in Section \ref{sec:cascade.pair_cr_spl}, these photons will split
before they reach the first threshold to produce pairs. Therefore,
regardless of whether the gap is dominated by CR or by ICS, Synchrotron
Radiation in IAR is not significant.

\subsubsection{Warm spot component \label{sec:radiation.thermal_reheating} }

Apart from the obvious X-ray component corresponding to the whole
surface radiation, the PSG model can explain both the hot and warm
spot radiation. The hot spot radiation is a natural consequence of
heating the actual polar cap region by the backstreaming particles
(see Section \ref{sec:x-ray.bgt1}). As was mentioned in Section \ref{sec:x-ray.blt},
the warm spot component can have two different sources: (I) the drastic
difference of the crustal transport process due to the non-dipolar
structure of the surface magnetic field (for young and middle-aged
pulsars), (II) and a mechanism of heating the surface adjacent to
the polar cap. In this section we present the second mechanism, i.e.
heating of the surface adjacent to the polar cap, which can be applied
to both young and old pulsars. 

\begin{figure}[H]
\begin{centering}
\includegraphics{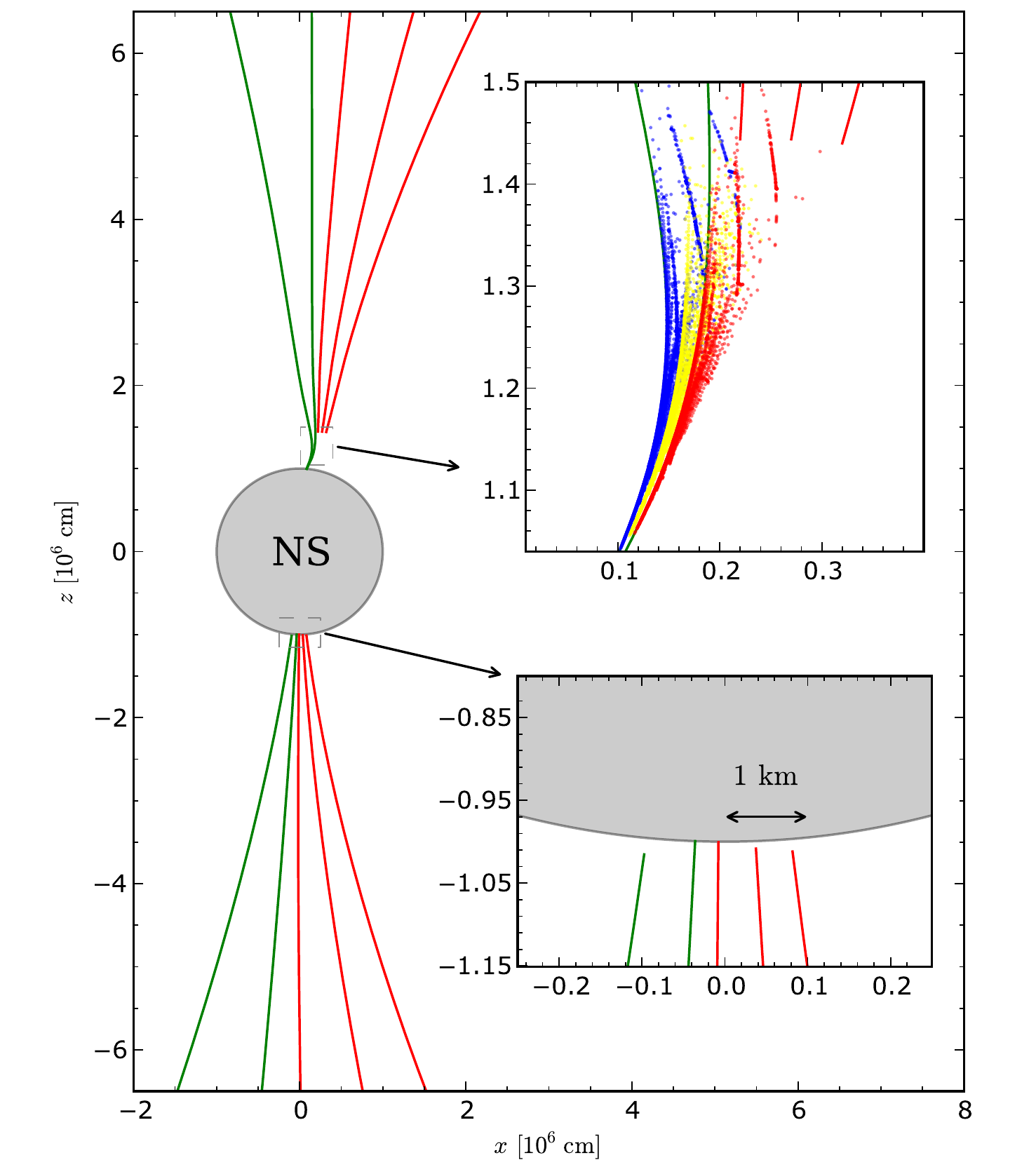}
\par\end{centering}

\caption[{The warm spot component {[}PSR B0950+08{]}}]{Global structure of magnetic field lines for PSR B0950+08. The structure
was obtained using two crust-anchored anomalies (see Section \ref{sec:model.0950}).
Green lines correspond to the outer open magnetic field lines, while
the red lines correspond to the closed magnetic field lines at which
secondary pairs are produced. Blue, yellow and red dots represent
the locations of secondary pair production for the outer left, the
middle and the outer right open field lines, respectively. \label{fig:radiation.warm_spot_0950}}
\end{figure}

Figure \ref{fig:radiation.warm_spot_0950} presents the mechanism
of heating the area adjacent to the polar cap for PSR B0950+08. When
the gap operates in the PSG-off mode the primary plasma (see Section
\ref{sec:radiation.primary_plasma}) will lose a significant part
of its energy via CR as the particles propagate through the region
of high curvature. For this particular magnetic line's configuration
the region of high CR extends up to an altitude about $4\,{\rm km}$
above the stellar surface. The most energetic CR photons emitted in
this region have a relatively short mean free path and they produce
electron-positron pairs in the region of open magnetic field lines.
However, both the less energetic CR photons and $\gamma$-photons
produced by SR have a large enough photon mean free path to produce
pairs in the region of the closed magnetic field lines. All newly
created pairs move along the closed magnetic field lines and heat
the surface beyond the polar cap on the opposite side of the star.

The fraction of energy transferred to the region of the closed field
lines highly depends on the region of open magnetic field lines considered
in CR/SR emission. In the Figure we use three different colours (blue,
yellow and red) to show the positions of pair creation for three characteristic
open magnetic field lines (the outer left, the middle and the outer
right). The simulation results in the following fractions of energy
transferred to the region of the closed field lines are: $0.02\%$,
$0.1\%$, $6\%$ for all three lines, respectively. For this specific
magnetic field configuration the transferred energy fraction increases
as we move towards the region with the highest curvature. We can roughly
estimate that for the proposed magnetic field configuration of PSR
B0950+08, about $1\%$ of the outflowing energy is responsible for
heating of the surface beyond the polar cap on the opposite side of
the star. Note that due to strong anisotropy of the outflowing and
backflowing stream of particles (see Section \ref{sec:radiation.primary_plasma}),
this fraction could be enough to obtain the warm spot component with
a luminosity equal or in some cases even higher than the luminosity
of the hot spot component. 

\begin{figure}[H]
\begin{centering}
\includegraphics{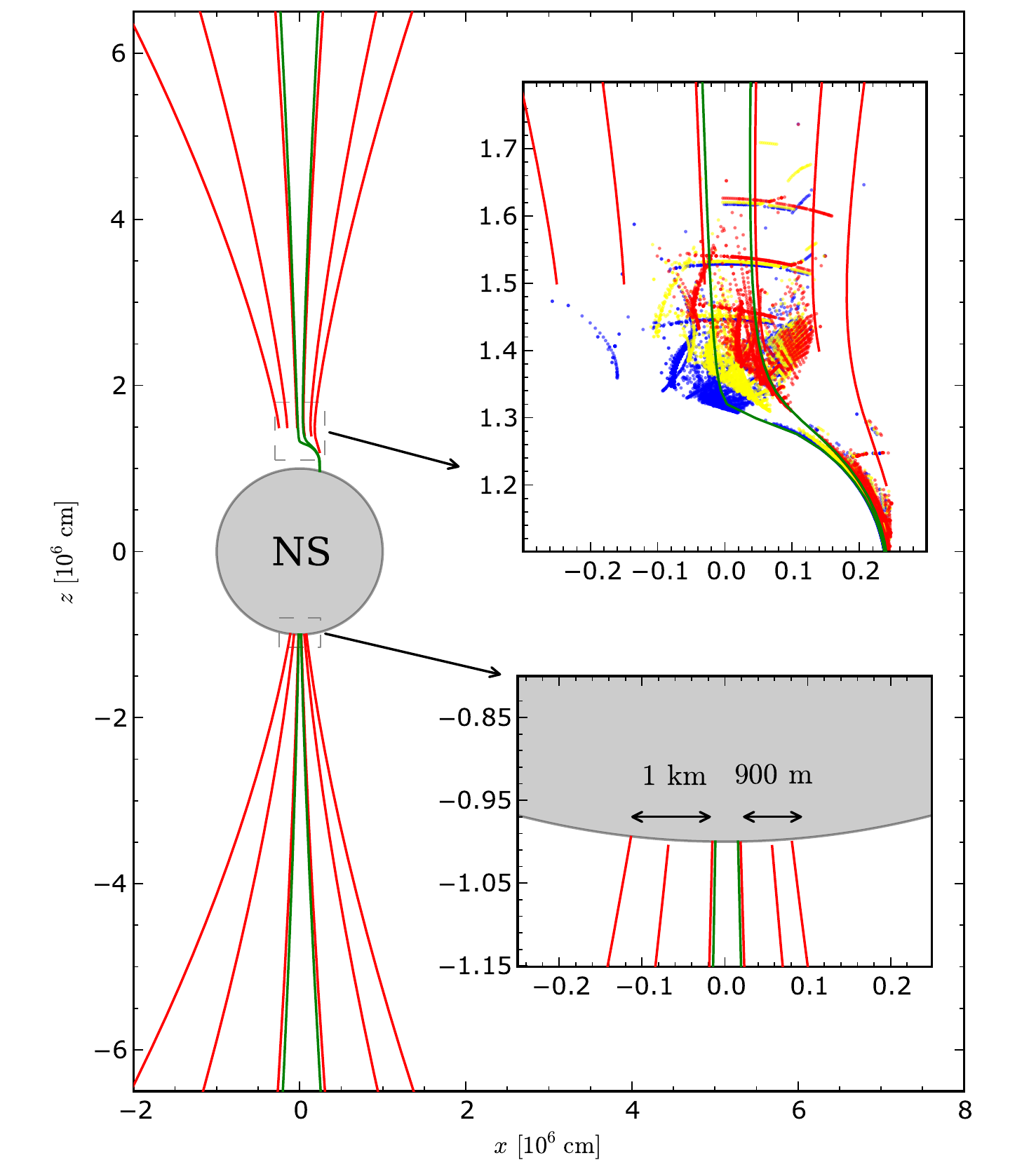}
\par\end{centering}

\caption[{The warm spot component {[}PSR B0943+10{]}}]{Global structure of magnetic field lines for PSR B0943+10. The structure
was obtained using two crust-anchored anomalies (see Section \ref{sec:model.b0943}).
Green lines correspond to the outer open magnetic field lines, while
the red lines correspond to the closed magnetic field lines at which
secondary pairs are produced. Blue, yellow and red dots represent
the locations of secondary pair creation for the outer left, the middle
and the outer right open field lines, respectively. \label{fig:radiation.warm_spot_0943}}
\end{figure}

The fraction of energy transferred to the region of closed field lines
highly depends on the magnetic field configuration. A more complicated
structure of the magnetic field lines proposed for PSR B0943+10 (see
Figure \ref{fig:radiation.warm_spot_0943}) results in a much wider
area of closed field lines at which pairs are created, and hence higher
fractions of energy transferred to the region of closed field lines.
For this specific structure of the magnetic field these fractions
are: $7\%$, $1\%$, $2\%$ for three characteristic lines, respectively.
We can roughly estimate that about $3-5\%$ of the outflowing energy
is responsible for the heating. Note that the magnetic field structure
of PSR B0950+08 results in the heating of only one side beyond the
polar cap, while in the case of PSR B0943+10 the whole surface around
the polar cap is heated. The actual size of the warm spot also depends
on the magnetic field configuration in the heating zone, and can either
be decreased or increased.

\subsection{Primary plasma \label{sec:radiation.primary_plasma}}

As we mentioned in Section \ref{sec:psg.multiplicity}, PSG-off and
PSG-on modes differ essentially by the Lorentz factor of primary particles
produced in the gap region. Furthermore, different scenarios of the
gap breakdown (due to surface overheating or due to production of
dense enough plasma) cause the evolution of primary particles in the
two modes to completely different.

We assume that in the PSG-off mode the gap breakdown is due to surface
overheating; hence the plasma cloud moving away from the stellar surface
is a mixture of ions and electron-positron plasma. In this scenario
the ions are the main source of charge density required to screen
the gap (see Equation \ref{eq:psg.overheating_parameter}). As the
plasma cloud moves away from the stellar surface both the spark height
and the spark width increase, which results in an increase of the
acceleration potential drop. When the particles gain the Lorentz factors
$\gamma\gtrsim10^{5}$, CR begins to produce $\gamma$-photons. In
the PSG-off mode most of the\linebreak{}
$\gamma$-photons are created near ZPF (see Figure \ref{fig:psg.cr_sol}).
All particles created by $\gamma$-photons above the ZPF do not contribute
to the heating of the surface. Furthermore, the acceleration in the
upper parts of the gap is relatively weak, and electrons produced
in this region will also escape from the gap, thus not contributing
to the surface heating. Depending on the details of the cascade formation,
the process described above may result in the creation of strong streaming
anisotropies, where the flux of the backstreaming particles is considerably
smaller than the flux of the outstreaming particles. Note that the
density of the backstreaming particles required to overheat the surface
is significantly lower than the co-rotational density $n_{{\rm CR}}\ll n_{{\rm GJ}}$
(see Table \ref{tab:psg.psg_top}).

In the PSG-on mode the quasi-equilibrium of the flux of backstreaming
particles and the flux of the polar cap radiation can cause the gap
to break only due to the production of dense enough plasma. Thus,
the surplus of positrons is the main source of the charge in the plasma
cloud moving away from the stellar surface. The ICS process responsible
for the cascade production of particles is effective only in the bottom
part of the gap. Hence, the backstreaming electrons will hit the surface
with a Lorentz factor $\gamma_{{\rm c}}$ well below the $\gamma_{{\rm max}}$.
As there is no strong pair production near (or above) ZPF, the backstreaming/outstreaming
anisotropy arises only due to the difference of the Lorentz factor
of electrons hitting the stellar surface and the Lorentz factor of
positrons accelerated in the gap $\gamma_{{\rm max}}/\gamma_{{\rm c}}\approx10$.
The actual density of newly created plasma to completely screen the
gap can be calculated only in the full cascade simulation. However,
as shown by \citet{2010_Timokhin}, this density should significantly
exceed the co-rotational Goldreich-Julian density $n_{{\rm ICS}}\gg n_{{\rm GJ}}$.
We describe the difference between the co-rotational density and the
actual density of primary plasma required to completely screen the
ICS-dominated gap by factor  $N_{{\rm ICS}}=n_{{\rm ICS}}/n_{{\rm GJ}}\gg1$.

\section{Inner magnetosphere of a pulsar}

\subsection{Gamma-ray emission\label{sec:radiation.im_gamma_ray}}

In general there are three processes which can produce $\gamma$-ray
emission in the inner magnetosphere of a pulsar ($R_{{\rm pc}}\ll z\ll R_{{\rm LC}}$):
CR, ICS and SR. Which of them produces the majority of $\gamma$-photons
depends on the parameters of the primary particles, and thus mainly
depends on the mode in which the gap operates. Additionally, the efficiency
of the ICS process strongly depends on the source of the background
photons.

\subsubsection{Curvature Radiation of primary particles\label{sec:radiation.cr_primary}}

When the gap operates in the PSG-off mode, high-energetic particles
are produced\linebreak{}
 $\gamma_{{\rm c}}\gtrsim10^{6}$. As they pass the region with high
curvature ($\Re_{6}\approx1$) they radiate a significant part of
their energy through CR (see Section \ref{sec:cascade.cr}). 

\begin{figure}[H]
\begin{centering}
\includegraphics[height=7cm]{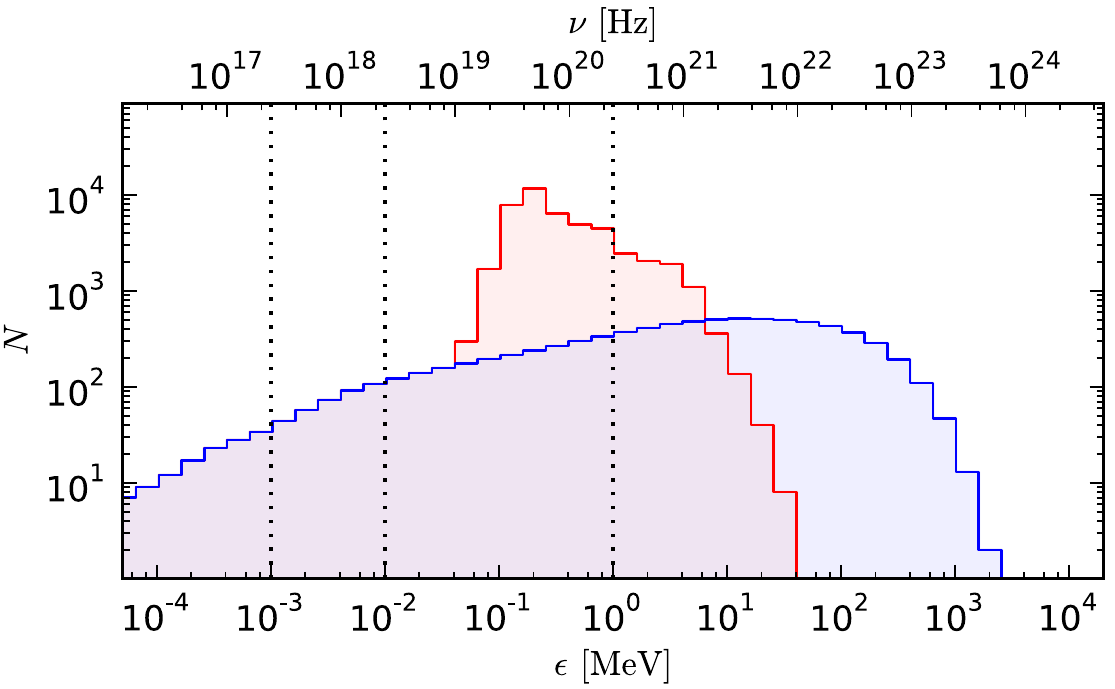}
\par\end{centering}

\caption[{Final photon distribution produced by a single primary particle {[}PSR
B1133+16{]}}]{Final photon distribution produced by a single primary particle for
PSR B1133+16. The blue line corresponds to the initial CR distribution,
while the red line presents the final distribution with the inclusion
of photon splitting, pair production and SR.\label{fig:radiation.cr_1133}}
\end{figure}

Figure \ref{fig:radiation.cr_1133} presents the distribution of CR
photons produced by a single primary particle moving along the open
magnetic field line of PSR B1133+16 (see Section \ref{sec:model.0950}
for the details of the magnetic field configuration). The initial
Lorentz factor of the particle $\gamma_{{\rm max}}=1.7\times10^{6}$
was set according to the value presented in Table \ref{tab:psg.psg_top}.
As the particle advanced through the region with high curvature, it
lost about $46\%$ of its initial energy, which was mainly converted
to high-energetic $\gamma$-photons with an energy up to about $2\,{\rm GeV}$.
The $\gamma$-photons are produced in a region of a strong magnetic
field, thus after passing a relatively short distance the most energetic
photons are absorbed by the magnetic field and electron-positron pairs
emerge. The red colour in the Figure corresponds to the final spectrum
(after photon splitting, pair production and SR) produced by a single
primary particle in the PSG-off mode. Most of the energy radiated
by the primary particle was converted into the secondary plasma (see
Section \ref{sec:radiation.secondary_plasma}) and only about $5\%$
of the particle's initial energy ended in the form of radiation with
a cut-off at about $30\,{\rm MeV}$. 

\begin{comment}
\textasciitilde{}/Programs/studies/phd/cascade\_plot/cascade\_plot.py 

(read\_data\_final, plot\_spectrum\_final, 341\_cr\_1e04\_10m\_new)
\end{comment}

\begin{figure}[H]
\begin{centering}
\includegraphics[height=7cm]{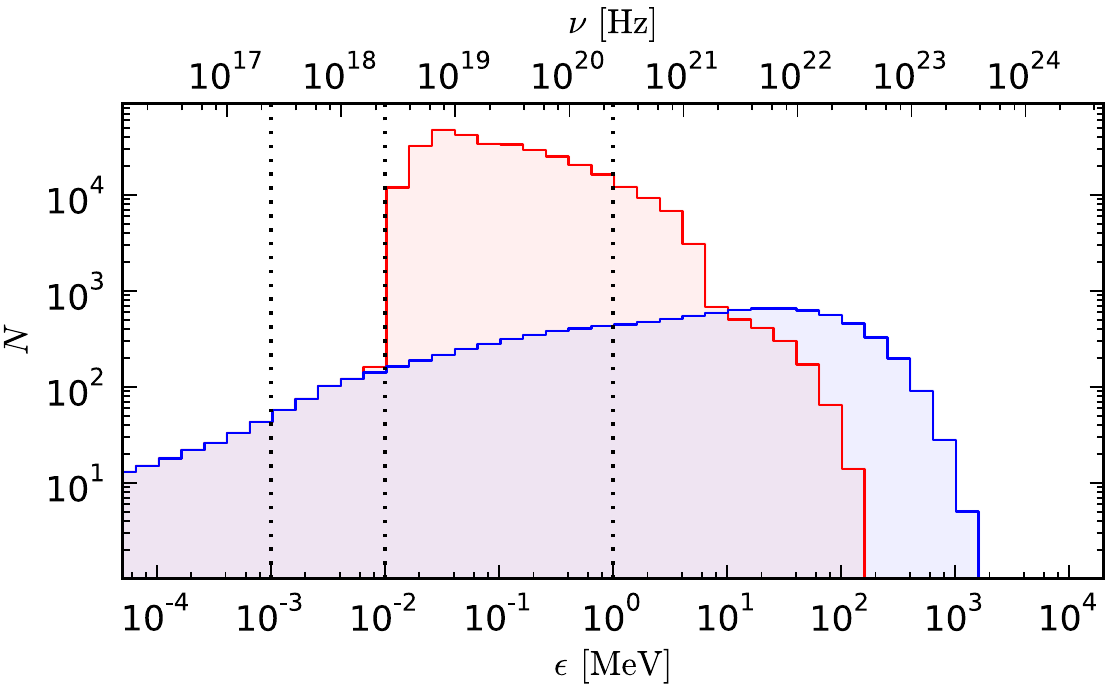}
\par\end{centering}

\caption[{Final photon distribution produced by a single primary particle {[}PSR
B0950+08{]}}]{Final photon distribution produced by a single primary particle for
PSR B0950+08. The blue line corresponds to the initial CR distribution,
while the red line presents the final distribution with the inclusion
of photon splitting, pair production and SR.\label{fig:radiation.cr_0950}}
\end{figure}

To increase the amount of photons reaching the observer, the emission
zone, i.e. the region with the highest curvature, should by located
in the area with a weaker magnetic field. Such a configuration allows
a photon to travel a longer distance before it is absorbed by the
magnetic field. As a result the electron-positron pairs are created
at higher Landau levels, which enhances SR. Figure \ref{fig:radiation.cr_0950}
presents the distribution of CR photons for PSR B0950+08. The calculations
were performed for the initial Lorentz factor of the particle $\gamma_{{\rm max}}=2.0\times10^{6}$
(see Table \ref{tab:psg.psg_top}).

\begin{comment}
\textasciitilde{}/Programs/studies/phd/cascade\_plot/cascade\_plot.py 

(read\_data\_final, plot\_spectrum\_final, 355\_cr\_1e04\_10m\_new)
\end{comment}

Due to CR the primary particle lost about $40\%$ of its initial energy.
In this case about a half of the energy radiated by the primary particle
was converted into the secondary plasma and the same amount of energy
(about $20\%$ of the particle's initial energy) ended in the form
of radiation. For both PSR B1133+16 and PSR B0950+08, the maximum
of the curvature is of the same order. However, the maximum of curvature
for \linebreak{}
PSR B1133+16 is located at an altitude of about $800\,{\rm m}$, while
for PSR B0950+08 it is located at an altitude of about $1.75\,{\rm km}$
(compare Figures \ref{fig:model.1133_curva} and \ref{fig:model.b0950_curva}).

\subsubsection{Inverse Compton Scattering of primary particles}

In the PSG-on mode the maximum Lorentz factor of primary particles
is in the range of $10^{4}-10^{5}$ (see Table \ref{tab:psg.psg_top}).
As it follows from Figures \ref{fig:cascade.lp_gamma} and \ref{fig:cascade.lp_b14_gamma},
the ICS process is most effective for particles with a Lorentz factor
in the range of $10^{3}-10^{4}$. Particles with high energies ($\gamma\gtrsim10^{5}$)
will upscatter thermal photons only just above the stellar surface,
where the density of the background photons is very high (see Section
\ref{sec:radiation.ics_iar}). Thus, if there is no additional source
of background photons, the most energetic particles ($\gamma\gtrsim10^{5}$)
will escape from the inner magnetosphere without losing their energy
by ICS. However, the plasma cloud produced by the ICS-dominated gap
has a density exceeding the co-rotational Goldreich-Julian density
even by a few orders of magnitude (see Section \ref{sec:radiation.primary_plasma}).
Such a high charge density reduces the acceleration \citep{2010_Timokhin}
and, consequently, the bulk of particles will escape from the IAR
with lower Lorentz factors. It is not possible to estimate the actual
Lorentz factor of particles in the plasma cloud at the moment of gap
breakdown without performing a full cascade simulation. Thus, in this
thesis we assume that at the moment of gap breakdown most of the particles
will have an energy that is about the characteristic value at which
the acceleration is stopped by ICS in the bottom parts of the IAR
$\gamma_{{\rm c}}$. To increase readability for cascade simulations
with very low surface temperature, in all the Figures of the ICS distribution
we present $\gamma$-photons produced by $50$ primary particles with
Lorentz factors in the range of $0.5\gamma_{{\rm c}}-2\gamma_{{\rm c}}$.

\begin{figure}[H]
\begin{centering}
\includegraphics[height=7cm]{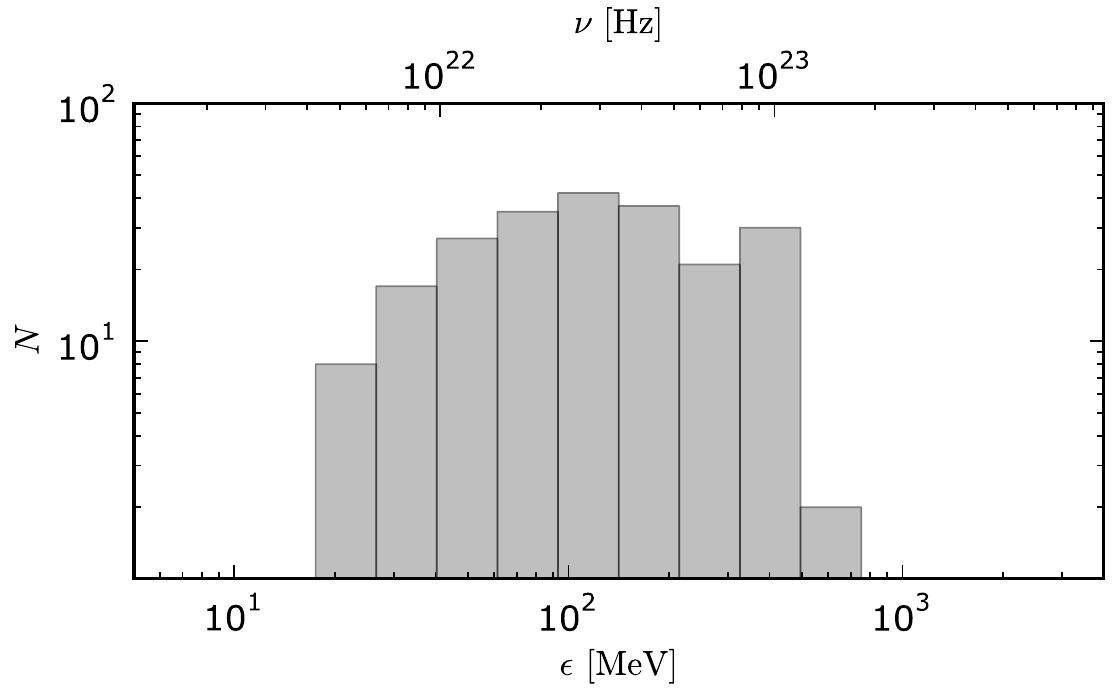}
\par\end{centering}

\caption[{Distribution of ICS photons for PSR B0834+06 {[}$T_{{\rm s}}=0.3\,{\rm MK}${]}}]{Distribution of ICS photons produced by an upscattering of surface
radiation \protect \linebreak{}
($T_{{\rm s}}=0.3\,{\rm MK}$) for PSR B0834+06. The plot includes
all $\gamma$-photons upscatterd by $50$ primary particles with Lorentz
factors in the range of $2.5\times10^{3}-10^{4}$. \label{fig:radiation.ics_0834}}
\end{figure}

In Figure \ref{fig:radiation.ics_0834} we present the distribution
of ICS photons produced by the upscattering of surface thermal radiation
with temperature $T_{{\rm s}}=0.3\,{\rm MK}$ for PSR B0834+06. Even
for such a low surface temperature the whole surface radiation is
the dominant source of background photons for ICS up to altitudes
of about one stellar radii. During the scattering the primary particles
lose about $30\%$ of their initial energy while producing $\gamma$-photons
with energy up to $1\,{\rm GeV}$. Since the $\gamma$-photons are
very energetic and are produced in a region with a strong magnetic
field, they will be absorbed by the magnetic field, thus giving rise
to the secondary plasma population (see Section \ref{sec:radiation.secondary_plasma}).
All pairs in the inner magnetosphere of a pulsar are created in the
nonzero Landau level, thus the pair production process is also accompanied
by strong SR (see the next section). 

\begin{comment}
\textasciitilde{}/Programs/studies/phd/cascade\_plot/cascade\_plot.py
t0834\_t03

(read\_data\_ics, plot\_hist\_ics, 384\_ics\_1e03\_10m\_ics2\_t03\_fifty)
\end{comment}

\begin{figure}[H]
\begin{centering}
\includegraphics[height=7cm]{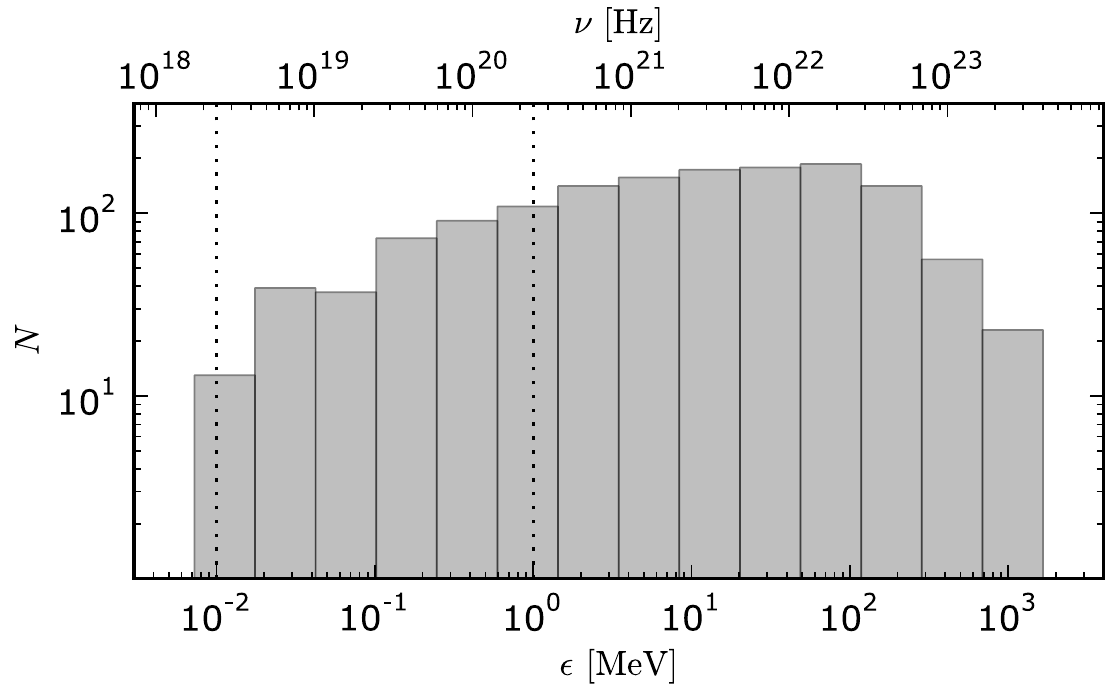}
\par\end{centering}

\caption[{Distribution of ICS photons for PSR B0834+06 {[}$T_{{\rm s}}=0.4\,{\rm MK}${]}}]{Distribution of ICS photons produced by an upscattering of the surface
radiation ($T_{{\rm s}}=0.4\,{\rm MK}$) for PSR B0834+06. The plot
includes all $\gamma$-photons upscatterd by $50$ primary particles
with Lorentz factors in the range of $2.5\times10^{3}-10^{4}$. \label{fig:radiation.ics_0834_t04}}
\end{figure}

A natural way of increasing the number of $\gamma$-photons produced
by ICS in the inner magnetosphere is to increase the number of background
photons. Figure \ref{fig:radiation.ics_0834_t04} presents the distribution
of ICS photons produced by an upscattering of the surface thermal
radiation with temperature $T_{{\rm s}}=0.4\,{\rm MK}$ for PSR B0834+06.
During the ICS process the primary particles lose about $65\%$ of
their initial energy. For higher surface temperatures the ICS produces
$\gamma$-photons up to higher altitudes (about two stellar radii),
thus photons with lower energy emerge $\epsilon_{{\rm min}}\approx3\,{\rm MeV}$.
These less energetic photons will reach the observer, but their total
energy is significantly lower than the total energy of the secondary
plasma created by more energetic $\gamma$-photons.

\begin{comment}
\textasciitilde{}/Programs/studies/phd/cascade\_plot/cascade\_plot.py
t0834\_t04c

(read\_data\_ics, plot\_hist\_ics, 384\_ics\_1e03\_10m\_ics2\_t04\_fifty)
\end{comment}

Note that although for PSR B0834+06 the X-ray spectral fit was performed
with only one BB component, the surface temperatures used in the calculations
($0.3\,{\rm MK}$ and $0.4\,{\rm MK}$) are in good agreement with
the predicted surface temperature of an old neutron star. 

Another source of background photons which could be relevant for ICS
in the inner magnetosphere is the warm spot component. As shown in
Section \ref{sec:radiation.thermal_reheating}, if the antipodal spot
operates in the PSG-off mode and if the magnetic field structure is
suitable then the warm spot is formed in the region adjacent to the
polar cap. With a temperature lower than the hot spot but a much larger
area, the warm spot is the main source of the background photons at
altitudes up to about half a stellar radius. 

\begin{figure}[H]
\begin{centering}
\includegraphics[height=7cm]{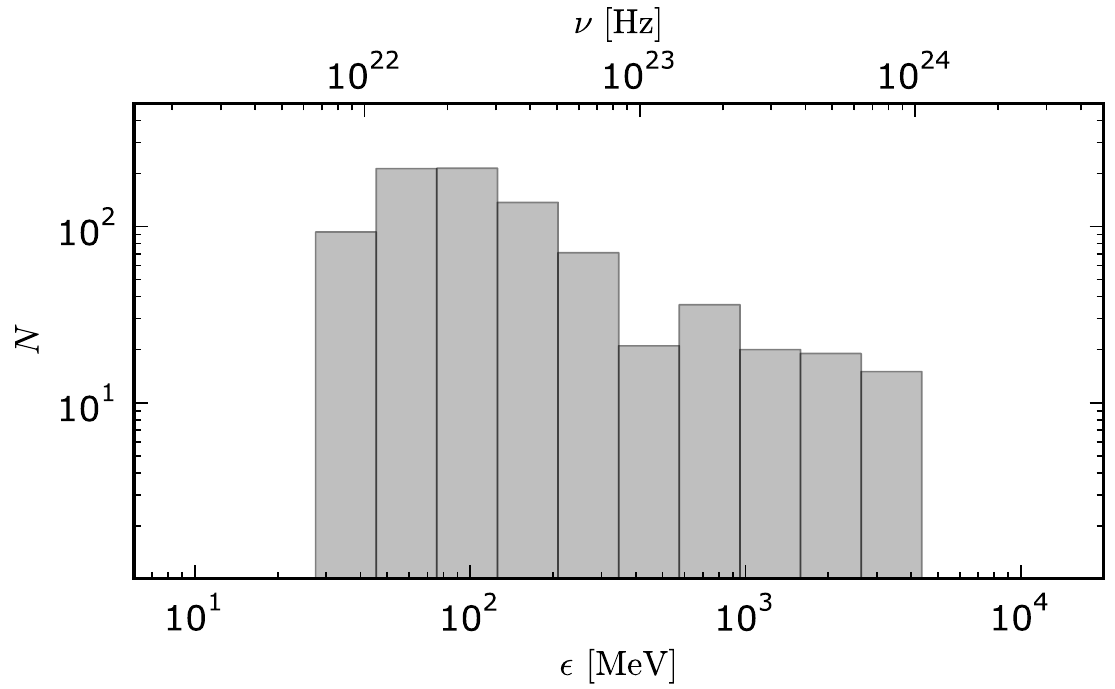}
\par\end{centering}

\caption[{Distribution of ICS photons for PSR B0834+06 {[}$T_{{\rm s}}=1.0\,{\rm MK}$,
$R_{{\rm ws}}=1\,{\rm km}${]}}]{Distribution of ICS photons produced by an upscattering of warm spot
radiation ($T_{{\rm s}}=1.0\,{\rm MK}$, $R_{{\rm ws}}=1\,{\rm km}$)
for PSR B0834+06. The plot includes all $\gamma$-photons upscatterd
by $50$ primary particles with Lorentz factors in the range of $2.5\times10^{3}-10^{4}$.
\label{fig:radiation.ics_0834_t10}}
\end{figure}

In Figure \ref{fig:radiation.ics_0834_t10} we present the distribution
of ICS photons produced by an upscattering of warm spot radiation
with temperature $T_{{\rm s}}=1.0\,{\rm MK}$ and radius $R_{{\rm ws}}=1\,{\rm km}$
for PSR B0834+06. When the warm spot is the main source of background
photons, the ICS process starts at lower altitudes. As a consequence,
the scattering produces photons with higher energy and the primary
particles lose up to $90\%$ of their initial energy. All these high
energetic $\gamma$-photons are absorbed by the magnetic field producing
electron-positron pairs. Note that for this specific pulsar the existence
of such a strong warm spot component is unlikely, but as mentioned
in Section \ref{sec:radiation.thermal_radiation} the X-ray spectral
fits should be extended to include more thermal components to put
better constraints on the X-ray emission of pulsars.

\begin{comment}
\textasciitilde{}/Programs/studies/phd/cascade\_plot/cascade\_plot.py
t0834\_t10

(read\_data\_ics, plot\_hist\_ics, 384\_ics\_1e03\_10m\_ics1\_t10\_fifty\_warm)
\end{comment}

\subsubsection{Synchrotron Radiation}

\begin{figure}[H]
\begin{centering}
\includegraphics{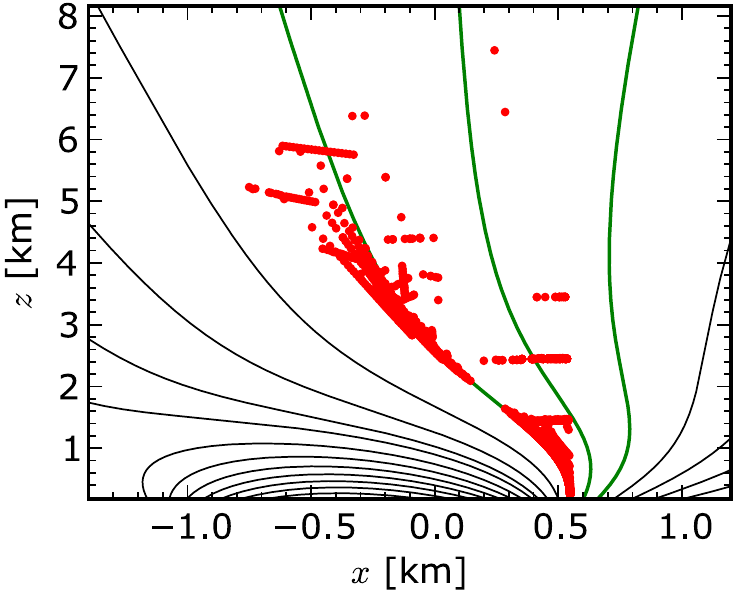}\hspace{0.5cm}\includegraphics{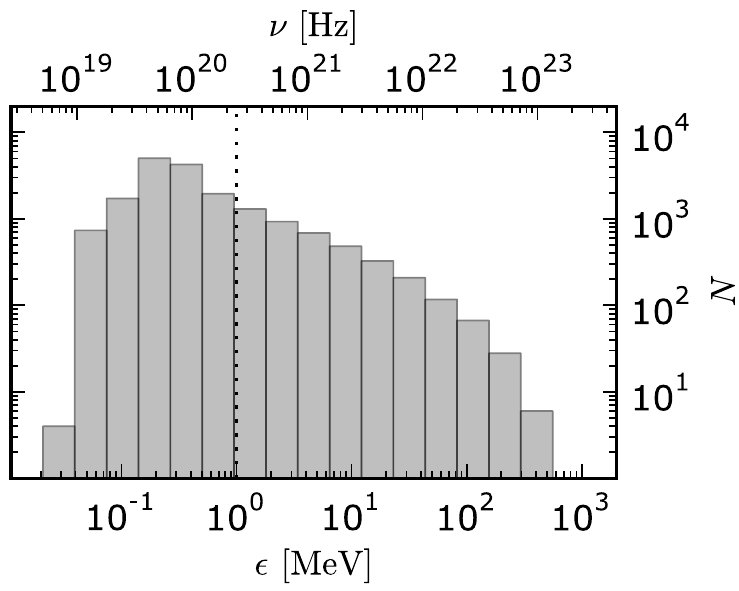}
\par\end{centering}

\caption[{Synchrotron Radiation in the PSG-off mode {[}PSR J0633+1746{]}}]{Synchrotron Radiation in the PSG-off mode for PSR J0633+1746. The
left panel presents the places at which SR-photons are generated,
while the right panel presents the SR-photons distribution. Plots
were obtained in a cascade simulation calculated for a single primary
particle moving along the extreme left open magnetic field line. \label{fig:radiation.sr_gamma_off_0633}}
\end{figure}

In both PSG-off and PSG-on modes SR plays a significant role in the
generation of soft $\gamma$-ray photons. Figure \ref{fig:radiation.sr_gamma_off_0633}
presents the places at which SR-photons are generated (left panel)
and the SR spectrum (right panel) in the PSG-off mode for Geminga.
The most energetic photons are generated close to the stellar surface
($z\approx500\,{\rm m}$), while the less energetic ones are produced
at altitudes $z>2\,{\rm km}$, where the magnetic field is weaker.
SR in the PSG-off mode produces photons with energy in the range of
from $30\,{\rm keV}$ to $1\,{\rm GeV}$. Again, the high energetic
$\gamma$-photons produce electron-positron pairs in a strong magnetic
field, thus its observation is not possible.

\begin{comment}
\textasciitilde{}/Programs/studies/phd/lines/lines.py t0633\_cr

\textasciitilde{}/Programs/studies/phd/cascade\_plot/cascade\_plot.py
t0633\_sr\_cr
\end{comment}

\begin{figure}[H]
\begin{centering}
\includegraphics{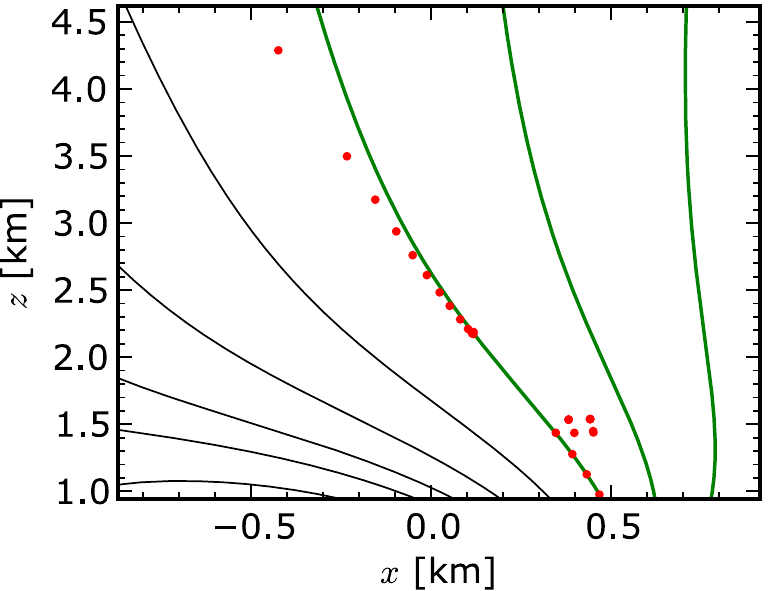}\hspace{0.5cm}\includegraphics{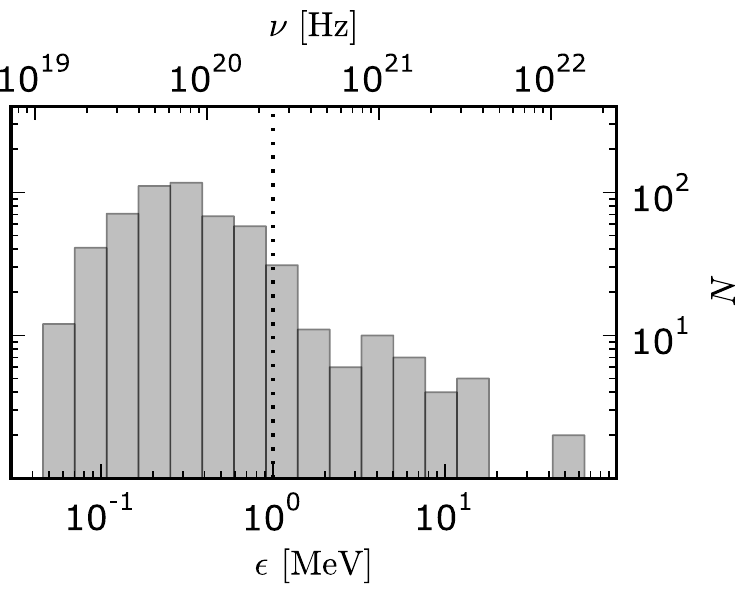}
\par\end{centering}

\caption[{Synchrotron Radiation in the PSG-on mode {[}PSR J0633+1746{]}}]{Synchrotron Radiation in the PSG-on mode for PSR J0633+1746. The
left panel presents places at which SR-photons are generated, while
the right panel presents the SR-photons distribution. Plots were obtained
in a cascade simulation calculated for a single primary particle moving
along the extreme left open magnetic field line. The ICS process was
calculated using the whole surface radiation with temperature $T_{{\rm s}}=0.5\,{\rm MK}$
(see Table \ref{tab:x-ray_thermal}). \label{fig:radiation.sr_gamma_on_0633}}
\end{figure}

In Figure \ref{fig:radiation.sr_gamma_on_0633} we present the places
of SR-photon generation (left panel) and the energy distribution of
photons (right panel) in the PSG-on mode for Geminga. 

\begin{comment}
\textasciitilde{}/Programs/studies/phd/lines/lines.py t0633\_ics

\textasciitilde{}/Programs/studies/phd/cascade\_plot/cascade\_plot.py
t0633\_sr\_ics
\end{comment}

The production of SR-photons in the PSG-off mode starts at altitudes
about $z\approx1\,{\rm km}$ and ends at altitudes $z\approx4.5\,{\rm km}$.
Thus, the energy range of SR-photons is narrower, with the minimum
and maximum photon energy $\epsilon_{{\rm min}}\approx40\,{\rm keV}$
and $\epsilon_{{\rm max}}\approx50\,{\rm MeV}$, respectively. Note
the significant difference in the number of photons produced by SR
in the PSG-off and PSG-on modes. The difference is a direct consequence
of low secondary plasma multiplicity in the PSG-on mode (see Section
\ref{sec:radiation.secondary_plasma}).

\subsection{X-ray emission\label{sec:radiation.x-ray_sr}}

The main source of X-ray photons produced in the inner magnetosphere
is SR. As mentioned in Section \ref{sec:radiation.cr_primary}, to
increase the amount of photons reaching the observer the emission
zone should by located in the area with a weaker magnetic field. In
this section we focus on the results of PSR B0943+10 and PSR 1929+10
for which the proposed configuration of a magnetic field satisfies
this requirement (see Sections \ref{sec:model.b0943} and \ref{sec:model.1929}). 

\begin{figure}[H]
\begin{centering}
\includegraphics[height=7cm]{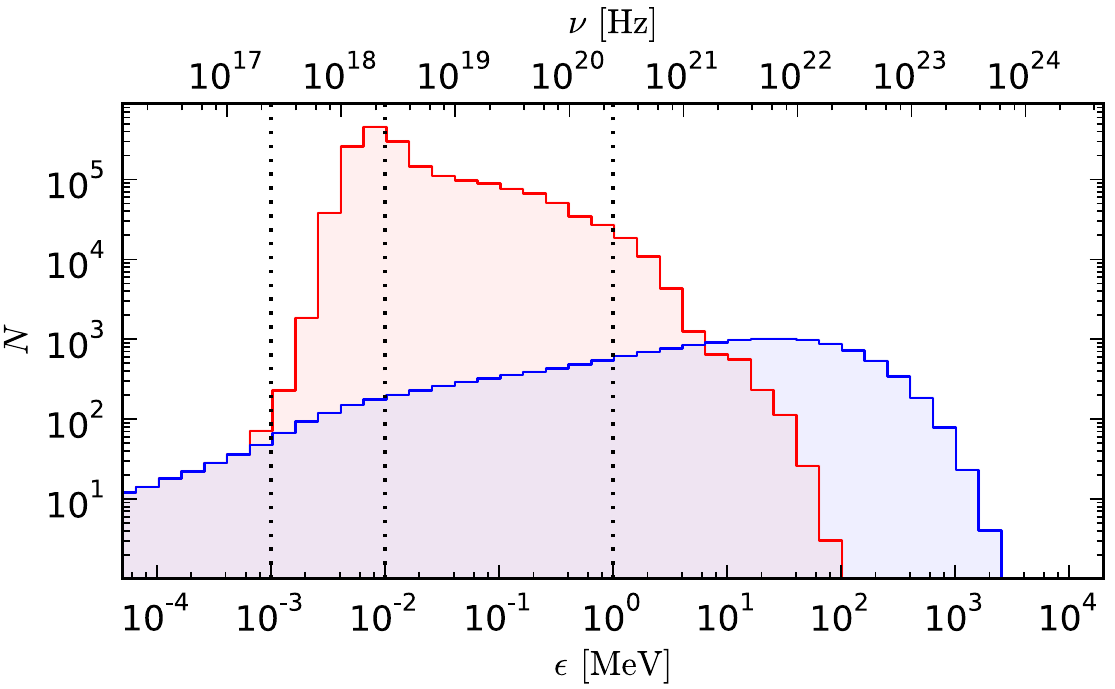}
\par\end{centering}

\caption[{Final photon distribution produced by a single primary particle {[}PSR
1929+10{]}}]{Final photon distribution produced by a single primary particle for
PSR 1929+10 in the PSG-off mode. The blue line corresponds to the
initial CR photons distribution, while the red line presents the final
distribution with the inclusion of photon splitting, pair production
and SR.\label{fig:radiation.cr_1929}}
\end{figure}

In the PSG-off mode most of the X-ray photons are produced by the
SR of newly created electron-positron pairs. Figure \ref{fig:radiation.cr_1929}
presents the final photon distribution produced by a single primary
particle of PSR B1929+10 in the PSG-off mode. For a single primary
particle we can estimate that only about $0.7\%$ of the total photon
energy is in the range of $1-10\,{\rm keV}$. The bulk of the energy
is carried away by newly created particles ($73\%$) and high energetic
photons ($27\%$). 

\begin{comment}
\textasciitilde{}/Programs/studies/phd/cascade\_plot/cascade\_plot.py 

t1929\_cr\_photons (read\_data\_final, plot\_spectrum\_final, 322\_cr\_1e04\_10m\_new)
\end{comment}

\begin{figure}[H]
\begin{centering}
\includegraphics[width=7.48cm,height=6cm]{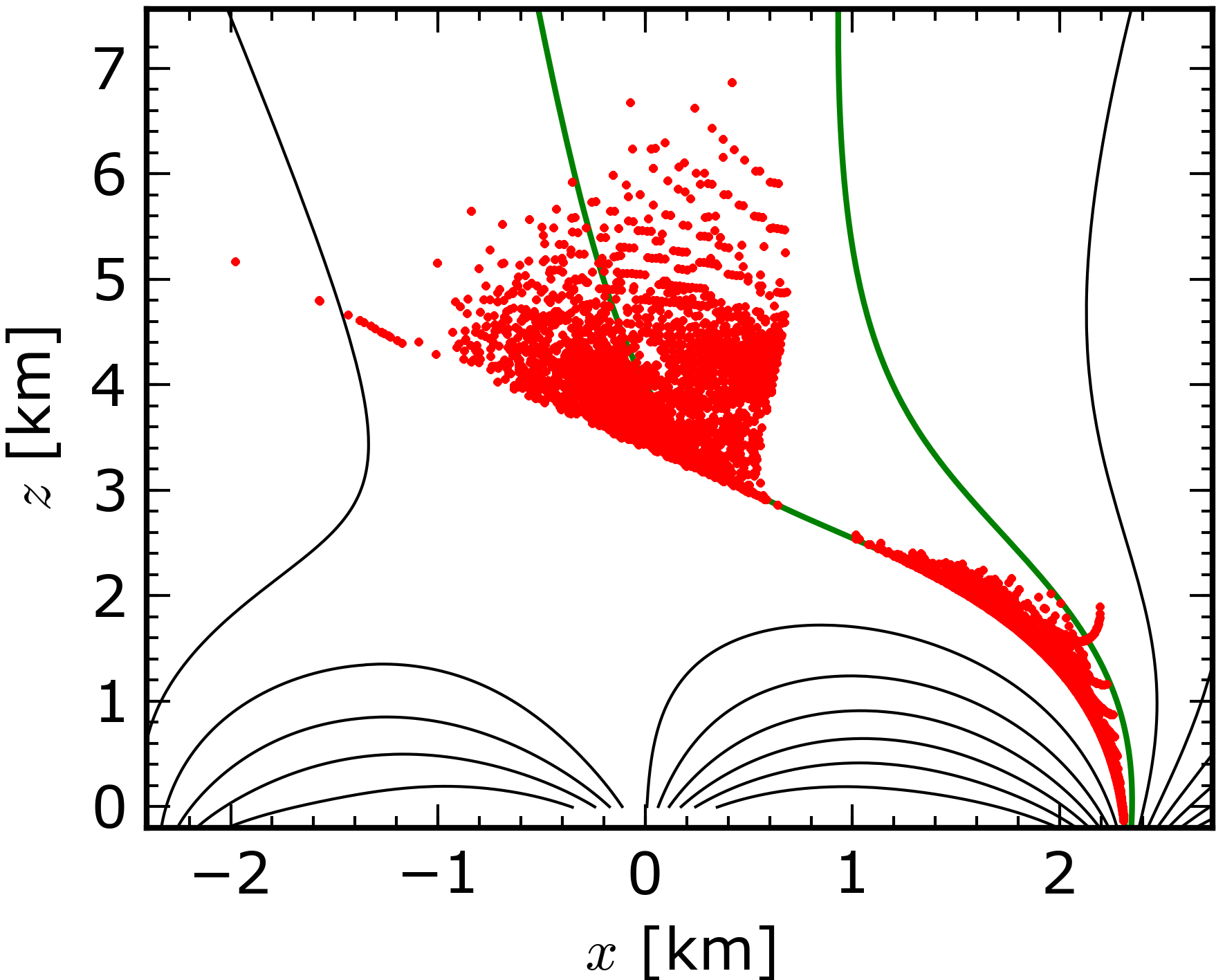}\hspace{0.5cm}\includegraphics{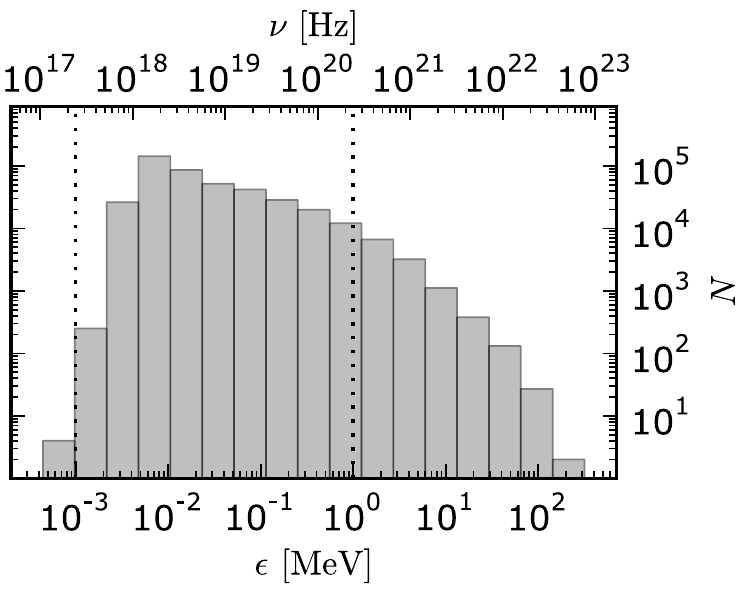}
\par\end{centering}

\caption[{Synchrotron Radiation in the PSG-off mode {[}PSR B1929+10{]}}]{Synchrotron Radiation in the PSG-off mode for PSR B1929+10. The left
panel presents the places at which SR-photons are generated, while
the right panel presents the SR-photons distribution. Plots were obtained
in a cascade simulation calculated for a single primary particle moving
along the extreme left open magnetic field line. \label{fig:radiation.sr_gamma_off_1929}}
\end{figure}

In Figure \ref{fig:radiation.sr_gamma_off_1929} we present the locations
and the photon distribution of SR for PSR 1929+10 in the PSG-off mode.
All SR-photons produced closer to the stellar surface will contribute
to $\gamma$-ray emission, while the SR-photons produced at higher
altitudes will produce photons in the X-ray band. As it results from
Figure \ref{fig:model.b1929_curva}, the curvature at an altitude
of $z\approx3.5\,{\rm km}$ is only $50\%$ higher than at $z\approx2\,{\rm km}$.
Furthermore, before the particle reaches the region with a relatively
low magnetic field ($z\approx3.5\,{\rm km}$), it radiates a significant
part of its energy at lower heights.

\begin{comment}
\textasciitilde{}/Programs/studies/phd/lines/lines.py t1929

\textasciitilde{}/Programs/studies/phd/cascade\_plot/cascade\_plot.py
t1929\_sr\_cr
\end{comment}

\begin{figure}[H]
\begin{centering}
\includegraphics[height=6.7cm]{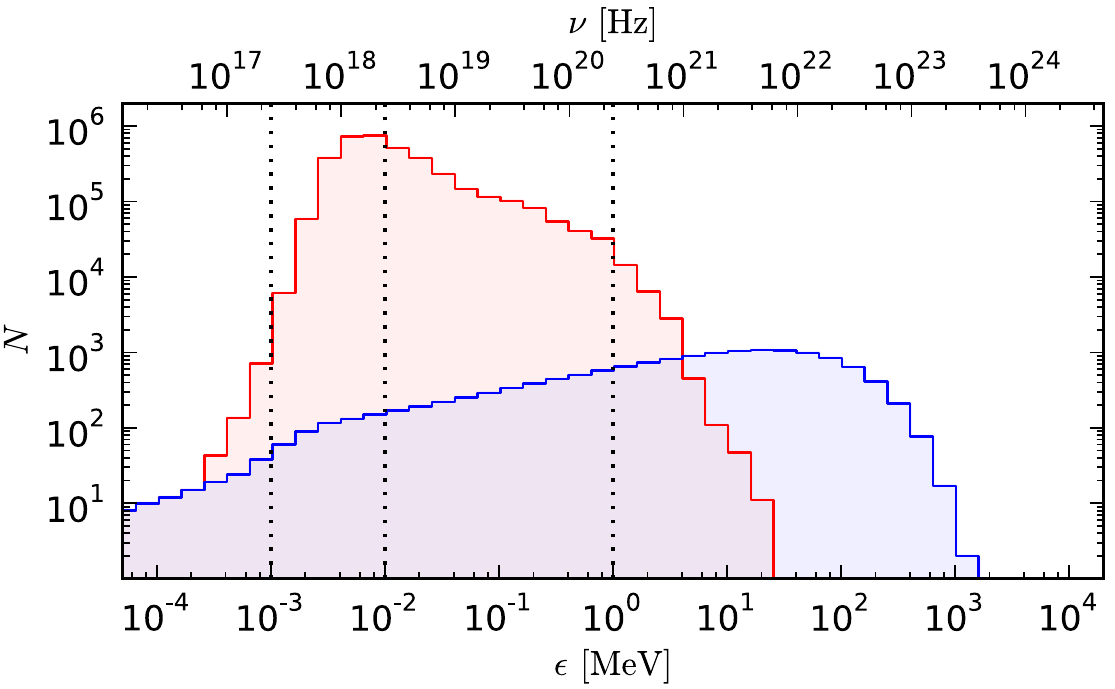}
\par\end{centering}

\caption[{Final photon distribution produced by a single primary particle {[}PSR
B0943+10{]}}]{Final photon distribution produced by a single primary particle for
PSR B0943+10 in the PSG-off mode. The blue line corresponds to the
initial CR photons distribution, while the red line presents the final
distribution with the inclusion of photon splitting, pair production
and SR.\label{fig:radiation.cr_0943}}
\end{figure}

To increase radiation in the $1-10\,{\rm keV}$ energy band we should
apply the magnetic field structure with considerably higher curvature
at altitudes where X-ray photons are generated. Although the curvature
will not directly affect the SR, it will enhance CR, and thus it will
increase the number of pairs produced in the region of a relatively
weak magnetic field. In Figure \ref{fig:radiation.cr_0943} we present
the final photon distribution produced by a single primary particle
for PSR B0943+10 calculated using the magnetic field configuration
as presented in Section \ref{sec:model.b0943}.

\begin{comment}
\textasciitilde{}/Programs/studies/phd/cascade\_plot/cascade\_plot.py 

t0943\_cr\_photons (read\_data\_final, plot\_spectrum\_final, 911\_cr\_1e04\_10m\_leftline)
\end{comment}

For this magnetic field structure about $3\%$ of the total photon
energy is in the range of $1-10\,{\rm keV}$. The newly created particles
carry away about $63\%$ of the energy radiated by the primary particle,
while about $37\%$ of the energy remains in the form of photons.

\begin{figure}[H]
\begin{centering}
\includegraphics[width=7.48cm,height=6cm]{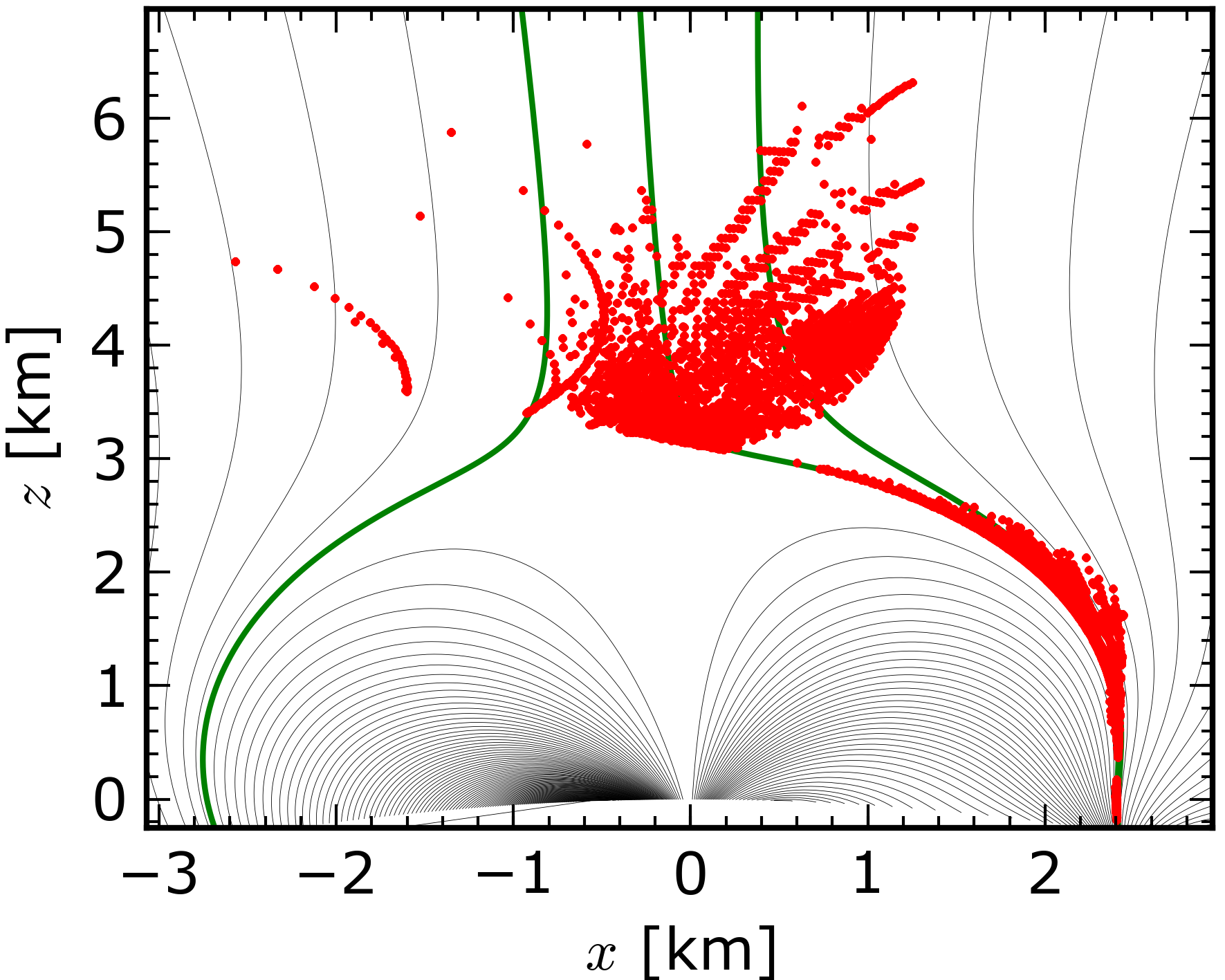}\hspace{0.5cm}\includegraphics{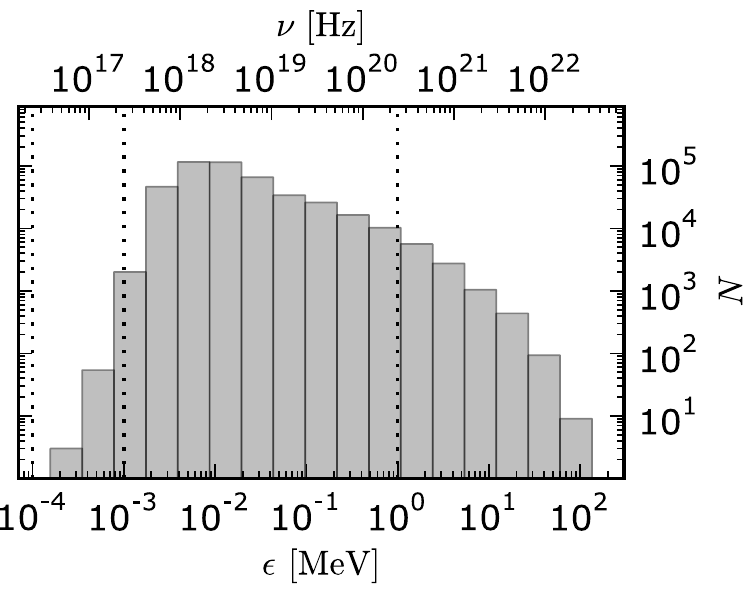}
\par\end{centering}

\caption[{Synchrotron Radiation in the PSG-off mode {[}PSR B0943+10{]}}]{Synchrotron Radiation in the PSG-off mode for PSR B0943+10. The left
panel presents the places at which SR-photons are generated, while
the right panel presents the SR-photons distribution. Plots were obtained
in a cascade simulation calculated for a single primary particle moving
along the extreme left open magnetic field line. \label{fig:radiation.sr_gamma_off_0943}}
\end{figure}

The structure of the magnetic field of PSR B0943+10 allows enhanced
pair production in a region of a weaker magnetic field (see Figure
\ref{fig:radiation.sr_gamma_off_0943}). The SR that accompanies pair
production at higher altitudes ($z>3\,{\rm km}$) essentially increases
the amount of energy radiated in the $1-10\,{\rm keV}$ energy band.
Note, however, that the fraction of energy radiated in this band is
still relatively low ($3\%$), and in order to be a substantial part
of the observed X-ray spectrum the strong anisotropy of backstreaming
and outstreaming plasma is required (see Section \ref{sec:radiation.primary_plasma}).

\begin{comment}
\textasciitilde{}/Programs/studies/phd/lines/lines.py t0943\_cr

\textasciitilde{}/Programs/studies/phd/cascade\_plot/cascade\_plot.py
t0943\_sr\_cr
\end{comment}

\begin{figure}[H]
\begin{centering}
\includegraphics[height=7cm]{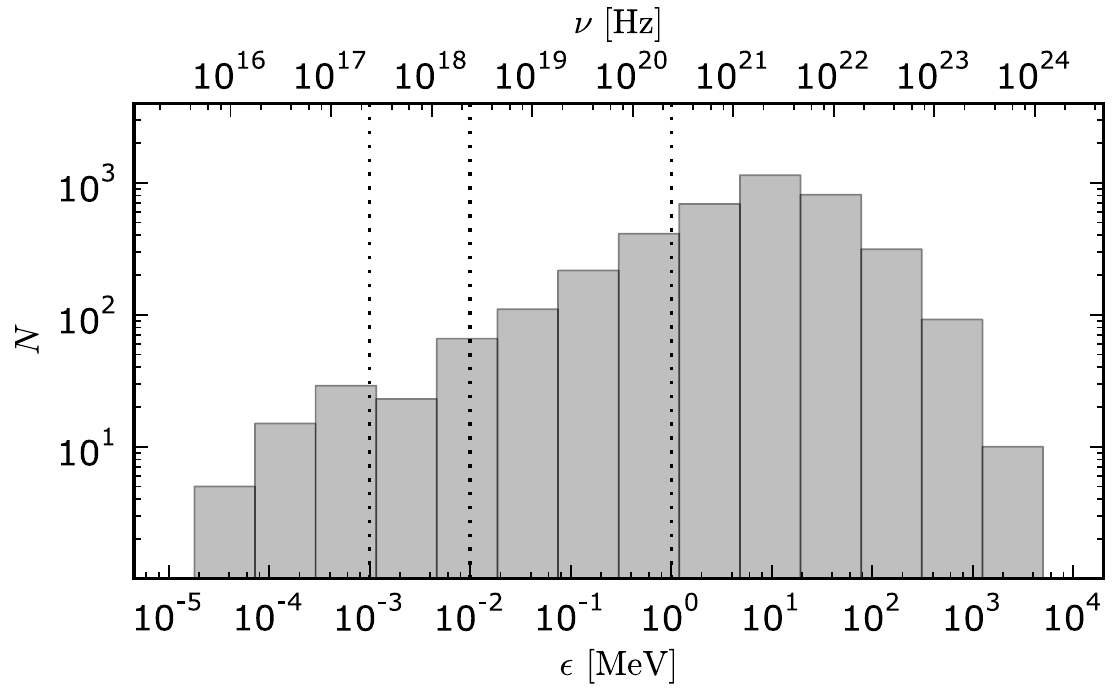}
\par\end{centering}

\caption[{Distribution of ICS photons for PSR B0628-28 {[}$T_{{\rm s}}=0.5\,{\rm MK}${]}}]{Distribution of ICS photons produced by an upscattering of the whole
surface radiation ($T_{{\rm s}}=0.5\,{\rm MK}$) for PSR B0628-28.
The plot includes all photons upscatterd by $50$ primary particles
with Lorentz factors in the range of $3\times10^{3}-1.2\times10^{4}$.
\label{fig:radiation.ics_0628}}
\end{figure}

In the PSG-on mode even for a complicated structure of the magnetic
field most of the outflowing energy is converted to secondary plasma.
Figure \ref{fig:radiation.ics_0628} presents the ICS-photons distribution
produced in the PSG-on mode for PSR B0628-28. The bulk of energy is
radiated in the form of high energetic $\gamma$-photons which are
responsible for pair production, and thus the formation of secondary
plasma. Taking into account not so high backstreaming/outstreaming
anisotropy in the PSG-on mode, the ICS process is not relevant for
the production of X-ray photons.

\begin{comment}
\textasciitilde{}/Programs/studies/phd/cascade\_plot/cascade\_plot.py
t0628\_t05\_ics
\end{comment}

\begin{figure}[H]
\begin{centering}
\includegraphics{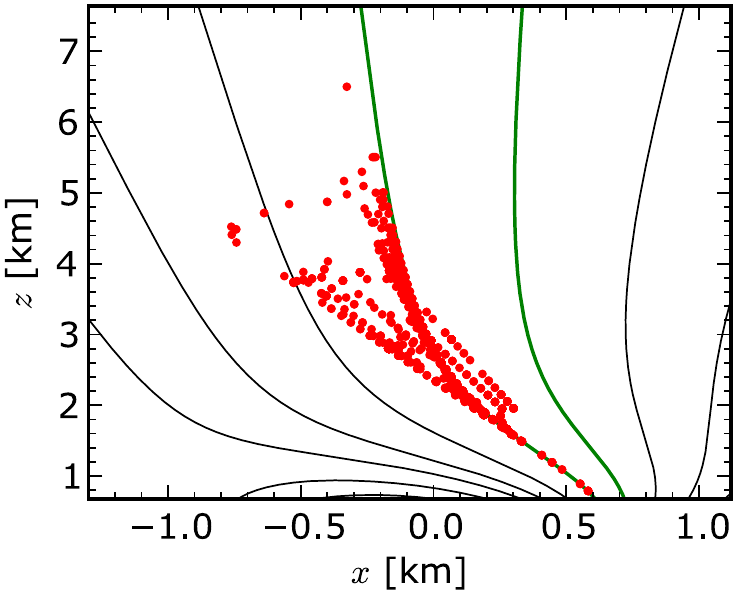}\hspace{0.5cm}\includegraphics{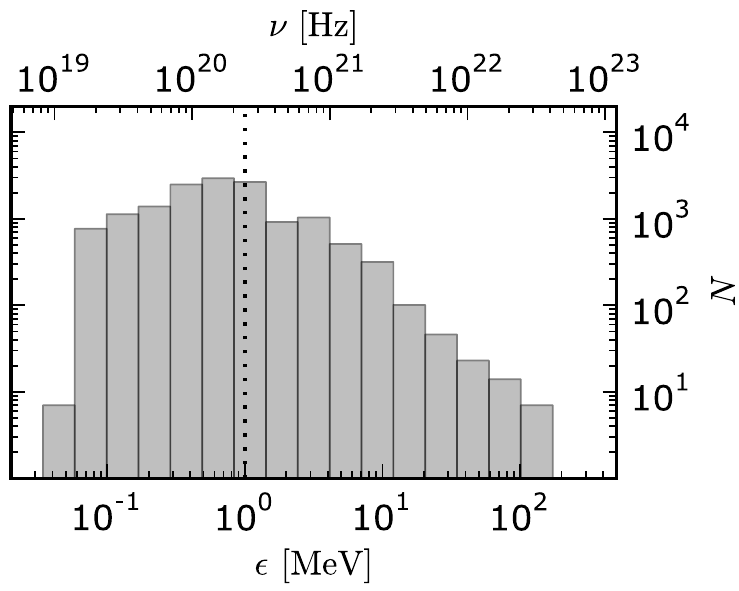}
\par\end{centering}

\caption[{Synchrotron Radiation in the PSG-on mode {[}PSR B0628-28{]}}]{Synchrotron Radiation in the PSG-on mode for PSR B0628-28. The left
panel presents places at which SR-photons are generated, while the
right panel presents the SR-photons distribution. Plots were obtained
in a cascade simulation calculated for $50$ primary particles with
Lorentz factors in the range of $3\times10^{3}-1.2\times10^{4}$ moving
along the extreme left open magnetic field line. \label{fig:radiation.sr_gamma_on_0628}}
\end{figure}

The SR which accompanies the pair creation process in the PSG-on mode
mostly produces soft $\gamma$-photons (see the right panel of Figure
\ref{fig:radiation.sr_gamma_on_0628}). Although the secondary pairs
are produced at similar altitudes in both modes, (compare the left
panels of Figures \ref{fig:radiation.sr_gamma_off_0943} and \ref{fig:radiation.sr_gamma_on_0628}),
the higher Lorentz factor of secondary plasma produced in the PSG-on
mode results in higher energy of the SR-photons. The results suggest
that when the gap operates in the PSG-on mode we should expect lower
efficiencies of nonthermal X-ray emission than in the PSG-off mode.
Note, however, that the final efficiency of X-ray radiation in the
PSG-off mode highly depends on the backstreaming/outstreaming anisotropy
and the structure of magnetic field lines.

\begin{comment}
\textasciitilde{}/Programs/studies/phd/lines/lines.py t0628\_ics\_xrays

\textasciitilde{}/Programs/studies/phd/cascade\_plot/cascade\_plot.py
t0628\_t05\_ics 
\end{comment}

\subsection{Secondary plasma \label{sec:radiation.secondary_plasma}}

\begin{figure}[H]
\begin{centering}
\includegraphics{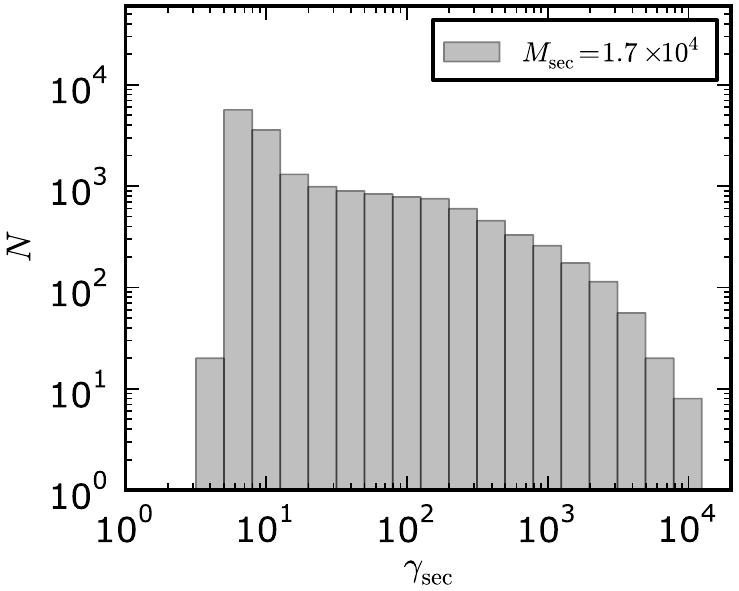}\hspace{0.5cm}\includegraphics{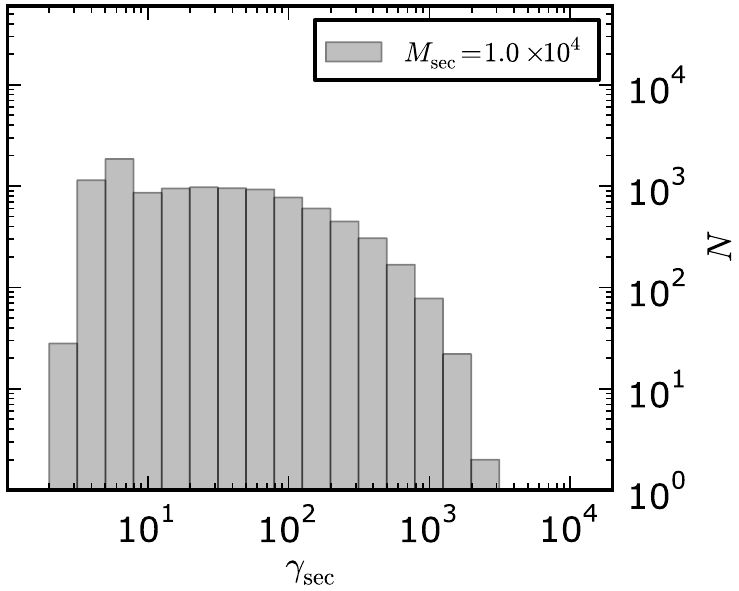}
\par\end{centering}

\caption[{Secondary plasma in the PSG-off mode {[}PSR J0633+1746, PSR B1133+16{]}}]{Energy histogram of secondary plasma in the PSG-off mode. The left
panel was obtained in a cascade simulation calculated for a single
primary particle moving along the extreme left open magnetic field
line of PSR J0633+1746, while the right panel corresponds to a cascade
simulation for PSR B1133+16.\label{fig:radiation.cr_pairs}}
\end{figure}

The multiplicity of secondary particles in the PSG-off mode is much
higher than in the PSG-on mode. However, the primary plasma produced
in the IAR of CR-dominated gaps has a density considerably lower than
the Goldreich-Julian co-rotational density (see Equation \ref{eq:psg.overheating_parameter}).
Figure \ref{fig:radiation.cr_pairs} presents the energy histogram
of secondary plasma for Geminga (left panel) and PSR B1133+16 (right
panel). Despite major differences in the magnetic field structure
and conditions in the IAR for both pulsars, the secondary plasma distribution
shows many similarities. The only significant difference is the maximum
Lorentz factor of secondary plasma, which for Geminga is about $\gamma_{{\rm sec}}^{{\rm max}}\approx10^{4}$,
while for PSR B1133+16 is is a few times smaller $\gamma_{{\rm sec}}^{{\rm max}}\approx3\times10^{3}$.

By using the overheating parameters presented in Table \ref{tab:psg.psg_top}
we can roughly estimate that the final multiplicity of particles in
the plasma cloud in the PSG-off mode ranges from $M=\kappa\cdot M_{{\rm sec}}\approx2$
(for Geminga) to $M=\kappa\cdot M_{{\rm sec}}\approx100$ (for PSR
B1133+16). Note, however, that these values do not take into account
the anticipated anisotropy of backstreaming and outstreaming particles.
The existence of such an anisotropy could further increase the final
multiplicity of particles in the plasma cloud leaving the inner magnetosphere.
Despite the fact that without a full cascade simulation in the IAR
we cannot unambiguously determine the final multiplicity in the plasma
cloud, it can be clearly seen that depending on the details of the
gap operating in the PSG-off mode, the produced plasma may be suitable
(e.g. PSR B1133+16) or unsuitable (e.g. Geminga) to generate radio
emission (see Section \ref{sec:radiation.radio_emission}). The main
factor determining the parameters of the CR-dominated gap, and thus
determining whether it is possible to effectively produce radio emission,
is the radius of curvature of the magnetic field lines (see Section
\ref{sec:psg.curvature_radius}).

\begin{comment}
\textasciitilde{}/Programs/studies/phd/cascade\_plot/cascade\_plot.py
t0633\_cr\_pairs, t1133\_cr\_pairs (change to right) 
\end{comment}

\begin{figure}[H]
\begin{centering}
\includegraphics{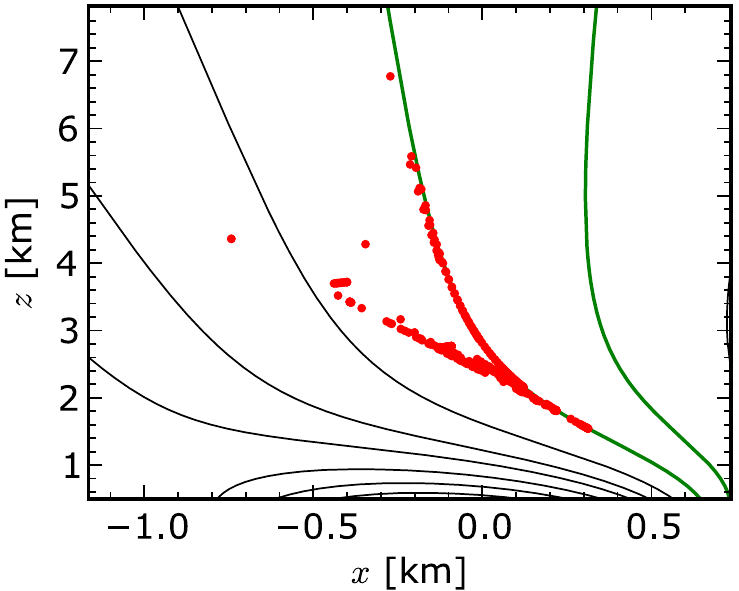}\hspace{0.5cm}\includegraphics{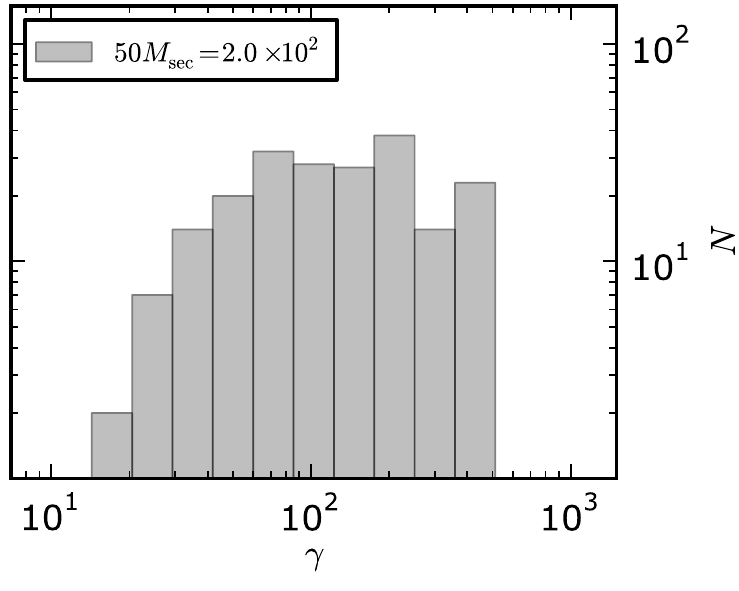}
\par\end{centering}

\caption[{Secondary plasma in the PSG-on mode {[}PSR B0628-28, $T_{{\rm s}}=0.3\,{\rm MK}${]}}]{Secondary plasma produced in the PSG-on mode for PSR B0628-28. The
left panel presents places at which pairs are produced, while the
right panel presents the histogram of particle energy. Plots were
obtained in a cascade simulation calculated for $50$ primary particles
with Lorentz factors in the range of $3\times10^{3}-1.2\times10^{4}$
moving along the extreme left open magnetic field line. The ICS process
was calculated using the whole surface radiation with temperature
$T_{{\rm s}}=0.3\,{\rm MK}$. \label{fig:radiation.plasma_sec_on_0628a}}
\end{figure}

In ICS-dominated gaps, on the other hand, the density of primary plasma
produced in IAR exceeds the co-rotational density. Thus, the development
of dense enough plasma for radio emission is much easier in the PSG-on
mode. In Figure \ref{fig:radiation.plasma_sec_on_0628a} we present
the locations of pair production and energy distribution of secondary
plasma in the PSG-on mode for PSR B0628-28. The final multiplicity
of particles in the plasma cloud in the PSG-on mode can be calculated
as $M=N_{{\rm ICS}}\times M_{{\rm sec}}$. As mentioned in Section
\ref{sec:radiation.primary_plasma}, the exact value of $N_{{\rm ICS}}$
can be found only by performing the full cascade simulation in IAR
but, as shown by \citet{2010_Timokhin}, we should expect a full screening
of the acceleration region when $N_{{\rm ICS}}$ reaches a value as
high as $20-100$. Thus we can roughly estimate that for the whole
surface radiation with temperature $T_{{\rm s}}=0.3\,{\rm MK}$, the
final multiplicity of secondary plasma in the PSG-on mode for PSR
B0628-28 is of the order of $M\approx100$.

\begin{comment}
\textasciitilde{}/Programs/studies/phd/lines/lines.py t0628\_ics\_a50

\textasciitilde{}/Programs/studies/phd/cascade\_plot/cascade\_plot.py
t0628\_ics\_a50
\end{comment}

In the PSG-on mode the main factor which determines the final multiplicity
of secondary plasma is the source of the background photons. As shown
in Section \ref{sec:cascade.background_photons}, the polar cap radiation
(the hot spot component) has a negligible impact on the ICS process
above the IAR. Figure \ref{fig:radiation.plasma_sec_on_0628a} presents
the location of pair production and energy distribution of secondary
plasma for PSR B0628-28 calculated assuming the whole surface radiation
with temperature $T_{{\rm s}}=0.5\,{\rm MK}$. The increase in the
number of background photons results in enhancement of the ICS process,
and thus an increase of the secondary multiplicity $M_{{\rm sec}}\approx60$.
For such conditions the final multiplicity of secondary plasma in
the PSG-on mode is of the order of $M\approx10^{3}-10^{5}$.

\begin{comment}
\textasciitilde{}/Programs/studies/phd/lines/lines.py t0628\_ics\_b

\textasciitilde{}/Programs/studies/phd/cascade\_plot/cascade\_plot.py
t0628\_ics\_b
\end{comment}

\begin{figure}[H]
\begin{centering}
\includegraphics{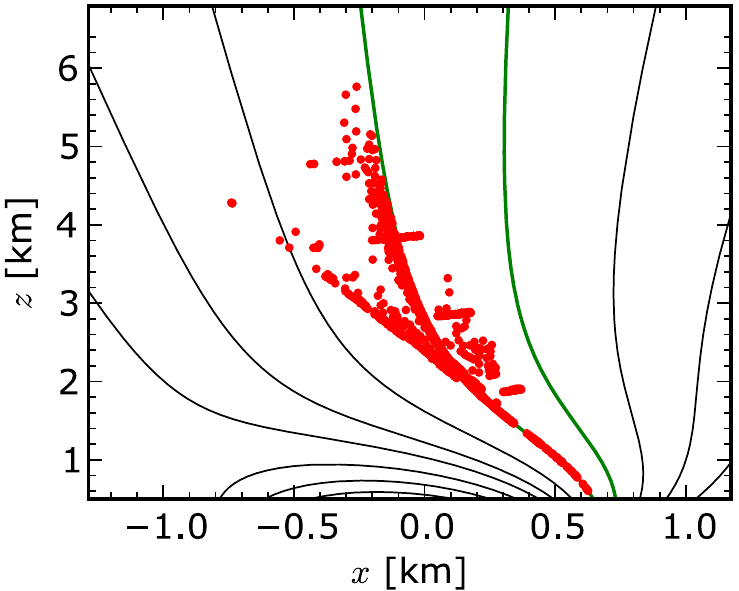}\hspace{0.5cm}\includegraphics{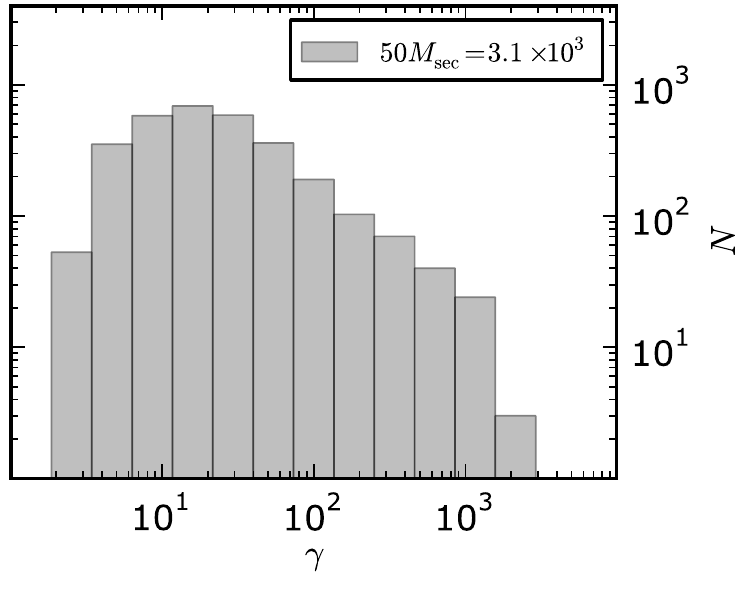}
\par\end{centering}

\caption[{Secondary plasma in the PSG-on mode {[}PSR B0628-28, $T_{{\rm s}}=0.5\,{\rm MK}${]}}]{Secondary plasma produced in the PSG-on mode for PSR B0628-28. The
left panel presents the places at which pairs are produced, while
the right panel presents the histogram of particle energy. Plots were
obtained in a cascade simulation calculated for $50$ primary particles
with Lorentz factors in the range of $3\times10^{3}-1.2\times10^{4}$
moving along the extreme left open magnetic field line. The ICS process
was calculated using the whole surface radiation with temperature
$T_{{\rm s}}=0.5\,{\rm MK}$. \label{fig:radiation.plasma_sec_on_0628b}}
\end{figure}

\chapter*{Conclusions}

\addcontentsline{toc}{chapter}{Conclusions}

The hot spot component identified in X-ray observations implies the
non-dipolar structure of surface magnetic field. We used the Partially
Screened Gap model to explain both the \linebreak{}
X-ray radiation of radio pulsars and production of secondary plasma
suitable for generation of radio emission.

\section*{A special case (PSR B0943+10)}

Our model predicts two additional sources of X-ray emission: (I) the
warm spot component and (II) enhanced SR radiation in the PSG-off
mode. The warm spot component is associated with particles originating
from the antipodal polar cap, while the high luminosity of X-ray photons
produced in the PSG-off mode is a result of strong anisotropy of backstreaming
and outstreaming particles.

Very recent results presented by \citet{2013_Hermsen} show the anti-correlation
of radio and X-ray emission of PSR B0943+10. The authors suggest an
unpulsed, non-thermal component in radio-bright mode and a $100\%$-pulsed
thermal component along with a nonthermal component in a radio-quiet
mode. In our model it is not possible to produce an unpulsed, nonthermal
X-ray component without the accompanying blackbody radiation of the
polar cap. Although it is possible to produce nonthermal X-ray radiation
which obscures the thermal component (strong SR in the PSG-off mode
with a high predominance of outstreaming particles), the resulting
radiation should be pulsed. We believe that the X-ray radiation of
PSR B0943+10 in the radio bright mode was misinterpreted as the nonthermal
one. As shown in Figure \ref{fig:x-ray.PSRS_quad} (panel d), for
a derived geometry of PSR B0943+10 the polar cap produces unpulsed,
thermal radiation. Furthermore, as reported by the authors, in the
radio-bright mode both the absorbed blackbody (BB) and the absorbed
power-law (PL) models fit the spectrum equally well (see Table S4
in \citealp{2013_Hermsen}). We believe that the observed radiation
modes of PSR B0943+10 correspond to a mode switch between the PSG-on
(radio-bright) and the PSG-off mode (radio-quiet). When pulsar is
in the PSG-on we observe both the radio emission and thermal radiation
which originates from the polar cap. In the PSG-off mode the secondary
plasma is not suitable to produce so strong radio emission as in the
PSG-on mode, but the polar cap radiation is accompanied by pulsed,
nonthermal emission produced by SR (see Sec. \ref{sec:radiation.x-ray_sr}).

\section*{Gamma-ray pulsars}

As was shown in Sections  \ref{sec:radiation.iar_gamma_ray} and \ref{sec:radiation.im_gamma_ray},
$\gamma$-rays produced in IAR and the inner magnetosphere cannot
reach the observer due to efficient pair production in those regions.
Current models of $\gamma$-ray emission propose that the emission
comes from outer magnetospheric gaps. The non-dipolar structure of
a magnetic field has two key implications on $\gamma$-ray emission
models: (I) the formation of slot gaps is not possible as pairs are
produced along all open magnetic field lines, (II) the high density
of electron-positron plasma ($n_{p}\gg n_{{\rm GJ}}$) produced in
the inner magnetosphere prevents the outer gap formation. The high-density
plasma which crosses the null line will screen the outer magnetospheric
region due to plasma separation (acceleration of electrons and deceleration
of positrons). Thus, the formation of outer gaps is possible only
in special cases when the pulsar operates in the PSG-off mode and
produces secondary plasma with low density $n_{p}\approx n_{{\rm GJ}}$.

As recently reported by \citet{2013_Arka}: ``It is possible for
relativistic populations of electrons and positrons in the current
sheet of a pulsar’s wind right outside the light cylinder to emit
synchrotron radiation that peaks in the ${\rm sub-GeV}$ to ${\rm GeV}$
regime, with $\gamma$-ray efficiencies similar to those observed
for the Fermi/LAT pulsars.'' We believe that the observed high-energetic
$\gamma$-rays are produced in the not yet well explored region right
outside the light cylinder.

\section*{Radio emission\label{sec:radiation.radio_emission}}

Pulsed radio emission remains one of the most intriguing puzzles of
astrophysics. It is remarkable that despite the large ranges in $P$,
$B_{{\rm d}}$, the variations in the pulse profile between different
classes of neutron stars (young, old, millisecond, magnetars) are
similar to those within classes \citep{2004_Melrose}. The radio emission
of most pulsars can be characterised by: a relatively narrow frequency
range, $\sim100\,{\rm MHz}$ to $\sim10\,{\rm GHz}$, and a high degree
of polarisation with a characteristic sweep of the position angle.
The extremely high brightness temperature of pulsar radio emission
(typically $T_{b}>10^{25}\,{\rm K}$) implies that a coherent emission
mechanism is involved. Many radio emission mechanisms have been proposed,
but no consensus on a specific emission mechanism has emerged. The
radio observations alone cannot identify the emission mechanism and,
hence, a model of the magnetosphere is needed to put constraints on
the radio emission model. An acceptable emission mechanism must involve
some form of instability to produce coherent radiation. The main difficulty
in finding a specific emission mechanism is that many of the predicted
features are common all proposed models. Furthermore, the polarisation
can also be regarded as generic rather than associated with a specific
emission mechanism \citep{2006_Melrose}. 

The X-ray observations have allowed us to put constraints on the polar
cap region of pulsars. The non-dipolar structure of the surface magnetic
field causes plasma to form under similar conditions regardless of
the global configuration of the magnetic field. We have showed that
depending on the details of IAR, the resulting plasma either meets
the requirements for efficient radio emission (suitable multiplicity
and energy distribution of secondary plasma) or is not suitable to
produce efficient radio emission (e.g. Geminga). Furthermore, the
proposed drift model allows to find a connection between radio and
X-ray emission processes (see Section \ref{sec:psg.drift}).

\section*{The mixed mode}

Although in the thesis we consider the PSG-on and PSG-off mode separately,
in a real case both of these modes can coexist either on two separate
polar caps or on the same polar cap occupying its different parts.
In the latter case the change of modes is associated with varying
degrees of intensity of the two modes. Furthermore, if specific conditions
are met, the ICS process can be a main source of $\gamma$-photons
in the lower parts of the gap, while the CR process can produce $\gamma$-photons
in the upper parts of the acceleration region. In such a case distinguishing
between the two modes is even more difficult.

\section*{Summary}

The main propositions associated with this thesis are as follows:
\begin{enumerate}
\item The size of the hot spots implies that the magnetic field configuration
just above the stellar surface differs significantly from a purely
dipole one. 
\item The analysis of X-ray observations shows that the temperature of the
actual polar cap is equal to the so-called critical value, i.e. the
temperature at which the outflow of thermal ions from the surface
screens the gap completely.
\item The non-dipolar structure of a surface magnetic field and the high
multiplicity of particles produced in IAR prevents the formation of
slot and outer gaps. 
\item The PSG model predicts the existence of two scenarios of gap breakdown:
the PSG-off mode for CR-dominated gaps and the PSG-on mode for ICS-dominated
gaps.
\item The two different scenarios of gap breakdown can in a natural way
explain the mode-changing phenomenon when both modes produce plasma
suitable to generate radio emission, and pulse nulling when the radio
emission is not generated in one of the modes.
\item The mode changes of the IAR may explain the anti-correlation of radio
and X-ray emission in very recent observations of PSR B0943+10 \citep{2013_Hermsen}.
\item The regular drift of subpulses can be expected only when the gap operates
in the PSG-on mode. The proposed model of drift allows to connect
the drift information obtained by radio observations with the X-ray
data of rotation-powered pulsars.
\end{enumerate}

\chapter*{Acknowledgements}

\addcontentsline{toc}{chapter}{Acknowledgements}

I would like to express my deep gratitude to Professor Giorgi Melikidze,
my research supervisor, for his patient guidance, enthusiastic encouragement
and useful critique of this research work. I would also like to thank
Professor Janusz Gil for his advice and support which allowed me to
complete this thesis. This research project would not have been possible
without the support of many people. I would like to thank all my colleagues
at the Institute of Astronomy who taught me a lot and never refused
to help: Professor Ulrich Geppert, Professor Dorota Gondek-Rosińska,
Professor Jarosław Kijak, Dr. Krzysztof Krzeszowki, Dr. Wojciech Lewandowski,
Professor Andrzej Maciejewski, Dr. Krzysztof Maciesiak, Dr. Olaf Maron,
Dr. Roberto Mignani, Dr. Marek Sendyk, and Dr. Agnieszka Słowikowska.
And a special thanks to Mrs Emilia Gil for her assistance in all the
administrative issues. 

I would also like to extend my thanks to friends and family for their
support, sacrifice, patience and wisdom. My special thanks are extended
to my parents for their support and encouragement throughout my studies.

\begin{flushright}
\textit{Thank you.}
\par\end{flushright}

\begin{center}
\includegraphics[height=9cm]{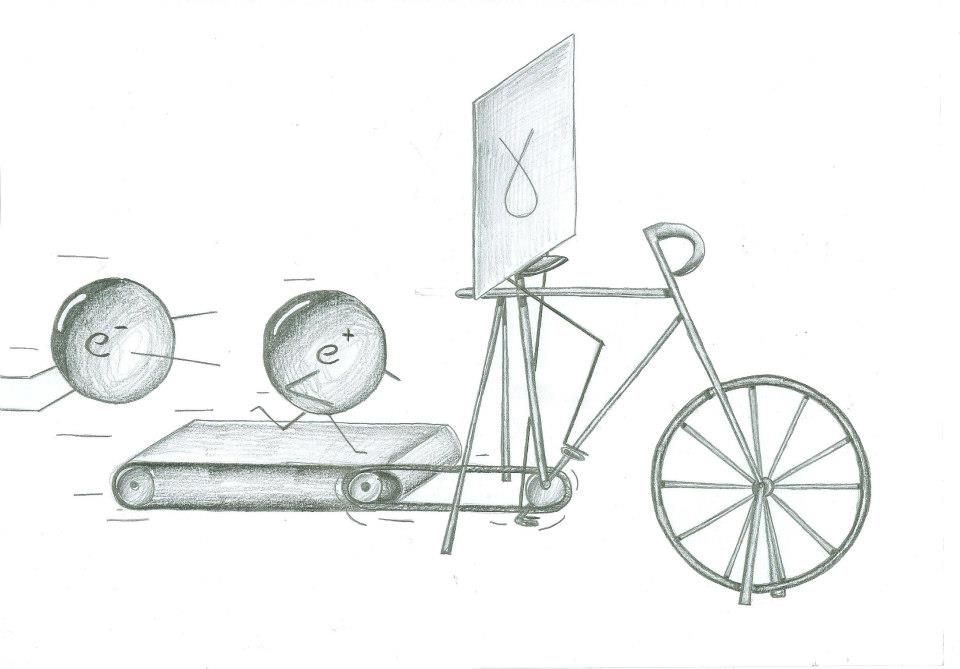}
\par\end{center}

\begin{center}
\textsc{Image by Karolina Rożko}
\par\end{center}

\clearpage{}

\clearpage{}

\listoftables

\addcontentsline{toc}{chapter}{List of tables}

\clearpage{}

\listoffigures

\addcontentsline{toc}{chapter}{List of figures}

\clearpage{}

\clearpage
\singlespacing

\bibliographystyle{apalike}
\bibliography{bibliography}

\addcontentsline{toc}{chapter}{\bibname}
\end{document}